\RequirePackage[thmmarks]{ntheorem}
\makeatletter
\renewtheoremstyle{plain} 
{\item[\hskip\labelsep \theorem@headerfont ##1\ \textup{##2}\theorem@separator]} 
{\item[\hskip\labelsep \theorem@headerfont ##1\ \textup{##2}\ (##3)\theorem@separator]}
\makeatother

\let\latexdocument\document
\let\latexarabic\arabic

\documentclass[final]{biometrika}

\let\document\latexdocument
\let\arabic\latexarabic

\def\rm{}
\usepackage{amsmath}

\usepackage{times}
\usepackage{bm}
\usepackage{natbib}

\usepackage{amssymb,subcaption}

\usepackage{graphicx}
\usepackage{epstopdf}
\usepackage{natbib} 
\usepackage[dvipsnames,table,dvipsnames*, svgnames*]{xcolor}
\usepackage{comment}

\allowdisplaybreaks

\definecolor{blue}{RGB}{000,000,200}
\definecolor{green}{RGB}{000,150,100}
\definecolor{purple}{RGB}{220,040,250}

\usepackage[plain,noend]{algorithm2e}

\makeatletter
\renewcommand{\algocf@captiontext}[2]{\quad #1\algocf@typo. \AlCapFnt{}#2} 
\def\@algocf@capt@plain{top}
\renewcommand{\algocf@makecaption}[2]{%
  \addtolength{\hsize}{\algomargin}%
  \sbox\@tempboxa{\algocf@captiontext{#1}{#2}}%
  \ifdim\wd\@tempboxa >\hsize
    \hskip .5\algomargin%
    \parbox[t]{\hsize}{\algocf@captiontext{#1}{#2}}
  \else%
    \global\@minipagefalse%
    \hbox to\hsize{\box\@tempboxa}
  \fi%
  \addtolength{\hsize}{-\algomargin}%
}
\makeatother

\newcommand{\be}{\begin{equation}}
\newcommand{\ee}{\end{equation}}

\newcommand{\bbR}{\mathbb{R}}

\newcommand{\A}{{A}}
\newcommand{\B}{{B}}

\newcommand{\I}{{I}}

\newcommand{\bbE}{{\mathbb{E}}}
\renewcommand{\P}{{P}}
\newcommand{\bbP}{{\mathbb{P}}}

\DeclareMathOperator{\Var}{Var}

\DeclareMathOperator{\Cov}{Cov}

\DeclareMathOperator{\var}{var}

\DeclareMathOperator{\diag}{diag}

\DeclareMathOperator*{\argmin}{arg\min}

\allowdisplaybreaks

\begin{document}

\markboth{P. Shi, Y. Zhou, and A. R. Zhang}{Log-Error-in-Variable Regression}

\title{High-dimensional Log-Error-in-Variable Regression with Applications to Microbial Compositional Data Analysis}

\author{Pixu Shi}
\affil{Department of Biostatistics \& Bioinformatics, Duke University, Durham, NC 27710, USA \email{pixu.shi@duke.edu}}

\author{Yuchen Zhou, \and Anru R. Zhang}
\affil{Department of Statistics, University of Wisconsin-Madison, Madison, WI 53706, USA \email{yuchenzhou@stat.wisc.edu} \email{anruzhang@stat.wisc.edu}}

\maketitle

\begin{abstract}
	In microbiome and genomic studies, the regression of compositional data has been a crucial tool for identifying microbial taxa or genes that are associated with clinical phenotypes. To account for the variation in sequencing depth, the classic log-contrast model is often used where read counts are normalized into compositions. However, zero read counts and the randomness in covariates remain critical issues. In this article, we introduce a surprisingly simple, interpretable, and efficient method for the estimation of compositional data regression through the lens of a novel high-dimensional log-error-in-variable regression model. The proposed method provides both corrections on sequencing data with possible overdispersion and simultaneously avoids any subjective imputation of zero read counts. We provide theoretical justifications with matching upper and lower bounds for the estimation error. The merit of the procedure is illustrated through real data analysis and simulation studies.
\end{abstract}


\footnotetext[1]{P. Shi and Y. Zhou contributed equally to this paper.}

\section{Introduction}
\label{sec.intro}

High-dimensional regression has attracted enormous attention in contemporary statistical research. The canonical model of high-dimensional regression can be written as $y=X\beta^\ast + \varepsilon$, where $y = (y_1,\ldots, y_n)^\top$ is the response vector, $X \in \mathbb{R}^{n\times p}$ is the covariate matrix, and $\beta^\ast\in \mathbb{R}^p$ is the unknown coefficient vector of interest. Much prior attention to high-dimensional regression focused on the clean data case where the covariates are accurately observed. However, in applications of econometrics, genomics, and engineering, we also frequently see covariates corrupted with noise. Previous literature referred to such scenarios as {\it error-in-variable} and showed that performing standard regression methods directly on the corrupted covariates may yield inaccurate inference results \citep{hausman2001mismeasured}. When the observable covariates are corrupted by additive Gaussian or sub-Gaussian noise, the methods and theories for error-in-variable regression have been widely considered in both the classic low-dimensional setting \citep{deming1943statistical} and more recent high-dimensional setting \citep{rosenbaum2010sparse,loh2012high,belloni2017linear,datta2017cocolasso}. 

The focus of high-dimensional error-in-variable regression has so far been mainly on homoskedastic Gaussian, sub-Gaussian, or bounded corruption setting. Motivated by applications in high-throughput sequencing in microbiome studies, we consider a new framework of high-dimensional error-in-variable regression that adapts to compositional covariates in this paper. 

The human microbiome is the aggregate of all microbes that reside on human bodies. It has attracted enormous recent attention due to its strong tie with human health \citep{methe2012framework}. For example, recent studies found that human microbiome may be closely related with various diseases, such as cancer \citep{schwabe2013microbiome}  and obesity \citep{turnbaugh2006obesity}. Modern next-generation sequencing technologies, such as 16S ribosomal RNA and shotgun metagenomics sequencing, provide quantification of the human microbiome by performing direct sequencing on either whole metagenomes or individual marker genes. By aligning sequencing reads to referential microbial genomes, we can organize the sequencing data into a count matrix with rows representing samples and columns representing microbial taxa or genes. Such data can be seen as the random realization of the relative abundance of bacteria in each sample.

To account for the difference in sequencing depth across samples, the read counts are often normalized into compositions \citep[see][for a survey and the references therein]{li2015microbiome}. The resulting data, also called compositional data, pose statistical challenges due to the collinearity and non-normality that come from their compositional nature. To address these issues, \cite{aitchison1984log} introduced the log-contrast model:
\begin{equation}\label{eq:log-contrast-aitchison}
	y_i=\textstyle\sum_{j=1}^{p-1}\log(Z_{ij}/Z_{ip})\beta_j^\ast+\varepsilon_i,\quad i=1,\dots,n.
\end{equation}
Here, $W_{ij}$ and $Z_{ij} = W_{ij} / (\sum_{j'=1}^p W_{ij'})$ are the absolute count and the relative abundance of the $j$th component (e.g., bacterial gene or taxon) in the $i$th sample, respectively. The analysis of the log-contrast model \eqref{eq:log-contrast-aitchison} is often dependent on the choice of reference component $Z_{ip}$, especially in high-dimensional settings. Thus, \cite{lin2014variable} reformulated \eqref{eq:log-contrast-aitchison} by introducing $\beta_p^\ast=-\sum_{j=1}^{p-1}\beta_j^\ast$,
\begin{equation}\label{eq:log-contrast-lin}
	y_i=\textstyle\sum_{j=1}^{p}\log(Z_{ij})\beta_j^\ast+\varepsilon_i,\quad i=1,\dots,n,\quad \mbox{ subject to } \textstyle\sum_{j=1}^p\beta_j^\ast=0
\end{equation}
and proposed to estimate $\beta^\ast$ through the constrained $l_1$ regularized estimator. More recently, \cite{shi2016regression} studied the statistical inference and confidence intervals for $\beta^\ast$; \cite{wang2017structured} considered the subcomposition selection in compositional data regression via a tree-guided regularization method.

The direct application of Models \eqref{eq:log-contrast-aitchison} and \eqref{eq:log-contrast-lin} by normalizing sequencing read counts, i.e., using $Z_{ij} = W_{ij}/(\sum_{j=1}^p W_{ij})$ as covariates, has several drawbacks. Firstly, it ignores the fact that the $Z_{ij}$'s are random realizations rather than true compositions of the components. In next-generation sequencing data, $Z_{ij}$ is the proportion of the read count of component $j$ among all components in sample $i$, and is thus a transformation of discrete random variables that reflect the underlying true compositions with measurement errors. As mentioned earlier, overlooking the measurement error in regressors may lead to inaccurate results. By treating $Z_{ij}$ as the true compositions, it is also overlooking the heteroskedasticity or overdispersion of $Z_{ij}$ caused by enormous uncontrollable factors of variation in sequencing, e.g., time, sampling location, or technical variability \citep{chen2013variable}. 
Secondly, the procedure requires $Z_{ij}>0$ while in reality, compositional data from next-generation sequencing often contain a lot of zeros due to the rarity of certain components. Strategies to deal with the zeros include replacing zero counts by a subjectively chosen small number, such as 0.5, before normalizing counts into compositions \citep{martin2000zero}, or imputing the entire composition matrix \citep{cao2020multisample} based on low-rank assumption. However, to the best of our knowledge, there is still no consensus on the best approach to deal with zero read counts in compositional data regression.

To address the aforementioned challenges in compositional data regression, we introduce a \emph{high-dimensional log-error-in-variable regression model} that directly handles count covariates without normalization into compositions or imputation of zeros. Recall $W_{ij}$ is the count of the $j$th component in the $i$th sample. We assume $W_i=(W_{i1},\ldots, W_{ip})^\top$ follows the Dirichlet-multinomial distribution \citep{mosimann1962compound} given the total count $N_i=\sum_{j=1}^pW_{ij}$ in the $i$th sample, and $N_i\sim\text{Poisson}(\nu_i)$ to account for the randomness of sequencing depth. That is,
\begin{equation}\label{eq:dir-multi}
	\begin{split}
		& {\mathrm{pr}}\left\{(W_{i1},\ldots, W_{ip}) = (k_{i1},\ldots, k_{ip})\Big|(N_i, q_{i1},\ldots, q_{ip})\right\} = N_i!(k_{i1}! \cdots k_{ip}!)^{-1}\textstyle\prod_{j=1}^p q_{ij}^{k_{ij}},\\
		& f(q_{i1}, \ldots, q_{ip}|N_i) = B(\alpha_i X_{i1},\ldots, \alpha_iX_{ip})^{-1} \textstyle\prod_{j=1}^p q_{ij}^{\alpha_i X_{ij}-1},
	\end{split}
\end{equation}
where $\sum_{j=1}^pk_{ij}=N_i, k_{i1},\ldots, k_{ip}\in  \{0,1,2,\ldots \}, \sum_{j=1}^pq_{ij} = 1, q_{i1},\ldots, q_{ip} \geq 0$. Here, $X_i=(X_{i1},\dots,X_{ip})^\top$ is the underlying true composition of the $p$ components, $B(\alpha_iX_{i1},\ldots, \alpha_iX_{ip}) = \{\prod_{j=1}^p\Gamma(\alpha_iX_{ij})\}/\Gamma(\alpha_i)$ is the Beta function, and $\alpha_i$ is the overdispersion parameter of the subject from which the $i$th sample is measured. The Dirichlet-multinomial distribution is a standard assumption and has been commonly used to model the multivariate count datasets with overdispersion. See, for example, \cite{chen2013variable,wadsworth2017integrative,mandal2015analysis,la2012hypothesis,holmes2012dirichlet,dai2019batch}. The Dirichlet-multinomial model has also been used in applications of econometrics \citep{guimaraes2007controlling}, single-cell mRNA studies \citep{qiu2017single}, text mining \citep{yin2014dirichlet} amongst others. When $\alpha_i\rightarrow +\infty$, the Dirichlet-multinomial distribution degenerates to the regular multinomial distribution.

Since the observable count, $W_{i}$, is merely a realization of the underlying composition, $X_{i}$, it is more reasonable to assume association between $y_i$ and $X_{i}$ rather than between $y_i$ and $W_i$. We thus assume the regression response $y_i$ to be dependent on $X_i$ through the following log-contrast model,
\begin{equation}\label{eq:log-contrast-EIV}
	y_i = \textstyle\sum_{j=1}^p \log(X_{ij})\beta_j^\ast + \varepsilon_i, \quad i=1,\ldots, n,\quad\mbox{subject to}\quad \textstyle\sum_{j=1}^p \beta_j^\ast=0.
\end{equation}
We refer to \eqref{eq:dir-multi} together with \eqref{eq:log-contrast-EIV} as the \emph{log-error-in-variable regression model}. Our aim is to estimate $\beta^\ast$ based on responses $y\in \mathbb{R}^n$ and error-in-covariates $W\in \mathbb{R}^{n\times p}$. Most of the results on error-in-variables regression deal with homoscedastic continuous variables and may not be directly applied here since the $W_i$'s are discrete random variables with heteroscedasticity depending on $X_i$ and $\alpha_i$. Therefore, new methods are in need. 

In this paper, we propose a surprisingly simple and straightforward estimation scheme, named \emph{variable correction regularized estimator}, for the high-dimensional log-error-in-variable regression. In particular, when the count observations are without overdispersion, we propose to \emph{add 0.5} to all counts $W_{ij}$, then estimate the regression parameters using constrained lasso with $\log(W_{ij}+0.5)$ as predictors; for overdispersed data, we propose to \emph{add an amount related to the overdispersion level} to $W_{ij}$ to alleviate the effect of any overly large or small counts due to overdispersion. We show that the proposed method achieves good performance in a general class of settings. In practice, we recommend to \emph{add 0.5 to all counts} if quantification of the overdispersion level is difficult, for example, when there is no repeated measurement, Dirichlet-multinomial distribution is largely violated, etc. 

In addition, we further generalize the proposed variable correction scheme in a more general log-error-in-variable regression model in Section \ref{sec:general}:
\begin{equation}\label{eq:general}
	\begin{gathered}
		y_i = \sum_{j=1}^p \log(\nu_{ij}) \beta_j^\ast + \varepsilon_i,\quad i=1,\ldots, n, \quad \text{ subject to }\quad \sum_{j=1}^p \beta_j^\ast = 0.
	\end{gathered}
\end{equation}
Here, the observed $y_i$ and independent random covariates $W_{ij}\geq 0$ are linked through $\log(\nu_{ij})$, where $\nu_{ij}=\mathbb{E}W_{ij}$ does not need to form compositions. $\beta^*$ is the parameter of interest and $\varepsilon_i$ is i.i.d. sub-Gaussian noise with mean zero and variance $\sigma^2$. We prove that the +0.5 variable correction regularized estimator for $\beta^\ast$ in \eqref{eq:general} achieves good performance. 

\section{Methods for Log-Error-in-Variable Regression}\label{sec:procedure}

\begin{figure}
	\begin{center}
		\includegraphics[width=0.6\linewidth]{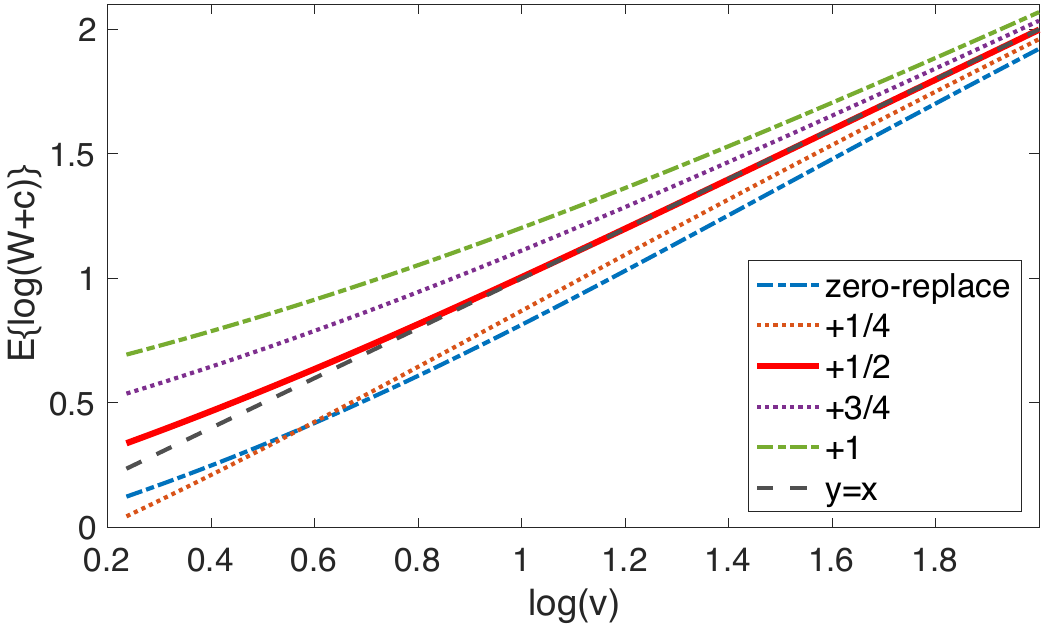}
		\caption{$\log(\nu)$ vs. $E\log(W\vee 0.5)$ (zero-replace) and $E\log(W+c)$ for $c = 1/4, 1/2, 3/4, 1$. Here, $W\sim \text{Poisson}(\nu)$. We can see $\log(W+1/2)$ has the smallest bias in estimating $\log(\nu)$.} 
		\label{fig:biasness}
	\end{center}
\end{figure}

To estimate $\beta^\ast$ in \eqref{eq:dir-multi} and \eqref{eq:log-contrast-EIV}, one classic method is simple normalization, i.e., using $W_{ij}/N_i$ as a surrogate for $X_{ij}$ and implement the classic high-dimensional regularized estimators with $\log(W_{ij}/N_i)$ as covariates. As discussed earlier, this idea has two critical issues: first, the zero-valued $W_{ij}$'s need to be replaced by a small value to make them positive in the log transformation. The choice of this value is often difficult but critical to the performance of the final estimates; secondly, even though $\bbE(W_{ij}/N_i|N_i)= X_{ij}$, $\log(W_{ij}/N_i)$ may be a biased estimator for $\log(X_{ij})$, which can cause additional inaccuracy to the regression analysis. To further illustrate the biasness of $\log(W_{ij}/N_i)$ and to introduce our fixing plan, we first focus on the non-overdispersion case, i.e., $\alpha_i=+\infty$ in Model \eqref{eq:dir-multi}, or equivalently $(W_{i1},\ldots, W_{ip})|N_i \sim \text{Multinomial}(N_i, X_{i1},\ldots, X_{ip})$.
In this case, $W_{ij}$ follows Poisson($\nu_iX_{ij}$) and
$$E (W_{ij})= \var(W_{ij})=\nu_iX_{ij}.$$
For any $z_i\geq 0$, the Taylor expansion of $\log(W_{ij}+z_i)$ at $\nu_iX_{ij}$ yields the following approximation,
\begin{equation*}
	\begin{split}
		E\{ \log(W_{ij}+z_i)\}&\approx \log(\nu_iX_{ij})+\dfrac{E(W_{ij}-\nu_iX_{ij}+z_i)}{\nu_iX_{ij}}-\dfrac{\var(W_{ij})+2z_iE(W_{ij}-\nu_iX_{ij})+z_i^2}{2\nu_i^2X_{ij}^2}\\
		&= \log(\nu_iX_{ij})+\dfrac{z_i-1/2}{\nu_iX_{ij}}-\dfrac{z_i^2}{2\nu_i^2X_{ij}^2}.
	\end{split}
\end{equation*}
Since $N_i$ is the total number of reads in sample $i$ and is generally large in practice (e.g., around $10^4$ to $10^5$ in our real data example), we can assume $\nu_i=E N_i$ to be large. Then, 
\begin{equation*}
	\begin{split}
		E\left[\log\left(\dfrac{W_{ij}+z_i}{N_i}\right)\right]-\log(X_{ij})\approx&\dfrac{z_i-1/2}{\nu_iX_{ij}}-\dfrac{z_i^2}{2\nu_i^2X_{ij}^2}+\log(\nu_i)-E\log(N_i) \approx \dfrac{z_i-1/2}{\nu_iX_{ij}}-\dfrac{z_i^2}{2\nu_i^2X_{ij}^2}.
	\end{split}
\end{equation*}
We can see the bias of $\log\{(W_{ij}+z_i)/N_i\}$ for estimating $\log(X_{ij})$ is approximately $-(1/2)\nu_iX_{ij}$ and $-(1/8)\nu_i^2X_{ij}^2$ when $z_i=0$ and $z_i=1/2$, respectively. For large $\nu_i$, one has $(1/8)\nu_i^2X_{ij}^2\ll (1/2)\nu_iX_{ij}$. Therefore, heuristically $\log\{(W_{ij}+1/2)/N_i\}$ is a significantly less biased estimator for $\log(X_{ij})$ compared to $\log(W_{ij}/N_i)$ (or $\log\{(W_{ij}+c)/N_i\}$ for any $c\neq 1/2$). Figure \ref{fig:biasness} illustrates the bias of $\log(W\vee 0.5)$ (i.e., replacing zeros by $1/2$) and $\log(W+c), c=1/4, 1/2, 3/4, 1$ for estimating $\log(\nu)$ when $W$ follows Poisson($\nu$). The plot suggests that $\log(W+1/2)$ achieves the minimum bias among these choices. In addition, by adding the positive value $1/2$ to all $W_{ij}$s, the earlier mentioned zero-replacement issue is simultaneously solved!

To account for higher variability in the count data, we also consider the overdispersed case where $\alpha_i<\infty$ in Model \eqref{eq:dir-multi}. In this case, we have 
$$
E (W_{ij})=\nu_iX_{ij}\quad \text{and}\quad \var(W_{ij})= \nu_iX_{ij}(1-X_{ij})(\nu_i+\alpha_i+1)/(\alpha_i+1)+\nu_iX_{ij}^2.$$ Similarly, by investigating the Taylor expansion of $E[\log\{(W_{ij}+z_i)/N_i\}]$, it can be shown that taking $z_i=(N_i+\alpha_i+1)/\{2(\alpha_i+1)\}$ will make $\log\{(W_{ij}+z_i)/N_i\}$ a better estimator for $\log(X_{ij})$ (more rigorous argument is postponed to the proof of Lemma \ref{lm:beta-binomial_1} in the supplementary materials). It is noteworthy that $z_i$ is an estimate for half of $(\nu_i+\alpha_i+1)/(\alpha_i+1)$, which quantifies the overdispersion rate of $W_{ij}$ compared with the multinomial distribution.

These heuristic arguments inspire us to the following \emph{variable correction regularized estimator} for the log-error-in-variable regression in \eqref{eq:dir-multi} and \eqref{eq:log-contrast-EIV}:
\begin{equation}\label{eq:hat_beta_original}
	\begin{split}
		& \hat{\beta} = \argmin_{\beta} \left\{\|y - B_W\beta\|_2^2/(2n) + \lambda\|\beta\|_1\right\} \quad \text{subject to}\quad\textstyle\sum_{j=1}^p \beta_j = 0,
	\end{split}
\end{equation}
where $B_W \in \mathbb{R}^{n\times p}$ and $(B_W)_{ij} = \log\left\{W_{ij} + (N_i+\alpha_i+1)/(2\alpha_i+2)\right\}$. Particularly if $\alpha_i = \infty$, i.e., $W_{i\cdot}|N_i$ satisfies the regular multinomial, $(B_W)_{ij} = \log(W_{ij} + 1/2)$.

\begin{remark}
	Different from the classic zero-replacement scheme that replaces only the zero covariates by a fixed value, we propose to add $1/2$ to all covariates in the non-overdispersion case. For overdispersed sample, we propose to correct $W_{ij}$ with a larger value: $z_i = (N_i + \alpha_i + 1)/\{2(\alpha_i+1)\}$. In particular, with larger total count $N_i$ or larger degree of overdispersion (i.e, smaller $\alpha_i$), the observable count covariate $W_{i}$ contains noisier information about the true underlying composition $X_{i}$. The larger added values can alleviate the effect of overly large or small counts due to overdispersion. 
	
	When there is evidence of overdispersion and there are multiple samples $W_i$ that share the same $X_i$ and $\alpha_i$ in practice, $\alpha_i$ can be estimated by method of moment \citep{la2012hypothesis} or maximum likelihood estimator \citep{tvedebrink2010overdispersion}. Otherwise, $\alpha_i=+\infty$ and $z_i=1/2$ are suggested. The more detailed discussions on the method of moment estimator of $\alpha_i$ is given in Section \ref{sec:overdispersion} of the supplementary materials.
\end{remark}

\begin{remark} 
	In the existing methods for high-dimensional error-in-variable regression, it is often a key step to construct good estimators for both the covariate $V$ and the Gram matrix $V^\top V$ (e.g., \cite{loh2012high,rosenbaum2013improved,datta2017cocolasso,rudelson2017errors,belloni2017linear}). Even though our construction targets at $(B_W)_{ij}$, a nearly unbiased estimator for $V_{ij} = \log(\nu_{ij})$, we can further show that $B_W^\top B_W$ is also a good estimator of $V^\top V$ based on some key properties of the log-error-in-variable model (e.g., Lemmas A1 and A2 in the supplementary materials). 
\end{remark}

\begin{remark}
	It is worth noting that the proposed variable correction scheme requires the knowledge of $N_i$, i.e., the total count from each subject. Due to the mechanism of sequencing techniques in microbiome studies, the raw data are usually counts as opposed to the normalized compositions and the total number of sequencing reads $N_i$ is usually available. $N_i$ is also available in a wide range of applications, e.g., the single-cell sequencing data analysis \citep{navin2011tumour}, text mining, etc. In addition, if the observation comes as a composition but the total count $N_i$ is not available, as long as the distribution of error-in-variable, i.e., $W_{ij}|X_{ij}$ can be parametrized as $\mathbb{P}_{X_{ij}}$ and $\mathbb{P}_{X_{ij}}$ is known for given $X_{ij}$, the compositional regression problem may be addressed by the general high-dimensional log-error-in-variable regression model discussed in Section 5. 
\end{remark}

\section{Theoretical Analysis}\label{sec:theory}

In this section, we investigate the theoretical performance of the proposed variable correction regularized estimator for the log-error-in-variable regression model. Denote $\bar{\nu}=\sum_{i=1}^n \nu_i/n$ and $\nu=(\nu_1,\dots,\nu_n)^\top$. We say a matrix $M$ satisfies the restricted isometry property (RIP) \citep{candes2007dantzig} with constant $\delta_s(M)\in(0,1)$ if 
\begin{equation}\label{ineq:RIP-condition}
	n\{1- \delta_s(M)\}\|\beta\|_2^2 \leq \|M\beta\|_2^2 \leq n\{1 + \delta_s(M)\}\|\beta\|_2^2 \quad \forall \text{ $s$-sparse vector $\beta$}.
\end{equation}
The restricted isometry property is commonly used in high-dimensional regression literature. Recall the corrected design matrix $B_W \in \mathbb{R}^{n\times p}$ and $(B_W)_{ij} = \log\left\{W_{ij} + (N_i+\alpha_i+1)/(2\alpha_i+2)\right\}$. We assume the centralized $\bar{B}_W = B_W\{I_p-(1/p)1_p1_p^\top\}$ satisfies the following condition:
\begin{condition}\label{con:RIP}
	$\bar{B}_W$ satisfies the restricted isometry property with constant $\delta_{2s}(\bar{B}_W)<1/10$ with probability $1-\epsilon'$ for some small quantity $\epsilon' > 0$.
\end{condition}

We first consider the case where the observable counts have no overdispersion, i.e., $\alpha=\infty$. We show the proposed variable correction regularized estimator \eqref{eq:hat_beta_original} satisfies the following upper bound.

\begin{theorem}[No overdispersion, upper bound]\label{th:upper_bound} 
	Consider \eqref{eq:dir-multi} and \eqref{eq:log-contrast-EIV} with $\alpha_i=\infty$, i.e., $W$ has no overdispersion. Suppose Condition \ref{con:RIP} holds, $n\geq Cs\log(p)$, and $a\bar{\nu} \leq \nu_i \leq b\bar{\nu}$, $a/p \leq X_{ij} \leq b/p$ for constants $0<a < 1 < b$. If for some large constant $C>0$, some $\epsilon>0$, and a constant $C_{\epsilon}$ that only depends on $\epsilon$, we have $(\bar{\nu}/p) \geq C(s + \log(np) + C_{\epsilon})$, then by choosing $\lambda = C[\left\{\log(p)/n\right\}\{\sigma^2 + (p/\bar{\nu})\|\beta^*\|_2^2\} + s\left(p/\bar{\nu}\right)^{3-2\epsilon}\|\beta^*\|_2^2]^{1/2}$ for some large constant $C > 0$, the variable correction regularized estimator \eqref{eq:hat_beta_original} satisfies
	\begin{equation}\label{ineq:upper_1}
		\|\hat{\beta} - \beta^*\|_2^2 \leq  C\left[(s\log p/n)\left\{\sigma^2 + (p/\bar{\nu})\|\beta^*\|_2^2\right\} + s^2\left(p/\bar{\nu}\right)^{3-2\epsilon}\|\beta^*\|_2^2\right]
	\end{equation}
	with probability at least $1 - 4p^{-C'}-\epsilon'$. Moreover, if $\bar{\nu} \geq p(sn/\log p)^{1/(2 - 2\epsilon)}$, then by choosing $\lambda = C[\left\{\log(p)/n\right\}\left\{\sigma^2 + (p/\bar{\nu})\|\beta^*\|_2^2\right\}]^{1/2}$ for constant $C > 0$, with probability at least $1 - 4p^{-C'}-\epsilon'$,
	\begin{equation}\label{ineq:upper_2}
		\|\hat{\beta} - \beta^*\|_2^2 \leq C(s\log p/n)\left\{\sigma^2 + (p/\bar{\nu})\|\beta^*\|_2^2\right\}.
	\end{equation}
\end{theorem}
We provide both the sketch and complete arguments in Section \ref{sec:proof-thm1} in the supplement. 
\begin{remark}
	Theorem \ref{th:upper_bound} shows that the estimation error gets smaller with the larger sample size $n$, smaller dimension $p$, smaller noise variance $\sigma^2$, higher $\bar{\nu}$, smaller signal amplitude $\|\beta^\ast\|_2^2$, or smaller sparsity level $s$. If $\bar{\nu}$ is large, a sufficient sample size to ensure consistency of $\beta$ is $n \geq C\max\{\sigma^2, 1\}s\log p$, which matches the classic result in high-dimensional sparse linear regression \citep{bickel2009simultaneous}. In addition, the error bound \eqref{ineq:upper_1} includes two components: $C(s\log p/n)\sigma^2$ corresponds to the error of $\varepsilon_i$ and $C\{(s\log p/n)(p/\bar{\nu})\|\beta^*\|_2^2 + s^2\left(p/\bar{\nu}\right)^{3-2\epsilon}\|\beta^*\|_2^2\}$ is incurred by the error in covariates.
	
	$\bar{\nu}$ is an important factor in both the condition and upper bound in Theorem \ref{th:upper_bound}. Since $\bar{\nu} = \sum_{i=1}^n \nu_i/n, \nu_i = \mathbb{E} \sum_{j=1}^p W_{ij}$, $(\bar{\nu}/p)$ quantifies the average count of all components among all subjects. In practice like microbiome studies, it is reasonable to assume $(\bar{\nu}/p)$ as a constant or logarithmic in scale as $p$ increases, because when we investigate bacteria with higher resolution at lower taxanomic levels, the average sequencing depth quantified by $\bar{\nu}$ should also be increased accordingly to maintain the accuracy of the analysis.
\end{remark}

Let $\bar V = \left\{\log(\nu_{i}X_{ij}) - p^{-1}\textstyle\sum_{l = 1}^p\log(\nu_{i}X_{il})\right\}_{1\leq i \leq n; 1\leq j \leq p}$ be the centralized population design matrix. We further consider the following class of covariate matrices and parameter vectors,
\begin{equation}\label{con:gaussian}
	\begin{split}
		\mathcal{F}_{p, n, s}(R, Q) = \big\{(\nu, X, \beta):& a\bar{\nu} \leq \nu_i \leq b\bar{\nu}, a/p \leq X_{ij} \leq b/p \text{ for constants } 0<a < 1 < b; \\ 
		& \delta_{2s}(\bar V) < 1/20, ||\beta||_2 \leq R, 1_p^\top\beta=0, e^{-3/2}Q \leq \bar\nu \leq e^{3/2}Q\big\}.\\
	\end{split}
\end{equation}
The constraints in $\mathcal{F}_{p, n, s}(R, Q)$ correspond to the regularization assumptions in Theorem \ref{th:upper_bound}. The upper bound in Theorem \ref{th:upper_bound} turns out to match the minimax lower bound in $\mathcal{F}_{p, n, s}(R, Q)$.
\begin{theorem}[Lower bound]\label{th:lower_bound} 
	Suppose $\varepsilon_i \overset{iid}{\sim} N(0, \sigma^2)$. If we have $n \geq Cs\log p$ for some large constant $C > 0$, $R \geq \bar{c}\surd\{s\log\left(p/s\right)\sigma^2/n\}$ for some constant $\bar{c} > 0$, $Q \geq p$, and $ s\geq 4$, then 
	\begin{equation}\label{eq: lower_bound}
		\inf_{\hat{\beta}} \sup_{(\nu, X, \beta) \in \mathcal{F}_{p, n, s}(R, Q)} E (\|\hat{\beta} - \beta\|_2^2) \geq c\{s\log (p/s)/n\}\left\{\sigma^2 + (p/Q)R^2\right\}.
	\end{equation}
\end{theorem}

The sketch and the complete proof of Theorem \ref{th:lower_bound} are provided in Section \ref{sec:lower-bound} in the supplement. Next, we consider the overdispersed case where $\alpha_i<\infty$ in Model \eqref{eq:dir-multi} and \eqref{eq:log-contrast-EIV}. We have the following upper bound for estimation accuracy of the proposed variable correction regularized estimator \eqref{eq:hat_beta_original}. 
\begin{theorem}[with overdispersion, upper bound]\label{th:upper_bound_Dirichlet multinomial} Suppose Condition \ref{con:RIP} holds and $a\bar{\nu} \leq \nu_i \leq b\bar{\nu}$, $a/p \leq X_{ij} \leq b/p$ for constants $0<a < 1 < b$. Set $\zeta_{\max} = \max_{i}\zeta_i$, where $\zeta_i = (\nu_i+\alpha_i+1)/\{2(\alpha_i+1)\}$ represents the level of overdispersion for $i$th sample. If for some $\delta>0$, some large constant $C$, and a large constant $C(\delta)$ that only depends on $\delta$, we have $\nu_{ij} \geq \zeta_i^{1 + \delta}$,  $n \geq Cs\log p$, and $\bar{\nu}/(p\zeta_{\max}) \geq \max\left[C\log(np), C(\delta)\right],$ then by choosing $\lambda = C\left(\{\log(p)/n\}^{1/2}[\sigma + \{(p/\bar\nu)\zeta_{\max}\}^{\frac{1}{2}}\|\beta^*\|_1] + \log(\bar\nu/p)(p/\bar\nu)\zeta_{\max}\|\beta^*\|_1\right)$, we have
	\begin{equation}
		\|\hat{\beta} - \beta\|_2^2 \leq C\left[(s\log p/n)\left\{\sigma^2 + (p/\bar{\nu})\zeta_{\max} \|\beta^*\|_1^2\right\} + s\log^2(\bar\nu/p)(p/\bar\nu)^{2}\zeta_{\max}^2\|\beta^*\|_1^2\right]
	\end{equation}
	with probability at least $1 - 6p^{-C'} - \epsilon'$, where $C'$ is a constant.
\end{theorem}
The proof is provided in Section \ref{sec:proof-thm3} in the supplement.

\section{General High-dimensional Log-Error-in-Variable Regression}\label{sec:general}

In this section, we extend the discussion to general high-dimensional log-error-in-variable regression that accommodates broader scenarios. This will also justify the +0.5 variable correction rule under a broader range of misspecified models. Specifically, let
\begin{equation}\label{eq:general2}
	\begin{gathered}
		y = V \beta^\ast + \varepsilon,\quad \text{or equivalently}\quad y_i = \sum_{j=1}^p \log(\nu_{ij}) \beta_j^* + \varepsilon_i \quad \text{ subject to }\quad \sum_{j=1}^{p}\beta_j^* = 0, \\ 
		\text{where}\quad W = (W_{ij}), \quad \mathbb{E} W_{ij}=\nu_{ij}, \quad W_{ij}\text{ are independent}, \quad i=1,\ldots, n, j=1,\ldots, p.
	\end{gathered}
\end{equation}
Here, $V = (\log(\nu_{ij}))_{\substack{1\leq i \leq n, 1\leq j \leq p}}$ are unknown underlying covariates, $\varepsilon_i$'s are i.i.d. sub-Gaussian noises with mean zero and variance $\sigma^2$, $\beta^\ast$ is the sparse parameter of interest, and $W_{ij}$ satisfies the following sub-exponential tail condition: 
\begin{equation}\label{ineq:sub_exponential_tail}
	\bbP(|W_{ij} - \nu_{ij}| \geq t) \leq C\exp\left\{-c(t^2/\nu_{ij}) \wedge t\right\}, \quad \forall t>0.
\end{equation}
In particular, Poisson distribution satisfies \eqref{ineq:sub_exponential_tail} and hence our log error-in-variable regression model without over-dispersion can be viewed as a special case of \eqref{eq:general2}.

We consider the following +0.5 variable correction regularized estimator for $\beta^*$ in the general high-dimensional log-error-in-variable regression model \eqref{eq:general2},
\begin{equation}\label{eq:hat_beta_original3}
	\begin{split}
		\hat{\beta} = \argmin_{\beta} \left(\frac{1}{2n}\|y - B_W\beta\|_2^2 + \lambda\|\beta\|_1\right) \quad \text{subject to } \sum_{j=1}^{p}\beta_j = 0.
	\end{split}
\end{equation}
Here, $B_W\in\mathbb{R}^{n\times p}$ with $(B_W)_{ij}=\log(W_{ij} + 0.5)$ and $\lambda$ is some tuning parameter.

The following theorem provides an upper bound for the variable correction regularized estimator \eqref{eq:hat_beta_original3} in the general high-dimensional log-error-in-variable regression model \eqref{eq:general2}. 
\begin{theorem}[General Upper Bound]\label{th:upper_bound3} Suppose Condition \ref{con:RIP} holds, $n\geq Cs\log(p)$, and $|\log(\nu_{ij}) - \log(\nu_{kl})| \leq a$ for some constant $a>0$ for all $1 \leq i, k \leq n, 1 \leq j, l \leq p$. Denote $\bar{\nu}=\frac{1}{n}\sum_{i=1}^n\sum_{j=1}^p\nu_{ij}$ and $F = \max_{ij}|\var(W_{ij})/\nu_{ij} - 1|$. If for some uniform constant $C>0$,some $\epsilon > 0$, and a constant $C_{\epsilon}$ that only relies on $\epsilon$, we have $\bar{\nu} \geq Cp(S + Fs\log(s) + \log(np) + C_{\epsilon})$, then by choosing  $\lambda = C[\{\log (p)/n\}\{\sigma^2 + (p/\bar{\nu})\|\beta^*\|_2^2\} + s\{\min\{F^2, C\}(p/\bar{\nu})^2 + (p/\bar{\nu})^{3-2\epsilon}\}\|\beta^*\|_2^2]^{1/2}$ for some large constant $C > 0$, we have
	\begin{equation}\label{ineq:general_rate1}
		\|\hat{\beta} - \beta^\ast\|_2^2 \leq \frac{Cs\log p}{n}\left\{\sigma^2 + (p/\bar{\nu})\|\beta^*\|_2^2\right\} + Cs^2\left\{\left(p/\bar{\nu}\right)^{3-2\epsilon} + \min\{F^2, C\}\left(p/\bar{\nu}\right)^2\right\}\|\beta^*\|_2^2
	\end{equation}
	with probability $1 - 3p^{-C'}-\epsilon'$. 
\end{theorem}
\begin{remark}
	Theorem \ref{th:upper_bound3} shows that for the log-error-in-variable model, when the observed variables $W_{ij}$'s with exponential tail probability are linked with the response $y$ in the form of model \eqref{eq:general2} through its first moment and $\var(W_{ij})$ is close to $\nu_{ij}$ (like the Poisson case), the +0.5 correction rule can achieve a reasonable estimation error under proper conditions.  If this is violated, such as $\var(W_{ij})$ is much larger than $\nu_{ij}$, the error upper bound for this correction can be large, which can be a potential limit of the method.
	
	Moreover, the estimation error upper bound of \eqref{ineq:general_rate1} includes two parts:  (a) $C(s\log p/n)\sigma^2$, which corresponds to the error of $\varepsilon_i$ and also appears in Theorem \ref{th:upper_bound}; (b) $C\{(s\log p/n)(p/\bar{\nu})\|\beta^*\|_2^2 + Cs^2\left\{\left(p/\bar{\nu}\right)^{3-2\epsilon} + \min\{F^2, C\}\left(p/\bar{\nu}\right)^2\right\}\|\beta^*\|_2^2\}$, which originates from the error-in-variable and is no smaller than the one in Theorem \ref{th:upper_bound}. When $W_{ij}$ is Poisson distributed, we have $\var(W_{ij}) = \mathbb{E} W_{ij}$, $F = 0$, and the upper bound \eqref{ineq:general_rate1} reduces to \eqref{ineq:upper_1} in Theorem \ref{th:upper_bound}.  
\end{remark}

\section{Simulation Studies}\label{sec:simu}

\begin{figure}
	\centering
	\begin{subfigure}[b]{0.48\textwidth}
		\includegraphics[page=2,width=1\linewidth]{sim_summary_with_pairs_dif_shape3.pdf}
		\caption{Estimation Error}
		\label{fig:esterr}
	\end{subfigure}
	\begin{subfigure}[b]{0.48\textwidth}
		\includegraphics[page=1,width=1\linewidth]{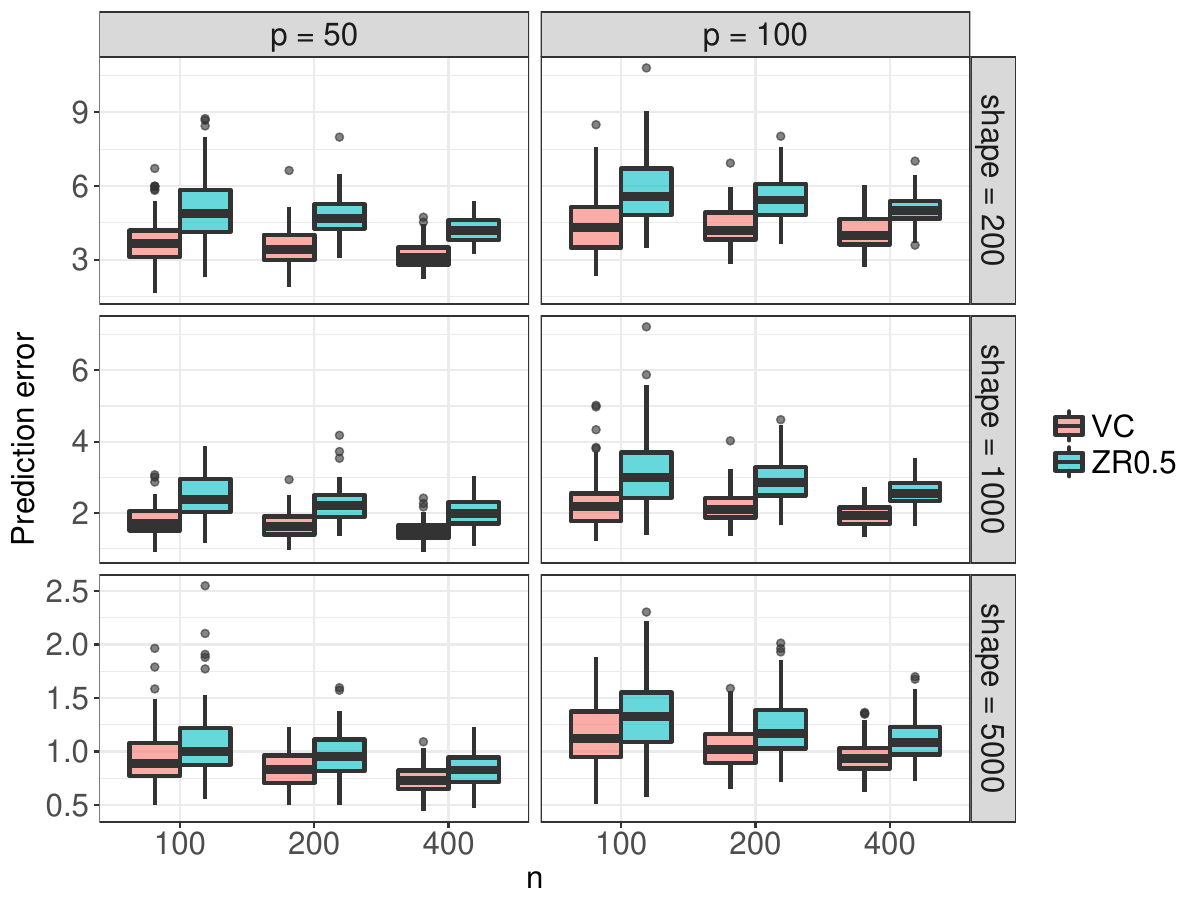}
		\caption{Prediction Error}
		\label{fig:prederr}
	\end{subfigure}
	\caption{Comparison between variable correction estimator (VC) and zero-replacement estimator by $0.5$ (ZR0.5) in simulation analysis. Noise terms $\varepsilon_i$ are all independent.}\label{fig:simu_summary}
\end{figure}

In this section, we evaluate the performance of our method using three simulation schemes for different purposes. In the first simulation scheme, we compare the prediction and estimation performance of the proposed variable correction regularized estimator with the classic method of zero-replacement under our model assumption with different parameter settings. To simulate the count matrix $W$ with $n=50, 100$ samples and $p=100,200,400$ covariates, we first generate $N_i$ from negative binomial distribution with mean $3\times 10^4$ and variance $3\times 10^6$. Here, the main purpose of choosing negative binomial instead of Poisson in previous theoretical analysis is to show that the Poisson assumption on $N_i$ is not crucial in real practice. Then we set $X_{ij}=X_{i+n/2,j}=\exp(\Phi_{ij})/\{\sum_{k=1}^p \exp(\Phi_{ij})\}$ for $j=1,\dots,p, i=1,\dots,n/2$ with $\{\Phi_{ij}\}$ generated independently from $N(\mu_j,1.5^2)$, where $\mu_1, \mu_2, \mu_3$ are drawn from Uniform[1,3], $\mu_4,\dots, \mu_7$ are drawn from Uniform[2,4], and $\mu_j$ for $j=8,\dots,p$ are drawn from Uniform[0,2]. With this setting, the average count of covariates will have a reasonable variation, with causal covariates slightly more abundant than non-causal ones. Then we generate $(W_{i1},\dots, W_{ip})$ from Dirichlet-Multinomial($N_i,\alpha X_{i1},\dots,\alpha X_{ip}$), where the overdispersion parameter $\alpha=200, 1000, 5000$. The $i$th and $(i+n/2)$th samples are designed to be from the same subject so they share the same $X_{ij}$ and can be used to estimate their shared overdispersion parameter. The response $y$ is generated as $y_i=\sum_{j=1}^p\log(X_{ij})\beta_j+\varepsilon_i$, where $\beta=(1, -0.8, -1.5, 0.6, -0.9, 1.2, 0.4, 0, \dots, 0)$ is the deterministic coefficient vector and $\varepsilon_i$ are i.i.d. noise generated from $N(0,0.5^2)$. The results are aggregated in Figure \ref{fig:simu_summary}. We can see variable correction significantly outperforms zero-replacement by 0.5 in all parameter configurations. 

\begin{figure}
	\centering
	\begin{subfigure}[b]{0.48\textwidth}
		\includegraphics[page=2,width=1\linewidth]{sim_summary_shared.pdf}
		\caption{Estimation Error}
		\label{fig:esterr_shared}
	\end{subfigure}
	\begin{subfigure}[b]{0.48\textwidth}
		\includegraphics[page=1,width=1\linewidth]{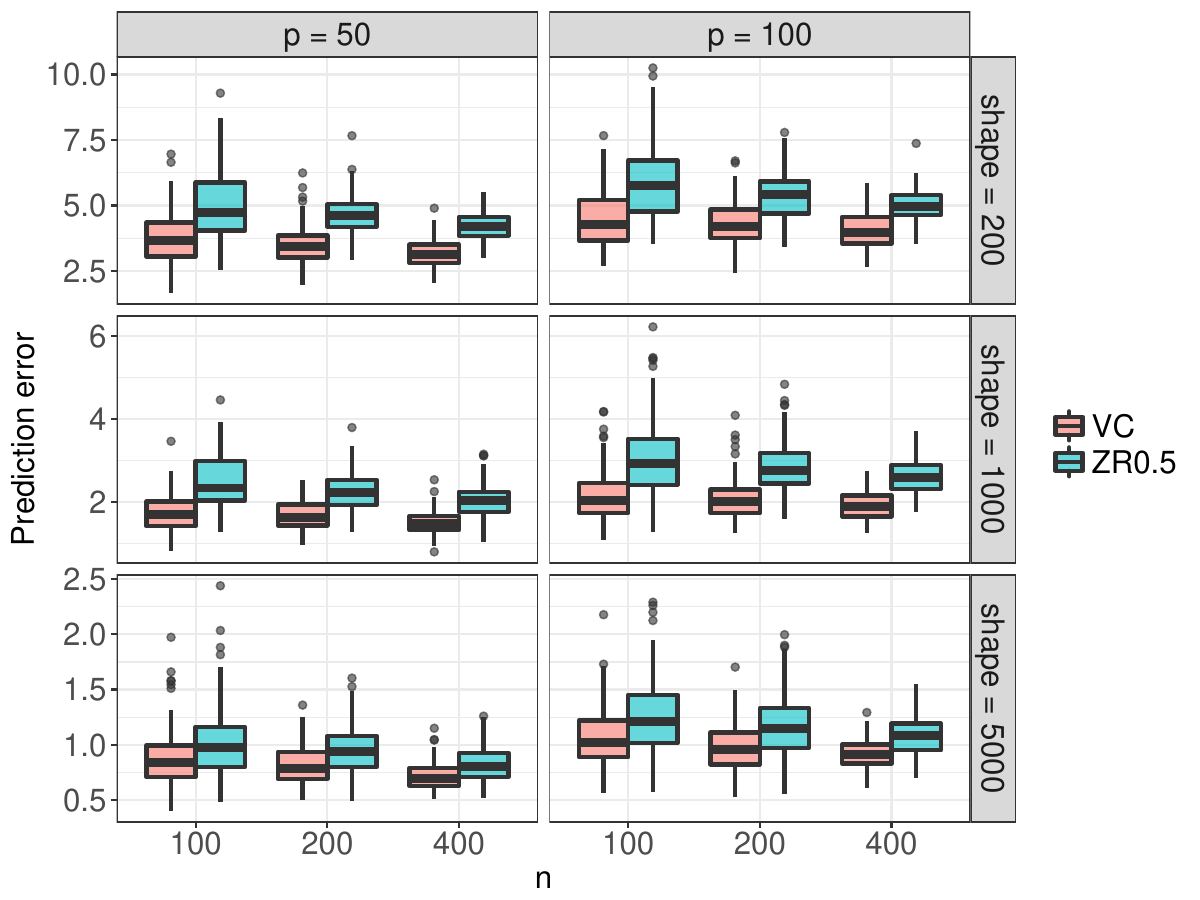}
		\caption{Prediction Error}
		\label{fig:prederr_shared}
	\end{subfigure}
	\caption{Comparison between variable correction estimator (VC) and zero-replacement estimator by $0.5$ (ZR0.5) in simulation analysis. $y$ is shared by samples from the same subject.}\label{fig:simu_summary_shared}
\end{figure}

To evaluate the performance of the proposed method when the response variable $y$ is shared by samples from the same subject like what we have in the real data analysis, we repeat the aforementioned simulation with one change: $y$ is generated with $\varepsilon_i=\varepsilon_{i+n/2}$ and $\varepsilon_i, i=1,\dots,n/2$ are i.i.d. from $N(0,0.5^2)$. The results are summarized in Figure \ref{fig:simu_summary_shared}. We can see the pattern of performance is similar to that of Figure \ref{fig:simu_summary} and the variable correction method still significantly achieves smaller estimation and prediction errors.

In the second simulation scheme, we compare different methods under misspecified models, i.e., when the data generation mechanism deviates from our model assumption. We generate $N_i$ and $X_{ij}$ in the same way as the first simulation scheme. $y_i$ is drawn in the same way as the first simulation scheme with independent error $\varepsilon_i$. To simulate $W_{ij}$, we first set $\alpha = 1000$ and generate $(\tilde{Q}_{i1}^{(1)},\dots, \tilde{Q}_{ip}^{(1)})$ from log-normal distribution such that $$\log(\tilde{Q}_{i1}^{(1)},\dots, Q_{ip}^{(1)}) \sim N\left(\log(\alpha X_{i1},\dots,\alpha X_{ip})-1/8,\Sigma\right), \quad  \Sigma_{ij}=0.5^{|i-j|}/4, 1\leq i,j\leq p.$$
Then $(\tilde{Q}_{i1}^{(1)},\dots, \tilde{Q}_{ip}^{(1)})$ are normalized into proportions: 
\begin{equation*}
	(Q_{i1}^{(1)},\dots, Q_{ip}^{(1)}) = (\tilde{Q}_{i1}^{(1)},\dots, \tilde{Q}_{ip}^{(1)}) / \textstyle\sum_{j=1}^p \tilde{Q}_{ij}^{(1)}.
\end{equation*}
Next, we generate $(Q_{i1}^{(2)},\dots, Q_{ip}^{(2)})$ from Dirichlet$(\alpha X_{i1},\dots,\alpha X_{ip})$ and take $Q_{ij}=wQ_{ij}^{(1)}+(1-w)Q_{ij}^{(2)}$, where $w$ takes a series of values between 0 and 0.5. Finally, $(W_{i1}, \dots, W_{ip})$ is drawn from multinomial$(N_i,Q_{i1},\dots, Q_{ip})$. When $w=0$, this simulation scheme is exactly Dirichlet-multinomial. The larger $w$ is, the more misspecified the model is. 

\begin{figure}
	\centering
	\includegraphics[page=2,width=0.8\linewidth]{misspec2_errorbar.pdf}\\
	\includegraphics[page=1,width=0.8\linewidth]{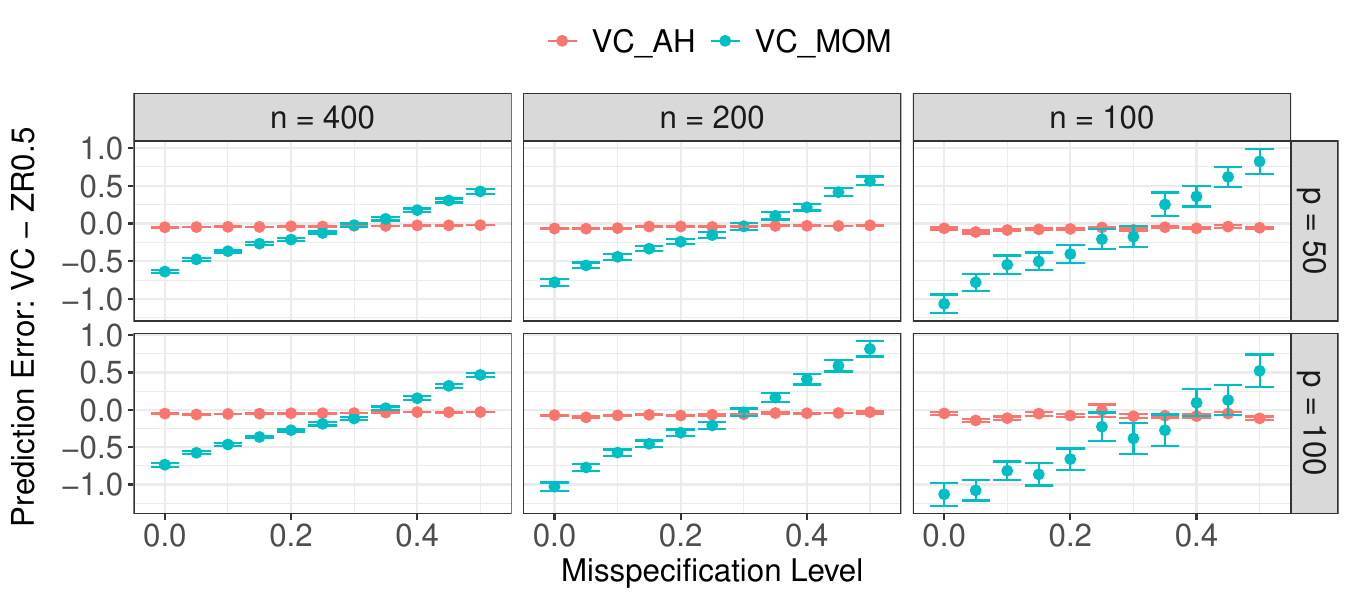}
	\caption{Estimation (upper panel) and prediction (lower panel) performance of VC\_AH, VC\_MOM in reference to ZR0.5 under the misspecified settings. The dots represent mean difference in performance and error bars represent mean $\pm$ standard deviation. }
	\label{fig:misspec}
\end{figure}

We compare the performance of (a) VC\_MOM: the variable correction estimator with the overdispersion parameters estimated via the method of moment described in Section \ref{sec:overdispersion} in the supplement; (b) VC\_AH: the variable correction estimator with the overdispersion parameters set to $\infty$ (which means to add all counts $W_{ij}$ by half); (c) ZR0.5: zero-replacement by $0.5$. We use ZR0.5 as the baseline and summarize the difference in the performance between each method and ZR0.5 for each round of simulation in Figure~\ref{fig:misspec}. We can see that the proposed VC\_MOM has significantly better performance than the classic ZR0.5 when the misspecification is moderate; the VR\_AH is always slightly better than the classic ZR0.5. 

\begin{figure}
	\centering
	\includegraphics[page=2,width=0.8\linewidth]{norep_errorbar.pdf}\\
	\includegraphics[page=1,width=0.8\linewidth]{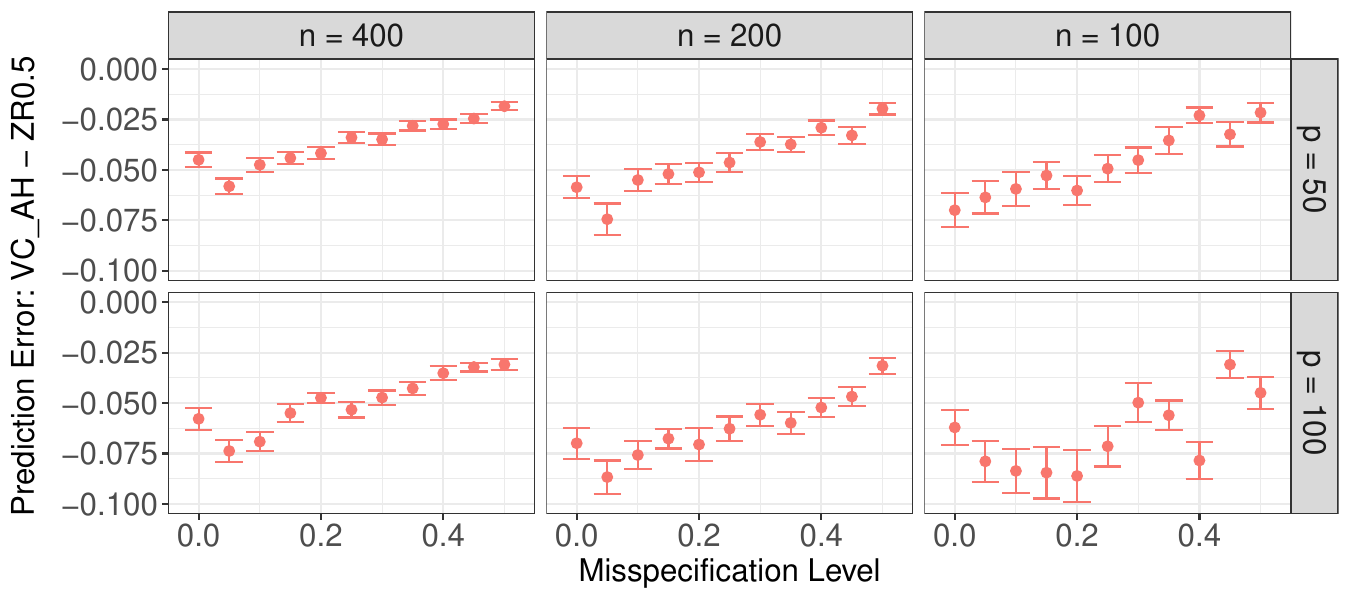}
	\caption{Estimation (upper panel) and prediction (lower panel) performance of VC\_AH in reference to ZR0.5 under the misspecified settings when no repeated measurement is available. The dots represent mean difference in performance and error bars represent mean $\pm$ standard deviation. }
	\label{fig:norep}
\end{figure}

Finally in the third simulation scheme, we consider a setting where only one measurement is available for each subject.  We generate $N_i$, $y_i$ and $W_{ij}$ in the same way as the second simulation scheme, but we no longer require $X_{ij}=X_{i+n/2, j}$ so that all the samples are from different subjects. We can see from Figure~\ref{fig:norep} that there is universal advantage of the proposed procedure (VC\_AH) than the classic zero-replacement (ZR0.5) scheme.

\section{Regression Analysis for Longitudinal Microbiome Studies}\label{sec:real}

In this section, we apply the proposed procedure to a longitudinal microbiome study reported by \cite{flores2014temporal}. We focus on the association between Body Mass Index (BMI) and gut microbiome composition at the genus level for healthy adults by excluding subjects with missing BMI, antibiotic disturbance, or other medication use. For the remaining 40 subjects each having 4 samples, 92 bacteria genera appear in more than 10\% of the samples and will be used for the analysis forward. We also adjust for the gender, age, race/ethnicity (caucasian, asian/pacific islander, hispanic/half white hispanic, or other), dietary preference (vegan, omnivore but no red meat, or omnivore), vitamin intake (yes or no), and exercise frequency (daily/regularly or occasionally/rarely) of the subjects in the regression analysis. 

We implement the proposed variable correction estimator. Specifically, we assume the samples of the same subject share the same unobserved composition $X_{ij}$ and overdispersion parameter $\alpha_i$, and estimate $\alpha_i$ for each subject respectively using the method of moment estimator ${\alpha}_{i,MOM}$ described in Section \ref{sec:overdispersion} in the supplementary materials. Then we apply the regression model \eqref{eq:hat_beta_original} 
with $y$ representing BMI, $W_{ij}$ representing read count of the $i$th sample and $j$th genus.
For comparison, we also perform the classic zero-replacement method in literature with zero counts changed to $c=0.1$ and $0.5$, respectively: 
\begin{equation}\label{eq:ori_beta}
	\hat{\beta}^{\mathrm{ZR}}=\argmin_{\beta: 1_p^\top\beta=0}\left[\textstyle\sum_{i=1}^n\left\{y_i-\textstyle\sum_{j=1}^p\log\left(W_{ij}\vee c\right)\beta_j\right\}^2/(2n)+\lambda\|\beta\|_1\right].
\end{equation}
To obtain stable variable selection, we generate 100 bootstrap samples of size $n/2$, repeat all methods with five-fold cross-validation choosing the tuning parameter $\lambda$ on each subsample, and record the frequency of each variable being selected among the 100 bootstrap fittings. For illustration purposes, we consider a variable to be selected if its selection frequency is no less than 0.7. 

\begin{figure}
	\centering
	\includegraphics[width=0.98\linewidth]{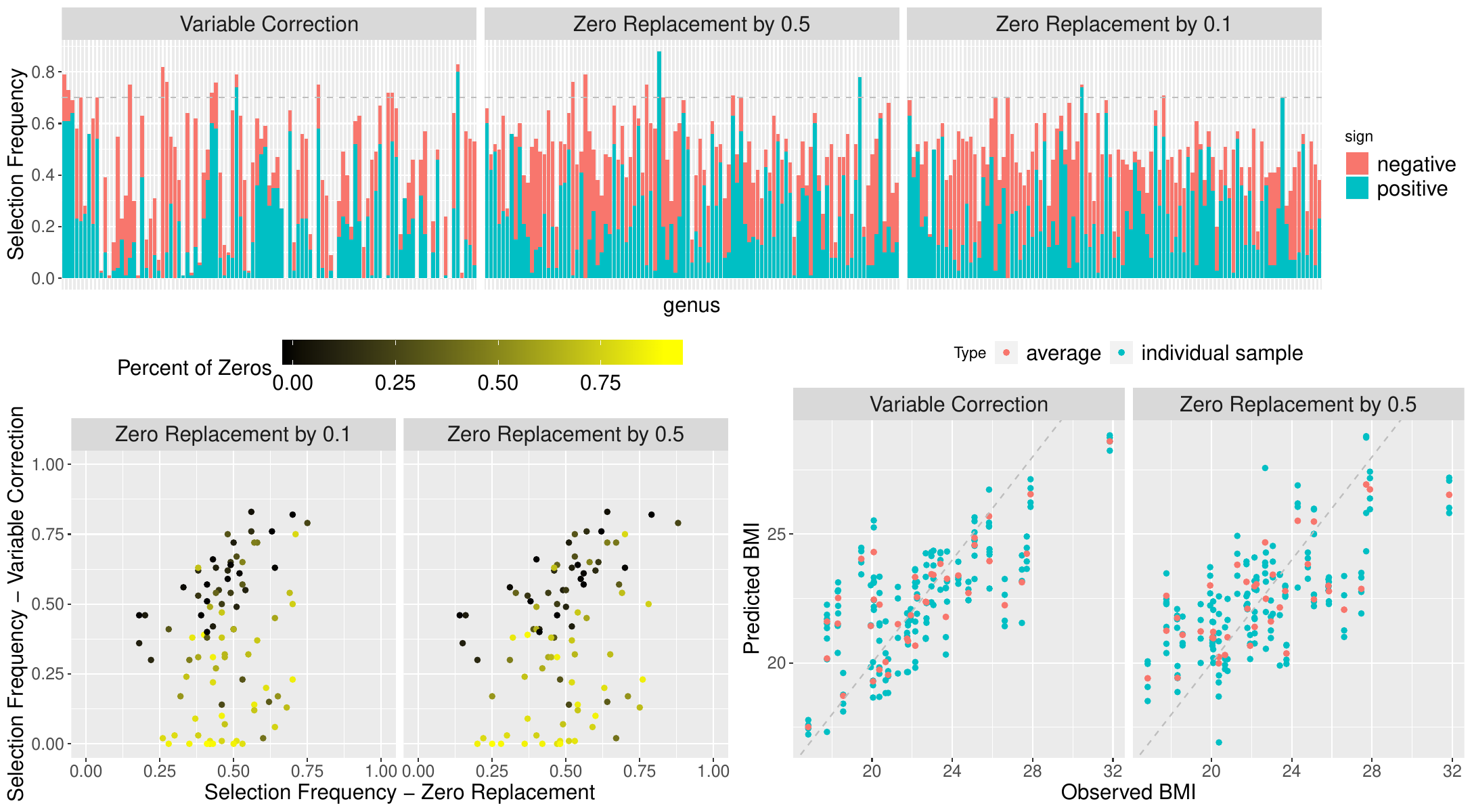}
	\caption{Top panel: Selection frequency of 92 genera using different methods. Bottom left panel: Comparison of selection frequency with regard to proportion of zero counts. Bottom right panel: Comparison of prediction performance.}\label{fig:allplots}
\end{figure}

The top panel of Figure \ref{fig:allplots} illustrates the selection frequency of the 9 covariates we adjusted for and 92 genera for each method. The dashed line corresponds to the selection frequency of 0.7. It can be observed that the variable correction estimator selects variables with either very high or very low frequency, while zero-replacement estimator has many more variables selected with a mid-range frequency. This comparison indicates that regularized estimator has much better stability in variable selection than zero-replacement. The variables selected by all three methods are Bacteroides(-), Dialister(+), and Megamonas(+). Variables selected by variable correction method only are being male(+),
age(+),
frequent exercise(+),
being hispanic/half white hispanic(-),
Akkermansia(-), Bifidobacterium(-),
Coprobacillus(+),
Coprococcus(+), Porphyromonas(-),
Prevotella(+), and Sutterella(+). Variables selected by zero-replacement with $c=0.1$ is a subset of that with $c=0.5$, where Arcanobacterium(-) and Slackia(+) are selected by both, and Dehalobacterium(-), 
Dorea(-), and Lactococcus(+) are selected by $c=0.5$ only. Here (+) and (-) are the signs of regression coefficients by plurality vote in the 100 bootstrap fittings. These selected genera correspond to the bars exceeding the dashed line in the top panel of Figure \ref{fig:allplots}. Among the genera selected by variable correction but missed by zero-replacement with $c=0.5$, Bifidobacterium has been widely studied for its lipid-lowering effect and negative association with obesity \citep{million2012obesity, an2011antiobesity}. Akkermansia has been reported to be negatively related with obesity in extensive literature 
\citep{everard2013cross, dao2016akkermansia, derrien2017akkermansia}. Coprococcus has also been reported to be positively related with obesity \citep{kasai2015comparison} and negatively related with weight loss induced by diet or gastric bypass surgery, as indicated by \cite{damms2015effects}. 

Bottom left panel of Figure \ref{fig:allplots} offers a closer look at the selection frequency with regard to the proportion of $W_{ij}=0$ for each variable.  Compared to zero-replacement method, the variable correction method tends to select variables with fewer zeros. This makes the variable correction method more desirable since the bacteria with large proportions of zeros are often possessed by only a few subjects, are far less reliable for prediction and interpretation purposes, and difficult to generalize to a larger population. Bottom right panel of Figure \ref{fig:allplots} compares the prediction performance of variable correction and zero-replacement with $c=0.5$, where the predicted BMI for each sample is obtained using refitted coefficients of the genera that have selection frequency no less than 0.7. The $R^2$ calculated using all the individual samples is 0.50 for variable correction and 0.42 for zero-replacement with $c=0.5$. Since each subject has multiple samples, we also provide the average predicted BMI of each subject in the figure. The $R^2$ using average predicted BMI is 0.56 for variable correction and 0.51 for zero-replacement. We can see the proposed method achieves much better prediction compared to zero-replacement using both the individual and average predicted BMI. 

\section{Discussions}\label{sec:discussion}

The proposed log-error-in-variable regression model method provides a new solution to deal with zero read counts in the high-dimensional regression analysis of microbiome studies. In contrast, many existing methods, such as EdgeR \citep{robinson2010edger}, DESeq2 \citep{love2014moderated}, and metagenomeSeq \citep{paulson2013metagenomeseq} (see \citep{mcmurdie2014waste} for an overview of procedures), model the zero read counts through negative binomial distribution or zero-inflated distributions. These methods can draw conclusion about the marginal effect of each component one at a time, while our method aims at the association between regression response and a component when other components are adjusted for. In another related line of work, \cite{de2018geometric} introduced a method to find a modified version of geometric mean that is close to the traditional geometric mean while being able to handle zeros. Their method can be modified to find a good alternative for a \textit{given} linear combination of the log read counts, but not for \textit{unknown} linear combinations like our regression equation.

One limitation of our regression model is that it does not discriminate between zero counts due to undersampling and actual zeros due to absence of the component. It assumes the underlying true composition to be positive, although it can be infinitely close to zero. Another limitation of our method is its ability to deal with rare components. Our variable correction is less accurate when the true bacterial abundance $X_{ij}$ is close to zero. This limitation is intrinsic to the log-error-in-variable model because of the large derivative of log function around zero. However, as is shown in our real data analysis, the $\ell_1$ penalized regression we used for variable selection tends to select the more abundant bacteria, possibly due to the fact that the small read counts are overshadowed by the correction added to it.

In addition to the aforementioned microbiome study, the proposed framework can be used for other applications on regression with count covariates. For example, in \emph{single-cell RNA-seq data  analysis}, high-throughput sequencing technique was performed on each of the single cells and the gene expressions can be measured as the total number of reads mapped to exonic regions. This resulting count matrix has rows and columns representing single cells and gene expressions, respectively. With the proposed method, we can perform regression analysis to study the association among the gene expressions of single cells and clinical phenotypes. Another potential application is in text mining, where one central task of topic modeling is to learn the topics of various documents when they share the same vocabulary of words. By counting the number of words or $n$-grams in each document, one can obtain count matrix data. Compared to the absolute counts of these words and $n$-grams, the relative abundances may be more predictive on the topic. Thus, our proposed method can be useful for building classifiers for topics of documents.

\section*{Acknowledgment}
We thank the Editor, the Associate Editor, and two anonymous referees for their helpful comments that help improve the presentation of this paper. We thank S\"und\"uz Kele\c{s} for helpful discussions. The research of Y. Zhou and A. R. Zhang was supported in part by NSF Grants CAREER-1944904, DMS-1811868, NIH R01 GM131399, and NIH Grant R01 GM131399. The research of P. Shi was supported in part by NIH Grants R01 HG003747 and R21 HG009744. 

\bibliographystyle{apa}
\bibliography{reference}

\newpage

\setcounter{page}{1}
\setcounter{section}{0}

\begin{center}
	{\Large\bf Supplement to ``High-dimensional Log-Error-in-Variable }
	
	\smallskip	
	{\Large\bf Regression with Applications to Microbial Compositional Data}
	
	\smallskip
	
	{\Large\bf Analysis Composition Matrix in Metagenomics Data"}
	
	\bigskip\medskip
	{Pixu Shi, ~ Yuchen Zhou, ~ and ~ Anru R. Zhang}
\end{center}

\begin{abstract}
	The supplementary materials contain three parts. Appendix A provides the estimation procedure for the overdispersion parameter. 
	Appendices B and C contain the proofs of main theorems and technical lemmas, respectively.
\end{abstract}

\appendix

\appendixone

\section{Estimation of Overdispersion Parameter}\label{sec:overdispersion}

When there exists repeated measurements, i.e., multiple samples with shared composition $X_i$, we can apply the following moment estimator for overdispersion parameter estimation. See \cite{mosimann1963compound} and \cite[Equation (3)]{la2012hypothesis} for more details.
\begin{proposition}\label{prop:MOM_alpha}
	Suppose we have $m$ samples from the Dirichlet-multinomial distribution,
	$$W_i=(W_{i1},\dots,W_{ip}) \sim {\mathrm{Dirichlet}}-{\mathrm{Multinomial}}(N_i, \alpha X_1,\dots, \alpha X_p), \quad i=1,\ldots, m.$$ 
	(See Section \ref{sec:procedure} for detailed explanations for Dirichlet-multinomial distribution) Then a moment estimator for $\alpha$ is given by
	\begin{equation}\label{eq:MOM_alpha}
		\widehat{\alpha}_{MOM}=\dfrac{\sum_{j=1}^p (S_j-G_j)}{\sum_{j=1}^p \{S_j+(N_c-1)G_j\}},
	\end{equation}
	where
	\begin{gather*}
		S_j=\dfrac{1}{m-1}\sum_{i=1}^m N_i\left(\dfrac{W_{ij}}{N_i}-\dfrac{\sum_{k=1}^m W_{kj}}{\sum_{k=1}^m N_k}\right),\quad
		G_j=\dfrac{\sum_{i=1}^m W_{ij}(N_i-W_{ij})/N_i}{\sum_{i=1}^m (N_i-1)},
	\end{gather*}
	$$N_c=\frac{1}{m-1}\left(\sum_{i=1}^m N_i -\dfrac{\sum_{i=1}^m N_i^2}{\sum_{i=1}^m N_i}\right).$$
\end{proposition}

\appendixtwo

\section{Proofs}\label{sec:proofs}

We collect the proofs of the main results in this section. For convenience, denote $\nu_{\min} = \min_{k, i}\nu_{ki}, \nu_{\max} = \max_{k, i}\nu_{ki}$, $\phi_1\left(\cdot\right) =  \log\left(\cdot + 1/2\right), \phi_2\left(\cdot\right) = \log^2\left(\cdot + 1/2\right)$ and
$$A_W \in \mathbb{R}^{p\times p},\quad A_W = B_W^\top B_W, \quad P = I_p - \frac{1}{p}1_p 1_p^\top.$$
Denote
$$ \quad \bar{A}_W = PA_WP, \quad \bar{B}_W = B_W P.$$
We also denote the Orlicz-$\psi_1$ and $-\psi_2$ norms as $\|X\|_{\psi_1} = \sup_{p \geq 1} p^{-1}(\mathbb{E}|X|^p)^{1/p}, \|X\|_{\psi_2} = \sup_{p \geq 1} p^{-1/2}(\mathbb{E}|X|^p)^{1/p}$ for any random variable $X$.

\subsection{Proof of Theorem \ref{th:upper_bound}}\label{sec:proof-thm1}

We first briefly discuss the sketch of proof here. First, we develop a series of inequalities on the corrected covariates in Lemmas \ref{lm:poisson_1}-\ref{lm:sub-Gaussian} based on the tail probability bounds of Poisson, multinomial, and sub-Gaussian distributions. Then we develop an upper bound for $\left\|\bar{B}_W^\top\left(\bar{B}_W\beta^* - y\right)\right\|_{\infty}$, a pivotal term in high-dimensional regression analysis, in Lemma \ref{lm:infinity_norm_bound}. Finally, we combine these inequalities with Condition \ref{con:RIP} and obtain the upper bound for estimation error. The biggest challenge in the proof is to bound the corrected covariates $\log(W_{ij} + 1/2)$. Although the aforementioned Taylor's expansion heuristically show $\log\left(W_{ij} + 1/2\right)$ are good estimators for $\log(\nu_{ij})$, more careful analysis is needed to obtain rigorous upper bounds on their biases and variances. To this end, we perform truncation on $W_{ij}$ as the direct application of Poisson tail bounds may not yield sharp enough results.

Now we provide the complete proof of this theorem. The model can be summarized as follows,
\begin{equation}
	y_i = \sum_{j=1}^n \log (X_{ij}) \beta_j^* + \varepsilon_i, \quad i=1,\ldots, n; 
\end{equation}
\begin{equation}
	W_{ij}\sim {\mathrm{Poisson}} (\nu_{ij}),\quad 1\leq i \leq n; 1\leq j \leq p.
\end{equation}
Here $y_i$ and $W_{ij}$ are observable and $\beta_i^\ast, \nu_{ij} = \nu_iX_{ij}$ are hidden parameters. The proposed estimator is as follows,
\begin{equation}\label{eq81}
	\hat{\beta} = \argmin_{\gamma} \left(\frac{1}{2n}\|y - B_WP\gamma\|_2^2 + \lambda\|\gamma\|_1\right) \quad \text{subject to}\quad \quad \sum_{j=1}^p \gamma_j = 0.
\end{equation}
In the following lemmas, suppose $W_0 \sim \text{Poisson}(v)$ and $W_0' = W_01_{\left\{v/10 \leq W \leq 10v\right\}} + v1_{\left\{W \notin \left[v/10, 10v\right]\right\}}$. 
\begin{lemma}[Bias of $\log(W_0'+1/2)$]\label{lm:poisson_1}
	For any $\epsilon > 0$, there exists $C_{\epsilon} > 0$ that only depends on $\epsilon$, such that for all $v \geq C_{\epsilon}$, we have $|E\log(W_0' + 1/2) - \log v| \leq 4v^{-\frac{3}{2}+\epsilon}$. 
\end{lemma}

\begin{lemma}[Bias of $\log^2(W_0'+1/2)$]\label{lm:poisson_2}
	There exists constant $C>0$ such that if $v \geq C$, then $|E\{\log^2(W_0' + 1/2)\} - \log^2v| \leq 4/v$. 
\end{lemma}

\begin{lemma}[Sub-Gaussianity]\label{lm:sub-Gaussian}
	There exist positive constants $K_0$ and $C$, such that for $v \geq C$, we have 
	$$\left\|v^{1/2}\left[\log\left(W_0' + 1/2\right) - E\left\{\log\left(W_0' + 1/2\right)\right\}\right]\right\|_{\psi_2} \leq K_0.$$
\end{lemma}

\begin{lemma}[Infinity norm bound]\label{lm:infinity_norm_bound}
	Under the setting of Theorem \ref{th:upper_bound}, there exist two constants $C$ and $C'$ such that
	$${\mathrm{pr}}\left[\left\|\bar{A}_W \beta^* - \bar{B}_W^\top y\right\|_\infty \leq C\left\{n\log p\left(\sigma^2 + \frac{p}{\bar\nu}\|\beta^*\|_2^2\right) + n^2s\left(\frac{p}{\bar{\nu}}\right)^{3-2\epsilon}\|\beta^*\|_2^2\right\}^{1/2}\right] \geq 1 - 4p^{-C'}.$$
\end{lemma}

Now let us move to the proof of Theorem \ref{th:upper_bound}. Denote $h = \hat{\beta} - \beta^*$. By lemma \ref{lm:infinity_norm_bound}, with probability at least $1 - 4p^{-C'}$,
\begin{equation}\label{eq15}
	\left\|\bar{A}_W\beta^* - \bar{B}_W^\top y\right\|_\infty \leq C\left\{n\log p\left(\sigma^2 + \frac{p}{\bar\nu}\|\beta^*\|_2^2\right) + n^2s\left(\frac{p}{\bar{\nu}}\right)^{3-2\epsilon}\|\beta^*\|_2^2\right\}^{1/2} = \frac{n}{2}\lambda.
\end{equation}
By the definition of the estimator and the fact that $P\beta^* = \beta^*, P\hat{\beta} = \hat{\beta}$, we have
$$\frac{1}{2n}\left\|y - \bar{B}_W\hat{\beta}\right\|_2^2 + \lambda\left\|\hat{\beta}\right\|_1 \leq \frac{1}{2n}\left\|y - \bar{B}_W\beta^*\right\|_2^2 + \lambda\left\|\beta^*\right\|_1, $$
which means
\begin{equation*}
	\frac{1}{2n}\left(2h^\top\left(\bar{A}_W\beta^* - \bar{B}_W^\top y\right) + h^\top\bar{A}_Wh\right) \leq \lambda\left(\|\beta^*\|_1 - \|\hat{\beta}\|_1\right).
\end{equation*}
Denote $S = \text{supp}\left(\beta^\ast\right)$. Noting that 
\begin{equation*}
	\begin{split}
		\|\beta^*\|_1 - \|\hat{\beta}\|_1 = \|\beta_{S}^\ast\|_1 - \|\hat{\beta}_S\|_1 - \|\hat{\beta}_{S^c}\|_1
		\leq \|\beta_{S}^\ast - \hat{\beta}_S\|_1 - \|h_{S^c}\|_1 \leq \|h_S\|_1 - \|h_{S^c}\|_1,
	\end{split}
\end{equation*}
we have 
\begin{equation}\label{eq47}
	\frac{1}{2n}\left(2h^\top\left(\bar{A}_W\beta^* - \bar{B}_W^\top y\right) + h^\top\bar{A}_Wh\right) \leq \lambda\left(\|h_S\|_1 - \|h_{S^c}\|_1\right).
\end{equation}
In addition, 
\begin{equation*}
	\begin{split}
		&\frac{1}{2n}\left(2h^\top\left(\bar{A}_W\beta^* - \bar{B}_W^\top y\right) + h^\top\bar{A}_Wh\right) = \frac{1}{2n}\left(2h^\top\left(\bar{A}_W\beta^* - \bar{B}_W^\top y\right) + \|\bar{B}_Wh\|_2^2\right)\\
		\geq& \frac{1}{n}h^\top\left(\bar{A}_W\beta^* - \bar{B}_W^\top y\right)
		\geq -\frac{1}{n}\|h\|_1\|\bar{A}_W\beta^* - \bar{B}_W^\top y\|_{\infty}\\
		=& -\frac{1}{n}\left(\|h_S\|_1 + \|h_{S^c}\|_1\right)\|\bar{A}_W\beta^* - \bar{B}_W^\top y\|_{\infty}.
	\end{split}
\end{equation*}
If (\ref{eq15}) holds, then (\ref{eq15}), (\ref{eq47}) and the previous inequality together imply
\begin{equation*}
	-\frac{1}{2}\left(\|h_S\|_1 + \|h_{S^c}\|_1\right) \leq \|h_S\|_1 - \|h_{S^c}\|_1, \quad \text{i.e.,}\quad \|h_S\|_1 \geq \frac{1}{3}\|h_{S^c}\|_1.
\end{equation*}
Therefore,
\begin{equation*}
	\|h_{\max(s)}\|_1 \geq \|h_S\|_1 \geq \frac{1}{3}\|h_{S^c}\|_1 \geq \frac{1}{3}\|h_{-\max(s)}\|_1,
\end{equation*}
where we set $h_{\max(s)}$ as $h$ with all but the largest $s$ entries in absolute value set to zero, and $h _{-\max(s)} = h - h_{\max(s)}$.
By the Karush-Kuhn-Tucker condition on \eqref{eq81}, we have
\begin{equation*}
	\left\|\bar{B}_W^\top\left(\bar{B}_W\hat{\beta} - y\right) + 1_p\kappa\right\|_{\infty} \leq n\lambda
\end{equation*}
for some $\kappa \in \bbR$. Since $\|P x\|_{\infty} = \|x - \frac{1}{p}\sum_{i=1}^{p}x_i\|_{\infty} \leq \|x\|_{\infty} + \left|\frac{1}{p}\sum_{i=1}^{p}x_i\right| \leq 2\|x\|_{\infty}$, 
\begin{equation*}
	\begin{split}
		\left\|\bar{A}_W\hat{\beta} - \bar{B}_W^\top y\right\|_{\infty} = \left\|P\left\{\bar{B}_W^\top\left(\bar{B}_W\hat{\beta} - y\right) + 1_p\kappa\right\}\right\|_{\infty}
		\leq 2\left\|\bar{B}_W^\top\left(\bar{B}_W\hat{\beta} - y\right) + 1_p\kappa\right\|_{\infty}
		\leq 2n\lambda.
	\end{split}
\end{equation*}
By \eqref{eq15} and the previous inequality, with probability at least $1 - 4p^{-C'}$,
\begin{equation}\label{eq70}
	\left\|\bar{A}_Wh\right\|_\infty = \left\|\bar{A}_W\left(\beta^\ast - \hat{\beta}\right)\right\|_{\infty} \leq \|\bar{A}_W\hat{\beta} - \bar{B}_W^\top y\|_{\infty} + \left\|\bar{A}_W\beta^* - \bar{B}_W^\top y\right\|_\infty \leq \frac{5}{2}n\lambda.
\end{equation}
Therefore, with probability at least $1 - 4p^{-C'}$,
\begin{equation}\label{eq17}
	h_{\max(s)}\bar{A}_Wh \leq \left\|h_{\max(s)}\right\|_1 \left\|\bar{A}_Wh\right\|_{\infty}\\
	\leq \frac{5}{2}n\lambda \cdot\sqrt{s}\left\|h_{\max(s)}\right\|_2.
\end{equation}
Define $\alpha = \left\|h_{\max(s)}\right\|_1/s$, then
\begin{equation*}
	\begin{split}
		\left\|h_{-max(s)} \right\|_{\infty} &\leq \alpha \leq 3\alpha, \quad 
		\left\|h_{-max(s)}\right\|_1  \leq 3\left\|h_{\max(s)}\right\|_1 = 3s\alpha.
	\end{split}
\end{equation*}
Apply Lemma 1.1 in \cite{cai2014sparse}, $h_{-\max(s)}$ can be expressed as a convex combinations of sparse vectors: $h_{-\max(s)} = \sum_{i = 1}^M \lambda_i u_i$, where $u_i$ is $s$-sparse and
\begin{equation*}
	\left\|u_i\right\|_1 = \left\|h^{(2)}\right\|_1, \quad \left\|u_i\right\|_{\infty} \leq 3\alpha, \quad supp(u_i) \subseteq supp(h_{-\max(s)}).
\end{equation*}   
Thus
\begin{equation}\label{eq49}
	\begin{split}
		\left\|u_i\right\|_2 &\leq \left\|u_i\right\|_0^{1/2}\left\|u_i\right\|_{\infty} \leq (\surd{s}) 3\alpha = 3\alpha\surd{s}.
	\end{split}
\end{equation}
Suppose $0 \leq \mu \leq 1$ is to be determined. Denote $\gamma_i = h_{\max(s)} + \mu u_i$, then
\begin{equation*}
	\begin{split}
		\sum_{j = 1}^{M} \lambda_j \gamma_j - \frac{1}{2}\gamma_i &= h_{\max(s)} + \mu h_{-\max(s)} - \frac{1}{2}\gamma_i = (1 - \mu - \frac{1}{2})h_{\max(s)} - \frac{1}{2}\mu u_i + \mu h. 
	\end{split}
\end{equation*}
Since $h_{\max(s)}$ and $u_i$ are $s$-sparse vectors, $\gamma_i = h_{\max(s)} + \mu u_i, \sum_{j = 1}^{M} \lambda_j \gamma_j - \gamma_i/2 - \mu h = \{(1/2) - \mu\}h_{\max(s)} - \mu u_i/2$ are all $2s$-sparse vectors. \\
Suppose $x = \left\|h_{\max(s)}\right\|_2$, by \eqref{eq49},
\begin{equation}\label{eq82}
	\begin{split}
		\left\|u_i\right\|_2 \leq 3\alpha\surd{s} \leq 3\left\|h_{\max(s)}\right\|_2 = 3x.
	\end{split}
\end{equation}
Also, we can check that
\begin{equation*}
	\begin{split}
		&\sum_{i=1}^M \lambda_i \left\|\bar{B}_W\left(\sum_{j=1}^M \lambda_j\gamma_j - \frac{1}{2}\gamma_i\right)\right\|_2^2 = \sum_{i = 1}^M \frac{\lambda_i}{4}\|\bar{B}_W\gamma_i\|_2^2.
	\end{split}
\end{equation*}
\begin{equation}\label{eq16}
	\begin{split}
		0 =& \sum_{i = 1}^M \lambda_i\left\|\bar{B}_W\left\{h_{\max(s)} + \mu h_{-\max(s)}-\frac{1}{2}(h_{\max(s)} + \mu u_i)\right\}\right\|_2^2 - \sum_{i = 1}^M \frac{\lambda_i}{4}\|\bar{B}_W \gamma_i\|_2^2 \\
		=& \sum_{i=1}^M \lambda_i\left\|\bar{B}_W\left\{\left(\frac{1}{2}-\mu\right)h_{\max(s)} - \frac{\mu}{2}u_i + \mu h\right\}\right\|_2^2 - \sum_{i = 1}^M \frac{\lambda_i}{4}\|\bar{B}_W \gamma_i\|_2^2 \\=&\sum_{i = 1}^M\lambda_i\left\|\bar{B}_W\left\{\left(\frac{1}{2}-\mu\right)h_{\max(s)}-\frac{\mu}{2}u_i\right\}\right\|_2^2+ 2\mu\left\{\left(\frac{1}{2}-\mu\right)h_{\max(s)}-\frac{\mu}{2}h_{-\max(s)}\right\}^{\top}\bar{A}_Wh\\ 
		&+\mu^2\|\bar{B}_Wh\|_2^2-\sum_{i = 1}^M \frac{\lambda_i}{4}\|\bar{B}_W \gamma_i\|_2^2\\
		=&\sum_{i = 1}^M\lambda_i\left\|\bar{B}_W\left\{\left(\frac{1}{2}-\mu\right)h_{\max(s)}-\frac{\mu}{2}u_i\right\}\right\|_2^2
		+\mu(1-\mu)h_{\max(s)}^{\top}\bar{A}_Wh-\sum_{i = 1}^M \frac{\lambda_i}{4}\|\bar{B}_W \gamma_i\|_2^2.
	\end{split}
\end{equation}
In the third equation above, we used the facts that $h_{-\max(s)} = \sum_{i = 1}^M\lambda_iu_i$ and $\bar{A}_W = \bar{B}_W^\top\bar{B}_W$. Noting that supp$(h_{\max(s)})$  $\cap$  supp$(u_i)$ $(i = 1, \dots, M)$ are empty sets and $\gamma_i, \{(1/2)-\mu\}h_{\max(s)}-\mu u_i/2$ are all $2s$-sparse vectors, we can apply Condition \ref{con:RIP}, \eqref{eq17}, and \eqref{eq82} to \eqref{eq16} and obtain
\begin{equation*}
	\begin{split}
		0 \leq& n\{1 + \delta_{2s}\left(\bar{B}_W\right)\}\sum_{i=1}^M \lambda_i \left\{\left(\frac{1}{2}-\mu\right)^2\|h_{\max(s)}\|_2^2+\frac{\mu^2}{4}\|u_i\|_2^2\right\}
		+ \frac{5}{2}n\mu(1-\mu)(\surd{s})\lambda\|h_{\max(s)}\|_2\\ &- n\left\{1 - \delta_{2s}\left(\bar{B}_W\right)\right\}\sum_{i = 1}^M \frac{\lambda_i}{4}\left(\|h_{\max(s)}\|_2^2+\mu^2\|u_i\|_2^2\right)\\
		=&n\sum_{i=1}^M \lambda_i\left(\left[\left\{1 + \delta_{2s}\left(\bar{B}_W\right)\right\}\left(\frac{1}{2}-\mu\right)^2 - \frac{1}{4}\left\{1 - \delta_{2s}\left(\bar{B}_W\right)\right\} \right]\|h_{\max(s)}\|_2^2+\frac{1}{2}\delta_{2s}\left(\bar{B}_W\right)\mu^2\|u_i\|_2^2\right)\\
		&+ \frac{5}{2}n\mu(1-\mu)(\surd{s})\lambda\|h_{\max(s)}\|_2\\
		\leq& n\sum_{i=1}^M \lambda_i\left(\left[\left\{1 + \delta_{2s}\left(\bar{B}_W\right)\right\}\left(\frac{1}{2}-\mu\right)^2 - \frac{1}{4}\left\{1 - \delta_{2s}\left(\bar{B}_W\right)\right\} \right]x^2+\left\{\frac{1}{2}\delta_{2s}\left(\bar{B}_W\right)\mu^2\right\}9x^2\right)\\
		&+ \frac{5}{2}n\mu(1-\mu)(\surd{s})\lambda x\\
		\leq& n\left[(\mu^2 - \mu)+\frac{1}{2}\delta_{2s}\left(\bar{B}_W\right)\left(1 - 2\mu+ 11\mu^2\right)\right]x^2 + \frac{5}{2}n\mu(1-\mu)(\surd{s})\lambda x
	\end{split}
\end{equation*}
holds with probability at least $1 - 4p^{-C'} - \epsilon'$. Set $\mu = \frac{1}{2}$ and notice that $\delta_{2s}\left(\bar{B}_W\right) \leq \frac{1}{10}$, we know that with probability at least $1 - 4p^{-C'} - \epsilon'$, 
\begin{equation*}
	x \leq\left(1 - \frac{11}{2}\delta_{2s}\left(\bar{B}_W\right)\right)^{-1}\frac{5}{2}(\surd{s})\lambda \leq \frac{50}{9}s^{1/2}\lambda.
\end{equation*}
Therefore, with probability at least $1 - 4p^{-C'} - \epsilon'$, 
\begin{equation*}
	\begin{split}
		\|h\|_2 = \left(\|h_{\max(s)}\|_2^2 + \|h_{-\max(s)}\|_2^2\right)^{1/2}
		\leq \left(\|h_{\max(s)}\|_2^2 + 9\|h_{\max(s)}\|_2^2\right)^{1/2}
		= 10^{1/2}x
		\leq 18s^{1/2}\lambda,
	\end{split}
\end{equation*}
which means with probability at least $1 - 4p^{-C'} - \epsilon'$, 
\begin{equation*}
	\|\hat{\beta} - \beta^*\|_2^2 \leq 324s\lambda^2 = \frac{Cs\log p}{n}\left(\sigma^2 + \frac{p}{\bar{\nu}}\|\beta^*\|_2^2\right) + Cs^2\left(\frac{p}{\bar{\nu}}\right)^{3-2\epsilon}\|\beta^*\|_2^2.
\end{equation*}
\qed

\subsection{Proof of Theorem \ref{th:lower_bound}}\label{sec:lower-bound}

We first provide a sketch of the proof here. The lower bound in Theorem \ref{th:lower_bound} consists of two terms: $cs\log (p/s)\sigma^2/n$ and $\{cs\log (p/s)/n\}(p/Q)R^2$, which originate from the uncertainty of $\varepsilon$ and $W$, respectively. We thus show $\inf \sup E(\|\hat{\beta} - \beta\|_2^2) \geq cs\log (p/s)\sigma^2/n$ and $\inf \sup E(\|\hat{\beta} - \beta\|_2^2) \geq \{cs\log (p/s)/n\}(p/Q)R^2$ separately in the proof of this theorem. While the first inequality develops from the classic high-dimensional regression literature (see, e.g., \cite{rigollet2011exponential}), the proof for the second one is far more complicated. In particular, we construct a series of instances of $(\nu^{(i)}, X^{(i)}, \beta^{(i)})_{i=1}^N$ satisfying the constraints in \eqref{con:gaussian} and $\diag(\nu^{(i)})X^{(i)}\beta^{(i)} = \diag(\nu^{(j)})X^{(j)}\beta^{(j)}$ for all $i, j$. By such the design, $y^{(i)} = \diag(\nu^{(i)})V^{(i)}\beta^{(i)} + \varepsilon^{(i)}$ becomes nullified for estimating $\beta^{(i)}$ and we show these instances are non-distinguishable based a sample of $(W^{(i)}, y^{(i)})$. Then we apply the generalized Fano's lemma and establish the desired error lower bound.

Then we provide a complete proof. In order to show the desired result, we only need to prove the following two lower bounds separately.
\begin{equation}\label{eq34}
	\inf_{\hat{\beta}} \sup_{(\nu, X, \beta) \in \mathcal{F}_{p, n, s}(R, Q)} E\left( \|\hat{\beta} - \beta\|_2^2\right) \geq \frac{c\sigma^2 s\log \left(\frac{p}{s}\right)}{n};
\end{equation}
\begin{equation}\label{ineq:lower2}
	\inf_{\hat{\beta}} \sup_{(\nu, X, \beta) \in \mathcal{F}_{p, n, s}(R, Q)} E\left( \|\hat{\beta} - \beta\|_2^2\right) \geq \frac{cs\log \left(\frac{p}{s}\right)}{n} \cdot \frac{p}{Q}R^2.
\end{equation}
First, we introduce the following lemma.
\begin{lemma}\label{lm: RIP fixed}
	If $n \times p$ random matrix $V$ i.i.d. entries that satisfy
	$$
	v_{ij} = \left\{\begin{array}{ll}
	v + 1, & \text{with probability } \frac{1}{2},\\
	v - 1, & \text{with probability } \frac{1}{2},
	\end{array}\right.
	$$
	where $v$ is a fixed positive number, then with probability $> 2/3$, for all $2s$-sparse $x \in \bbR^p$,
	\begin{equation}\label{eq45}
		n\left(1 - \frac{1}{40}\right)\|x\|_2^2 \leq \|VP x\|_2^2 \leq n\left(1 + \frac{1}{40}\right)x^2,
	\end{equation}
	where $P = I_p - (1/p)1_p1_p^\top$.
\end{lemma}
To prove \eqref{eq34}, we only need to show that for some fixed $\nu, X$ satisfying $e^{-3}\bar{\nu} \leq \nu_{i} \leq e^{3}\bar{\nu}$ and $e^{-3}/p \leq X_{ij} \leq e^{3}/p$, $V \in \bbR^{n \times p}$ with $(V)_{ij} = \log(\nu_iX_{ij})$ satisfying \eqref{eq45}, and $e^{-\frac{3}{2}}Q \leq \bar{\nu} \leq e^{\frac{3}{2}}Q$,
\begin{equation*}
	\inf_{\hat{\beta}} \sup_{\beta \in B_0(s) \cap B_2(R) \cap \{\beta: 1^\top\beta = 0\}} E (\|\hat{\beta} - \beta\|_2)^2 \geq \frac{c\sigma^2 s\log \left(\frac{p}{s}\right)}{n}.
\end{equation*}
Denote 
\begin{equation*}
	\mathcal{M} = \left\{x \in \{0, 1\}^{\lfloor\frac{p}{2}\rfloor}: \|x\|_0 = \left\lfloor
	\frac{s}{2}\right\rfloor\right\}.
\end{equation*} 
By Lemma 2.9 in \cite{tsybakov2009introduction} and Lemma A.3 in \cite{rigollet2011exponential}, there exists a subset $\mathcal{M}'$ of $\mathcal{M}$ such that $\forall x, x' \in \mathcal{M}', x' \neq x$, we have $\|x - x'\|_0 \geq s/32$ and 
\begin{equation*}
	\log\left|\mathcal{M}'\right| \geq \bar{c}s\log\left(\frac{p}{s}\right).
\end{equation*}
Let the elements in $\mathcal{M}'$ be $x^{(1)}, x^{(2)}, \dots, x^{(\left|\mathcal{M}'\right|)}$. For all $1 \leq j \leq \left|\mathcal{M}'\right|$, set 
\begin{equation*}
	\beta^{(j)} = \left\{\begin{array}{ll}
		\frac{\tau}{\surd(2\left\lfloor\frac{s}{2}\right\rfloor)}\left(x^{(j)\top}, -x^{(j)\top}\right)^\top, & p \text{ is even},\\
		\frac{\tau}{\surd(2\left\lfloor\frac{s}{2}\right\rfloor)}\left(x^{(j)\top}, 0, -x^{(j)\top}\right)^\top, & p \text{ is odd},
	\end{array}\right.
\end{equation*}
where $\tau > 0$ is to be determined. Then we have $\|\beta^{(j)}\|_0 = 2\lfloor s/2\rfloor \leq s, \|\beta^{(j)}\|_2 = \tau$ and 
\begin{equation*}
	\|\beta^{(i)} - \beta^{(j)}\|_2 = \frac{\tau}{\surd(2\left\lfloor\frac{s}{2}\right\rfloor)}\left(\|\beta^{(i)} - \beta^{(j)}\|_0\right)^{\frac{1}{2}} = \frac{\tau}{\surd(2\left\lfloor\frac{s}{2}\right\rfloor)}(\frac{s}{32})^{1/2}\geq \frac{\tau}{4\sqrt{2}}, \quad 1 \leq i, j \leq |\mathcal{M}|', i \neq j.
\end{equation*}
Moreover, given $\left(V^{(i)}, \beta^{(i)}\right)$, the Kullback-Leibler divergence
\begin{equation}\label{eq: KL}
	D_{KL}\left\{\left(y^{(i)}, W^{(i)}\right), \left(y^{(j)}, W^{(j)}\right)\right\} = D_{KL}\left(y^{(i)}, y^{(j)}\right) + D_{KL}\left(W^{(i)}, W^{(j)}\right), \quad \forall i \neq j.
\end{equation}
Actually, given $V^{(i)}$ and $\beta^{(i)}$, $y^{(i)}$ and $W^{(i)}$ are independent. In addition, $y^{(i)}$ and $y^{(j)}$ are independent. Therefore \eqref{eq: KL} holds.

Here $V^{(i)} = V^{(j)} = V$, thus $W^{(i)}$ and $W^{(j)}$ have the same distribution, and
\begin{equation}\label{eq400}
	D_{KL}\left(W^{(i)}, W^{(j)}\right) = 0.
\end{equation}
Here, $D_{KL}(\cdot, \cdot)$ is the Kullback–Leibler divergence between two distributions. Note that $P\beta^{(i)} = P\beta^{(j)} = 0$,
\begin{equation*}
	\begin{split}
		D_{KL}(y^{(i)}, y^{(j)}) =& \frac{1}{2\sigma^2}\|V(\beta^{(i)} - \beta^{(j)})\|_2^2 = \frac{1}{2\sigma^2}\|VP(\beta^{(i)} - \beta^{(j)})\|_2^2 \leq \frac{n(1 + \delta)}{2\sigma^2}\|\beta^{(i)} - \beta^{(j)}\|_2^2\\
		\leq& \frac{n(1 + \delta)}{2\sigma^2}2\left(\|\beta^{(i)}\|_2^2 + \|\beta^{(j)}\|_2^2\right) = \frac{2n(1 + \delta)\tau^2}{\sigma^2},
	\end{split}
\end{equation*}
where $\delta = 1/20$.The first inequality follows from \eqref{eq45}; the second inequality is due to Cauchy-Schwarz inequality. Combine \eqref{eq: KL}, \eqref{eq400} and the previous inequality together, we have
\begin{equation*}
	D_{KL}\left\{\left(y^{(i)}, W^{(i)}\right), \left(y^{(j)}, W^{(j)}\right)\right\} \leq \frac{2n(1 + \delta)\tau^2}{\sigma^2}.
\end{equation*}
By generalized Fano's lemma, 
\begin{equation*}
	\inf_{\hat{\beta}} \sup_{\beta \in B_0(s) \cap B_2(R) \cap \{\beta: 1^\top\beta = 0\}}E\left(\|\hat{\beta} - \beta\|_2\right) \geq \frac{\tau}{8\sqrt{2}}\left\{1 - \frac{\frac{2n(1 + \delta)\tau^2}{\sigma^2} + \log 2}{\bar{c}s\log\left(\frac{p}{s}\right)}\right\}.
\end{equation*}
Set $\tau = \bar{c}_1\surd\{s\log(p/s)\sigma^2/n\}$ with a small constant $\bar{c}_1 > 0$, we have
\begin{equation*}
	\begin{split}
		\inf_{\hat{\beta}} \sup_{\beta \in B_0(s) \cap B_2(R) \cap \{\beta: 1^\top\beta = 0\}}E\left(\|\hat{\beta} - \beta\|_2\right)^2 \geq& \left\{\inf_{\hat{\beta}} \sup_{\beta \in B_0(s) \cap B_2(R) \cap \{\beta: 1^\top\beta = 0\}}E_j\left(\|\hat{\beta} - \beta\|_2\right)\right\}^2\\ \geq& \frac{c\sigma^2s\log\left(\frac{p}{s}\right)}{n}.
	\end{split}
\end{equation*}
\ \par
For the second part, denote $s_0 = \lfloor s/4\rfloor, p_0 = \left\lfloor p/2\right\rfloor$. Set $\tilde{\beta}^{(0)} \in \bbR^{p_0}, \beta^{(0)} \in \bbR^p$ such that
$$\tilde\beta^{(0)}_j = \left\{\begin{array}{ll}
1, & 1 \leq j \leq s_0,\\
0, & \text{otherwise}
\end{array}\right.$$
and
$$\beta^{(0)} = \left\{\begin{array}{ll}
\frac{R}{\surd\{2s_0(1 + \theta^2)\}}\left\{\left(\tilde{\beta}^{(0)}\right)^\top, -\left(\tilde{\beta}^{(0)}\right)^\top\right\}^\top, & p \text{ is even}, \\
\frac{R}{\surd\{2s_0(1 + \theta^2)\}}\left\{\left(\tilde{\beta}^{(0)}\right)^\top, 0, -\left(\tilde{\beta}^{(0)}\right)^\top\right\}^\top, & \text{otherwise}.
\end{array}\right.$$
Moreover, set $\tilde\beta^{(i)} \in \bbR^{p_0}, \beta^{(i)} \in \bbR^{p} (i = 1, 2, \dots, M)$ such that
$$\tilde\beta^{(i)}_j = \left\{\begin{array}{ll}
\theta, & \text{if } j \in \Omega_i,\\
0, & \text{if } j \notin \Omega_i,
\end{array}\right.$$
and
$$\beta^{(j)} = \left\{\begin{array}{ll}
\frac{R}{\surd\{2s_0(1 + \theta^2)\}}\left\{\left(\tilde{\beta}^{(0)} + \tilde\beta^{(i)}\right)^\top, -\left(\tilde{\beta}^{(0)} + \tilde\beta^{(i)}\right)^\top\right\}^\top, & p \text{ is even}, \\
\frac{R}{\surd\{2s_0(1 + \theta^2)\}}\left\{\left(\tilde{\beta}^{(0)} + \tilde\beta^{(i)}\right)^\top, 0, -\left(\tilde{\beta}^{(0)} + \tilde\beta^{(i)}\right)^\top\right\}^\top, & \text{otherwise},
\end{array}\right.$$
where $\Omega_1, \Omega_2, \dots, \Omega_M$ are uniformly random subsets from $\{s_0 + 1, s_0 + 2, \dots, p_0\}$ such that $\left|\Omega_1\right| = \cdots = \left|\Omega_M\right| = s_0$, $\theta$ and $M$ are to be determined. It is clear that 
\begin{equation*}
	\|\beta^{(i)}\|_{2} = \frac{R}{\surd\{2s_0(1 + \theta^2)\}}\left\{2\left(s_0 + s_0\theta^2\right)\right\}^{\frac{1}{2}} = R,
\end{equation*}
and
\begin{equation*}
	\begin{split}
		\|\beta^{(i)} - \beta^{(j)}\|_2^2 =& \left\{\frac{R}{\surd\{2s_0(1 + \theta^2)\}}\right\}^2 2\|\tilde\beta^{(i)} - \tilde\beta^{(j)}\|_2^2 \leq  \left\{\frac{R}{\surd\{2s_0(1 + \theta^2)\}}\right\}^2 4\theta^2s_0 = \frac{2R^2\theta^2}{1 + \theta^2}.
	\end{split}
\end{equation*}
Additionally,
\begin{equation*}
	\begin{split}
		&{\mathrm{pr}}\left(\|\beta^{(i)} - \beta^{(j)}\|_2^2 \leq \frac{R^2\theta^2}{1 + \theta^2}\right) = {\mathrm{pr}}\left(\|\tilde\beta^{(i)} - \tilde\beta^{(j)}\|_2^2 \leq s_0\theta^2\right) = {\mathrm{pr}}\left(\left|\Omega_i \cap \Omega_j\right| \geq \frac{s_0}{2}\right)\\ =& \sum_{t = \left\lceil\frac{s_0}{2}\right\rceil}^{s_0}{\mathrm{pr}}\left(\left|\Omega_i \cap \Omega_j\right| = t\right)
		= \sum_{t = \left\lceil\frac{s_0}{2}\right\rceil}^{s_0}\frac{\binom{s_0}{t}\binom{p_0 - 2s_0}{s_0 - t}}{\binom{p_0 - s_0}{s_0}} = \sum_{t = \left\lceil\frac{s_0}{2}\right\rceil}^{s_0}\binom{s_0}{t}\frac{\frac{s_0!}{\left(s_0 - t\right)!}}{\frac{(p_0 - s_0)\cdots(p_0 - 2s_0 + 1)}{(p_0 - 2s_0)\cdots(p_0 - 3s_0 + t + 1)}}\\ \leq& \sum_{t = \left\lceil\frac{s_0}{2}\right\rceil}^{s_0}\binom{s_0}{t}\left(\frac{s_0}{p_0 - 2s_0 + 1}\right)^t
		\leq \sum_{t = \left\lceil\frac{s_0}{2}\right\rceil}^{s_0}\binom{s_0}{t}\left(\frac{s_0}{p_0 - 2s_0 + 1}\right)^{\frac{s_0}{2}} \leq \left(\frac{4s_0}{p_0 - 2s_0 + 1}\right)^{\frac{s_0}{2}}.
	\end{split}
\end{equation*}  
The first inequality comes from $s_0!/\left(s_0 - t\right)! \leq s_0^t$ and 
\begin{equation*}
	\frac{(p_0 - s_0)\cdots(p_0 - 2s_0 + 1)}{(p_0 - 2s_0)\cdots(p_0 - 3s_0 + t + 1)} \geq (p_0 - 2s_0 + 1)^t;
\end{equation*}
the third inequality holds since 
\begin{equation*}
	\sum_{t = \left\lceil\frac{s_0}{2}\right\rceil}^{s_0}\binom{s_0}{t} \leq 2^{s_0}.
\end{equation*}

Set $M = \lfloor  \{(p_0 - 2s_0 + 1)/(4s_0)\}^{s_0/4}\rfloor$, we have
\begin{equation*}
	\begin{split}
		{\mathrm{pr}}\left(\|\beta^{(i)} - \beta^{(j)}\|_2^2 > \frac{R^2\theta^2}{1 + \theta^2}, \forall i \neq j\right) \geq& 1 - \frac{M(M - 1)}{2}\left(\frac{4s_0}{p_0 - 2s_0 + 1}\right)^{\frac{s_0}{2}}\\
		>& 1 - M^2\left(\frac{4s_0}{p_0 - 2s_0 + 1}\right)^{\frac{s_0}{2}} > 0,
	\end{split}
\end{equation*}
which means there exist fixed $\beta^{(1)}, \dots, \beta^{(M)}$ satisfying
\begin{equation}\label{eq37}
	\frac{R^2\theta^2}{1 + \theta^2} < \min_{i \neq j}\|\beta^{(i)} - \beta^{(j)}\|_2^2 \leq \frac{2R^2\theta^2}{1 + \theta^2}.
\end{equation}
Consider an $n \times p$ random matrix $V$ with entries
$$
v_{ij} = \left\{\begin{array}{ll}
v + 1, & \text{with probability } \frac{1}{2},\\
v - 1, & \text{with probability } \frac{1}{2},
\end{array}\right.
$$
where $v$ is to be determined. By Lemma \ref{lm:epsilon-net}, with probability $> 2/3$, for all $2s$-sparse $x \in \bbR^p$,
\begin{equation}\label{eq76}
	n\left(1 - \frac{1}{40}\right)\|x\|_2^2 \leq \|VP x\|_2^2 \leq n\left(1 + \frac{1}{40}\right)x^2.
\end{equation}
We would like to construct $\widetilde{V}^{(1)}, \dots, \widetilde{V}^{(M)}$ such that
\begin{equation*}
	(V + \widetilde{V}^{(i)})\beta^{(1)} = \cdots = (V + \widetilde{V}^{(M)})\beta^{(M)}.
\end{equation*}
Actually, we can set
\begin{equation*}
	\left(\widetilde{V}^{(i)}\right)_{kj} = \left\{\begin{array}{ll}
		0, & j \notin \{1, 2, \dots, s_0\}\cup\{p - p_0 +1, p - p_0 + s_0\}, \\
		\theta v - \frac{\theta}{s_0}\sum_{l \in \Omega_i}v_{kl}, & j \in \{1, 2, \dots, s_0\},\\
		\theta v - \frac{\theta}{s_0}\sum_{l \in \Omega_i'}v_{kl}, & j \in \{p - p_0 +1, \dots, p - p_0 + s_0\},
	\end{array}\right.
\end{equation*}
where $\Omega_j' = \{i + p - p_0: i \in \Omega_j\}$, $\theta > 0$ is a fixed number need to be determined. 
Denote 
\begin{equation*}
	\gamma^{(i)} = \beta^{(i)} - \beta^{(0)} = \left\{\begin{array}{ll}
		\frac{R}{\surd\{2s_0(1 + \theta^2)\}}\left\{\left(\tilde\beta^{(i)}\right)^\top, -\left(\tilde\beta^{(i)}\right)^\top\right\}^\top, & p \text{ is even}, \\
		\frac{R}{\surd\{2s_0(1 + \theta^2)\}}\left\{\left(\tilde\beta^{(i)}\right)^\top, 0, -\left(\tilde\beta^{(i)}\right)^\top\right\}^\top, & \text{otherwise}.
	\end{array}\right.
\end{equation*}
Then for any $1 \leq i \leq M$, 
\begin{equation*}
	\begin{split}
		\left(V + \widetilde V^{(i)}\right)\beta^{(i)} =& \left(V + \widetilde V^{(i)}\right)\left(\beta^{(0)} + \gamma^{(i)}\right) = V\beta^{(0)} + \left(V\gamma^{(i)} + \widetilde{V}^{(i)}\beta^{(0)}\right) + \widetilde{V}^{(i)}\gamma^{(i)}\\
		=& V\beta^{(0)} + 0 + 0 = V\beta^{(0)}.
	\end{split}
\end{equation*}
The third equation holds since for all $1 \leq j \leq n$,
\begin{equation*}
	\begin{split}
		&\left(V\gamma^{(i)} + \widetilde{V}^{(i)}\beta^{(0)}\right)_j = \sum_{k = 1}^{p}\left(V\right)_{jk}\gamma^{(i)}_k + \sum_{j = 1}^{p}\left(\widetilde{V}^{(i)}\right)_{jk}\beta^{(0)}_{k}\\
		=& \frac{R\theta}{\surd\{2s_0(1 + \theta^2)\}}\left(\sum_{k \in \Omega_i}v_{jk} - \sum_{k' \in \Omega_i'}v_{jk'}\right) + \frac{R}{\surd\{2s_0(1 + \theta^2)\}}\left\{\sum_{l = 1}^{s_0}\left(\widetilde{V}^{(i)}\right)_{jl} - \sum_{l' = p - p_0 + 1}^{p - p_0 + s_0}\left(\widetilde{V}^{(i)}\right)_{jl'}\right\}\\
		=& \frac{R\theta}{\surd\{2s_0(1 + \theta^2)\}}\left(\sum_{k \in \Omega_i}v_{jk} - \sum_{k' \in \Omega_i'}v_{jk'}\right) + \frac{R}{\surd\{2s_0(1 + \theta^2)\}} s_0\left(\theta v - \frac{\theta}{s_0}\sum_{l \in \Omega_i}v_{jl}\right)\\ &- \frac{R}{\surd\{2s_0(1 + \theta^2)\}}s_0\left(\theta v - \frac{\theta}{s_0}\sum_{l' \in \Omega_i'}v_{jl'}\right) = 0,
	\end{split}
\end{equation*}
and
\begin{equation*}
	\left[\widetilde{V}^{(i)}\gamma^{(i)}\right]_j = \sum_{k \in \Omega_i}\left(\widetilde{V}^{(i)}\right)_{jk}\gamma^{(i)}_k = \sum_{k \in \Omega_i}0\cdot\gamma^{(i)}_k = 0.
\end{equation*}
Moreover, we have $\widetilde{V}^{(i)} = V_0U^{(i)}$, where $V_0$ is a random matrix with $\left(V_0\right)_{ij} = v_{ij} - v$ and $U_i \in \bbR^{p \times p}$,
\begin{equation*}
	\left(U^{(i)}\right)_{kl} = \frac{\theta}{s_0}\left(1_{\{k \in \Omega_i, l \in \{1, 2, \dots, s_0\}\}} + 1_{\{k \in \Omega_i', l \in \{p - p_0 + 1, p - p_0 + 2, \dots, p - p_0 + s_0\}\}}\right).
\end{equation*}
For any $\gamma \in \bbR^p$, by Cauchy-Schwarz inequality,
\begin{equation*}
	\begin{split}
		\|U^{(i)}P\gamma\|_2^2 =& \sum_{k = 1}^{n}\left\{\sum_{l=1}^{p}\left(U^{(i)}\right)_{kl}\left(P\gamma\right)_l\right\}^2 = \sum_{k \in \Omega_i}\left\{\sum_{l=1}^{s_0}\frac{\theta}{s_0}\left(P\gamma\right)_l\right\}^2 + \sum_{k \in \Omega_i'}\left\{\sum_{l=p - p_0 + 1}^{p - p_0 + s_0}\frac{\theta}{s_0}\left(P\gamma\right)_l\right\}^2\\
		=& \frac{\theta^2}{s_0}\left[\left\{\sum_{l=1}^{s_0}\left(P\gamma\right)_l\right\}^2 + \left\{\sum_{l=p-p_0+1}^{p-p_0+s_0}\left(P\gamma\right)_l\right\}^2\right] \leq \theta^2\left\{\sum_{l=1}^{s_0}\left(P\gamma\right)_l^2 + \sum_{l=p-p_0+1}^{p-p_0+s_0}\left(P\gamma\right)_l^2\right\}\\
		\leq& \theta^2\|P\gamma\|_2^2 \leq \theta^2\|\gamma\|_2^2.
	\end{split}
\end{equation*}
Recall that $\widetilde{V}^{(i)} = V_0U^{(i)}$ and the $n \times p$ matrix $V_0$ satisfies
$$
v_{ij} = \left\{\begin{array}{ll}
+ 1, & \text{with probability } \frac{1}{2},\\
- 1, & \text{with probability } \frac{1}{2},
\end{array}\right.
$$
by Lemma 5 in \cite{achlioptas2001database}, $\forall \epsilon > 0$, we have
\begin{equation*}
	\begin{split}
		{\mathrm{pr}}\left(\left|\frac{1}{n}\|\widetilde{V}^{(i)}P\gamma\|_2^2 - \|U^{(i)}P\gamma\|_2^2\right| \geq \epsilon\|U^{(i)}P\gamma\|_2^2\right) \leq 2\exp\left\{-n\left(\frac{\epsilon^2}{4} - \frac{\epsilon^3}{6}\right)\right\},
	\end{split}
\end{equation*}
which means
\begin{equation*}
	\begin{split}
		{\mathrm{pr}}\left\{\frac{1}{n}\|\widetilde{V}^{(i)}P\gamma\|_2^2 \geq \left(1 + \epsilon\right)\theta^2\|\gamma\|_2^2\right\} \leq& {\mathrm{pr}}\left\{\frac{1}{n}\|\tilde{V}^{(i)}P\gamma\|_2^2 \geq \left(1 + \epsilon\right)\|U^{(i)}P\gamma\|_2^2\right\}\\
		\leq& 2\exp\left\{-n\left(\frac{\epsilon^2}{4} - \frac{\epsilon^3}{6}\right)\right\}.
	\end{split}
\end{equation*}
For any $S_0 \subset \{1, \dots, p\}, |S_0| = 2s$, denote $\mathcal{A}_{S_0} = \{x: x \in S^{p-1}, \text{supp}(x) \subseteq S_0\}$. By Lemma 5.2 in \cite{vershynin2010introduction}, there exists an $\epsilon/4$-net of $\mathcal{A}_{S_0}$, say $\mathcal{N}_{\frac{\epsilon}{4}}$, with $\left|\mathcal{N}_{\epsilon}\right| \leq \{1 + 2/(\epsilon/4)\}^{2s} = \left(1 + 8/\epsilon\right)^{2s}$.
Thus
\begin{equation*}
	\begin{split}
		{\mathrm{pr}}\left\{\exists \gamma \in \mathcal{N}_{\frac{\epsilon}{4}}, \frac{1}{n}\|\widetilde{V}^{(i)}P\gamma\|_2^2 \geq \left(1 + \epsilon\right)\theta^2\|\gamma\|_2^2\right\} \leq
		\left(1 + \frac{8}{\epsilon}\right)^{2s}\times 2\exp\left\{-n\left(\frac{\epsilon^2}{4} - \frac{\epsilon^3}{6}\right)\right\}.
	\end{split}
\end{equation*}
By Lemma \ref{lm:epsilon-net} in the supplementary materials, 
\begin{equation*}
	\sup_{\gamma \in \mathcal{A}_{S_0}} \|\widetilde{V}^{(i)}P\gamma\|_2^2 \leq \left(1 - \frac{\epsilon}{2}\right)^{-1} \sup_{x \in \mathcal{N}_{\frac{\epsilon}{4}}} \|\widetilde{V}^{(i)}P\gamma\|_2^2,
\end{equation*}
For any $0 <\epsilon \leq 1/2$, if $\sup_{\gamma \in \mathcal{A}_{S_0}} \|\widetilde{V}^{(i)}P\gamma\|_2^2 \geq (1 + 2\epsilon)\theta^2\|\gamma\|^2$, then
\begin{equation*}
	\sup_{\gamma \in \mathcal{N}_{\frac{\epsilon}{4}}} \|\widetilde{V}^{(i)}P\gamma\|_2^2 \geq \left(1 - \frac{\epsilon}{2}\right)(1 + 2\epsilon)\theta^2\|\gamma\|_2^2 \geq (1 + \epsilon)\theta^2\|\gamma\|_2^2.
\end{equation*}
Thus for fixed $S_0 \subset \bbR^p, |S_0| = 2s$, 
\begin{equation}\label{eq71}
	\begin{split}
		&{\mathrm{pr}}\left\{\exists \gamma \in \bbR^p, \text{supp}(\gamma) \subset S_0, \frac{1}{n}\|\widetilde{V}^{(i)}P\gamma\|_2^2 \geq \left(1 + 2\epsilon\right)\theta^2\|\gamma\|_2^2\right\} \\ =& {\mathrm{pr}}\left\{\exists \gamma \in \mathcal{A}_{S_0}, \frac{1}{n}\|\widetilde{V}^{(i)}P\gamma\|_2^2 \geq \left(1 + 2\epsilon\right)\theta^2\|\gamma\|_2^2\right\}\\ \leq& {\mathrm{pr}}\left\{\exists \gamma \in \mathcal{N}_{\frac{\epsilon}{4}}, \frac{1}{n}\|\widetilde{V}^{(i)}P\gamma\|_2^2 \geq \left(1 + \epsilon\right)\theta^2\|\gamma\|_2^2\right\}\\
		\leq&  2\left(1 + \frac{8}{\epsilon}\right)^{2s}\exp\left\{-n\left(\frac{\epsilon^2}{4} - \frac{\epsilon^3}{6}\right)\right\}.
	\end{split}
\end{equation}
In addition, 
\begin{equation*}
	\begin{split}
		&{\mathrm{pr}}\left\{\exists 1 \leq i \leq M, \gamma \in \bbR^p \text{ s.t. } \|\gamma\|_0 \leq 2s, \frac{1}{n}\|\widetilde{V}^{(i)}P\gamma\|_2^2 \geq \left(1 + 2\epsilon\right)\theta^2\|\gamma\|_2^2\right\}\\
		=& {\mathrm{pr}}\left\{\exists 1 \leq i \leq M, S_0, \gamma \text{ s.t. } S_0 \subset \bbR^p, |S_0| = 2s, \gamma \in \mathcal{A}_{S_0}, \frac{1}{n}\|\widetilde{V}^{(i)}P\gamma\|_2^2 \geq \left(1 + 2\epsilon\right)\theta^2\|\gamma\|_2^2\right\}\\
		\leq& \sum_{i = 1}^{M}\sum_{S_0 \subset \bbR^p, |S_0| = 2s}{\mathrm{pr}}\left\{\exists \gamma \in \bbR^p, \text{supp}(\gamma) \subset S_0, \frac{1}{n}\|\widetilde{V}^{(i)}P\gamma\|_2^2 \geq \left(1 + 2\epsilon\right)\theta^2\|\gamma\|_2^2\right\}.
	\end{split}
\end{equation*}
Combining \eqref{eq71} and the previous inequality together, we have
\begin{equation*}
	\begin{split}
		&{\mathrm{pr}}\left\{\exists 1 \leq i \leq M, \gamma \in \bbR^p \text{ s.t. } \|\gamma\|_0 \leq 2s, \frac{1}{n}\|\widetilde{V}^{(i)}P\gamma\|_2^2 \geq \left(1 + 2\epsilon\right)\theta^2\|\gamma\|_2^2\right\}\\ \leq& \left\lfloor\left(\frac{p_0 - 2s_0 + 1}{4s_0}\right)^{\frac{s_0}{4}}\right\rfloor\binom{p}{2s}\times 2\left(1 + \frac{8}{\epsilon}\right)^{2s}\exp\left\{-n\left(\frac{\epsilon^2}{4} - \frac{\epsilon^3}{6}\right)\right\}\\
		\leq& p^s\times p^{2s}\times 2\left(1 + \frac{8}{\epsilon}\right)^{2s}\exp\left\{-n\left(\frac{\epsilon^2}{4} - \frac{\epsilon^3}{6}\right)\right\}\\
		=& \exp\left\{3s\log p + 2s\log\left(1 + \frac{8}{\epsilon}\right) + \log (2) -n\left(\frac{\epsilon^2}{4} - \frac{\epsilon^3}{6}\right)\right\}.
	\end{split}
\end{equation*}
Notice that $n \geq Cs\log p$ for a large constant $C > 0$. By setting $\epsilon = 1/2$ in the previous inequality, we have
\begin{equation*}
	{\mathrm{pr}}\left(\forall 1 \leq i \leq M, \gamma \in \bbR^p \text{ s.t. } \|\gamma\|_0 \leq 2s, \frac{1}{n}\|\widetilde{V}^{(i)}P\gamma\|_2^2 \leq 2\theta^2\|\gamma\|_2^2\right) > \frac{2}{3}.
\end{equation*}
Set $\theta = c\surd\{s\log(p/s)/(ne^v)\}$ with a small constant $c$, since $v > 0$ and $n \geq Cs\log p$, $2\theta^2 \leq 2c^2s\log \left(p/s\right)/(ne^v) \leq c^2$. Therefore, 
\begin{equation}\label{eq46}
	{\mathrm{pr}}\left(\forall 1 \leq i \leq M, \gamma \in \bbR^p \text{ s.t. } \|\gamma\|_0 \leq 2s, \|\widetilde{V}^{(i)}P\gamma\|_2^2 \leq nc^2\|\gamma\|_2^2\right) > \frac{2}{3}.
\end{equation}
Denote $\delta_0 = 1/40, \delta = 1/20$, by \eqref{eq76} and \eqref{eq46}, with probability $>1/3$, for all $1 \leq i \leq M, \gamma \in \bbR^p, \|\gamma\|_0 \leq 2s$,
\begin{equation*}
	\begin{split}
		&\left\|\left(V + \widetilde V^{(i)}\right)P\gamma\right\|_2^2 \leq \|\widetilde V^{(i)}P\gamma\|_2^2 + \|VP\gamma\|_2^2 + 2\|\widetilde V^{(i)}P\gamma\|_2\|VP\gamma\|_2\\ \leq& nc^2\|\gamma\|_2^2 + n(1 + \delta_0)\|\gamma\|_2^2 + 2nc\surd(1 + \delta_0)\|\gamma\|_2^2 \leq n(1 + 2\delta_0)\|\gamma\|_2^2 = n(1 + \delta)\|\gamma\|_2^2,
	\end{split}
\end{equation*}
and
\begin{equation*}
	\begin{split}
		\left\|\left(V + \widetilde V^{(i)}\right)P\gamma\right\|_2^2 \geq& \|\widetilde V^{(i)}P\gamma\|_2^2 + \|VP\gamma\|_2^2 - 2\|\widetilde V^{(i)}P\gamma\|_2\|VP\gamma\|_2\\ \geq& n(1 - \delta_0)\|\gamma\|_2^2 - 2nc\surd(1 + \delta_0)\|\gamma\|_2^2\\
		\geq& n(1 - 2\delta_0)\|\gamma\|_2^2 = n(1 - \delta)\|\gamma\|_2^2.
	\end{split}
\end{equation*}
Denote $V^{(i)} = V + \widetilde V^{(i)}$, then
\begin{equation}\label{eq77}
	{\mathrm{pr}}\left(\forall 1 \leq i \leq M, \gamma \in \bbR^p, \|\gamma\|_0 \leq 2s, n(1 - \delta)\|\gamma\|_2^2 \leq \|V^{(i)}P\gamma\|_2^2 \leq n(1 + \delta)\|\gamma\|_2^2\right) > \frac{1}{3}.
\end{equation}
i.e., with probability $> 1/3$, for all $1 \leq i \leq M, \gamma \in \bbR^p, \|\gamma\|_0 \leq 2s$, $\bar{V}^{(i)} = V^{(i)}P  (i = 1, 2, \dots, M)$ satisfy RIP condition with constant $\delta_{2s}(\bar{V}^{(i)}) = 1/20$. Also by definition, for all $1 \leq k \leq n, 1 \leq j \leq p$,
\begin{equation*}
	-\theta = \theta v - \frac{\theta}{s_0}s_0(v + 1) \leq \left(\widetilde{V}^{(i)}\right)_{kj} \leq \theta v - \frac{\theta}{s_0}s_0(v - 1) = \theta.
\end{equation*}
Thus
\begin{equation}\label{eq73}
	v - 1 - \theta \leq \left(V^{(i)}\right)_{kj} = v_{kj} + \left(\widetilde{V}^{(i)}\right)_{kj} \leq v + 1 + \theta.
\end{equation}
Since $\theta = c\surd\{s\log(p/s)/(ne^v)\} \leq c\surd\{s\log(p/s)/(Cs\log p)\} \leq 1/2$,
$v - 3/2 \leq \left(V^{(i)}\right)_{kj} \leq v + 3/2$,
which means
\begin{equation}\label{eq72}
	\left|\left(V^{(i)}\right)_{kl} - \left(V^{(i)}\right)_{k'l'}\right| \leq 3, \quad \forall 1 \leq i \leq M, 1 \leq k, k' \leq n, 1 \leq l, l' \leq p. 
\end{equation}
Denote $\nu^{(i)} \in \bbR^n, X^{(i)} \in \bbR^{n \times p}$ such that
\begin{equation*}
	\nu^{(i)}_{k} = \sum_{j = 1}^{p}\exp\left\{\left(V^{(i)}\right)_{kj}\right\}, \quad \forall 1 \leq k \leq n,
\end{equation*}
\begin{equation*}
	\left\{X^{(i)}\right\}_{kj} = \frac{\exp\left\{\left(V^{(i)}\right)_{kj}\right\}}{\sum_{l = 1}^{p}\exp\left\{\left(V^{(i)}\right)_{kl}\right\}}, \quad \forall 1 \leq k \leq n, 1 \leq j \leq p.
\end{equation*}
\eqref{eq72} implies
\begin{equation}\label{eq79}
	e^{-3}\bar{\nu}^{(i)} \leq \nu_k^{(i)} \leq e^3\bar{\nu}^{(i)}, \frac{e^{-3}}{p} \leq X_{kl}^{(i)} \leq \frac{e^3}{p}, \quad \forall 1 \leq k \leq n, 1 \leq l \leq p.
\end{equation}
Using the fact that $D_{KL}(W_1, W_2) = \lambda_1\log(\lambda_1/\lambda_2) + \lambda_2 - \lambda_1$ for $W_1 \sim $ Poisson($\lambda_1$) and $W_2 \sim $ Poisson($\lambda_2$), we have
\begin{equation}\label{eq35}
	\begin{split}
		D_{KL}\left(W^{(i)}, W^{(j)}\right) =& \sum_{k = 1}^{n}\sum_{l = 1}^{p}\left[e^{\left(V^{(i)}\right)_{kl}}\log\left\{\frac{e^{\left(V^{(i)}\right)_{kl}}}{e^{\left(V^{(j)}\right)_{kl}}}\right\} + e^{\left(V^{(j)}\right)_{kl}} - e^{\left(V^{(i)}\right)_{kl}}\right]\\ =& \sum_{k=1}^{n}\sum_{l =1}^{s_0}\left[e^{\left(V^{(i)}\right)_{kl}}\log\left\{\frac{e^{\left(V^{(i)}\right)_{kl}}}{e^{\left(V^{(j)}\right)_{kl}}}\right\} + e^{\left(V^{(j)}\right)_{kl}} - e^{\left(V^{(i)}\right)_{kl}}\right]\\ &+ \sum_{k=1}^{n}\sum_{l = p -p_0 + 1}^{p - p_0 + s_0}\left[e^{\left(V^{(i)}\right)_{kl}}\log\left\{\frac{e^{\left(V^{(i)}\right)_{kl}}}{e^{\left(V^{(j)}\right)_{kl}}}\right\} + e^{\left(V^{(j)}\right)_{kl}} - e^{\left(V^{(i)}\right)_{kl}}\right]. 
	\end{split}
\end{equation}
The second equation holds since $\exp\left(\left(V^{(i)}\right)_{kl}\right) = \exp\left(\left(V^{(j)}\right)_{kl}\right) = 1$ for all $1 \leq k \leq n, l \notin \{1, \dots, s_0\} \cup \{p - p_0 + 1, \dots, p - p_0 + s_0\}$. 
For any $l \in \{1, 2, \dots, s_0\}$, 
\begin{equation}\label{eq36}
	\begin{split}
		&e^{\left(V^{(i)}\right)_{kl}}\log\left\{\frac{e^{\left(V^{(i)}\right)_{kl}}}{e^{\left(V^{(j)}\right)_{kl}}}\right\} + e^{\left(V^{(j)}\right)_{kl}} - e^{\left(V^{(i)}\right)_{kl}}\\
		=& e^{\left(V^{(i)}\right)_{kl}}\left[\exp\left\{\left(V^{(j)}\right)_{kl} - \left(V^{(i)}\right)_{kl}\right\} - \left\{\left(V^{(j)}\right)_{kl} - \left(V^{(i)}\right)_{kl}\right\} - 1\right]\\
		=& e^{\left(V^{(i)}\right)_{kl}}\left[\exp\left\{\left(\widetilde V^{(j)}\right)_{kl} - \left(\widetilde V^{(i)}\right)_{kl}\right\} - \left\{\left(\widetilde V^{(j)}\right)_{kl} - \left(\widetilde V^{(i)}\right)_{kl}\right\} - 1\right]\\
		=&e^{\left(V^{(i)}\right)_{kl}}\left[\exp\left\{\frac{\theta}{s_0}\left(\sum_{m \in \Omega_j}v_{km}' - \sum_{m \in \Omega_i}v_{km}'\right)\right\} - \frac{\theta}{s_0}\left(\sum_{m \in \Omega_j}v_{km}' - \sum_{m \in \Omega_i}v_{km}'\right) - 1\right]\\
		=&e^{\left(V^{(i)}\right)_{kl}}\left(e^{z_{k, i:j}} - z_{k, i:j} - 1\right),
	\end{split}
\end{equation}
where 
\begin{equation*}
	v_{km}' = v_{km} - v =  \left\{\begin{array}{ll}
		1, & \text{with probability} 1/2, \\
		-1, & \text{with probability} 1/2,
	\end{array}\right.
\end{equation*}
\begin{equation*}
	z_{k, i:j} = \frac{\theta}{s_0}\left(\sum_{m \in \Omega_j\backslash\Omega_i}v_{km}' - \sum_{m \in \Omega_i\backslash\Omega_j}v_{km}'\right).
\end{equation*}
Noting that
\begin{equation*}
	|z_{k, i:j}| \leq \frac{\theta}{s_0}\left(\left|\Omega_j\backslash\Omega_i\right| + \left|\Omega_i\backslash\Omega_j\right|\right) \leq \frac{\theta}{s_0}2s_0 = 2\theta \leq 2c\surd\left\{\frac{sp\log\left(\frac{p}{s}\right)}{n\bar{\nu}}\right\} \leq 2c\surd\left\{\frac{sp\log\left(\frac{p}{s}\right)}{Cs\log(p)\cdot p}\right\}< 1,
\end{equation*}
we have 
\begin{equation}\label{eq74}
	e^{z_{k, i:j}} - z_{k, i:j} - 1 \leq z_{k, i:j}^2.
\end{equation}
Since $v_{km}'$ are i.i.d. Rademacher random variables, by Hoeffding's inequality,
\begin{equation*}
	{\mathrm{pr}}\left(\left|\sum_{m \in \Omega_j\backslash\Omega_i}v_{km}' - \sum_{m \in \Omega_i\backslash\Omega_j}v_{km}'\right|  > t\right) \leq 2\exp\left\{-\frac{2t^2}{4\left(\left|\Omega_i\backslash\Omega_j\right| + \left|\Omega_i\backslash\Omega_j\right|\right)}\right\} \leq 2\exp\left(-\frac{t^2}{4s_0}\right),
\end{equation*}
which means
\begin{equation*}
	{\mathrm{pr}}\left\{\left(\sum_{m \in \Omega_j\backslash\Omega_i}v_{km}' - \sum_{m \in \Omega_i\backslash\Omega_j}v_{km}'\right)^2  > t\right\} \leq 2\exp\left(-\frac{t}{4s_0}\right) \leq 2\exp\left(-\frac{t}{s}\right).
\end{equation*}
For $q \geq 1$,
\begin{equation*}
	\begin{split}
		E\left|\left(\frac{s_0}{\theta}z_{k, i:j}\right)^2\right|^q = &E\left\{\left(\sum_{m \in \Omega_j\backslash\Omega_i}v_{km}' - \sum_{m \in \Omega_i\backslash\Omega_j}v_{km}'\right)^2\right\}^p\\
		=& \int_{0}^{\infty}qt^{q - 1}{\mathrm{pr}}\left\{\left(\sum_{m \in \Omega_j\backslash\Omega_i}v_{km}' - \sum_{m \in \Omega_i\backslash\Omega_j}v_{km}'\right)^2 > t\right\}dt\\ \leq& \int_{0}^{\infty}qt^{q - 1}\times 2\exp\left(-\frac{t}{s}\right)dt = 2q!s^q \leq 2\left(\frac{q + 1}{2}\right)^qs^q.
	\end{split}
\end{equation*}
Thus
\begin{equation*}
	\left\|\left(\frac{s_0}{\theta}z_{k, i:j}\right)^2\right\|_{\psi_1} = \sup_{q \geq 1}\left[\frac{1}{q}\left\{E\left|\left(\frac{s_0}{\theta}z_{k, i:j}\right)^2\right|^q\right\}^{\frac{1}{q}}\right] \leq \sup_{q \geq 1}\left(2^{1/q}\frac{q + 1}{2q}s\right) \leq 2s.
\end{equation*}
Therefore
\begin{equation*}
	\|z_{k, i:j}^2 - E z_{k, i:j}^2\|_{\psi_1} \leq 2\|z_{k, i:j}^2\|_{\psi_1} \leq 2\frac{2\theta^2 s}{s_0^2} \leq 2\frac{2\theta^2 s}{\left\lfloor\frac{s}{4}\right\rfloor^2} \leq \frac{200\theta^2}{s}.
\end{equation*}
By Bernstein-type inequality, 
\begin{equation}\label{eq39}
	{\mathrm{pr}}\left\{\left|\sum_{k = 1}^{n}z_{k, i:j}^2 - E\left(\sum_{k = 1}^{n}z_{k, i:j}^2\right)\right| \geq t\right\} \leq 2\exp\left\{-c'\min\left(\frac{s^2t^2}{n\theta^4}, \frac{st}{\theta^2}\right)\right\}.
\end{equation}
Set $t = \frac{n\theta^2}{s_0}$ in \eqref{eq39}, since
\begin{equation*}
	E\left(\sum_{k = 1}^{n}z_{k, i:j}^2\right) = n\frac{\theta^2}{s_0^2}\left(\left|\Omega_i\backslash\Omega_j\right| + \left|\Omega_j\backslash\Omega_i\right|\right) \leq \frac{2n\theta^2}{s_0},
\end{equation*}
we have
\begin{equation}\label{eq75}
	{\mathrm{pr}}\left(\sum_{k = 1}^{n}z_{k, i:j}^2 > 3\frac{n\theta^2}{s_0}\right) \leq 2\exp\left(-c'n\right).
\end{equation}
In addition, by \eqref{eq73}, for all $1 \leq k \leq n, 1 \leq l \leq p$,
\begin{equation}\label{eq80}
	\begin{split}
		e^{v - \frac{3}{2}} \leq e^{v - 1 - \theta} \leq e^{\left(V^{(i)}\right)_{kl}} \leq e^{v + 1 + \theta} \leq e^{v + \frac{3}{2}}.
	\end{split}
\end{equation}
\eqref{eq36}, \eqref{eq74} and the previous inequality imply that
\begin{equation*}
	e^{\left(V^{(i)}\right)_{kl}}\log\left\{\frac{e^{\left(V^{(i)}\right)_{kl}}}{e^{\left(V^{(j)}\right)_{kl}}}\right\} + e^{\left(V^{(j)}\right)_{kl}} - e^{\left(V^{(i)}\right)_{kl}} \leq e^{v + \frac{3}{2}}z_{k, i:j}^2.
\end{equation*}
for all $1 \leq k \leq n, 1 \leq l \leq s_0$. 
Thus
\begin{equation*}
	\sum_{k=1}^{n}\sum_{l =1}^{s_0}\left[e^{\left(V^{(i)}\right)_{kl}}\log\left\{\frac{e^{\left(V^{(i)}\right)_{kl}}}{e^{\left(V^{(j)}\right)_{kl}}}\right\} + e^{\left(V^{(j)}\right)_{kl}} - e^{\left(V^{(i)}\right)_{kl}}\right] \leq s_0\cdot e^{v + \frac{3}{2}}\sum_{k = 1}^{n}z_{k, i:j}^2,
\end{equation*}
By \eqref{eq75} and the previous inequality, we have
\begin{equation}\label{eq40}
	{\mathrm{pr}}\left(\sum_{k=1}^{n}\sum_{l =1}^{s_0}\left[e^{\left(V^{(i)}\right)_{kl}}\log\left\{\frac{e^{\left(V^{(i)}\right)_{kl}}}{e^{\left(V^{(j)}\right)_{kl}}}\right\} + e^{\left(V^{(j)}\right)_{kl}} - e^{\left(V^{(i)}\right)_{kl}}\right] > 3ne^{v + \frac{3}{2}}\theta^2\right) \leq 2\exp\left(-c'n\right).
\end{equation}
Similarly, 
\begin{equation}\label{eq41}
	{\mathrm{pr}}\left( \sum_{k=1}^{n}\sum_{l = p -p_0 + 1}^{p - p_0 + s_0}\left[e^{\left(V^{(i)}\right)_{kl}}\log\left\{\frac{e^{\left(V^{(i)}\right)_{kl}}}{e^{\left(V^{(j)}\right)_{kl}}}\right\} + e^{\left(V^{(j)}\right)_{kl}} - e^{\left(V^{(i)}\right)_{kl}}\right] > 3ne^{v + \frac{3}{2}}\theta^2\right) \leq 2\exp\left(-c'n\right).
\end{equation}
By \eqref{eq35}, (\ref{eq40}) and (\ref{eq41}),
\begin{equation*}
	{\mathrm{pr}}\left\{D_{KL}\left(W^{(i)}, W^{(j)}\right) > 6ne^{v + \frac{3}{2}}\theta^2\right\} \leq 4\exp\left(-c'n\right).
\end{equation*}
Thus,
\begin{equation*}
	{\mathrm{pr}}\left\{\exists 1 \leq i \neq j \leq M, D_{KL}\left(W^{(i)}, W^{(j)}\right) > 6ne^{v + \frac{3}{2}}\theta^2\right\} \leq 2M(M - 1)\exp\left(-c'n\right).
\end{equation*}
Since $\log\left(2M\left(M - 1\right)\right) \asymp s\log(p/s)$ and $n \geq Cs\log p$ for sufficient large $C$, we know that
\begin{equation*}
	{\mathrm{pr}}\left\{\exists 1 \leq i \neq j \leq M, D_{KL}\left(W^{(i)}, W^{(j)}\right) > 6ne^{v + \frac{3}{2}}\theta^2\right\} < \frac{1}{3},
\end{equation*}
which means
\begin{equation}\label{eq78}
	{\mathrm{pr}}\left\{\forall 1 \leq i \neq j \leq M, D_{KL}\left(W^{(i)}, W^{(j)}\right) \leq 6ne^{v + \frac{3}{2}}\theta^2\right\} > \frac{2}{3}.
\end{equation}
By \eqref{eq77}, \eqref{eq79} and \eqref{eq78}, there exist fixed $\nu^{(i)} \in \bbR^n, X^{(i)} \in \bbR^{n \times p}$ ($1 \leq i \leq M$) satisfying $\bar{\nu}/e^3 \leq \nu_k^{(i)} \leq e^3\bar{\nu}$ and $e^{-3}p \leq X_{kl} \leq e^3p$,  $\bar{V}^{(i)} = V^{(i)}P \in \bbR^{n \times p}$($1 \leq i \leq M$) satisfying RIP condition with constant $\delta_{2s}(\bar{V}^{(i)}) = 1/20$, and $\forall 1 \leq i \neq j \leq M$, 
\begin{equation}\label{eq402}
	D_{KL}\left(W^{(i)}, W^{(j)}\right) \leq  6ne^{v + \frac{3}{2}}\theta^2 = 6ne^{v + \frac{3}{2}}\cdot c^2\frac{s\log\left(\frac{p}{s}\right)}{ne^v} = 6e^{\frac{3}{2}}c^2s\log\left(\frac{p}{s}\right) \leq cs\log\left(\frac{p}{s}\right).
\end{equation} 
Since $V^{(1)}\beta^{(1)} = \cdots = V^{(M)}\beta^{(n)}$, $y^{(i)}$ and $y^{(j)}$ have the same distribution, and $D_{KL}(y^{(i)}, y^{(j)}) = 0.$
Combine \eqref{eq: KL}, \eqref{eq402} and the previous inequality,
\begin{equation*}
	D_{KL}\left((y^{i}, W^{(i)}), (y^{(i)}, W^{(j)})\right) \leq cs\log\left(\frac{p}{s}\right).
\end{equation*}
Set $v = \log (Q/p)$, by \eqref{eq80} and $\nu_{kl}^{(i)} = e^{(V^{(i)})_{kl}}$, $e^{-3/2}Q \leq \bar{\nu} \leq e^{3/2}Q$. By (\ref{eq37}) and \eqref{eq80}, 
\begin{equation*}
	\min_{i \neq j}\|\beta^{(i)} - \beta^{(j)}\|_2 > \frac{1}{2}R\theta \geq c\left\{\frac{s\log\left(\frac{p}{s}\right)}{ne^v}\right\}^{1/2}R = c\left\{\frac{sp\log\left(\frac{p}{s}\right)}{nQ}\right\}^{1/2}R. 
\end{equation*}
By the precious two inequalities and generalized Fano's lemma, we have
\begin{equation*}
	\inf_{\hat{\beta}} \sup_{(\nu, X, \beta) \in \mathcal{F}_{p, n, s}(R, Q)} E \left(\|\hat{\beta} - \beta\|_2\right) \geq c\left\{\frac{sp\log\left(\frac{p}{s}\right)}{nQ}\right\}^{1/2}R\left\{1 - \frac{cs\log\left(\frac{p}{s}\right) + \log 2}{\log M}\right\}.
\end{equation*}
Note that $\log M \asymp s\log\left(\frac{p}{s}\right)$, there exists a small constant $c > 0$ such that
\begin{equation*}
	\inf_{\hat{\beta}} \sup_{(\nu, X, \beta) \in \mathcal{F}_{p, n, s}(R, Q)} E \left(\|\hat{\beta} - \beta\|_2^2\right) \geq \left\{\inf_{\hat{\beta}} \sup_{(\nu, X, \beta) \in \mathcal{F}_{p, n, s}(R, Q)} E\left(\|\hat{\beta} - \beta\|_2\right)\right\}^2 \geq c\frac{sp\log\left(\frac{p}{s}\right)}{nQ}R^2.
\end{equation*}
In summary, we have proved our assertion. \qed

\subsection{Proof of Theorem \ref{th:upper_bound_Dirichlet multinomial}}\label{sec:proof-thm3}
$W_{ij}$ can be seen as $W_{ij}|q_i \sim \text{Poisson}(\nu_iq_{ij})$ and $q_{ij} \sim \text{Beta}(\alpha_iX_{ij}, \alpha_i - \alpha_iX_{ij})$. We introduce the truncated version of $W_{ij}$ as
$$W'_{ij}|W_{ij}, q_i = \left\{\begin{array}{l}
W_{ij}, \quad q_{ij} \in \left[\frac{X_{ij}}{10}, 10X_{ij}\right] \text{and } W \in [\frac{\nu_{i}q_{ij}}{10}, 10\nu_{i} q_{ij}],\\
\nu_{ij}, \quad \text{otherwise}.
\end{array}\right.$$
Under the setting of Theorem \ref{th:upper_bound_Dirichlet multinomial}, we introduce the following lemmas.
\begin{lemma}\label{lm:beta-binomial_1}
	$\left|E\{\log(W_{ij}' + z_i)\} - \log\left(\nu_{ij}\right)\right| \leq 10\zeta_i/\nu_{ij}$ holds if $\nu_{ij} \geq C(\delta)$.
\end{lemma}
\begin{lemma}\label{lm:beta-binomial_2}
	$\left|E\{\log^2(W_{ij}' + z_i)\} - \log^2\left(\nu_{ij}\right)\right| \leq C\log(\nu_{ij})\zeta_{i}/\nu_{ij}$ holds if $\nu_{ij} \geq C(\delta)$.
\end{lemma}
\begin{lemma}[Sub-Gaussianity]\label{lm:sub-Gaussian_diri-mul}
	$\left\|\surd{(\nu_{ij}/\zeta_{\max})}[\log\left(W'_{ij} + z_i\right) - E\{\log(W'_{ij} + z_i)\}]\right\|_{\psi_2} \leq K_0$ for a constant $K_0$.
\end{lemma}
\begin{lemma}[Infinity norm bound]\label{lm:infinity_norm_bound_Dirichlet Multinomial}
	Under the setting of Theorem \ref{th:upper_bound_Dirichlet multinomial}, there exists constants $C, C' > 0$, such that
	$${\mathrm{pr}}\left(\left\|\bar{A}_W \beta^* - \bar{B}_W^\top y\right\|_\infty \leq C\left[\surd{(n\log p)}\left\{\sigma + \left(\frac{p}{\bar\nu}\zeta_{\max}\right)^{\frac{1}{2}}\|\beta^*\|_1\right\} + n\frac{\log(\bar\nu/p)}{\bar\nu/p}\zeta_{\max}\|\beta^*\|_1\right]\right) \geq 1 - 6p^{-C'}.$$
\end{lemma} 
Applying these lemmas, we can prove Theorem \ref{th:upper_bound_Dirichlet multinomial} by essentially the same method as the proof of Theorem \ref{th:upper_bound} with $\lambda = C\left(\{\log(p)/n\}^{1/2}[\sigma + \{(p/\bar\nu)\zeta_{\max}\}^{\frac{1}{2}}\|\beta^*\|_1] + \log(\bar\nu/p)(\bar\nu/p)^{-1}\zeta_{\max}\|\beta^*\|_1\right)$.\qed

\subsection{Proof of Theorem \ref{th:upper_bound3}} 
In the following lemmas, suppose $W_{ij}' = W_{ij}1_{\left\{\nu_{ij}/10 \leq W_{ij} \leq 10\nu_{ij}\right\}} + \nu_{ij}1_{\left\{W_{ij} \notin \left[\nu_{ij}/10, 10\nu_{ij}\right]\right\}}$. 
\begin{lemma}[Bias of $\log(W_0'+1/2)$]\label{lm:general_1}
	For any $\varepsilon > 0$, there exists a constant $C_{\epsilon} > 0$ that only relies on $\epsilon$ such that for all $\nu_{ij} \geq C_{\epsilon}$, we have $|E\log(W_{ij}' + 1/2) - \log\nu_{ij}| \leq \nu_{ij}^{-3/2 + \epsilon} + F/(2\nu_{ij})$ for some constant $C$. 
\end{lemma}

\begin{lemma}[Bias of $\log^2(W_0'+1/2)$]\label{lm:general_2}
	There exists a constant $C>0$ such that if $\nu_{ij} \geq C$, then $|E\{\log^2(W_{ij}' + 1/2)\} - \log^2\nu_{ij}| \leq F\log(\nu_{ij})/\nu_{ij} + C/\nu_{ij}$ for some constant $C$. 
\end{lemma}

\begin{lemma}[Sub-Gaussianity]\label{lm:sub-Gaussian_general}
	There exist positive constants $K_0$ and $C$, such that for $\nu_{ij} \geq C$, we have 
	$$\left\|\nu_{ij}^{1/2}\left[\log\left(W_{ij}' + 1/2\right) - E\left\{\log\left(W_{ij}' + 1/2\right)\right\}\right]\right\|_{\psi_2} \leq K_0.$$
\end{lemma}

\begin{lemma}\label{lm:infinity_norm_bound3}
	Given the setting in Theorem \ref{th:upper_bound3}, there exist two constants $C$ and $C'$, such that with probability at least $1 - 3p^{-C'}$,
	$$\left\|\bar{\A}_W \beta^* - \bar{\B}_W^\top y\right\|_\infty \leq C\left(n\log p\left(\sigma^2 + \frac{p}{\bar\nu}\|\beta^*\|_2^2\right) + n^2s\left(\left(\frac{p}{\bar{\nu}}\right)^{3 - 2\epsilon} + \min\{F^2, C\}\left(\frac{p}{\bar{\nu}}\right)^{2}\right)\|\beta^*\|_2^2\right)^{1/2}.$$
\end{lemma}

By these lemmas, we can prove Theorem \ref{th:upper_bound3} essentially the same as the proof of Theorem \ref{th:upper_bound}. \qed

\appendixthree

\section{Technical Lemmas}

We collect the technical lemmas used in the theoretical analysis in this section.

\begin{lemma}[Poisson tail bound (\cite{boucheron2013concentration}, Pages 22 - 23)]\label{lm:poisson tail bound}
	For any $x \geq 0$,
	\begin{equation*}
		{\mathrm{pr}}\left(W \geq v + x\right) \leq \exp\left(-\frac{x^2}{2v}\psi_{Benn}\left(x/v\right)\right).
	\end{equation*}
	For any $0 \leq x \leq v$, 
	\begin{equation*}
		{\mathrm{pr}}\left(W \leq v - x\right) \leq \exp\left(-\frac{x^2}{2v}\psi_{Benn}\left(-x/v\right)\right) \leq \exp\left(-\frac{x^2}{2v}\right),
	\end{equation*}
	where $\phi_{Benn}$ is the Bennett function: $\psi_{Benn}(t) = \frac{(1 + t)\log(1 + t) - t}{t^2/2}$ for $t > -1, t \neq 0$, and $\psi_{Benn}(0) = 1$.
\end{lemma}

\begin{lemma}[Truncated Poisson tail bound]\label{lm:trancated poisson tail bound}
	There exists constant $c > 0$ such that for any $x \geq 0$, 
	\begin{equation*}
		{\mathrm{pr}}\left(W' \geq v + x\right) \leq \exp\left(-c\frac{x^2}{v}\right),
	\end{equation*}
	for any $0 \leq x \leq \lambda$, 
	\begin{equation*}
		{\mathrm{pr}}\left(W' \leq v - x\right) \leq \exp\left(-c\frac{x^2}{v}\right).
	\end{equation*}
\end{lemma}

{\noindent Proof of Lemma \ref{lm:trancated poisson tail bound}.} 
By definition of $W'$ and Lemma \ref{lm:poisson tail bound}, for any $0 \leq x \leq 9v$,
\begin{equation*}
	{\mathrm{pr}}\left(W' \geq v + x\right) \leq {\mathrm{pr}}\left(W \geq v + x\right) \leq \exp\left(-\frac{x^2}{2v}\psi_{Benn}\left(x/v\right)\right).
\end{equation*}
Noting that $\psi_{Benn}$ is a positive continuous function, denote $c > 0$ is a constant such that $\min_{0 \leq t \leq 9}\psi_{Benn}(t) = 2c$, then 
\begin{equation*}
	{\mathrm{pr}}\left(W' \geq v + x\right) \leq \exp\left(-c\frac{x^2}{v}\right).
\end{equation*}
For any $x > 9v$, we have
\begin{equation*}
	{\mathrm{pr}}\left(W' \geq v + x\right) = 0 < \exp\left(-c\frac{x^2}{v}\right).
\end{equation*}
Therefore, for any $x > 0$, 
\begin{equation*}
	{\mathrm{pr}}\left(W' \leq v - x\right) \leq {\mathrm{pr}}\left(W \leq v - x\right) \leq \exp\left(-\frac{x^2}{2v}\right).
\end{equation*}
For $x = 0$, 
\begin{equation*}
	{\mathrm{pr}}\left(W' \leq v - x\right) \leq 1 = \exp\left(-\frac{x^2}{2v}\right).
\end{equation*}
\qed\medskip

Consider the regression model \eqref{eq:dir-multi} and \eqref{eq:log-contrast-EIV}. Denote $V, \bar{V} \in \mathbb{R}^{n\times p}$ where $V_{ij}=\log(\nu_iX_{ij})$ and $\bar{V}=V\{I_p-(1/p)1_p1_p^\top\}$. The following Lemma \ref{lm:RIP} shows that the RIP condition (Condition \ref{con:RIP}) holds if the deterministic matrix $\bar{V}$ satisfies RIP condition with constant $\delta_{2s}(\bar{V})<1/20$ in the case without overdispersion. 
\begin{lemma}[RIP condition]\label{lm:RIP}
	Suppose $n\geq Cs\log(p)$, $\bar{\nu} \geq p$, and $a\bar{\nu} \leq \nu_i \leq b\bar{\nu}$, $a/p \leq X_{ij} \leq b/p$ for constants $0<a < 1 < b$. If for some large constant $C>0$, some $\epsilon>0$, we have $\bar{\nu} \geq Cp(s + \log(np))$, and $\bar{V}$ satisfies RIP condition with constant $\delta_{2s}(\bar{V})$, then with probability $1 - 4p^{-C'}$, $\bar{\B}_W$ satisfies RIP condition with constants $\delta_{2s}(\bar{\B}_W) = 2\delta_{2s}(\bar{V})$, i.e.,
	\begin{equation}\label{eq14}
		n\left(1 - 2\delta_{2s}\left(\bar{V}\right)\right)\|x\|_2^2 \leq x^\top \bar{\A}_W x = \|\bar{\B}_W x\|_2^2 \leq n\left(1 + 2\delta_{2s}\left(\bar{V}\right)\right)\|x\|_2^2
	\end{equation}
	holds for all $2s$-sparse vector $x \in \bbR^p$, with probability $1 - 4p^{-C'}$.
\end{lemma}

{\noindent Proof of Lemma \ref{lm:RIP}.}
Define $\|X\|_{\psi_1} = \sup_{p \geq 1} p^{-1}(E |X|^p)^{1/p}, \|X\|_{\psi_2} = \sup_{p \geq 1} p^{-1/2}(E |X|^p)^{1/p}$.
For fixed $x$ such that $\|x\|_2 = 1$, supp$(x) \subseteq S_0 \subset \{1, \dots, p\}$ and $\left|S_0\right| = 2s$, denote $\widetilde{x} = \P x$, $\bar{x} = \frac{1}{p}\sum_{i = 1}^{p}x_i$. Then
\begin{equation}\label{eq27}
	\sum_{i = 1}^{p}\widetilde{x}_i^2 = \sum_{i = 1}^{p}\left(x_i - \bar{x}\right)^2 = \sum_{i=1}^{p}x_i^2 - p\bar{x}^2 \leq \sum_{i = 1}^{p}x_i^2 = 1.
\end{equation}

First consider $x^{\top}\left(\bar{\A}_{W'}-  E \bar{\A}_{W'}\right)x$. Note that $\A_{W'}= \sum_{k=1}^{n} \A_{W',k}$, where
$$\A_{W', k} \in \mathbb{R}^{p\times p},\quad (A_{W', k})_{ij} = \left\{\begin{array}{ll}
\phi_1(W'_{ki})\phi_1(W'_{kj}), & i \neq j\\
\phi_2(W'_{ki}), & i = j,
\end{array}\right.$$
and $\phi_1(\cdot) = \log\left(\cdot + \frac{1}{2}\right), \phi_2(\cdot) = \log^2\left(\cdot + \frac{1}{2}\right)$,
we have
\begin{equation*}
	x^\top\left(\bar{\A}_{W'}-  E \bar{\A}_{W'}\right)x = \sum_{k=1}^{n} x^{\top}\left(\bar{\A}_{W', k}-  E \bar{\A}_{W', k}\right)x,
\end{equation*}
where $\bar\A_{W',k} = \P\A_{W',k}\P$. If we can bound $\|x^{\top}\left(\bar{\A}_{W', k}-  E \bar{\A}_{W', k}\right)x\|_{\psi_{1}}$, then Bernstein-type inequality could give us an upper bound for ${\mathrm{pr}}\left(\left|x^\top \left(\bar{\A}_{W'} -  E \bar{\A}_{W'}\right) x\right| \geq t\right)$.
Actually,
\begin{equation}\label{eq99}
	\begin{split}
		&{\mathrm{pr}} (|x^{\top} (\bar{\A}_{W',k} - E  \bar{\A}_{W',k}) x| > t)
		= {\mathrm{pr}} (|\widetilde{x}^{\top} (\A_{W',k} - E  \A_{W',k}) \widetilde{x}| > t)\\   
		= &{\mathrm{pr}} (|\sum_{i = 1}^p \sum_{j = 1}^p ((A_{W',k})_{ij} - E  (A_{W',k})_{ij}) \widetilde{x}_i \widetilde{x}_j| > t).
	\end{split}      
\end{equation}
Denote $D_{ki} = \phi_1\left(W'_{ki}\right) -  E \phi_1\left(W'_{ki}\right)$, by Lemma \ref{lm:sub-Gaussian}, there exist two constants $C$ and $K_0 > 0$, if $\nu_{\min} \geq C$, then 
\begin{equation*}
	\|\sqrt{\nu_{ki}}D_{ki}\|_{\psi_2} \leq K_0.
\end{equation*}
Thus,
\begin{equation}\label{eq95}
	\|D_{ki}\|_{\psi_{2}} \leq \frac{1}{\sqrt{\nu_{ki}}}K_0 \leq \frac{1}{\sqrt{\nu_{\min}}}K_0.
\end{equation}
In addition, 
\begin{equation}
	\begin{split}
		&\sum_{i = 1}^p \sum_{j = 1}^p \left(\left(A_{W',k}\right)_{ij} - E  \left(A_{W',k}\right)_{ij}\right) \widetilde{x}_i \widetilde{x}_j\\
		=&\sum_{1 \leq i, j \leq p, i \neq j} \left(\phi_{1}\left(W'_{ki}\right) \phi_{1}\left(W'_{kj}\right) -  E  \phi_{1}\left(W'_{ki}\right) E \phi_{1}\left(W'_{kj}\right) \right) \widetilde{x}_i \widetilde{x}_j+ \sum_{i = 1}^p\left(\phi_{2}\left(W'_{ki}\right) - E \phi_{2}\left(W'_{ki}\right)\right) \widetilde{x}_i^2\\
		=&\sum_{i = 1}^p \sum_{j = 1}^p \left(\phi_{1}\left(W'_{ki}\right) \phi_{1}\left(W'_{kj}\right) -  E  \phi_{1}\left(W'_{ki}\right) E \phi_{1}\left(W'_{kj}\right) \right) \widetilde{x}_i \widetilde{x}_j- \sum_{i = 1}^p\left(E \phi_{1}^2\left(W'_{ki}\right) - \left(E \phi_{1}\left(W'_{ki}\right)\right)^2\right) \widetilde{x}_i^2\\
		=&\sum_{i = 1}^p \sum_{j = 1}^p \left(\left(D_{ki} + E  \phi_{1}\left(W'_{ki}\right)\right)\left(D_{kj} + E  \phi_{1}\left(W'_{kj}\right)\right) -  E  \phi_{1}\left(W'_{ki}\right) E \phi_{1}\left(W'_{kj}\right)\right) \widetilde{x}_i \widetilde{x}_j - \sum_{i = 1}^pE D_{ki}^2 \widetilde{x}_i^2\\
		=&\sum_{i = 1}^p \sum_{j = 1}^p D_{ki}D_{kj}\widetilde{x}_i \widetilde{x}_j -  \sum_{i = 1}^pE D_{ki}^2 \widetilde{x}_i^2 + 2\sum_{i = 1}^p D_{ki} \sum_{j = 1}^p E \phi_{1}\left(W'_{kj}\right)\widetilde{x}_i \widetilde{x}_j.
	\end{split}
\end{equation}
Therefore, 
\begin{equation}\label{eq97}
	\begin{split}
		&{\mathrm{pr}} \left(\left|\sum_{i = 1}^p \sum_{j = 1}^p \left(\left(A_{W',k}\right)_{ij} - E  \left(A_{W',k}\right)_{ij}\right) \widetilde{x}_i \widetilde{x}_j\right| > t\right)\\
		\leq&{\mathrm{pr}}\left(\left|\sum_{i = 1}^p \sum_{j = 1}^p D_{ki}D_{kj}\widetilde{x}_i \widetilde{x}_j -  \sum_{i = 1}^pE D_{ki}^2 \widetilde{x}_i^2\right| > \frac{t}{2}\right)\\
		&+ {\mathrm{pr}}\left(\left|2\sum_{i = 1}^p D_{ki} \sum_{j = 1}^pE \phi_{1}\left(W'_{kj}\right)\widetilde{x}_i \widetilde{x}_j\right| > \frac{t}{2}\right).
	\end{split}
\end{equation}
Denote
\begin{displaymath}
	\A \in \mathbb{R}^{p \times p},  
	A_{ij} = \widetilde{x}_i \widetilde{x}_j,
	D_k = \left(D_{k1}, \dots, D_{kp}\right)^\top,
\end{displaymath}
then
\begin{displaymath}
	\sum_{i = 1}^p \sum_{j = 1}^p D_{ki}D_{kj}\widetilde{x}_i \widetilde{x}_j = D_{k}^{\top}\A D_{k},
\end{displaymath}
and
\begin{equation*}
	\begin{split}
		E D_{k}^{\top}\A D_{k} =& \sum_{1 \leq i, j \leq p, i \neq j}\widetilde{x}_i \widetilde{x}_jE D_{ki}E D_{kj} + \sum_{i = 1}^pE D_{ki}^2 \widetilde{x}_i^2\\
		=& 0 + \sum_{i = 1}^p E D_{ki}^2 \widetilde{x}_i^2 = \sum_{i = 1}^p E D_{ki}^2 \widetilde{x}_i^2.
	\end{split}
\end{equation*}
Denote $K = \frac{1}{\sqrt{C}}K_0$, \eqref{eq95} tells us 
\begin{displaymath}
	\max_{i \in S} \|D_{ki}\|_{\psi_{2}} \leq K, \quad \forall \nu_{\min} \geq C.
\end{displaymath}
By Hanson-Wright inequality, for every $t \geq 0$, we know that there exists a constant $c > 0$,
\begin{equation*}
	\begin{split}
		{\mathrm{pr}}\left(\left|\sum_{i = 1}^p \sum_{j = 1}^p D_{ki}D_{kj}\widetilde{x}_i \widetilde{x}_j -  \sum_{i = 1}^pE D_{ki}^2 \widetilde{x}_i^2\right| > \frac{t}{2}\right) \leq 2\exp\left[-c\cdot\min\left(\frac{t^2}{4 K^4 \|\A\|_{HS}^{2}}, \frac{t}{2\|\A\|}\right)\right],
	\end{split}
\end{equation*}
where 
\begin{equation*}
	\begin{split}
		\|\A\|^2_{HS} &= \sum_{i = 1}^p \sum_{j = 1}^p\left|A_{ij}\right|^2 = \sum_{i = 1}^p \sum_{j = 1}^p \widetilde{x}_i^{2} \widetilde{x}_j^{2}, \quad \|\A\| = \max_{\|u\|_2 \leq 1} \|\A u\|_2.
	\end{split}	
\end{equation*}
By \eqref{eq27},
\begin{equation*}
	\|\A\|^2_{HS} = \left(\sum_{i = 1}^p\widetilde{x}_i^2\right)^2 \leq 1,
\end{equation*}
and
\begin{equation}\label{eq96}
	\begin{split}
		&\|\A\| = \max_{\substack{\|u\|_2 = 1\\\|v\|_2 = 1}} \left|u^T \A v\right| = \max_{\substack{\|u\|_2 = 1\\\|v\|_2 = 1}} \left|\sum_{i = 1}^p \sum_{j = 1}^p u_i A_{ij} v_j\right|\leq \max_{\substack{\|u\|_2 = 1\\\|v\|_2 = 1}}\sum_{i = 1}^p \sum_{j = 1}^p\left|u_i \widetilde{x}_i \widetilde{x}_j v_j\right|\\
		\leq& \max_{\substack{\|u\|_2 = 1\\\|v\|_2 = 1}}\left(\sum_{i = 1}^p \left|u_i \widetilde{x}_i\right|\right)\left(\sum_{j = 1}^p \left|v_j \widetilde{x}_j\right|\right) \leq \max_{\substack{\|u\|_2 = 1\\\|v\|_2 = 1}}\left(\sum_{i = 1}^p \widetilde{x}_i^2\sum_{i = 1}^p u_i^2\right)^{\frac{1}{2}}\left(\sum_{i = 1}^p \widetilde{x}_i^2\sum_{i = 1}^p v_i^2\right)^{\frac{1}{2}} \leq 1.
	\end{split}
\end{equation}
The third inequality of \eqref{eq96} comes from Cauchy-Schwarz inequality.\\
Thus,
\begin{equation}\label{eq28}
	\begin{split}
		{\mathrm{pr}}\left(\left|\sum_{i = 1}^p \sum_{j = 1}^p D_{ki}D_{kj}\widetilde{x}_i \widetilde{x}_j -  \sum_{i = 1}^pE D_{ki}^2 \widetilde{x}_i^2\right| > \frac{t}{2}\right) \leq 2\exp\left[-c\min\left(\frac{t^2}{4 K^4}, \frac{t}{2}\right)\right].
	\end{split}
\end{equation}
Now we consider ${\mathrm{pr}}\left(\left|2\sum_{i = 1}^p D_{ki} \sum_{j = 1}^pE \phi_{1}(W'_{kj})\widetilde{x}_i \widetilde{x}_j\right| > \frac{t}{2}\right)$.
Since
\begin{equation*}
	\frac{a}{b}\nu_{ki'} \leq \nu_{ki} \leq \frac{b}{a}\nu_{ki'}, \quad \forall 1 \leq i, i' \leq p,
\end{equation*}
we have
\begin{equation*}
	\log\nu_{ki} \leq \log\left(\frac{b}{a}\right) + \log\nu_{ki'}, \quad \forall 1 \leq i, i' \leq p.
\end{equation*}
By Lemma \ref{lm:poisson_1}, there exists a constant $C > 0$, for $\nu_{\min} \geq C$, 
\begin{equation*}
	|E \phi_{1}(W'_{ki}) - \log\nu_{ki}| \leq \frac{4}{\nu_{ki}}, \quad \forall 1 \leq i, i' \leq p, 1 \leq k \leq n.
\end{equation*}
Therefore, if $\nu_{\min} \geq C$, 
\begin{equation*}
	\left| E \phi_1\left(W'_{ki}\right) -  E \phi_1\left(W'_{ki'}\right)\right| \leq \log\left(\frac{b}{a}\right) + \frac{8}{\nu_{\min}}, \quad \forall 1 \leq i, i' \leq p.
\end{equation*}
Moreover,
\begin{equation*}
	\begin{split}
		&\sum_{i = 1}^p\left(\sum_{j = 1}^pE \phi_{1}\left(W'_{kj}\right)\widetilde{x}_i\widetilde{x}_j\right)^2
		= \sum_{i = 1}^p\widetilde{x}_i^2\left(\sum_{j = 1}^pE \phi_{1}\left(W'_{kj}\right)\widetilde{x}_j\right)^2\\
		=& \sum_{i = 1}^p\widetilde{x}_i^2\left(\sum_{j = 1}^pE \phi_{1}(W'_{kj})\left(x_j - \bar{x}\right)\right)^2
		= \sum_{i = 1}^p\widetilde{x}_i^2\left(\sum_{j = 1}^p\left(E \phi_{1}(W'_{kj}) - \overline{E \phi_{1}(W'_{k\cdot})}\right)x_j\right)^2\\
		=&\sum_{i = 1}^p\widetilde{x}_i^2\left(\sum_{j \in S_0}\left(E \phi_{1}(W'_{kj}) - \overline{E \phi_{1}(W'_{k\cdot})}\right)x_j\right)^2
		\leq \sum_{i = 1}^p\widetilde{x}_i^2\sum_{j \in S_0}\left(E \phi_{1}(W'_{kj}) - \overline{E \phi_{1}(W'_{k\cdot})}\right)^2\sum_{j \in S_0}x_j^2\\
		=& \sum_{i = 1}^p\widetilde{x}_i^2\sum_{j \in S_0}\left(E \phi_{1}(W'_{kj}) - \overline{E \phi_{1}(W'_{k\cdot})}\right)^2,
	\end{split}
\end{equation*}
where $\overline{E \phi_{1}(W'_{k\cdot})} = \frac{1}{p}\sum_{i = 1}^{p} E \phi_{1}\left(W'_{kj}\right)$. The inequality comes from Cauchy-Schwarz Inequality.
Since 
\begin{equation*}
	\left(E \phi_{1}(W'_{kj}) - \overline{E \phi_{1}(W'_{k\cdot})}\right)^2 \leq \max_{i', i}\left| E \phi_1\left(W'_{ki}\right) -  E \phi_1\left(W'_{ki'}\right)\right|^2 \leq \left(\log\left(\frac{b}{a}\right) + \frac{8}{\nu_{\min}}\right)^2, 
\end{equation*}
for $\nu_{\min} \geq C$,
\begin{equation}\label{eq98}
	\begin{split}
		&\sum_{i = 1}^p\left(\sum_{j = 1}^pE \phi_{1}\left(W'_{kj}\right)\widetilde{x}_i\widetilde{x}_j\right)^2
		\leq \sum_{i = 1}^{p}\widetilde{x}_i^2\cdot 2s\left(\log\left(\frac{b}{a}\right) + \frac{8}{\nu_{\min}}\right)^2
		\leq 2s\left(\log\left(\frac{b}{a}\right) + \frac{8}{C}\right)^2.
	\end{split}
\end{equation}
Denote $K_1 = \frac{1}{\sqrt{\nu_{\min}}}K_0$, due to \eqref{eq95}, \eqref{eq98} and Hoeffding-type inequality, 
\begin{equation*}
	\begin{split}
		&{\mathrm{pr}}\left(\left|2\sum_{i = 1}^p D_{ki} \sum_{j = 1}^pE \phi_{1}(W'_{kj})\widetilde{x}_i \widetilde{x}_j\right| > \frac{t}{2}\right)\\ \leq& e\cdot \exp\left(-\frac{ct^2}{\left(\|\max_{i}D_{ki}\|_{\psi_{2}}\right)^2\sum_{i = 1}^p\left(\sum_{j = 1}^pE \phi_{1}\left(W'_{kj}\right)\widetilde{x}_i\widetilde{x}_j\right)^2}\right)\\ \leq& e\cdot\exp\left(-\frac{ct^2}{sK_1^2}\right).
	\end{split}
\end{equation*}
\eqref{eq99}, \eqref{eq97}, \eqref{eq28} and the previous inequality together imply that
\begin{equation*}
	{\mathrm{pr}}\left(\left|x^\top(\bar\A_{W',k} -  E  \bar\A_{W',k})x\right| > t\right) \leq 2\exp\left[-c\min\left(\frac{t^2}{4 K^4}, \frac{t}{2}\right)\right] + e\cdot\exp\left(-\frac{ct^2}{sK_1^2}\right).
\end{equation*}

For $q \geq 1$, we have
\begin{equation*}
	\begin{split}
		& E  \left|x^\top(\bar\A_{W',k} -  E  \bar\A_{W',k})x\right|^q\\
		=&\int_{0}^{\infty}{\mathrm{pr}}\left(\left|x^\top(\bar\A_{W',k} -  E  \bar\A_{W',k})x\right| > t\right)q t^{q-1}dt\\
		\leq& \int_{0}^{\infty}\left(2\exp\left[-c\min\left(\frac{t^2}{4 K^4}, \frac{t}{2}\right)\right] + e\cdot\exp\left(-\frac{ct^2}{sK_1^2}\right)\right)qt^{q - 1}dt\\
		\leq& 2\int_{0}^{\infty}\exp\left(-\frac{ct^2}{4K^4}\right)qt^{q-1}dt +e\int_{0}^{\infty}\exp\left(-\frac{ct^2}{sK_1^2}\right)qt^{q-1}dt\\
		&+ 2\int_{0}^{\infty}\exp\left(-\frac{ct}{2}\right)qt^{q-1}dt\\
		=& q\left(\frac{4K^4}{c}\right)^{\frac{q}{2}}\Gamma\left(\frac{q}{2}\right) + \frac{eq}{2}\left(\frac{sK_1^2}{c}\right)^{\frac{q}{2}}\Gamma\left(\frac{q}{2}\right) + 2\left(\frac{2}{c}\right)^{q}q!\\
		\leq& q\left(\frac{4K^4}{c}\right)^{\frac{q}{2}}\left(\frac{q}{2}\right)^{\frac{q}{2}} + \frac{eq}{2}\left(\frac{sK_1^2}{c}\right)^{\frac{q}{2}}\left(\frac{q}{2}\right)^{\frac{q}{2}} + 2\left(\frac{2}{c}\right)^{q}\left(\frac{q+1}{2}\right)^q.
	\end{split}
\end{equation*}
Note that
\begin{equation*}
	x^q + y^q + z^q \leq (x + y + z)^q, \quad \forall q \geq 1, x, y, z \geq 0,
\end{equation*}
for any $q \geq 1$, 
\begin{equation*}
	\begin{split}
		&\frac{1}{q}\left( E  \left|x^\top(\bar\A_{W',k} -  E  \bar\A_{W',k})x\right|^q\right)^{\frac{1}{q}}\\
		\leq& \frac{1}{q}\left(q^{\frac{1}{q}}\left(\frac{4K^4}{c}\right)^{\frac{1}{2}}\left(\frac{q}{2}\right)^{\frac{1}{2}} + \left(\frac{eq}{2}\right)^{\frac{1}{q}}\left(\frac{sK_1^2}{c}\right)^{\frac{1}{2}}\left(\frac{q}{2}\right)^{\frac{1}{2}} + 2^{\frac{1}{q}}\frac{q+1}{c}\right)\\
		\leq& \frac{4K^2}{\sqrt{2c}} + eK_1\sqrt{\frac{s}{2c}} + \frac{4}{c}.
	\end{split}	
\end{equation*}
The first inequality holds since $q^{\frac{1}{q}}q^{\frac{1}{2}} \leq 2q$ and $2^{\frac{1}{q}}\frac{q + 1}{q} \leq 2 \times 2 = 4$ for $q \geq 1$.\\
Thus for any $1 \leq k \leq n$,
\begin{equation}\label{eq100}
	\|x^\top(\bar{\A}_{W',k} -  E  \bar{\A}_{W',k})x\|_{\psi_1} \leq \frac{4K^2}{\sqrt{2c}} + eK_1\sqrt{\frac{s}{2c}} + \frac{4}{c}.
\end{equation}
Note that $\nu_{\min} \gtrsim \bar{\nu}/p \geq Cs$, $K_1\sqrt{s} \leq C\frac{K_0}{\sqrt{s}}\sqrt{s} = CK_0$ can be bounded by a constant. Therefore
for any $1 \leq k \leq n$, 
\begin{equation*}
	\|x^\top(\bar{\A}_{W',k} -  E  \bar{\A}_{W',k})x\|_{\psi_1} \leq C.
\end{equation*}
can be bounded by a constant.
By Bernstein-type inequality,
\begin{equation*}
	\begin{split}
		{\mathrm{pr}}\left(\left|x^\top(\bar{\A}_{W'} -  E  \bar{\A}_{W'})x\right| \geq t\right) =& {\mathrm{pr}}\left(\left|\sum_{k = 1}^{n}x^\top(\bar{\A}_{W',k} -  E  \bar{\A}_{W',k})x\right| \geq t\right)\\ \leq& 2\exp\left[-c\min\left\{\frac{t^2}{n}, t\right\}\right].
	\end{split}
\end{equation*}
Thus there exists a constant $C > 0$ such that for $1 \leq t \lesssim n$,
\begin{equation*}
	{\mathrm{pr}}\left(\left|x^\top(\bar{\A}_{W'} -  E  \bar{\A}_{W'})x\right| \geq C\sqrt{nt}\right) \leq 2e^{-t},
\end{equation*}
which means there exists a constant $c > 0$,
\begin{equation}\label{eq4}
	{\mathrm{pr}}\left(\left|x^\top(\bar{\A}_{W'} -  E  \bar{\A}_{W'})x\right| \geq \frac{1}{4}\delta_{2s}\left(\bar{V}\right)n\right) \leq 2e^{-cn}.
\end{equation}
For any $S_0 \subset \{1, \dots, p\}, |S_0| = 2s$, by Lemma \ref{lm:epsilon-net} with $\widetilde{\A} = \bar{\A}_{W'} -  E \bar{\A}_{W'}$ and $\epsilon = \frac{1}{4}$, we have
\begin{equation}\label{eq12}
	\sup_{x \in \mathcal{A}_{S_0}} \left|x^{\top}(\bar{\A}_{W'} -  E \bar{\A}_{W'})x\right| \leq 2\sup_{x \in \mathcal{N}_{\frac{1}{4}}} \left|x^{\top}(\bar{\A}_{W'} -  E \bar{\A}_{W'})x\right|,
\end{equation}
where $\mathcal{A}_{S_0} = \{x: x \in \bbR^p, \|x\|_2 = 1, \text{supp}(x) \subset S_0\}$ and $\mathcal{N}_{\frac{1}{4}}$ is an $\epsilon$-net of $\mathcal{A}_{S_0}$. By Lemma 5.2 in \cite{vershynin2010introduction}, we can choose $\mathcal{N}_{\frac{1}{4}}$ such that card$\left(\mathcal{N}_{\frac{1}{4}}\right) \leq 9^{2s}$. By \eqref{eq4},
\begin{equation}\label{eq13}
	\begin{split}
		&{\mathrm{pr}}\left(\sup_{x \in \mathcal{N}_{\frac{1}{4}}} \left|x^{\top}(\bar{\A}_{W'} -  E \bar{\A}_{W'})x\right| \geq \frac{1}{4}\delta_{2s}\left(\bar{V}\right)n\right)\\
		=&{\mathrm{pr}}\left(\exists x \in \mathcal{N}_{\frac{1}{4}}, \left|x^{\top}(\bar{\A}_{W'} -  E \bar{\A}_{W'})x\right| \geq \frac{1}{4}\delta_{2s}\left(\bar{V}\right)n\right)\\
		\leq& 9^{2s}\times 2e^{-cn}.
	\end{split}
\end{equation}
Combine (\ref{eq12}) and (\ref{eq13}) together, we have 
\begin{equation*}
	{\mathrm{pr}}\left(\sup_{x \in \mathcal{A}_{S_0}} \left|x^{\top}(\bar{\A}_{W'} -  E \bar{\A}_{W'})x\right| \geq \frac{1}{2}\delta_{2s}\left(\bar{V}\right)n\right) \leq 9^{2s}\times 2e^{-cn}.
\end{equation*}
Therefore,
\begin{equation}\label{eq52}
	\begin{split}
		&{\mathrm{pr}}\left(\exists x \in \bbR^p, \|x\|_0 \leq s, \|x\|_2 = 1, \left|x^{\top}(\bar{\A}_{W'} -  E \bar{\A}_{W'})x\right|\geq \frac{1}{2}\delta_{2s}\left(\bar{V}\right)n\right)\\
		\leq& {\mathrm{pr}}\left(\exists S_0 \subset \{1, 2, \dots, p\}, \left|S_0\right| = 2s, \sup_{x \in \mathcal{A}_{S_0}} \left|x^{\top}(\bar{\A}_{W'} -  E \bar{\A}_{W'})x\right|\geq \frac{1}{2}\delta_{2s}\left(\bar{V}\right)n\right)\\
		\leq& \binom{p}{2s} \times 9^{2s} \times 2e^{-cn} \leq e^{2s\log p} \times 9^{2s} \times 2e^{-cn}.
	\end{split}
\end{equation}
Next, we need to bound $x^\top (E  \bar{\A}_{W'} - \bar{V}^\top\bar{V}) x$. \\
By Lemma \ref{lm:poisson_1} and Lemma \ref{lm:poisson_2}, $\exists C_{\epsilon'} > 0$(only depends on $\epsilon'$), if $\nu_{\min} \geq C_{\epsilon'} \geq C$, then
\begin{equation*}
	\begin{split}
		\left(\log(\nu_{ki}) - \frac{4}{\nu_{ki}^{\frac{3}{2}-\epsilon'}}\right)\left(\log(\nu_{kj}) - \frac{4}{\nu_{ki}^{\frac{3}{2}-\epsilon'}}\right) &\leq  E \phi_{1}(W'_{ki})\phi_{1}(W'_{kj})\\ &\leq \left(\log(\nu_{ki}) + \frac{4}{\nu_{ki}^{\frac{3}{2}-\epsilon'}}\right)\left(\log(\nu_{kj}) + \frac{4}{\nu_{ki}^{\frac{3}{2}-\epsilon'}}\right),
	\end{split}
\end{equation*}
and
\begin{equation*}
	\log^2(\nu_{ki}) - \frac{4}{\nu_{ki}} \leq  E \phi_{2}\left(W'_{ki}\right) \leq \log^2(\nu_{ki}) + \frac{4}{\nu_{ki}}.
\end{equation*}
Thus 
\begin{equation*}
	\left| E \phi_{1}(W'_{ki})\phi_{1}(W'_{kj}) - \log(\nu_{ki})\log(\nu_{kj})\right| \leq 8\frac{\log\nu_{\max}}{\nu_{\min}^{\frac{3}{2}-\epsilon'}} + \frac{16}{\nu_{\min}^{3 - 2\epsilon'}}.
\end{equation*}
Since $\nu_{\max} \asymp \nu_{\min}$,  for $\nu_{\min} \geq C$,
\begin{equation*}
	\left| E \phi_{1}(W'_{ki})\phi_{1}(W'_{kj}) - \log(\nu_{ki})\log(\nu_{kj})\right| \leq C'\frac{\log\nu_{\min}}{\nu_{\min}^{\frac{3}{2}-\epsilon'}} + \frac{16}{\nu_{\min}^{3 - 2\epsilon'}}.
\end{equation*}
\begin{equation*}
	\left| E \phi_{2}(W'_{ki}) - \log^2\nu_{ki}\right| \leq \frac{4}{\nu_{\min}}.
\end{equation*}
Hence, 
\begin{equation}\label{eq514}
	\begin{split}
		\max_{i,j}\left|\left(E \A_{W'} - V^\top V\right)_{ij}\right| =& \max_{i,j}\left|\sum_{k = 1}^{n}\left( E \phi_{1}(W'_{ki})\phi_{1}(W'_{kj}) - \log\nu_{ki}\log\nu_{kj}\right)\right|\\ \leq& n\cdot\max\left\{C'\frac{\log\nu_{\min}}{\nu_{\min}^{\frac{3}{2}-\epsilon'}} + \frac{16}{\nu_{\min}^{3 - 2\epsilon'}}, \frac{4}{\nu_{\min}}\right\}.
	\end{split}
\end{equation}
Also, for any $p \times p$ matrix $\A$, one can show that
\begin{equation}\label{eq515}
	\max_{i,j}\left|\left(\P^\top \A\P\right)_{ij}\right| \leq 4\max_{i,j}\left|A_{ij}\right|.
\end{equation}
In fact, 
\begin{equation*}
	\begin{split}
		\left|\left(\P^\top \A\P\right)_{ij}\right| =& \left|A_{ij} - \frac{1}{p}\sum_{i' =1}^pA_{i'j} - \frac{1}{p}\sum_{j'=1}^{p}A_{ij'} + \frac{1}{p^2}\sum_{i'=1}^{p}\sum_{j'=1}^{p}A_{i'j'}\right|\\
		\leq& \left|A_{ij}\right| + \left|\frac{1}{p}\sum_{i' =1}^pA_{i'j}\right| + \left|\frac{1}{p}\sum_{j'=1}^{p}A_{ij'}\right| +\left|\frac{1}{p^2}\sum_{i'=1}^{p}\sum_{j'=1}^{p}A_{i'j'}\right|\\
		\leq& 4\max_{i,j}\left|A_{ij}\right|.
	\end{split}	
\end{equation*}
Combine \eqref{eq514} and \eqref{eq515} together, we have
\begin{equation}\label{ineq52}
	\begin{split}
		\left|x^\top (E  \bar{\A}_{W'} - \bar{V}^\top\bar{V}) x\right|
		=&\left|\sum_{i \in S_0}\sum_{j \in S_0}x_i\left(\P\left( E  \A_{W'} - V^\top V\right)\P\right)_{ij}x_j\right|\\
		\leq& \sum_{i \in S_0}\sum_{j \in S_0}\left|x_i\right|\left|\left(\P\left( E  \A_{W'} - V^\top\ V\right)\P\right)_{ij}\right|\left|x_j\right|\\
		\leq& 4\max_{i,j}\left|\left(E \A_{W'} - V^\top V\right)_{ij}\right|\sum_{i \in S_0}\sum_{j \in S_0}\left|x_i\right|\left|x_j\right|\\
		\leq& 4\max_{i,j}\left|\left(E \A_{W'} - V^\top V\right)_{ij}\right|\left(\sum_{i \in S_0}|x_i|\right)^2\\
		\leq& 4\max_{i,j}\left|\left(E \A_{W'} - V^\top V\right)_{ij}\right|\cdot 2s\sum_{i \in S_0}|x_i|^2\\
		=&8s\cdot n\cdot\max\left\{C'\frac{\log\nu_{\min}}{\nu_{\min}^{\frac{3}{2}-\epsilon'}} + \frac{16}{\nu_{\min}^{3 - 2\epsilon'}}, \frac{4}{\nu_{\min}}\right\}.
	\end{split}
\end{equation}
Set $\epsilon' = \frac{1}{5}$, and note that $\lim_{t \to \infty}\frac{\log t}{t^{\frac{1}{4}}} = 0$, we know that there exists two constants $C$ and $C'$ such that for any $\nu_{\min} \geq C$,
\begin{equation*}
	\left|x^\top (E  \bar{\A}_{W'} - \bar{V}^\top\bar{V}) x\right| \leq \frac{C'sn}{\nu_{\min}}.
\end{equation*}
Since $\nu_{\min} \gtrsim \bar\nu/p \geq Cs$, for any $2s$-sparse $x \in \bbR^p$ such that $\|x\|_2 = 1$, we have
\begin{equation}\label{eq5}
	\left|x^\top (E  \bar{\A}_{W'} - \bar{V}^\top\bar{V}) x\right| \leq \frac{1}{2}n\cdot\delta\left(\mathbf\Gamma\right).
\end{equation}
Combine \eqref{eq52} and \eqref{eq5} together, also note that $\bar{V}^\top\bar{V}$ satisfies RIP condition with constant $\delta_{2s}(\bar{V})$, we know that
\begin{equation}\label{eq505}
	n\left(1 - 2\delta_{2s}\left(\bar{V}\right)\right)\|x\|_2^2 \leq x^\top \bar{\A}_{W'} x \leq n\left(1 + 2\delta_{2s}\left(\bar{V}\right)\right)\|x\|_2^2
\end{equation}
holds for all $2s$-sparse $x$ with probability at least $1 - 2\times9^{2s} e^{2s\log p-cn}$. \\
By \eqref{eq503} and \eqref{eq504},
\begin{equation}\label{eq506}
	{\mathrm{pr}}\left(W_{ij} \neq W_{ij}'\right) = {\mathrm{pr}}\left(W_{ij} \leq \frac{1}{10}\nu_{ij}\right) + {\mathrm{pr}}\left(W_{ij} \geq 10\nu_{ij}\right) \leq 2e^{-c\nu_{ij}}.
\end{equation}
\eqref{eq505} and \eqref{eq506} tell us
\begin{equation*}
	\begin{split}
		&{\mathrm{pr}}\left(\forall 2s\text{-sparse } x, n\left(1 - 2\delta_{2s}\left(\bar{V}\right)\right)\|x\|_2^2 \leq x^\top \bar{\A}_{W} x \leq n\left(1 + 2\delta_{2s}\left(\bar{V}\right)\right)\|x\|_2^2\right)\\
		\geq& {\mathrm{pr}}\left(\forall 2s\text{-sparse } x, n\left(1 - 2\delta_{2s}\left(\bar{V}\right)\right)\|x\|_2^2 \leq x^\top \bar{\A}_{W'} x \leq n\left(1 + 2\delta_{2s}\left(\bar{V}\right)\right)\|x\|_2^2\right)\\
		&- {\mathrm{pr}}\left(\exists 1 \leq i \leq n, 1 \leq j \leq p, W_{ij} \neq W_{ij}'\right)\\
		\geq& 1 - 2\times 9^{2s} e^{2s\log p-cn} - \sum_{i = 1}^{n}\sum_{j = 1}^{p}{\mathrm{pr}}\left(W_{ij} \neq W_{ij}'\right)\\
		\geq& 1 - 2\times 9^{2s} e^{2s\log p-cn} - 2npe^{-c\nu_{\min}}.	
	\end{split}
\end{equation*}
Since $\bar\nu \geq Cp\log(np)$, we know that $\nu_{\min} \geq C\log(np)$ with a large constant $C$. Also notice that $n \geq Cs\log p$, there exists a large constant $C' > 0$,
\begin{equation*}
	{\mathrm{pr}}\left(\forall s\text{-sparse } x, n\left(1 - 2\delta\left(\Gamma\right)\right)\|x\|_2^2 \leq x^\top \bar{\A}_{W} x \leq n\left(1 + 2\delta\left(\Gamma\right)\right)\|x\|_2^2\right) \geq 1 - 4p^{-C'}.
\end{equation*} 
\qed
\medskip
\begin{lemma}[Truncated Beta tail bound]\label{lm:trancated Beta tail bound}
	Suppose $q \sim \text{Beta}(\alpha, \beta)$, define 
	\begin{equation*}
		q' = \left\{\begin{array}{ll}
			q, & \frac{\alpha}{10\left(\alpha + \beta\right)} \leq q \leq \frac{10\alpha}{\alpha + \beta},\\
			\frac{\alpha}{\alpha + \beta}, & \text{otherwise}.
		\end{array}
		\right.
	\end{equation*}
	If $\frac{\beta}{\alpha} \asymp p$, then $\exists c > 0$, $\forall t \geq 0$,
	\begin{equation*}
		{\mathrm{pr}}\left(q' \geq \frac{\alpha}{\alpha + \beta} + t\right) \vee {\mathrm{pr}}\left(q' \leq \frac{\alpha}{\alpha + \beta} - t\right) \leq 2e^{-cp(\alpha + \beta)t^2}.
	\end{equation*}
\end{lemma}
{\noindent Proof of Lemma \ref{lm:trancated Beta tail bound}.}
Since $\beta > \alpha$, by Theorem 9 in \cite{zhang2018non}, for any $0 < t < \frac{\beta}{\alpha + \beta}$,
\begin{equation*}
	{\mathrm{pr}}\left(q \geq \frac{\alpha}{\alpha + \beta} + t\right) \leq 2\exp\left(-c\min\left\{\frac{\beta^2t^2}{\alpha}, \beta t\right\}\right).
\end{equation*}
If $\frac{\beta}{\alpha} \asymp p$ and $0 < t \leq 9\frac{\alpha}{\alpha + \beta}$, 
\begin{equation*}
	p(\alpha + \beta)t^2 \asymp \frac{\beta^2t^2}{\alpha} \gtrsim \beta t.
\end{equation*}
Thus for any $t \in [0, 9\frac{\alpha}{\alpha + \beta}]$, $\exists c > 0$, 
\begin{equation*}
	{\mathrm{pr}}\left(q \geq \frac{\alpha}{\alpha + \beta} + t\right) \leq 2e^{-cp(\alpha + \beta)t^2}.	
\end{equation*}
By the definition of $q'$, $\forall t \geq 0$, 
\begin{equation*}
	{\mathrm{pr}}\left(q' \geq \frac{\alpha}{\alpha + \beta} + t\right) \leq 2e^{-cp(\alpha + \beta)t^2}.	
\end{equation*}
Moreover, apply Theorem 9 in \cite{zhang2018non},  for any $0 < t < \frac{\alpha}{\alpha + \beta}$, 
\begin{equation*}
	{\mathrm{pr}}\left(q \leq \frac{\alpha}{\alpha + \beta} - t\right) \leq 2e^{-c\frac{\beta^2t^2}{\alpha}}.
\end{equation*}
If $\frac{\beta}{\alpha} \asymp p$, then $\frac{\beta^2}{\alpha} \asymp p(\alpha + \beta)$, thus $\exists c> 0$ such that for all $t \in [0, \frac{9\alpha}{10\left(\alpha + \beta\right)}]$,
\begin{equation*}
	{\mathrm{pr}}\left(q \leq \frac{\alpha}{\alpha + \beta} - t\right) \leq 2e^{-cp\left(\alpha + \beta\right)t^2}.
\end{equation*}
Thus for any $t \geq 0$, 
\begin{equation*}
	{\mathrm{pr}}\left(q' \leq \frac{\alpha}{\alpha + \beta} - t\right) \leq 2e^{-cp\left(\alpha + \beta\right)t^2}.
\end{equation*}
\qed
\medskip

\begin{lemma}[Truncated Beta-Poisson tail bound]\label{lm:tail bound}
	$W|q \sim \text{Poisson}(\mu q)$, and $q \sim \text{Beta}(\alpha, \beta)$, define
	\begin{equation*}
		W'|W, q = \left\{\begin{array}{ll}
			W, & q \in \left[\frac{\alpha}{10\left(\alpha + \beta\right)}, \frac{10\alpha}{\alpha + \beta}\right] \text{and } W \in [\frac{\mu q}{10}, 10\mu q],\\
			\mu\frac{\alpha}{\alpha + \beta}, & \text{otherwise}.
		\end{array}\right.
	\end{equation*} If $\frac{\beta}{\alpha} \asymp p$, then we have
	\begin{equation*}
		\begin{split}
			&{\mathrm{pr}}\left(W' \geq \mu\frac{\alpha}{\alpha + \beta} + t\right) \vee {\mathrm{pr}}\left(W' \leq \mu\frac{\alpha}{\alpha + \beta} - t\right) \leq 3e^{-ct^2\min\left\{\frac{\alpha + \beta}{\mu\alpha}, \frac{(\alpha + \beta)^2}{\mu^2\alpha}\right\}}.
		\end{split}
	\end{equation*} 
\end{lemma}
{\noindent Proof of Lemma \ref{lm:tail bound}.}
\begin{equation*}
	\begin{split}
		&{\mathrm{pr}}\left(W' \geq \mu(1 + \delta)\frac{\alpha}{\alpha + \beta}\right) =  E \left({\mathrm{pr}}\left(W' \geq \mu(1 + \delta)\frac{\alpha}{\alpha + \beta}\bigg|q\right)\right)\\
		=& \int_{0}^{1}{\mathrm{pr}}\left(W' \geq \mu(1 + \delta)\frac{\alpha}{\alpha + \beta}\bigg|q\right)\frac{1}{\text{B}(\alpha, \beta)}q^{a - 1}(1 - q)^{b - 1}dq\\
		=& \int_0^{\min\{(1 + \frac{\delta}{2})\frac{\alpha}{\alpha + \beta}, 1\}}{\mathrm{pr}}\left(W' \geq \mu(1 + \delta)\frac{\alpha}{\alpha + \beta}\bigg|q\right)\frac{1}{\text{B}(\alpha, \beta)}q^{a - 1}(1 - q)^{b - 1}dq\\
		& + \int_{\min\{(1 + \frac{\delta}{2})\frac{\alpha}{\alpha + \beta}, 1\}}^1{\mathrm{pr}}\left(W' \geq \mu(1 + \delta)\frac{\alpha}{\alpha + \beta}\bigg|q\right)\frac{1}{\text{B}(\alpha, \beta)}q^{a - 1}(1 - q)^{b - 1}dq.
	\end{split}
\end{equation*} 
By the definition of $W'$, for $q \geq 10\frac{\alpha}{\alpha + \beta}$, we know that
\begin{equation*}
	{\mathrm{pr}}\left(W' \geq \mu(1 + \delta)\frac{\alpha}{\alpha + \beta}\bigg|q\right) = 0.
\end{equation*}
Therefore,
\begin{equation}\label{eq349}
	\begin{split}
		{\mathrm{pr}}\left(W' \geq \mu(1 + \delta)\frac{\alpha}{\alpha + \beta}\right) \leq  \sup_{q \in (0, (1 + \frac{\delta}{2})\frac{\alpha}{\alpha + \beta}]}{\mathrm{pr}}\left(W' \geq \mu(1 + \delta)\frac{\alpha}{\alpha + \beta}\bigg|q\right) + {\mathrm{pr}}\left(q' \geq \left(1 + \frac{\delta}{2}\right)\frac{\alpha}{\alpha + \beta}\right).
	\end{split}
\end{equation}
By Lemma \ref{lm:trancated poisson tail bound}, for any $t \geq 0$,
\begin{equation*}
	\begin{split}
		{\mathrm{pr}}\left(W' \geq \mu(q + t)\right|q) \leq \exp\left(-c\frac{(\mu t)^2}{\mu q}\right).
	\end{split}
\end{equation*}
Hence, for all $q \in (0, (1 + \frac{\delta}{2})\frac{\alpha}{\alpha + \beta}]$, 
\begin{equation*}
	\begin{split}
		{\mathrm{pr}}\left(W' \geq \mu(1 + \delta)\frac{\alpha}{\alpha + \beta}\bigg|q\right) \leq& \exp\left(-c\frac{\left(\mu \left((1 + \delta)\frac{\alpha}{\alpha + \beta} - q\right)\right)^2}{\mu q}\right)\\
		\leq& \exp\left(-c\mu\frac{\delta^2}{1 + \frac{\delta}{2}}\frac{\alpha}{\alpha + \beta}\right).
	\end{split}
\end{equation*}
If $\delta \leq 99, \exists c > 0$,
\begin{equation*}
	{\mathrm{pr}}\left(W' \geq \mu(1 + \delta)\frac{\alpha}{\alpha + \beta}\bigg|q\right) \leq \exp\left(-c\mu\delta^2\frac{\alpha}{\alpha + \beta}\right).
\end{equation*}
By the definition of $W'$, for $\delta \geq 99$,
\begin{equation*}
	{\mathrm{pr}}\left(W' \geq \mu\left(1 + \delta\right)\frac{\alpha}{\alpha + \beta}\bigg|q\right) = 0.
\end{equation*}
Thus for all $\delta > 0$ and $q \in (0, (1 + \frac{\delta}{2})\frac{\alpha}{\alpha + \beta}]$.
\begin{equation}\label{eq231}
	{\mathrm{pr}}\left(W' \geq \mu(1 + \delta)\frac{\alpha}{\alpha + \beta}\bigg|q\right) \leq \exp\left(-c\mu\delta^2\frac{\alpha}{\alpha + \beta}\right).
\end{equation}
In addition, by Lemma \ref{lm:trancated Beta tail bound},
\begin{equation*}
	{\mathrm{pr}}\left(q' \geq \left(1 + \frac{\delta}{2}\right)\frac{\alpha}{\alpha + \beta}\right) \leq 2e^{-c\delta^2\alpha}.
\end{equation*}
Combine \eqref{eq349}, \eqref{eq231} and the previous inequality together,
\begin{equation*}
	\begin{split}
		&{\mathrm{pr}}\left(W' \geq \mu(1 + \delta)\frac{\alpha}{\alpha + \beta}\right) \leq 3e^{-c\alpha\delta^2\min\left\{1, \frac{\mu}{\alpha + \beta}\right\}}.
	\end{split}
\end{equation*}
Similarly, we have
\begin{equation*}
	\begin{split}
		&{\mathrm{pr}}\left(W' \leq \mu(1 - \delta)\frac{\alpha}{\alpha + \beta}\right) \leq 3e^{-c\alpha\delta^2\min\left\{1, \frac{\mu}{\alpha + \beta}\right\}}.
	\end{split}
\end{equation*}
Thus
\begin{equation*}
	\begin{split}
		&{\mathrm{pr}}\left(W' \geq \mu\frac{\alpha}{\alpha + \beta} + t\right) \leq 3e^{-ct^2\min\left\{\frac{\alpha + \beta}{\mu\alpha}, \frac{(\alpha + \beta)^2}{\mu^2\alpha}\right\}},\\
		&{\mathrm{pr}}\left(W' \leq \mu\frac{\alpha}{\alpha + \beta} - t\right) \leq 3e^{-ct^2\min\left\{\frac{\alpha + \beta}{\mu\alpha}, \frac{(\alpha + \beta)^2}{\mu^2\alpha}\right\}}.
	\end{split}
\end{equation*} 
\qed
\medskip

{\noindent Proof of Lemma \ref{lm:poisson_1}.} In the following proof, $C_i\left(\epsilon\right)$ and $C_{\epsilon}$ only depend on $\epsilon$, and for fixed $\epsilon$, they are large constant.\\ According to the Lagrange form of Taylor's Theorem, $\forall x > -1$, 
\begin{equation*}
	\log(x + 1) = x - \frac{x^2}{2} + \frac{x^3}{3(1 + \xi)^3}
\end{equation*}
for some number $\xi$ between 0 and $x$. Take $x = (t - v + \frac{1}{2})/v$, we know that
\begin{equation*}
	\log\left(t + \frac{1}{2}\right) = \log v + \frac{t - v + \frac{1}{2}}{v} - \frac{\left(t - v + \frac{1}{2}\right)^2}{2v^2} + \frac{\left(t - v + \frac{1}{2}\right)^3}{3(1+\xi)^3v^3}
\end{equation*}
for some number $\xi$ between 0 and $ (t - v + \frac{1}{2})/v$.\\
Denote $f(t) = \log(t + \frac{1}{2}) - \log v - \frac{t - v + \frac{1}{2}}{v} + \frac{(t - v + \frac{1}{2})^2}{2v^2}$, $\exists C_1(\epsilon_1) > 0$, if $v \geq C_1\left(\epsilon_1\right)$, for any $v - v^{1/2 + \epsilon_1} < t < v + v^{1/2 + \epsilon_1}$, 
\begin{equation*}
	\left|f(t)\right| \leq v^{-\frac{3}{2} + 3\epsilon_1}.
\end{equation*}
Therefore for all $v \geq C_1\left(\epsilon_1\right)$, 
\begin{equation}\label{eq18}
	E \left|f(W_0)1_{\{v - v^{1/2 + \epsilon_1} < W_0 < v + v^{1/2 + \epsilon_1}\}}\right| \leq v^{-\frac{3}{2} + 3\epsilon_1}.
\end{equation} 
If $t \leq v - v^{1/2 + \epsilon_1}, v \geq 1$, then 
\begin{equation*}
	\begin{split}
		\left|f(t)\right| \leq& \max\left\{\log\left(t + \frac{1}{2}\right), \log v\right\} + \left|\frac{t - v + \frac{1}{2}}{v}\right| + \left|\frac{\left(t - v + \frac{1}{2}\right)^2}{2v^2}\right|
		\leq \log v + \frac{3}{2}.
	\end{split}
\end{equation*}
Set $x = v^{1/2 + \epsilon_1}$ in Lemma \ref{lm:poisson tail bound}, we have
\begin{equation*}
	{\mathrm{pr}}\left(W_0 \leq v - v^{1/2 + \epsilon_1}\right) \leq \exp\left(- \frac{v^{2\epsilon_1}}{2}\right).
\end{equation*}
Therefore, $\exists C_2\left(\epsilon_1\right) > 0$, for $v \geq C_2\left(\epsilon_1\right)$,
\begin{equation}\label{eq19}
	\begin{split}
		E \left(\left|f(W_0)1_{\{W_0 \leq v - v^{1/2 + \epsilon_1}\}}\right|\right) \leq& \left(\log v + \frac{3}{2}\right){\mathrm{pr}}\left(W_0 \leq v - v^{1/2 + \epsilon_1}\right)\\
		\leq& \left(\log v + \frac{3}{2}\right)\exp\left(- \frac{v^{2\epsilon_1}}{2}\right)
		\leq v^{-\frac{3}{2}}. 
	\end{split}
\end{equation} 

$\forall t > v + v^{1/2 + \epsilon_1}, v \geq 1$, 
\begin{equation}\label{eq87}
	\begin{split}
		&\left|\log\left(t + \frac{1}{2}\right) - \log v - \frac{t - v + \frac{1}{2}}{v} + \frac{\left(t - v + \frac{1}{2}\right)^2}{2v^2}\right|\\ 
		\leq& \left|\log\left(t + \frac{1}{2}\right) - \log v\right| + \left|\frac{\left(t - v + \frac{1}{2}\right)^2}{2v^2} - \frac{t - v + \frac{1}{2}}{v}\right| \\
		=& \left|\log\left(t + \frac{1}{2}\right) - \log v\right| + \frac{t - v + \frac{1}{2}}{v}\left|\frac{t + \frac{1}{2}}{2v} - \frac{3}{2}\right| \\
		\leq& \log\left(t + \frac{1}{2}\right) + \frac{3\left(t + \frac{1}{2}\right)^2}{2v^2}
		\leq \log\left(t + \frac{1}{2}\right) + 3\frac{t^2}{v^2}.
	\end{split}
\end{equation}
Set $x = v^{\frac{1}{2}+ \epsilon_1}$ in Lemma \ref{lm:poisson tail bound}, since $\lim_{y \to 0}\psi_{Benn}(y) = 1$, for $v \geq C_3\left(\epsilon_1\right)$, we have 
\begin{equation}\label{eq86}
	{\mathrm{pr}}\left(W_0 \geq v + v^{1/2 + \epsilon_1}\right) \leq \exp\left(-\frac{\left(v^{1/2 + \epsilon_1}\right)^2}{2v}\psi_{Benn}\left(\frac{v^{1/2 + \epsilon_1}}{v}\right)\right) \leq \exp\left(-\frac{v^{2\epsilon_1}}{4}\right).
\end{equation}
By the previous inequality and \eqref{eq87}, for $v \geq C_3\left(\epsilon_1\right)$, 
\begin{equation}\label{eq20}
	\begin{split}
		&\left|E f(W_0)1_{\{W_0 \geq v + v^{1/2 + \epsilon_1}\}}\right|\\
		\leq& E \left(\log\left(W_0 + \frac{1}{2}\right)1_{\{W_0 \geq v + v^{1/2 + \epsilon_1}\}}\right) + 3E \left(\frac{W^2}{v^2}1_{\{W_0 \geq v + v^{1/2 + \epsilon_1}\}}\right)\\
		\leq& E \left(W_01_{\{W_0 \geq v + v^{1/2 + \epsilon_1}\}}\right) + \frac{3}{v^2}E \left(W_0^21_{\{W_0 \geq v + v^{1/2 + \epsilon_1}\}}\right)\\
		\leq& \left[ E  W_0^2\right]^{\frac{1}{2}}\left[{\mathrm{pr}}\left(W_0 \geq v + v^{1/2 + \epsilon_1}\right)\right]^{\frac{1}{2}}
		+ \frac{3}{v^2}\left[ E  W_0^4\right]^{\frac{1}{2}}\left[{\mathrm{pr}}\left(W_0 \geq v + v^{1/2 + \epsilon_1}\right)\right]^{\frac{1}{2}}\\
		\leq& \left(2v^2\right)^{\frac{1}{2}}\exp\left(-\frac{v^{2\epsilon_1}}{8}\right) + \frac{3}{v^2}\left(2v^4\right)^{\frac{1}{2}}\exp\left(-\frac{v^{2\epsilon_1}}{8}\right)
		\leq v^{-\frac{3}{2}}.
	\end{split}
\end{equation}
The third inequality comes from Cauchy-Schwarz Inequality.\\
Set $\epsilon_1 = \frac{\epsilon}{3}$, combine (\ref{eq18}), (\ref{eq19}) and (\ref{eq20}) together, for $\nu \geq C_{\epsilon} = \max\{C_1\left(\epsilon_1\right), C_2\left(\epsilon_1\right), C_3\left(\epsilon_1\right)\}$, 
\begin{equation}\label{ineq40}
	\begin{split}
		\left|E f(W_0)\right| =&E \left|f(W_0)1_{\{v - v^{1/2 + \epsilon_1} < W_0 < v + v^{1/2 + \epsilon_1}\}}\right|\\ &+  \left|E f(W_0)1_{\{ W_0 \leq v - v^{1/2 + \epsilon_1}\}}\right| + \left|E f( W_0)1_{ W_0 \geq v + v^{1/2 + \epsilon_1}}\right|\\
		\leq& v^{-\frac{3}{2}+\epsilon} + v^{-\frac{3}{2}} + v^{-\frac{3}{2}}\\
		\leq& 3v^{-\frac{3}{2}+\epsilon}.
	\end{split}
\end{equation}
Note that  $E f( W_0) = E \log( W_0+\frac{1}{2}) - \log v + \frac{1}{8v^2}$, for $v \geq C_{\epsilon}$, 
\begin{equation}\label{eq88}
	|E \log ( W_0 + \frac{1}{2}) - \log v| \leq 3v^{-\frac{3}{2}+\epsilon} + \frac{1}{8v^2}.
\end{equation}
Moreover, by Lemma \ref{lm:poisson tail bound},
\begin{equation}\label{eq503}
	{\mathrm{pr}}\left( W_0 \leq \frac{1}{10}v\right) = {\mathrm{pr}}\left( W_0 \leq v - \frac{9}{10}v\right) \leq \exp\left(-\frac{\left(\frac{9}{10}v\right)^2}{v}\right) = \exp\left(-\frac{81}{100}v\right),
\end{equation}
\begin{equation}\label{eq504}
	{\mathrm{pr}}\left( W_0 \geq 10v\right) \leq \exp\left(-\frac{(9v)^2}{v}\psi_{Benn}\left(\frac{9v}{v}\right)\right) = \exp\left(-9\psi_{Benn}(9)v\right).
\end{equation}
By the previous two inequalities, there exist constants $C, c > 0$, for $v \geq C_{\epsilon} > C$,
\begin{equation}\label{eq90}
	\begin{split}
		&\left| E \log\left( W_0' + \frac{1}{2}\right) -  E \log\left( W_0 + \frac{1}{2}\right)\right|\\
		\leq&\left|\log\left(v + \frac{1}{2}\right){\mathrm{pr}}\left( W_0 \leq \frac{1}{10}v\right) -  E \log\left( W_0 + \frac{1}{2}\right)1_{\{ W_0 \leq \frac{1}{10}v\}}\right|\\
		&+ \left|\log\left(v + \frac{1}{2}\right){\mathrm{pr}}\left( W_0 \geq 10v\right) -  E \log\left( W_0 + \frac{1}{2}\right)1_{\{ W_0 \geq 10v\}}\right|\\
		\leq& \max\left\{\log\left(v + \frac{1}{2}\right){\mathrm{pr}}\left( W_0 \leq \frac{1}{10}v\right),  E \left|\log\left( W_0 + \frac{1}{2}\right)\right|1_{\{ W_0 \leq \frac{1}{10}v\}}\right\}\\
		&+ \max\left\{\log\left(v + \frac{1}{2}\right){\mathrm{pr}}\left( W_0 \geq 10v\right),  E \log\left|\left( W_0 + \frac{1}{2}\right)\right|1_{\{ W_0 \geq 10v\}}\right\}\\
		\leq& \max\left\{\log\left(v + \frac{1}{2}\right)e^{-cv}, \left( E \log^2\left( W_0 + \frac{1}{2}\right)\right)^{\frac{1}{2}}\left({\mathrm{pr}}\left( W_0 \leq \frac{1}{10}v\right)\right)^{\frac{1}{2}}\right\}\\
		&+ \max\left\{\log\left(v + \frac{1}{2}\right)e^{-cv}, \left( E \log^2\left( W_0 + \frac{1}{2}\right)\right)^{\frac{1}{2}}\left({\mathrm{pr}}\left( W_0 \geq 10v\right)\right)^{\frac{1}{2}}\right\}\\
		\leq& 2\max\left\{\log\left(v + \frac{1}{2}\right)e^{-cv}, \left( E   W_0\right)^{\frac{1}{2}}e^{-cv}\right\}
		\leq 2\sqrt{v}e^{-cv}.
	\end{split}
\end{equation}
The third inequality is due to Cauchy-Schwarz Inequality.

\eqref{eq88} and the previous inequality imply that for $v \geq C_{\epsilon}$,
\begin{equation*}
	\begin{split}
		\left| E \log( W_0' + \frac{1}{2}) - \log v\right| \leq& \left| E \log( W_0 + \frac{1}{2}) - \log v\right| + \left| E \log( W_0' + \frac{1}{2}) -  E \log(  W_0 + \frac{1}{2})\right|\\
		\leq& 3v^{-\frac{3}{2}+3\epsilon_1} + \frac{1}{8v^2} + 2\sqrt{v}e^{-cv}\\
		\leq& 4v^{-\frac{3}{2}+3\epsilon_1}.
	\end{split}
\end{equation*}
\qed
\medskip

{\noindent Proof of Lemma \ref{lm:poisson_2}.}
By Taylor's expansion, 
\begin{equation*}
	\log^2(v + x) = \log^2v + 2\frac{\log v}{v}x + \frac{1-\log v}{v^2}x^2 + \frac{2\log(v +\xi)-3}{3(v+\xi)^3}x^3,
\end{equation*}
where $\xi$ is  a real number between 0 and $x$.\\
Let $x = t - v + \frac{1}{2}$, we have
\begin{equation*}
	\begin{split}
		\log^2\left(t + \frac{1}{2}\right) =& \log^2 v + 2\frac{\log v}{v}\left(t - v + \frac{1}{2}\right)+ \frac{1-\log v}{v^2}\left(t - v + \frac{1}{2}\right)^2\\ &+ \frac{2\log(v +\xi) - 3}{3(v+\xi)^3}\left(t - v + \frac{1}{2}\right)^3,
	\end{split}
\end{equation*}
where $\xi$ is  a real number between 0 and $t - v + \frac{1}{2}$.\\
Let $g(t) = \log^2(t + \frac{1}{2}) - \log^2v - 2\frac{\log v}{v}(t - v + \frac{1}{2}) - \frac{1-\log v}{v^2}(t - v + \frac{1}{2})^2$, then there exists a constant $C > 0$, for all $v - v^{5/8} < t < v + v^{5/8}$, $v \geq C$, we have
\begin{equation*}
	\begin{split}
		|g(t)| \leq& \frac{2\log(v + \xi)}{3(v + \xi)^3} \left|t - v + \frac{1}{2}\right|^3
		\leq \frac{2\log(3v)}{3\left(v - v^{5/8}\right)^3}\left|v^{5/8}+\frac{1}{2}\right|^3
		\leq \frac{\log v}{v^3}v^{15/8}\\
		\leq& \frac{v^{\frac{1}{8}}}{v^3}v^{15/8} = \frac{1}{v}.
	\end{split}
\end{equation*}
Therefore, for $v \geq C$,
\begin{equation}\label{eq21}
	\left|E g(  W_0)1_{\{v - v^{5/8} <  W_0 < v + v^{5/8}\}}\right| \leq \frac{1}{v}.
\end{equation}
For $t \leq v - v^{5/8}$, $v \geq 1$, we have
\begin{equation*}
	\begin{split}
		\left|g(t)\right| \leq& \max\{\log^2\left(t + \frac{1}{2}\right), \log^2 v\} + 2\log v\left|\frac{t - v + \frac{1}{2}}{v}\right| + \left|\frac{1 - \log v}{v^2}\right| \left(t - v + \frac{1}{2}\right)^2\\
		\leq& \log^2 v + 2\log v + \frac{\log v}{v^2}\left(v - \frac{1}{2}\right)^2
		\leq \log^2 v + 2\log v + \log v = \log^2 v + 3\log v.
	\end{split}
\end{equation*}
By Lemma \ref{lm:poisson tail bound}, for $v \geq C$,
\begin{equation}\label{eq22}
	\begin{split}
		E (\left|g( W_0)1_{\{ W_0 \leq v - v^{1/2 + \epsilon_2}\}}\right|) \leq& \left(\log^2 v + 3\log v\right){\mathrm{pr}}\left( W_0 \leq v - v^{5/8}\right)\\
		\leq& \left(\log^2 v + 3\log v\right)\exp\left(- \frac{v^{1/4}}{2}\right)
		\leq v^{-\frac{3}{2}}. 
	\end{split}
\end{equation}
$\forall t > v + v^{5/8}$ and $v \geq C$,
\begin{equation}\label{eq23}
	\begin{split}
		g(t) \leq& \left|\log^2 \left(t + \frac{1}{2}\right) - \log^2 v\right| + \left|2\frac{\log v}{v}\left(t - v + \frac{1}{2}\right)+ \frac{1-\log v}{v^2}\left(t - v + \frac{1}{2}\right)^2\right|\\
		\leq& \log^2 \left(t + \frac{1}{2}\right) + \frac{t - v + \frac{1}{2}}{v}\left|2\log v + \frac{1 - \log v}{v}\left(t - v + \frac{1}{2}\right)\right|\\
		\leq& \log^2 \left(t + \frac{1}{2}\right) + \frac{t - v + \frac{1}{2}}{v}\left(2\log v + \frac{\log v}{v}\left(t - v + \frac{1}{2}\right)\right)\\
		\leq& \log^2 \left(t + \frac{1}{2}\right) + 2(\log v) \frac{t}{v} + (\log v) \frac{t^2}{v^2}
		\leq 3t + (\log v) \frac{t^2}{v^2}.
	\end{split}
\end{equation}
By Lemma \ref{lm:poisson tail bound}, for all $v \geq C$ with a large constant $C$,
\begin{equation*}
	{\mathrm{pr}}( W_0 \geq v + v^{5/8}) \leq \exp\left(-\frac{\left(v^{\frac{5}{8}}\right)^2}{v}\psi_{Benn}\left(\frac{v^{\frac{5}{8}}}{v}\right)\right) \leq \exp\left(-\frac{1}{2}v^{\frac{1}{4}}\right).
\end{equation*}
Combine \eqref{eq23} and the previous inequality together, for all $v \geq C$,
\begin{equation}\label{eq89}
	\begin{split}
		&\left|E g( W_0)1_{\{ W_0 \geq v + v^{5/8}\}}\right|\\
		\leq&E \left(3 W_0 + (\log v) \frac{ W_0^2}{v^2}\right)1_{\{ W_0 \geq v + v^{5/8}\}}\\
		=& 3E  W_01_{\{ W_0 \geq v + v^{5/8}\}} + \frac{\log v}{v^2}E  W_0^21_{\{ W_0 \geq v + v^{5/8}\}}\\
		\leq& 3\left( E   W_0^2\right)^{\frac{1}{2}}\left[{\mathrm{pr}}\left( W_0 \geq v + v^{5/8}\right)\right]^{\frac{1}{2}} + \frac{\log v}{v^2}\left( E   W_0^4\right)^{\frac{1}{2}}\left[{\mathrm{pr}}\left( W_0 \geq v + v^{5/8}\right)\right]^{\frac{1}{2}}\\
		\leq& 3\left(2v^2\right)^{\frac{1}{2}}\exp\left(-\frac{1}{4}v^{\frac{1}{4}}\right) + \frac{\log v}{v^2}\left(2v^4\right)^{\frac{1}{2}}\exp\left(-\frac{1}{4}v^{\frac{1}{4}}\right)\\
		\leq& (6v + 2\log v)\exp\left(-\frac{v^{1/4}}{4}\right)\leq v^{-\frac{3}{2}}.
	\end{split}
\end{equation}
By \eqref{eq21}, \eqref{eq22} and \eqref{eq89}, for $v \geq C$, we have
\begin{equation}\label{ineq41}
	\begin{split}
		\left|E g( W_0)\right| =&E \left|g( W_0)1_{\{v - v^{5/8} <  W_0 < v + v^{5/8}\}}\right|\\ &+  \left|E g( W_0)1_{\{ W_0 \leq v - v^{5/8}\}}\right| + \left|E g( W_0)1_{\{ W_0 \geq v + v^{5/8}\}}\right|\\
		\leq& v^{-1} + v^{-\frac{3}{2}} + v^{-\frac{3}{2}} = \frac{1}{v} + \frac{2}{v^{\frac{3}{2}}}.
	\end{split}
\end{equation}
Note that $E g( W_0) = E \log^2 ( W_0 + \frac{1}{2}) - \log^2 v - \frac{1}{v} + \frac{\log v - 1}{4v^2}$, for $v \geq C$,
\begin{equation}\label{eq91}
	\left|E \log^2( W_0 + \frac{1}{2}) - \log^2 v\right| \leq \left| E  g( W_0)\right| + \left|\frac{1}{v} - \frac{\log v - 1}{4v^2}\right| \leq \frac{1}{v} + \frac{2}{v^{\frac{3}{2}}} + \max\{\frac{1}{v}, \frac{\log v - 1}{4v^2}\} = \frac{2}{v} + \frac{2}{v^{\frac{3}{2}}}.
\end{equation} 
Similarly to \eqref{eq90}, there exists a constant $c > 0$, for all $v \geq C$, 
\begin{equation}\label{ineq42}
	\begin{split}
		&\left| E \log^2\left( W_0' + \frac{1}{2}\right) -  E \log^2\left( W_0 + \frac{1}{2}\right)\right|\\
		=&\left|\log^2\left(v + \frac{1}{2}\right){\mathrm{pr}}\left( W_0 \leq \frac{1}{10}v\right) -  E \log^2\left( W_0 + \frac{1}{2}\right)1_{\{ W_0 \leq \frac{1}{10}v\}}\right|\\
		&+ \left|\log^2\left(v + \frac{1}{2}\right){\mathrm{pr}}\left( W_0 \geq 10v\right) -  E \log^2\left( W_0 + \frac{1}{2}\right)1_{\{ W_0 \geq 10v\}}\right|\\
		\leq& \max\left\{\log^2\left(v + \frac{1}{2}\right){\mathrm{pr}}\left( W_0 \leq \frac{1}{10}v\right),  E \log^2\left( W_0 + \frac{1}{2}\right)1_{\{ W_0 \leq \frac{1}{10}v\}}\right\}\\
		&+ \max\left\{\log^2\left(v + \frac{1}{2}\right){\mathrm{pr}}\left( W_0 \geq 10v\right),  E \log^2\left( W_0 + \frac{1}{2}\right)1_{\{ W_0 \geq 10v\}}\right\}\\
		\leq& \max\left\{\log^2\left(v + \frac{1}{2}\right)e^{-cv}, \left( E \log^4\left( W_0 + \frac{1}{2}\right)\right)^{\frac{1}{2}}\left({\mathrm{pr}}\left( W_0 \leq \frac{1}{10}v\right)\right)^{\frac{1}{2}}\right\}\\
		&+ \max\left\{\log^2\left(v + \frac{1}{2}\right)e^{-cv}, \left( E \log^4\left( W_0 + \frac{1}{2}\right)\right)^{\frac{1}{2}}\left({\mathrm{pr}}\left( W_0 \geq 10v\right)\right)^{\frac{1}{2}}\right\}\\
		\leq& 2\max\left\{\log^2\left(v + \frac{1}{2}\right)e^{-cv}, \left( E   W_0\right)^{\frac{1}{2}}e^{-cv}\right\}\\
		\leq& 2\sqrt{v}e^{-cv}.
	\end{split}
\end{equation}
By \eqref{eq91} and the previous inequality,
\begin{equation*}
	\begin{split}
		& \left| E \log^2\left( W_0' + \frac{1}{2}\right) - \log^2v\right| \\
		\leq& \left| E \log^2\left( W_0' + \frac{1}{2}\right) -  E \log^2\left( W_0 + \frac{1}{2}\right)\right| + \left|E \log^2( W_0 + \frac{1}{2}) - \log^2 v\right|\\
		\leq& \frac{2}{v} + \frac{2}{v^{\frac{3}{2}}} + 2\sqrt{v}e^{-cv} \leq \frac{4}{v}.
	\end{split}
\end{equation*}
\qed
\medskip

{\noindent Proof of Lemma \ref{lm:sub-Gaussian}.} For convenience, denote $D = \phi_{1}( W_0') - E \phi_{1}( W_0')$. By Lemma \ref{lm:poisson_1}, there exists a constant $C_1 > 0$ such that $\left| E \phi_1( W_0') - \log v\right| \leq \frac{4}{v}$. For $t > 0$ and $v \geq C_1$, we have
\begin{equation}\label{eq1}
	\begin{split}
		&{\mathrm{pr}}\left(\left|\sqrt{v}D\right| > t\right) = {\mathrm{pr}}\left(\sqrt{v}D > t\right) + {\mathrm{pr}}\left(\sqrt{v}D < - t\right)\\ =& {\mathrm{pr}}\left(\log\left( W_0' +\frac{1}{2}\right) > E \log\left( W_0'+\frac{1}{2}\right) + \frac{t}{\sqrt{v}}\right)\\ &+  {\mathrm{pr}}\left(\log\left( W_0'+\frac{1}{2}\right) < E \log\left( W_0'+\frac{1}{2}\right) - \frac{t}{\sqrt{v}}\right)\\
		\leq& {\mathrm{pr}}\left(\log\left( W_0'+\frac{1}{2}\right) > \log v - \frac{4}{v}  + \frac{t}{\sqrt{v}}\right) + {\mathrm{pr}}\left(\log\left( W_0'+\frac{1}{2}\right) < \log v + \frac{4}{v} - \frac{t}{\sqrt{v}}\right)\\
		=& {\mathrm{pr}}\left( W_0' > v e^{\frac{t}{\sqrt{v}} - \frac{4}{v}}-\frac{1}{2}\right) + {\mathrm{pr}}\left( W_0' < v e^{-\frac{t}{\sqrt{v}} + \frac{4}{v}} - \frac{1}{2}\right).
	\end{split}	
\end{equation}
$\forall t \geq 1$ and $v \geq C_2 = \max\{C_1, 100\}$,
\begin{equation}\label{eq24}
	\begin{split}
		&v \left(e^{\frac{t}{\sqrt{v}} - \frac{4}{v}} -1\right) -\frac{1}{2} \geq v\left(e^{\frac{t}{\sqrt{v}} - \frac{4}{v}} -1 - \frac{1}{2v}\right)\geq v\left(1 + \frac{t}{\sqrt{v}} - \frac{4}{v} -1 - \frac{1}{2v}\right)\\
		\geq& v\frac{t}{2\sqrt{v}} + v\left(\frac{t}{2\sqrt{v}} - \frac{9}{2v}\right) \geq v\frac{t}{2\sqrt{v}}> 0.
	\end{split}	
\end{equation}
In the second inequality, we used $e^x \geq 1 + x$ for all $x \in \bbR$.\\
By Lemma \ref{lm:trancated poisson tail bound} and \eqref{eq24}, 
\begin{equation}\label{eq2}
	\begin{split}
		&{\mathrm{pr}}\left( W_0' > ve^{\frac{t}{\sqrt{v}} - \frac{4}{v}}-\frac{1}{2}\right) = {\mathrm{pr}}\left( W_0' > v + v\left(e^{\frac{t}{\sqrt{v}} - \frac{4}{v}} -1\right) -\frac{1}{2}\right)\\
		\leq& \exp\left(-c\frac{\left(v\left(e^{\frac{t}{\sqrt{v}} - \frac{4}{v}} -1\right) -\frac{1}{2}\right)^2}{v}\right) \leq \exp\left(-c\frac{\left(v\frac{t}{2\sqrt{v}}\right)^2}{v}\right) = \exp\left(-\frac{1}{4}ct^2\right). 
	\end{split}	
\end{equation} 
Now consider the left tail. For any $v \geq C_2$ and $t \geq 1$,
\begin{equation}\label{eq25}
	\begin{split}
		&{\mathrm{pr}}\left( W_0' < ve^{-\frac{t}{\sqrt{v}} + \frac{4}{v}} - \frac{1}{2}\right)\leq {\mathrm{pr}}\left( W_0' < ve^{-\frac{t}{2\sqrt{v}}} - \frac{1}{2}\right).
	\end{split}	
\end{equation}
For any $t \geq 6\sqrt{v}$, we have $ve^{-\frac{t}{2\sqrt{v}}} - \frac{1}{2} < \frac{1}{10}v$, by the definition of $ W_0'$, immediately we have
\begin{equation}\label{eq92}
	\begin{split}
		&{\mathrm{pr}}\left( W_0' < ve^{-\frac{t}{\sqrt{v}} + \frac{4}{v}} - \frac{1}{2}\right) = {\mathrm{pr}}\left( W_0' < ve^{-\frac{t}{2\sqrt{v}}} - \frac{1}{2}\right) = 0.
	\end{split}
\end{equation}
For any $t < 6\sqrt{v}$, by Lemma \ref{lm:poisson tail bound}, we have
\begin{equation}\label{eq93}
	\begin{split}
		&{\mathrm{pr}}\left( W_0' < ve^{-\frac{t}{2\sqrt{v}}} - \frac{1}{2}\right)
		\leq {\mathrm{pr}}\left( W_0' < ve^{-\frac{t}{2\sqrt{v}}}\right)
		= {\mathrm{pr}}\left( W_0' < v - v\left(1 - e^{-\frac{t}{2\sqrt{v}}}\right)\right)\\
		\leq& \exp\left(-\frac{\left(v\left(1 - e^{-\frac{t}{2\sqrt{v}}}\right)\right)^2}{2v}\right)
		= \exp\left(-\frac{v}{2}\left(1 - e^{-\frac{t}{2\sqrt{v}}}\right)^2\right).
	\end{split}
\end{equation}
Note that for $x > 0$,
\begin{equation*}
	\left(\frac{1 - e^{-x}}{x}\right)' = \frac{(x + 1)e^{-x} - 1}{x^2} < 0,
\end{equation*}
therefore for $t < 6\sqrt{v}, v \geq C_2$, $1 - e^{-\frac{t}{2\sqrt{v}}} \geq \frac{1 - e^{-3}}{3}\frac{t}{2\sqrt{v}}$, which means that
\begin{equation*}
	\begin{split}
		& \exp\left(-\frac{v}{2}\left(1 - e^{-\frac{t}{2\sqrt{v}}}\right)^2\right) \leq \exp\left(-\frac{v}{2}\frac{(1 - e^{-3})^2}{9}\frac{t^2}{4v}\right)\leq \exp\left(-\frac{1}{100}t^2\right).
	\end{split}
\end{equation*}
\eqref{eq25}, \eqref{eq93} and the previous inequality imply that for any $v \geq C_2, t < 6\sqrt{v}$,
\begin{equation*}
	{\mathrm{pr}}\left( W_0' < ve^{-\frac{t}{\sqrt{v}} + \frac{4}{v}} - \frac{1}{2}\right) \leq \exp\left(-\frac{1}{100}t^2\right).
\end{equation*}
By \eqref{eq92} and the previous inequality, if $v \geq C_2$, then for all $t \geq 1$, we have 
\begin{equation}\label{eq3}
	\begin{split}
		{\mathrm{pr}}\left( W_0' < ve^{-\frac{t}{\sqrt{v}} + \frac{4}{v}} - \frac{1}{2}\right)
		\leq \exp\left(-\frac{1}{100}t^2\right).
	\end{split}
\end{equation}
combine \eqref{eq1}, \eqref{eq2} and \eqref{eq3}, we know that $\exists C, c > 0$, for all $t \geq 1$, $v \geq C$,
\begin{equation*}
	\begin{split}
		{\mathrm{pr}}\left(\left|\sqrt{v}D\right| > t\right)
		\leq 2\exp\left(-ct^2\right).
	\end{split}
\end{equation*}

For $p \geq 1$, we have
\begin{equation*}
	\begin{split}
		&E \left|\sqrt{v}D\right|^p = \int_{0}^{\infty}{\mathrm{pr}}\left(\left|\sqrt{v}D\right| > t\right)pt^{p-1}dt \leq \int_{0}^{1}pt^{p-1}dt + \int_{1}^{\infty}2\exp\left(-ct^2\right)pt^{p-1}dt\\
		\leq& 1 + \int_{0}^{\infty}2\exp\left(-ct^2\right)pt^{p-1}dt
		= 1 + c^{-\frac{p}{2}} p\Gamma\left(\frac{p}{2}\right)
		\leq 1 + pc^{-\frac{p}{2}}\left(\frac{p}{2}\right)^{\frac{p}{2}}\\
		\leq& \left(1 + p^{\frac{1}{p}}c^{-\frac{1}{2}}\left(\frac{p}{2}\right)^{\frac{1}{2}}\right)^p.
	\end{split}
\end{equation*}
In the last step, we use $x^p + y^p \leq (x + y)^p$ for all $x, y \geq 0, p \geq 1$.\\
Therefore, for $p \geq 1$, we have
\begin{equation*}
	\begin{split}
		&p^{-\frac{1}{2}}\left(E \left|\sqrt{v}D\right|^p\right)^{\frac{1}{p}}
		\leq p^{-\frac{1}{2}}\left(1 + p^{\frac{1}{p}}c^{-\frac{1}{2}}\left(\frac{p}{2}\right)^{\frac{1}{2}}\right)
		\leq 1 + \frac{e^{\frac{1}{e}}c^{-\frac{1}{2}}}{\sqrt{2}},
	\end{split}
\end{equation*}
which means 
\begin{equation*}
	\left\|\sqrt{v}D\right\|_{\psi_{2}} \leq K_0,
\end{equation*} 
where $K_0 = 1 + \frac{e^{\frac{1}{e}}c^{-\frac{1}{2}}}{\sqrt{2}}$. 
\qed
\medskip

{\noindent Proof of Lemma \ref{lm:infinity_norm_bound}.} Since 
\begin{equation*}
	\begin{split}
		y_i = \sum_{j = 1}^p\log\left(X_{ij}\right)\beta_j^* + \varepsilon_i = \sum_{j = 1}^p\log\left(X_{ij}\right)\beta_j^* + \log\nu_i \sum_{j = 1}^{p}\beta_j^* + \varepsilon_i = \sum_{j = 1}^{p}\log\left(\nu_{ij}\right)\beta_j^* + \varepsilon_i,
	\end{split}
\end{equation*}
i.e., $y = V\beta^* + \varepsilon$ (where $V_{ij} = \log\nu_{ij}$). Also note that $\P\beta^* = \beta^*$, 
\begin{equation*}
	\begin{split}
		& \left\|\bar{\A}_{W'}\beta^* - \bar{\B}_{W'}^\top y\right\|_\infty = \left\|\P\A_{W'}\beta^* - \P\B_{W'}^\top (V\beta^* + \varepsilon)\right\|_\infty \leq \left\|\P(\A_{W'} - \B_{W'}^\top V)\beta^*\right\|_\infty + \|\bar \B_{W'}^\top \varepsilon\|_\infty. \\
	\end{split}
\end{equation*}
More specifically,
\begin{equation*}
	\left(\bar \B_{W'}^\top\varepsilon\right)_i = \sum_{k=1}^{n}\left(\bar{\B}_{W'}\right)_{ki}\varepsilon_k =  \sum_{k=1}^{n}\left(\phi_1(W'_{ki}) - \frac{1}{p}\sum_{j=1}^{p}\phi_1(W'_{kj})\right)\varepsilon_k.
\end{equation*}
For any $t \geq 0$, by Hoeffding-type inequality,
\begin{equation}\label{eq510}
	{\mathrm{pr}} \left(\left|\sum_{k=1}^{n}\left(\bar{\B}_{W'}\right)_{ki}\varepsilon_k\right| > t\Big|W'_{kj}, 1 \leq k \leq n, 1 \leq j \leq p\right) \leq 2e^{-\frac{t^2}{2\sigma^2\sum_{k=1}^n \left(\bar{\B}_{W'}\right)_{ki}^2}}.
\end{equation}
Denote $D_{ki} = \phi_1(W'_{ki}) -  E \phi_1(W'_{ki}), Q_{ki} = D_{ki}^2 -  E  D_{ki}^2, M_{ki} = \left(\bar{\B}_W\right)_{ki} -  E \left(\bar{\B}_W\right)_{ki}$, $R_{ki} = M_{ki}^2 -  E  M_{ki}^2$, we have shown that there exists a constant $K$ such that $\|D_{ki}\|_{\psi_{2}} \leq K$ in Lemma \ref{lm:sub-Gaussian}. According to Lemma 5.9 in \cite{vershynin2010introduction}, 
\begin{equation}\label{eq6}
	\begin{split}
		\|M_{ki}\|_{\psi_2}^2 =& \left\|(1 - \frac{1}{p})D_{ki} + \frac{1}{p}\sum_{j \neq i}D_{kj}\right\|_{\psi_2}^2 \leq C\left(\left(1 - \frac{1}{p}\right)^2\|D_{ki}\|_{\psi_{2}}^2 + \frac{1}{p^2}\sum_{j \neq i}\|D_{kj}\|_{\psi_2}^2\right)\\ \leq& C\max_{j}\|D_{kj}\|_{\psi_2}^2 \leq C K^2.
	\end{split}
\end{equation}
Thus $\|M_{ki}^2\|_{\psi_{1}} \leq 2\|M_{ki}\|_{\psi_{2}}^2 \leq CK^2$, which leads to $\|R_{ki}\|_{\psi_{1}} \leq 2\|M_{ki}^2\|_{\psi_{1}} \leq CK^2$. Note that
\begin{equation*}
	\left(\bar{\B}_{W'}\right)_{ki}^2 -  E \left(\bar{\B}_{W'}\right)_{ki}^2 = R_{ki} + 2\left( E \left(\bar{\B}_{W'}\right)_{ki}\right)M_{ki},
\end{equation*}
$\forall t \geq 0$, we have
\begin{equation}\label{eq507}
	\begin{split}
		&{\mathrm{pr}} \left(\left| \sum_{k=1}^n\left(\bar{\B}_{W'}\right)_{ki}^2 - \sum_{k=1}^{n} E \left(\bar{\B}_{W'}\right)_{ki}^2\right| > t\right)\\
		\leq&{\mathrm{pr}}\left(\left|\sum_{k=1}^{n}R_{ki}\right| > \frac{t}{2}\right) + {\mathrm{pr}}\left(\left|\sum_{k = 1}^{n}\left( E \left(\bar{\B}_{W'}\right)_{ki}\right)M_{ki}\right| > \frac{t}{4}\right).
	\end{split}
\end{equation}
By Bernstein-type inequality, we have
\begin{equation*}
	{\mathrm{pr}}\left(\left|\sum_{k=1}^{n}R_{ki}\right| > \frac{t}{2}\right) \leq 2\exp\left[-c\min\left(\frac{t^2}{nK^4}, \frac{t}{K^2}\right)\right],
\end{equation*}
By Hoeffding-type inequality, 
\begin{equation*}
	{\mathrm{pr}}\left(\left|\sum_{k = 1}^{n}\left( E \left(\bar{\B}_{W'}\right)_{ki}\right)M_{ki}\right| > \frac{t}{4}\right) \leq e\cdot\exp\left(-\frac{ct^2}{K^2\sum_{k=1}^{n} E \left(\bar{\B}_{W'}\right)_{ki}^2}\right),
\end{equation*}
Set $t = n$, combine \eqref{eq507} and the previous two inequalities together, we have
\begin{equation}\label{eq509}
	\begin{split}
		&{\mathrm{pr}}\left(\sum_{k=1}^{n}\left(\bar{\B}_{W'}\right)_{ki}^2 > \sum_{k=1}^{n} E \left(\bar{\B}_{W'}\right)_{ki}^2 + n\right)\\
		\leq&{\mathrm{pr}} \left(\left| \sum_{k=1}^n\left(\bar{\B}_{W'}\right)_{ki}^2 - \sum_{k=1}^{n} E \left(\bar{\B}_{W'}\right)_{ki}^2\right| > n\right)\\
		\leq&{\mathrm{pr}}\left(\left|\sum_{k=1}^{n}R_{ki}\right| > \frac{n}{2}\right) + {\mathrm{pr}}\left(\left|\sum_{k = 1}^{n}\left( E \left(\bar{\B}_{W'}\right)_{ki}\right)M_{ki}\right| > \frac{n}{4}\right)\\
		\leq&2\exp\left[-c\min\left(\frac{n}{K^4}, \frac{n}{K^2}\right)\right]
		+ e\cdot\exp\left(-\frac{cn^2}{K^2\sum_{k=1}^{n} E \left(\bar{\B}_{W'}\right)_{ki}^2}\right).
	\end{split}
\end{equation}
By Lemma \ref{lm:poisson_2}, $ E \phi_1^2\left(W'_{ki}\right) =  E \phi_{2}\left(W'_{ki}\right) \leq \log^2\nu_{ki} + \frac{4}{\nu_{ki}}$ for all $\nu_{ki} \geq C$. Since $a\bar{\nu} \leq \nu_k \leq b\bar{\nu}$ and $a/p \leq X_{ki} \leq b/p$, we know that $\left(\frac{a}{b}\right)^2\nu_{ki} \leq \nu_{\min}$ for all $1 \leq k \leq n, 1 \leq i \leq p$. Thus $c \log^2\nu_{\min} \leq  E  \phi_{1}^2\left(W'_{ki}\right) \leq C\log^2\nu_{\min}$.\\
If $\nu_{\min} \geq C$, Lemma \ref{lm:poisson_1} and the entry-wise upper and lower bounds of $\nu_i$ and $X_{ij}$ together imply that
\begin{equation}\label{eq511}
	\begin{split}
		&\left| E \left(\bar{\B}_{W'}\right)_{ki}\right| = \left| E \phi_1\left(W'_{ki}\right) - \frac{1}{p}\sum_{j=1}^{p} E \phi_1\left(W'_{kj}\right)\right|\\
		\leq& \left|\log\nu_{ki} - \frac{1}{p}\sum_{j=1}^{n}\log\nu_{kj}\right| + \frac{4}{\nu_{ki}} + \frac{1}{p}\sum_{j = 1}^{p}\frac{4}{\nu_{ki}}
		\leq \log\left(\frac{b}{a}\right) + \frac{8}{\nu_{\min}}.
	\end{split}
\end{equation}
By the previous inequality and the independence of $W'_{kj} (j = 1, 2, \dots, p)$, for $\nu_{\min} \geq C$, we have
\begin{equation}\label{eq508}
	\begin{split}
		E \left(\bar{\B}_{W'}\right)_{ki}^2 =& \left( E \left(\bar{\B}_{W'}\right)_{ki}\right)^2 + \Var\left(\left(\bar{\B}_{W'}\right)_{ki}\right)\\
		\leq& \left(\log\left(\frac{b}{a}\right) + \frac{8}{\nu_{\min}}\right)^2 + \Var\left(\phi_1(W'_{ki}) - \frac{1}{p}\sum_{j=1}^{p}\phi_1(W'_{kj})\right)\\
		\leq& \left(\log\left(\frac{b}{a}\right) + \frac{8}{\nu_{\min}}\right)^2 +  \left(1 - \frac{1}{p}\right)^2\Var\left(\phi_1\left(W'_{ki}\right)\right) + \frac{1}{p^2}\sum_{j \neq i}\Var\left(\phi_1\left(W'_{kj}\right)\right).
	\end{split}
\end{equation}
Notice that for all $\nu_{kj} \geq C$, 
\begin{equation}\label{eq513}
	\begin{split}
		\Var\left(\phi_1\left(W'_{kj}\right)\right) =&  E \phi_1^2\left(W'_{kj}\right) - \left( E \phi_1\left(W'_{kj}\right)\right)^2\\
		\leq& \log^2(\nu_{kj}) + \frac{4}{\nu_{kj}} - \left(\log(\nu_{kj}) - \frac{4}{\nu_{kj}^{\frac{3}{2} - \frac{1}{4}}}\right)^2
		\leq \frac{8}{\nu_{kj}}.
	\end{split}
\end{equation}
Combine \eqref{eq508} and the previous inequality, if $\nu_{\min} \geq C$, then
\begin{equation*}
	\begin{split}
		E \left(\bar{\B}_{W'}\right)_{ki}^2 \leq& \left(\log\left(\frac{b}{a}\right) + \frac{8}{\nu_{\min}}\right)^2 + \left(\left(1 - \frac{1}{p}\right)^2 + p\cdot \frac{1}{p^2}\right)\frac{8}{\nu_{\min}}\\ \leq& \left(\log\left(\frac{b}{a}\right) + \frac{8}{\nu_{\min}}\right)^2 + \frac{8}{\nu_{\min}} \leq \left(\log\left(\frac{b}{a}\right) + \frac{8}{C}\right)^2 + \frac{8}{C}
	\end{split}
\end{equation*}
can be bounded by a constant. Thus
\begin{equation*}
	\sum_{k = 1}^{n} E \left(\bar{\B}_{W'}\right)_{ki}^2 \leq Cn.
\end{equation*}
Therefore, if $\nu_{\min} \geq C$, $\forall t \geq 0$, by \eqref{eq510}, \eqref{eq509} and the previous inequality, there exists an absolute constant $c > 0$, 
\begin{equation*}
	\begin{split}
		&{\mathrm{pr}} \left(\left|\sum_{k=1}^{n}\left(\bar{\B}_{W'}\right)_{ki}\varepsilon_k\right| > t\right)\\
		=&{\mathrm{pr}}\left(\left|\sum_{k=1}^{n}\left(\bar{\B}_{W'}\right)_{ki}\varepsilon_k\right| > t, \sum_{k=1}^{n}\left(\bar{\B}_{W'}\right)_{ki}^2 \leq \sum_{k=1}^{n} E \left(\bar{\B}_{W'}\right)_{ki}^2 + n\right)\\
		&+{\mathrm{pr}}\left(\left|\sum_{k=1}^{n}\left(\bar{\B}_{W'}\right)_{ki}\varepsilon_k\right| > t, \sum_{k=1}^{n}\left(\bar{\B}_{W'}\right)_{ki}^2 > \sum_{k=1}^{n} E \left(\bar{\B}_{W'}\right)_{ki}^2 + n\right)\\
		\leq& {\mathrm{pr}}\left(\left|\sum_{k=1}^{n}\left(\bar{\B}_{W'}\right)_{ki}\varepsilon_k\right| > t, \sum_{k=1}^{n}\left(\bar{\B}_{W'}\right)_{ki}^2 \leq Cn\right)
		+ {\mathrm{pr}}\left( \sum_{k=1}^{n}\left(\bar{\B}_{W'}\right)_{ki}^2 > \sum_{k=1}^{n} E \left(\bar{\B}_{W'}\right)_{ki}^2 + n\right)\\
		\leq& 2\exp\left(-\frac{ct^2}{n\sigma^2}\right) + (2 + e)\exp\left(-cn\right).
	\end{split}
\end{equation*}
Thus
\begin{equation}\label{eq9}
	\begin{split}
		&{\mathrm{pr}}\left(\|\bar{\B}_{W'}^\top\varepsilon\|_{\infty} > t\right)
		= {\mathrm{pr}}\left(\exists 1 \leq i \leq p, \left(\bar{\B}_{W'}^\top\varepsilon\right)_i > t\right)
		\leq \sum_{i = 1}^{p}{\mathrm{pr}} \left(\left|\sum_{k=1}^{n}\left(\bar{\B}_{W'}\right)_{ki}\varepsilon_k\right| > t\right)\\
		\leq& p\left(2e^{-\frac{ct^2}{n\sigma^2}} + (2 + e)e^{-cn}\right)
		\leq 5p\left(e^{-\frac{ct^2}{n\sigma^2}} + e^{-cn}\right).
	\end{split}	
\end{equation}
Now we consider $\|\P\left(\A_{W'} - \B_{W'}^\top V\right)\beta^*\|_{\infty}$. Note that
\begin{equation*}
	(\B_{W'}^\top  V)_{ij} = \sum_{k=1}^{n}\phi_1\left(W'_{ki}\right)\log\nu_{kj},
\end{equation*}
we have
\begin{equation*}
	\begin{split}
		\left[\left(\A_{W'} - \B_{W'}^\top  V\right)\beta^*\right]_i =& \sum_{j=1}^{p}\left(\A_{W'} - \B_{W'}^\top  V\right)_{ij}\beta_j^*\\
		=& \sum_{k=1}^{n}\sum_{j=1}^{p}\phi_{1}(W'_{ki})\left(\phi_{1}\left(W'_{kj}\right) - \log\nu_{kj}\right)\beta_j^*\\
		=& \sum_{k=1}^{n}\phi_{1}(W'_{ki})\sum_{j \in S}\left(\phi_{1}\left(W'_{kj}\right) - \log\nu_{kj}\right)\beta_j^*,
	\end{split}
\end{equation*}
where $S$ is the support set of $\beta^*$.

Thus
\begin{equation}\label{eq31}
	\begin{split}
		\left[\P\left(\A_{W'} - \B_{W'}^\top  V\right)\beta^*\right]_i =& \sum_{k=1}^{n}\left(\phi_{1}(W'_{ki}) - \frac{1}{p}\sum_{l = 1}^{p}\phi_1\left(W'_{kl}\right)\right)\sum_{j \in S}\left(\phi_{1}\left(W'_{kj}\right) - \log\nu_{kj}\right)\beta_j^*\\
		=& \sum_{k = 1}^{n}\left(\bar{\B}_{W'}\right)_{ki}\sum_{j \in S}\left(\phi_{1}\left(W'_{kj}\right) - \log\nu_{kj}\right)\beta_j^*.
	\end{split}
\end{equation}
Consider $\left(\bar{\B}_{W'}\right)_{ki}\sum_{j \in S}\left(\phi_{1}\left(W'_{kj}\right) - \log\nu_{kj}\right)\beta_j^*$ first. 
By Lemma \ref{lm:poisson_1}, $\forall 1 > \epsilon > 0$, $\exists C_{\epsilon} \geq C > 0$(only depends on $\epsilon$) such that for all $k, j$, if $\nu_{\min} \geq C_{\epsilon}$,
\begin{equation*}
	\left| E \phi_1\left(W'_{kj}\right) - \log\nu_{kj}\right| \leq \frac{4}{\nu_{kj}^{\frac{3}{2} - \epsilon}} \leq \frac{4}{\nu_{\min}^{\frac{3}{2} - \epsilon}}.
\end{equation*}
Apply Cauchy-Schwarz inequality, 
\begin{equation*}
	\begin{split}
		&\left(\sum_{j \in S} E \left(\phi_{1}\left(W'_{kj}\right) - \log\nu_{kj}\right)\beta_j^*\right)^2 \leq \sum_{j \in S}\left( E \phi_1\left(W'_{kj}\right) - \log\nu_{kj}\right)^2\|\beta^*\|_2^2 \leq \frac{16s}{\nu_{\min}^{3 - 2\epsilon}}\|\beta^*\|_2^2,
	\end{split}	
\end{equation*} 
i.e., 
\begin{equation}\label{eq32}
	\left|\sum_{j \in S} E \left(\phi_{1}\left(W'_{kj}\right) - \log\nu_{kj}\right)\beta_j^*\right| \leq \frac{4\sqrt{s}}{\nu_{\min}^{\frac{3}{2} - \epsilon}}\|\beta^*\|_2.
\end{equation}
For convenience, denote $U_{kj} = \phi_{1}\left(W'_{kj}\right) - \log\nu_{kj}$. Note that $\forall t \geq 0$, if $|XY -  E  X E  Y| > t$, then for any $\alpha > 0$, we have $|X -  E  X| > \min\left(\frac{t}{3| E  Y|}, \sqrt{\frac{\alpha t}{3}}\right)$ or $|Y -  E  Y| > \min\left(\frac{t}{3| E  X|}, \sqrt{\frac{t}{3\alpha}}\right)$. Otherwise
\begin{equation*}
	\begin{split}
		\left|XY -  E  X E  Y\right| =& \left|(X -  E  X)(Y -  E  Y) +  E  Y(X -  E  X) +  E  X(Y -  E  Y)\right|\\
		\leq& \left|(X -  E  X)(Y -  E  Y)\right| + \left| E  Y(X -  E  X)\right| + \left| E  X(Y -  E  Y)\right|\\
		\leq& \sqrt{\frac{\alpha t}{3}}\sqrt{\frac{t}{3\alpha}} + \left| E  X\right|\frac{t}{3| E  X|} + \left| E  Y\right|\frac{t}{3| E  Y|} = t.
	\end{split}
\end{equation*}
Contradiction! By \eqref{eq511} and \eqref{eq32}, for any $t \geq 0$, 
\begin{equation}\label{eq512}
	\begin{split}
		&{\mathrm{pr}}\left(\left|\left(\bar{\B}_{W'}\right)_{ki}\sum_{j \in S}U_{kj}\beta_j^* -  E \left(\bar{\B}_{W'}\right)_{ki} E \left(\sum_{j \in S}U_{kj}\beta_j^*\right)\right| > t\right)\\
		\leq& {\mathrm{pr}}\left(\left|\left(\bar{\B}_{W'}\right)_{ki} -  E \left(\bar{\B}_{W'}\right)_{ki}\right| > \min\left(\frac{t}{3\left| E \sum_{j \in S}U_{kj}\beta_j^*\right|}, \sqrt{\frac{t}{3\|\beta^*\|_2}}\right)\right)\\
		&+ {\mathrm{pr}}\left(\left|\sum_{j \in S}U_{kj}\beta_j^* -  E \left(\sum_{j \in S}U_{kj}\beta_j^*\right)\right| >  \min\left(\frac{t}{3\left| E \left(\bar{\B}_{W'}\right)_{ki}\right|}, \sqrt{\frac{\|\beta^*\|_2t}{3}}\right)\right)\\
		\leq&  {\mathrm{pr}}\left(\left|M_{ki}\right| > \min\left(\frac{t}{\frac{12\sqrt{s}}{\nu_{\min}^{\frac{3}{2} - \epsilon}}\|\beta^*\|_2}, \sqrt{\frac{t}{3\|\beta^*\|_2}}\right)\right)\\ &+ {\mathrm{pr}}\left(\left|\sum_{j \in S}D_{kj}\beta_j^*\right| > \min\left(\frac{t}{3\log\left(\frac{b}{a}\right) + \frac{24}{\nu_{\min}}}, \sqrt{\frac{\|\beta^*\|_2t}{3}}\right)\right).
	\end{split}
\end{equation}
By Lemma \ref{lm:sub-Gaussian} and (\ref{eq6}), 
\begin{equation*}
	\|D_{kj}\|_{\psi_2} \leq \frac{K_0}{\sqrt{\nu_{\min}}}, \quad \|M_{ki}\|_{\psi_2}^2 \leq C\max_{j}\|D_{kj}\|_{\psi_2}^2 \leq C\frac{K_0^2}{\nu_{\min}}.
\end{equation*}
Due to Hoeffding-type inequality, 
\begin{equation*}
	{\mathrm{pr}}\left(\left|M_{ki}\right| > \min\left(\frac{t}{\frac{12}{\nu_{\min}}\|\beta^*\|_2}, \sqrt{\frac{t}{3\|\beta^*\|_2}}\right)\right) \leq e\cdot\exp\left(-\frac{c}{\frac{1}{\nu_{\min}}K_0^2}\min\left(\frac{t^2}{\frac{s}{\nu_{\min}^{3 - 2\epsilon}}\|\beta^*\|_2^2}, \frac{t}{\|\beta^*\|_2}\right)\right).
\end{equation*}
\begin{equation*}
	{\mathrm{pr}}\left(\left|\sum_{j \in S}D_{kj}\beta_j^*\right| > \min\left(\frac{t}{6\frac{b}{a} + \frac{24}{\nu_{\min}}}, \sqrt{\frac{\|\beta^*\|_2t}{3}}\right)\right) \leq e\cdot \exp\left(-\frac{c}{\frac{1}{\nu_{\min}}K_0^2\|\beta^*\|_2^2}\min\left(t^2, \|\beta^*\|_2t\right)\right).
\end{equation*}
Combine \eqref{eq512} and the previous two inequalities together, we have
\begin{equation*}
	\begin{split}
		&{\mathrm{pr}}\left(\left|\left(\bar{\B}_{W'}\right)_{ki}\sum_{j \in S}U_{kj}\beta_j^* -  E \left(\bar{\B}_{W'}\right)_{ki} E \left(\sum_{j \in S}U_{kj}\beta_j^*\right)\right| > t\right)\\ \leq& e\cdot\exp\left(-\frac{c}{\frac{1}{\nu_{\min}}K_0^2}\min\left(\frac{t^2}{\frac{s}{\nu_{\min}^{3 - 2\epsilon}}\|\beta^*\|_2^2}, \frac{t}{\|\beta^*\|_2}\right)\right) + e\cdot \exp\left(-\frac{c}{\frac{1}{\nu_{\min}}K_0^2\|\beta^*\|_2^2}\min\left(t^2, \|\beta^*\|_2t\right)\right).
	\end{split}
\end{equation*}
For $q \geq 1$, the previous inequality implies that
\begin{equation*}
	\begin{split}
		& E \left|\left(\bar{\B}_{W'}\right)_{ki}\sum_{j \in S}U_{kj}\beta_j^* -  E \left(\bar{\B}_{W'}\right)_{ki} E \left(\sum_{j \in S}U_{kj}\beta_j^*\right)\right|^q\\
		=&\int_{0}^{\infty}{\mathrm{pr}}\left(\left|\left(\bar{\B}_{W'}\right)_{ki}\sum_{j \in S}U_{kj}\beta_j^* -  E \left(\bar{\B}_{W'}\right)_{ki} E \left(\sum_{j \in S}U_{kj}\beta_j^*\right)\right| > t\right)qt^{q - 1}dt\\
		\leq& \int_{0}^{\infty}e\cdot\exp\left(-\frac{ct^2}{\frac{1}{\nu_{\min}}K_0^2\frac{s}{\nu_{\min}^{3 - 2\epsilon}}\|\beta^*\|_2^2}\right)qt^{q-1}dt + \int_{0}^{\infty}e\cdot\exp\left(-\frac{ct}{\frac{1}{\nu_{\min}}K_0^2\|\beta^*\|_2}\right)qt^{q-1}dt\\
		&+ \int_{0}^{\infty}e\cdot\exp\left(-\frac{ct^2}{\frac{1}{\nu_{\min}}K_0^2\|\beta^*\|_2^2}\right)qt^{q-1}dt + \int_{0}^{\infty}e\cdot\exp\left(-\frac{ct}{\frac{1}{\nu_{\min}}K_0^2\|\beta^*\|_2}\right)qt^{q-1}dt\\
		=& \frac{eq}{2}\Gamma\left(\frac{q}{2}\right)\left(\frac{K_0\sqrt{s}\|\beta^*\|_2}{\sqrt{c}\nu_{\min}^{2 - \epsilon}}\right)^q + 2e\cdot\left(\frac{K_0^2\|\beta^*\|_2}{c\nu_{\min}}\right)^q q! + \frac{eq}{2}\Gamma\left(\frac{q}{2}\right)\left(\frac{K_0\|\beta^*\|_2}{\sqrt{c}\nu_{\min}^{\frac{1}{2}}}\right)^q\\
		\leq& \frac{eq}{2}\left(\frac{q}{2}\right)^{\frac{q}{2}}\left(\frac{K_0\sqrt{s}\|\beta^*\|_2}{\sqrt{c}\nu_{\min}^{2 - \epsilon}}\right)^q + 2e\cdot\left(\frac{K_0^2\|\beta^*\|_2}{c\nu_{\min}}\right)^q\left(\frac{q + 1}{2}\right)^q + \frac{eq}{2}\left(\frac{q}{2}\right)^{\frac{q}{2}}\left(\frac{K_0\|\beta^*\|_2}{\sqrt{c}\nu_{\min}^{\frac{1}{2}}}\right)^q.
	\end{split}
\end{equation*}
Thus 
\begin{equation*}
	\begin{split}
		&\left( E \left|\left(\bar{\B}_{W'}\right)_{ki}\sum_{j \in S}U_{kj}\beta_j^* -  E \left(\bar{\B}_{W'}\right)_{ki} E \left(\sum_{j \in S}U_{kj}\beta_j^*\right)\right|^q\right)^{\frac{1}{q}}\\
		\leq& \left(\frac{eq}{2}\left(\frac{q}{2}\right)^{\frac{q}{2}}\left(\frac{K_0\sqrt{s}\|\beta^*\|_2}{\sqrt{c}\nu_{\min}^{2 - \epsilon}}\right)^q\right)^{\frac{1}{q}} + \left(2e\cdot\left(\frac{K_0^2\|\beta^*\|_2}{c\nu_{\min}}\right)^q\left(\frac{q + 1}{2}\right)^q\right)^{\frac{1}{q}}\\ &+ \left(\frac{eq}{2}\left(\frac{q}{2}\right)^{\frac{q}{2}}\left(\frac{K_0\|\beta^*\|_2}{\sqrt{c}\nu_{\min}^{\frac{1}{2}}}\right)^q\right)^{\frac{1}{q}}\\
		\leq& q\left(2\frac{K_0\sqrt{s}\|\beta^*\|_2}{\sqrt{c}\nu_{\min}^{2 - \epsilon}} + 2e\cdot\frac{K_0^2\|\beta^*\|_2}{c\nu_{\min}} + 2\frac{K_0\|\beta^*\|_2}{\sqrt{c}\nu_{\min}^{\frac{1}{2}}}\right).
	\end{split}
\end{equation*}
The second inequality holds since $\left(2e\right)^{\frac{1}{q}}\frac{q + 1}{2} \leq 2eq$ and $\left(\frac{eq}{2}\right)^{\frac{1}{q}}\left(\frac{q}{2}\right)^{\frac{1}{2}} \leq \frac{e}{2\sqrt{2}}\cdot 2q \leq 2q$ for $q \geq 1$.
Set $\epsilon = \frac{1}{2}$ in the previous inequality, we know that if $\nu_{\min} \geq Cs$,
\begin{equation*}
	\left\|\left(\bar{\B}_{W'}\right)_{ki}\sum_{j \in S}U_{kj}\beta_j^* -  E \left(\bar{\B}_{W'}\right)_{ki} E \left(\sum_{j \in S}U_{kj}\beta_j^*\right)\right\|_{\psi_1} \leq \frac{C}{\sqrt{\nu_{\min}}}\|\beta^*\|_2.
\end{equation*}
Consider the centering, we have
\begin{equation*}
	\begin{split}
		&\left\|\left(\bar{\B}_{W'}\right)_{ki}\sum_{j \in S}U_{kj}\beta_j^* -  E \left(\left(\bar{\B}_{W'}\right)_{ki}\sum_{j \in S}U_{kj}\beta_j^*\right)\right\|_{\psi_1}\\
		\leq&2\left\|\left(\bar{\B}_{W'}\right)_{ki}\sum_{j \in S}U_{kj}\beta_j^* -  E \left(\bar{\B}_{W'}\right)_{ki} E \left(\sum_{j \in S}U_{kj}\beta_j^*\right)\right\|_{\psi_1}
		\leq \frac{C}{\sqrt{\nu_{\min}}}\|\beta^*\|_2
	\end{split}	
\end{equation*}
for all $1 \leq k \leq n$. By Bernstein-type inequality,
\begin{equation*}
	{\mathrm{pr}}\left(\left|\sum_{k = 1}^{n}\left(\left(\bar{\B}_{W'}\right)_{ki}\sum_{j \in S}U_{kj}\beta_j^* -  E \left(\left(\bar{\B}_{W'}\right)_{ki}\sum_{j \in S}U_{kj}\beta_j^*\right)\right)\right| \geq t\right) \leq 2\exp\left[-c\min\left\{\frac{\nu_{\min}t^2}{n\|\beta^*\|_2^2}, \frac{\sqrt{\nu_{\min}}t}{\|\beta^*\|_2}\right\}\right].
\end{equation*}
Recall that $\left[\P\left(\A_{W'} - \B_{W'}^\top  V\right)\beta^*\right]_i = \sum_{k=1}^{n}\left(\bar{\B}_{W'}\right)_{ki}\sum_{j \in S}U_{kj}\beta_j^*$,  set $t = C\frac{\sqrt{n}}{\sqrt{\nu_{\min}}}\|\beta^*\|_2\sqrt{x}$ with $1 \leq x \lesssim n$ and large constant $C$, 
\begin{equation}\label{eq7}
	{\mathrm{pr}}\left(\left|\left[\P\left(\A_{W'} - \B_{W'}^\top  V\right)\beta^*\right]_i -  E \left[\P\left(\A_{W'} - \B_{W'}^\top  V\right)\beta^*\right]_i\right| \geq \sqrt{n}\frac{C}{\sqrt{\nu_{\min}}}\|\beta^*\|_2\sqrt{x}\right) \leq e^{-x}.
\end{equation}
Since $\left|\Cov\left(X, Y\right)\right| \leq \sqrt{\Var\left(X\right)\Var\left(Y\right)}$, we have
\begin{equation*}
	\begin{split}
		&\left| E \left[\sum_{k=1}^{n}\left(\bar{\B}_{W'}\right)_{ki}\sum_{j \in S}U_{kj}\beta_j^*\right] -  E \sum_{k=1}^{n}\left(\bar{\B}_{W'}\right)_{ki} E \sum_{j \in S}U_{kj}\beta_j^*\right|\\
		\leq& \left(\Var\left(\sum_{k=1}^{n}\left(\bar{\B}_{W'}\right)_{ki}\right)\Var\left(\sum_{j \in S}U_{kj}\beta_j^*\right)\right)^{\frac{1}{2}}.
	\end{split}	
\end{equation*}
By \eqref{eq513}, if $\nu_{\min} \geq C$,
\begin{equation*}
	\begin{split}
		& \Var\left(\sum_{k=1}^{n}\left(\bar{\B}_{W'}\right)_{ki}\right) = \sum_{k=1}^{n}\Var\left(\bar{\B}_{W'}\right)_{ki}\\ 
		=& \sum_{k=1}^{n}\left[\left(1 - \frac{1}{p}\right)^2\Var\left(\phi_1(W_{ki}')\right)+ \frac{1}{p^2}\sum_{j \neq i}\Var\left(\phi_1(W_{kj}')\right)\right] \leq \frac{8n}{\nu_{\min}}.
	\end{split}
\end{equation*}
\begin{equation*}
	\Var\left(\sum_{j \in S}U_{kj}\beta_j^*\right) = \sum_{j \in S}\beta_j^{*2}\Var\left(U_{kj}\right) = \sum_{j \in S}\beta_j^{*2}\Var\left(\phi_1\left(W'_{kj}\right)\right) \leq \frac{8}{\nu_{\min}}\|\beta^*\|_2^2.
\end{equation*}
Hence
\begin{equation}\label{eq33}
	\left| E \left[\left(\bar{\B}_{W'}\right)_{ki}\sum_{j \in S}U_{kj}\beta_j^*\right] -  E \left(\bar{\B}_{W'}\right)_{ki} E \sum_{j \in S}U_{kj}\beta_j^*\right| \leq \frac{8\sqrt{n}}{\nu_{\min}}\|\beta^*\|_2.
\end{equation}
By (\ref{eq31}), (\ref{eq32}) and (\ref{eq33}),
\begin{equation}\label{eq8}
	\begin{split}
		\left| E \left[\P\left(\A_{W'} - \B_{W'}^\top  V\right)\beta^*\right]_i\right| =& \left|\sum_{k=1}^{n} E \left(\left(\bar{\B}_{W'}\right)_{ki}\sum_{j \in S}U_{kj}\beta_j^*\right)\right|\\
		\leq& \left| E \sum_{k=1}^{n}\left(\bar{\B}_{W'}\right)_{ki} E \sum_{j \in S}U_{kj}\beta_j^*\right| + \frac{8\sqrt{n}}{\nu_{\min}}\|\beta^*\|_2\\
		\leq& \sum_{k=1}^{n}\left|\left(\bar{\B}_{W'}\right)_{ki}\right|\left| E \sum_{j \in S}U_{kj}\beta_j^*\right| + \frac{8\sqrt{n}}{\nu_{\min}}\|\beta^*\|_2\\
		\leq& n\left(\log\left(\frac{b}{a}\right) + \frac{8}{\nu_{\min}}\right)\cdot \frac{4\sqrt{s}}{\nu_{\min}^{\frac{3}{2} - \epsilon}}\|\beta^*\|_2 + \frac{8\sqrt{n}}{\nu_{\min}}\|\beta^*\|_2\\
		\leq& Cn\frac{\sqrt{s}}{\nu_{\min}^{\frac{3}{2} - \epsilon}}\|\beta^*\|_2 + 8\frac{\sqrt{n}}{\sqrt{\nu_{\min}}}\|\beta^*\|_2.
	\end{split}
\end{equation}
Combine (\ref{eq7}) and (\ref{eq8}) together, we know that there exists a large constant $C$, for $1 \leq x \lesssim n$,
\begin{equation*}
	{\mathrm{pr}}\left(\left|\left[\P\left(\A_{W'} - \B_{W'}^\top  V\right)\beta^*\right]_i\right| \geq \sqrt{n}\frac{C}{\sqrt{\nu_{\min}}}\|\beta^*\|_2\sqrt{x} + Cn\frac{\sqrt{s}}{\nu_{\min}^{\frac{3}{2} - \epsilon}}\|\beta^*\|_2\right) \leq e^{-x}.
\end{equation*}
Since $\nu_{\min} \geq \frac{a^2}{b^2}\frac{\bar\nu}{p}$,
\begin{equation*}
	{\mathrm{pr}}\left(\left|\left[\P\left(\A_{W'} - \B_{W'}^\top  V\right)\beta^*\right]_i\right| \geq C\sqrt{n}\left(\frac{p}{\bar\nu}\right)^{\frac{1}{2}}\|\beta^*\|_2\sqrt{x} + Cn\sqrt{s}\left(\frac{p}{\bar{\nu}}\right)^{\frac{3}{2} - \epsilon}\|\beta^*\|_2\right) \leq e^{-x}.
\end{equation*}
Therefore,
\begin{equation*}
	\begin{split}
		&{\mathrm{pr}}\left(\|\P\left(\A_{W'} - \B_{W'}^\top  V\right)\beta^*\|_{\infty} \geq C\sqrt{n}\left(\frac{p}{\bar\nu}\right)^{\frac{1}{2}}\|\beta^*\|_2\sqrt{x} + Cn\sqrt{s}\left(\frac{p}{\bar{\nu}}\right)^{\frac{3}{2} - \epsilon}\|\beta^*\|_2\right)\\
		= & {\mathrm{pr}}\left(\exists 1 \leq i \leq p, \left|\left[\P\left(\A_{W'} - \B_{W'}^\top  V\right)\beta^*\right]_i\right| \geq C\sqrt{n}\left(\frac{p}{\bar\nu}\right)^{\frac{1}{2}}\|\beta^*\|_2\sqrt{x} + Cn\sqrt{s}\left(\frac{p}{\bar{\nu}}\right)^{\frac{3}{2} - \epsilon}\|\beta^*\|_2\right)\\	
		\leq& pe^{-x}.
	\end{split}
\end{equation*}
Set $x = \left(C' + 1\right)\log p$, where $C'$ is an absolute constant, of course $x \lesssim n$, we have
\begin{equation}\label{eq10}
	{\mathrm{pr}}\left(\|\P\left(\A_{W'} - \B_{W'}^\top  V\right)\beta^*\|_{\infty} \geq C\sqrt{n\log p}\left(\frac{p}{\bar\nu}\right)^{\frac{1}{2}}\|\beta^*\|_2 + Cn\sqrt{s}\left(\frac{p}{\bar{\nu}}\right)^{\frac{3}{2} - \epsilon}\|\beta^*\|_2\right) \leq e^{-C'\log p}.
\end{equation}
Choose $t = C\sqrt{n\log p\cdot\sigma^2}$ in (\ref{eq9}) and notice that $n \geq Cs\log p$,
\begin{equation*}
	{\mathrm{pr}}\left(\|\bar{\B}_{W'}^\top\epsilon\|_{\infty} > C\sqrt{n\log p\cdot\sigma^2}\right) \leq 2e^{-C'\log p}.
\end{equation*}
Since $\sqrt{x} + \sqrt{y} \leq \sqrt{2}\sqrt{x + y}$, we know that
\begin{equation}\label{ineq48}
	\begin{split}
		&{\mathrm{pr}}\left(\|\bar{\A}_{W'}\beta^* - \bar{\B}_{W'}^\top y\|_{\infty} > C\sqrt{n\log p\left(\sigma^2 + \frac{p}{\bar\nu}\|\beta^*\|_2^2\right) + Cn^2s\left(\frac{p}{\bar{\nu}}\right)^{3-2\epsilon}\|\beta^*\|_2^2}\right)\\
		\leq& {\mathrm{pr}}\left(\|\P\left(\A_{W'} - \B_{W'}^\top  V\right)\beta^*\|_{\infty} \geq C\sqrt{n\log p}\left(\frac{p}{\bar\nu}\right)^{\frac{1}{2}}\|\beta^*\|_2 + Cn\sqrt{s}\left(\frac{p}{\bar{\nu}}\right)^{\frac{3}{2} - \epsilon}\|\beta^*\|_2\right)\\
		& + {\mathrm{pr}}\left(\|\bar{\B}_{W'}^\top\epsilon\|_{\infty} > C\sqrt{n\log p\cdot\sigma^2}\right)\\
		\leq& 3e^{-C'\log p} = 3p^{-C'}.
	\end{split}	
\end{equation}
Note that $\nu_{\min} \gtrsim \frac{\bar{\nu}}{p} \geq C\log(np)$, by \eqref{eq506} and the previous inequality, immediately we have
\begin{equation}\label{ineq49}
	\begin{split}
		&{\mathrm{pr}}\left(\|\bar{\A}_{W}\beta^* - \bar{\B}_{W}^\top y\|_{\infty} > C\sqrt{n\log p\left(\sigma^2 + \frac{p}{\bar\nu}\|\beta^*\|_2^2\right) + Cn^2s\left(\frac{p}{\bar{\nu}}\right)^{3-2\epsilon}\|\beta^*\|_2^2}\right)\\
		\leq& {\mathrm{pr}}\left(\|\bar{\A}_{W'}\beta^* - \bar{\B}_{W'}^\top y\|_{\infty} > C\sqrt{n\log p\left(\sigma^2 + \frac{p}{\bar\nu}\|\beta^*\|_2^2\right) + Cn^2s\left(\frac{p}{\bar{\nu}}\right)^{3-2\epsilon}\|\beta^*\|_2^2}\right)\\
		&+ {\mathrm{pr}}\left(\exists 1 \leq i \leq n, 1 \leq j \leq p, W_{ij} \neq W'_{ij}\right)\\
		\leq& 3p^{-C'} + 2npe^{-c\nu_{\min}} \leq 3p^{-C'} + 2np\exp\left(-(C' + 1)\log(2np)\right)\leq 4p^{-C'}.
	\end{split}
\end{equation}
\qed
\medskip

{\noindent Proof of Lemma \ref{lm:beta-binomial_1}.}
By Taylor's expansion, for $x > -1$,
\begin{equation*}
	\log(x + 1) = x - \frac{x^2}{2} + \frac{x^3}{3(1 + \xi)^3}
\end{equation*}
for some number $\xi$ between 0 and $x$. 
Take $x = \frac{t + z_i - \nu_{ij}}{\nu_{ij}}$, where $t \geq 0, z_i = \frac{N_i + \alpha_i + 1}{2(1 + \alpha_i)}> 0$, $\nu_{ij} = \nu_iX_{ij}$, we have
\begin{equation*}
	\log(t + z_i) = \log \nu_{ij} + \frac{t + z_i - \nu_{ij}}{\nu_{ij}}- \frac{\left(t + z_i - \nu_{ij}\right)^2}{2\nu_{ij}^2} + \frac{\left(t + z_i - \nu_{ij}\right)^3}{3\left(1 + \xi_t\right)^3\nu_{ij}^3},
\end{equation*}
where $\xi_t$ is  a real number between 0 and $\frac{t - \nu_{ij} + z_i}{\nu_{ij}}$.
Denote $f(t) = \log(t + z_i) - \log \nu_{ij} - \frac{t + z_i - \nu_{ij}}{\nu_{ij}} + \frac{\left(t + z_i - \nu_{ij}\right)^2}{2\nu_{ij}^2}$. For $\nu_{ij} - \left(\nu_{ij}\zeta_i\right)^{\frac{1}{2} + \epsilon} \leq t \leq \nu_{ij} + \left(\nu_{ij}\zeta_i\right)^{\frac{1}{2} + \epsilon}$, $\exists C(\epsilon) > 0$, if $\nu_{ij} \geq C(\epsilon)$,
\begin{equation*}
	\left|\frac{(t - \nu_{ij})^3}{3(1 + \xi_t)^3\nu_{ij}^3}\right| \leq \frac{\left(\nu_{ij}\zeta_i\right)^{\frac{3}{2} + 3\epsilon}}{2\nu_{ij}^3}.
\end{equation*} 
Since $\nu_{ij} \geq \zeta_i^{1 + \delta}$, set $\epsilon = \frac{\delta}{24 + 12\delta} < \frac{\delta}{12 + 6\delta}$, we have
\begin{equation}\label{eq232}
	\left|\frac{(t - \nu_{ij})^3}{3(1 + \xi_t)^3\nu_{ij}^3}\right| \leq \frac{\left(\nu_{ij}\zeta_i\right)^{\frac{3}{2} + 3\frac{\delta}{12 + 6\delta}}}{2\nu_{ij}^3} = \frac{\zeta_{i}\cdot \zeta_{i}^{\frac{1 + \delta}{2 + \delta}}}{2\nu_{ij}^{\frac{3}{2} - 3\frac{\delta}{12 + 6\delta}}}\leq \frac{\zeta_{i}\cdot \nu_{ij}^{\frac{1}{2 + \delta}}}{2\nu_{ij}^{\frac{3}{2} - 3\frac{\delta}{12 + 6\delta}}} = \frac{\zeta_i}{2\nu_{ij}}.
\end{equation}
Denote $r_{ij} = \left(\nu_{ij}\zeta_i\right)^{\frac{1}{2} + \frac{\delta}{24 + 12\delta}}$, then $r_{ij} \leq \left(\nu_{ij}^{\frac{2 + \delta}{1 + \delta}}\right)^{\frac{12 + 7\delta}{24 + 12\delta}} = \nu_{ij}^{\frac{12 + 7\delta}{12 + 12\delta}} < \nu_{ij}$. If $\nu_i \geq C(\delta) \geq C$, then $N_i \sim \nu_{i}$ implies $ E  N_i^k \leq 2\nu_i^k$ for $k = 1, 2, 3$, and 
\begin{equation*}
	E  z_i^3 =  E  \left(\frac{N_i + \alpha_i + 1}{2(\alpha_i + 1)}\right)^3 \leq 2\left(\frac{\nu_i + \alpha_i + 1}{2\left(\alpha_i + 1\right)}\right)^3 = 2\zeta_i^3.
\end{equation*}
Also note that $\nu_{ij} \geq \zeta_i^{1 + \delta}$,
\begin{equation*}
	E \frac{z_i^3}{3\left(1 + \xi_t\right)^3\nu_{ij}^3}1_{\{\nu_{ij} - r_{ij} < W_{ij}' < \nu_{ij} + r_{ij}\}} \leq  E \frac{z_i^3}{3\left(1 + \xi_t\right)^3\nu_{ij}^3} \leq \frac{2\zeta_i^3}{2\nu_{ij}^3} \leq \frac{\zeta_i}{\nu_{ij}}.
\end{equation*}
Combine \eqref{eq232} and the previous inequality together, 
\begin{equation}\label{eq201}
	\begin{split}
		&E \left|f(W_{ij}')1_{\{\nu_{ij} - r_{ij} < W_{ij}' \leq \nu_{ij} + r_{ij}\}}\right| =  E \left|\frac{\left(t + z_i - \nu_{ij}\right)^3}{3\left(1 + \xi_t\right)^3\nu_{ij}^3}1_{\{\nu_{ij} - r_{ij} < W_{ij}' \nu_{ij} + r_{ij}\}}\right| \\ \leq& 4\left( E 	\left|\frac{(t - \nu_{ij})^3}{3(1 + \xi_t)^3\nu_{ij}^3}1_{\{\nu_{ij} - r_{ij} < W_{ij}' < \nu_{ij} + r_{ij}\}}\right| +  E \frac{z_i^3}{3\left(1 + \xi_t\right)^3\nu_{ij}^3}1_{\{\nu_{ij} - r_{ij} < W_{ij}' < \nu_{ij} + r_{ij}\}} \right)\\  \leq& 4\left(\frac{\zeta_i}{2\nu_{ij}} + \frac{\zeta_i}{\nu_{ij}}\right) = 6\frac{\zeta_i}{\nu_{ij}}.
	\end{split}
\end{equation} 
The first inequality holds since $\left|(t + z_i - \nu_{ij})^3\right| \leq 4\left(|t - \nu_{ij}|^3 + z_i^3\right).$\\ 
For $t \leq \nu_{ij} - r_{ij}, \nu_{ij} \geq 1$, we have 
\begin{equation*}
	\begin{split}
		\left|f(t)\right| \leq& \max\{\log\left(t + z_i\right), \log \nu_{ij}\} + \left|\frac{t - \nu_{ij} + z_i}{\nu_{ij}}\right| + \left|\frac{\left(t - \nu_{ij} + z_i\right)^2}{2\nu_{ij}^2}\right|\\
		\leq& \log(\nu_{ij} + z_i) + \left|\frac{t - \nu_{ij}}{\nu_{ij}}\right| + \frac{z_i}{\nu_{ij}} + 2\frac{(t - \nu_{ij})^2}{2\nu_{ij}^2} + 2\frac{z_i^2}{2\nu_{ij}^2}\\
		\leq& \left(\log(\nu_{ij}) + \frac{z_i}{\nu_{ij}}\right) + 1 + \frac{z_i}{\nu_{ij}} + 1 + \frac{z_i^2}{\nu_{ij}^2}\\
		\leq& \log \nu_{ij} + 2 + \frac{z_i}{\nu_{ij}} + \frac{z_i}{\nu_{ij}} + \frac{z_i^2}{\nu_{ij}^2}.
	\end{split}
\end{equation*}
The third inequality holds since $\log(u + v) \leq \log(u) + \frac{v}{u}$ for $u, v > 0$, and $|t - \nu_{ij}| \leq \nu_{ij}$ for $0 \leq t \leq \nu_{ij} - r_{ij}$.
By Lemma \ref{lm:tail bound}, we have
\begin{equation*}
	{\mathrm{pr}}\left(W_{ij}' \leq \nu_{ij} - t\right) \leq 3\exp\left(-ct^2\min\left\{\frac{\alpha_i}{\nu_{i}\cdot \alpha_iX_{ij}}, \frac{\alpha_i^2}{\nu_{i}^2\cdot \alpha_{i}X_{ij}}\right\}\right) \leq 3\exp\left(-ct^2\frac{1}{\nu_{ij}}\frac{\alpha_i}{\alpha_i + \nu_{i}}\right).
\end{equation*}
Set $t = r_{ij}$ and note that
\begin{equation*}
	r_{ij}^2\frac{1}{\nu_{ij}}\frac{\alpha_i}{\alpha_i + \nu_{i}} \geq \left(\nu_{ij}\zeta_{i}\right)^{2(1 + \epsilon)}\frac{1}{4\nu_{ij}\zeta_{i}} = \frac{\left(\nu_{ij}\zeta_{i}\right)^{2\epsilon}}{4},
\end{equation*}
where $\epsilon = \frac{\delta}{24 + 12\delta}$. Also note that $\zeta_i \geq \frac{1}{2}$, we have
\begin{equation}\label{eq214}
	\begin{split}
		{\mathrm{pr}}\left(W_{ij}' \leq \nu_{ij} - r_{ij}\right) 
		&\leq 3\exp\left(-c\left(\nu_{ij}\zeta_i\right)^{2\epsilon}\right) \leq 3\exp\left(-c\nu_{ij}^{2\epsilon}\right).
	\end{split}
\end{equation}
Therefore $\exists C(\delta)$, if $\nu_{ij} \geq C(\delta)$,
\begin{equation}\label{eq202}
	\begin{split}
		&\left| E  f(W_{ij}')1_{\{W_{ij}' \leq \nu_{ij} - r_{ij}\}}\right|\\ \leq& (\log \nu_{ij} + 2){\mathrm{pr}}\left(W_{ij}' \leq \nu_{ij} - r_{ij}\right) + \frac{2}{\nu_{ij}} E  z_i1_{\{W_{ij}' \leq \nu_{ij} - r_{ij}\}} + \frac{1}{\nu_{ij}^2} E  z_i^21_{\{W_{ij}' \leq \nu_{ij} - r_{ij}\}}\\
		\leq& (\log \nu_{ij} + 2)3\exp\left(-c\nu_{ij}^{2\epsilon}\right) + \frac{2}{\nu_{ij}}( E  z_i^2)^{\frac{1}{2}}( E  1_{\{W_{ij}' \leq \nu_{ij} - r_{ij}\}})^{\frac{1}{2}} + \frac{1}{\nu_{ij}^2}( E  z_i^4)^{\frac{1}{2}}( E  1_{\{W_{ij}' \leq \nu_{ij} - r_{ij}\}})^{\frac{1}{2}}\\
		\leq& (\log \nu_{ij} + 2)3\exp\left(-c\nu_{ij}^{2\epsilon}\right) + \frac{2}{\nu_{ij}}\cdot 2\zeta_i\cdot 3\exp\left(-c\nu_{ij}^{2\epsilon}\right) + \frac{1}{\nu_{ij}^2}2\zeta_i^2\cdot 3\exp\left(-c\nu_{ij}^{2\epsilon}\right)\\
		\leq& (\log \nu_{ij} + 2)3\exp\left(-c\nu_{ij}^{2\epsilon}\right) + 6\left(2\frac{\zeta_i}{\nu_{ij}} + \frac{\zeta_i}{\nu_{ij}^2}\right)\exp\left(-c\nu_{ij}^{2\epsilon}\right) \leq\frac{\zeta_i}{\nu_{ij}}.
	\end{split}
\end{equation}
In the second inequality, we used Cauchy-Schwarz inequality; the third inequality comes from $ E  z_i^2 \leq 2\zeta_i^2,  E  z_i^4 \leq 2\zeta_i^4$ for $z_i \geq C(\delta) \geq C$.\\
For $t \geq \nu_{ij} + r_{ij}, \nu_{ij} \geq 1$, 
\begin{equation*}
	\begin{split}
		\left|f(t)\right| 
		\leq& \left|\log\left(t + z_i\right) - \log \nu_{ij}\right| + \left|\frac{\left(t - \nu_{ij} + z_i\right)^2}{2\nu_{ij}^2} - \frac{t - \nu_{ij} + z_i}{\nu_{ij}}\right| \\
		=& \left|\log\left(t + z_i\right) - \log \nu_{ij}\right| + \frac{t - \nu_{ij} + z_i}{\nu_{ij}}\left|\frac{t + z_i}{2\nu_{ij}} - \frac{3}{2}\right| \\
		\leq& \log\left(t + z_i\right) + \frac{3\left(t + z_i\right)^2}{2\nu_{ij}^2} 
		\leq t + z_i + \frac{3\left(t + z_i\right)^2}{2\nu_{ij}^2}.
	\end{split}
\end{equation*}
Therefore, for $\nu_{ij} \geq C_3\left(\delta\right)$, by Cauchy-Schwarz inequality,
\begin{equation}\label{eq216}
	\begin{split}
		&\left|E f(W_{ij}')1_{\{W_{ij}' \geq \nu_{ij} + r_{ij}\}}\right|\\
		\leq&E (W_{ij}' + z_i)1_{\{W_{ij}' \geq \nu_{ij} + r_{ij}\}} + 3E \frac{(W_{ij}' + z_i)^2}{\nu_{ij}^2}1_{\{W_{ij}' \geq \nu_{ij} + r_{ij}\}}\\
		\leq&E W_{ij}'\cdot1_{\{W_{ij}' \geq \nu_{ij} + r_{ij}\}} + 6E \frac{W_{ij}'^2}{\nu_{ij}^2} 1_{\{W_{ij}' \geq \nu_{ij} + r_{ij}\}} + 6E \frac{z_i^2}{\nu_{ij}^2} 1_{\{W_{ij}' \geq \nu_{ij} + r_{ij}\}} + E z_i 1_{\{W_{ij}' \geq \nu_{ij} +r_{ij}\}}.
	\end{split}
\end{equation}
The last inequality is due to Cauchy-Schwarz inequality.

Notice that $\exists$ constant $C > 0$ such that $ E  q_{ij}^k \leq CX_{ij}^k$ and $ E \left(W_{ij}^k|q_{ij}\right) \leq C\left(\nu_iq_{ij}\right)^k$ for $1 \leq k \leq 4$, we have
\begin{equation}\label{eq245}
	E  W_{ij}^k \leq C\nu_{ij}^k
\end{equation}
for $k = 1, 2, 3, 4$.
Similar to \eqref{eq214}, 
\begin{equation}\label{eq215}
	\begin{split}
		{\mathrm{pr}}\left(W_{ij}' \geq \nu_{ij} + r_{ij}\right) 
		&\leq 3\exp\left(-c\nu_{ij}^{2\epsilon}\right).
	\end{split}
\end{equation}
By Cauchy-Schwarz inequality,
\begin{equation}\label{eq241}
	\begin{split}
		E W_{ij}1_{\left\{W_{ij}' \geq \nu_{ij} + r_{ij}\right\}} \leq \left( E  W_{ij}^2 E 1_{\left\{W_{ij}' \geq \nu_{ij} + r_{ij}\right\}}\right)^{\frac{1}{2}}
		\leq \left(C\nu_{ij}^2 \cdot 3\exp\left(-c\nu_{ij}^{2\epsilon}\right)\right)^{\frac{1}{2}}
		\leq C\nu_{ij}\exp\left(-c\nu_{ij}^{2\epsilon}\right).
	\end{split}
\end{equation}
\begin{equation}\label{eq242}
	\begin{split}
		E W_{ij}^2 1_{\left\{W_{ij}' \geq \nu_{ij} + r_{ij}\right\}} \leq \left( E  W_{ij}^4 E  1_{\left\{W_{ij} \geq \nu_{ij} + r_{ij}\right\}}\right)^{\frac{1}{2}}
		\leq \left(C\nu_{ij}^4 \cdot 3\exp\left(-c\nu_{ij}^{2\epsilon}\right)\right)^{\frac{1}{2}}
		\leq C\nu_{ij}^2\exp\left(-c\nu_{ij}^{2\epsilon}\right).
	\end{split}	    	
\end{equation}
In addition,  Cauchy-Schwarz inequality tells us
\begin{equation}\label{eq243}
	\begin{split}
		E  \left|W_{ij} - W_{ij}'\right|1_{\{W_{ij}' \geq \nu_{ij} + r_{ij}\}} \leq& \left( E \left(W_{ij} - W_{ij}'\right)^2{\mathrm{pr}}\left(W_{ij}' \geq \nu_{ij} + r_{ij}\right)\right)^{\frac{1}{2}}.
	\end{split}
\end{equation}
Since 
\begin{equation*}
	\begin{split}
		E \left(W_{ij} - W_{ij}'\right)^2 =&  E \left(W_{ij} - W_{ij}'\right)^21_{\{W_{ij} \neq W_{ij}'\}}
		\leq 2 E  \left[(W_{ij}^2 + W_{ij}'^2)1_{\{W_{ij} \neq W_{ij}'\}}\right]\\
		\leq& 2 E  \left[(W_{ij}^2 + \nu_{ij}^2)1_{\{W_{ij} \neq W_{ij}'\}}\right] \leq 2 E  W_{ij}^2 + 2\nu_{ij}^2 \leq C\nu_{ij}^2 + 2\nu_{ij}^2,	
	\end{split}
\end{equation*}
By \eqref{eq215} and \eqref{eq243}, we have
\begin{equation*}
	\begin{split}
		E  \left|W_{ij} - W_{ij}'\right|1_{\{W_{ij}' \geq \nu_{ij} + r_{ij}\}} \leq& \left[C\nu_{ij}^2\cdot 3\exp\left(-c\nu_{ij}^{2\epsilon}\right)\right]^{\frac{1}{2}} \leq C\nu_{ij}\exp\left(-c\nu_{ij}^{2\epsilon}\right).
	\end{split}
\end{equation*}
Combine \eqref{eq241} and the previous inequality together,
\begin{equation}\label{eq217}
	E  W_{ij}'1_{\{W_{ij}' \geq \nu_{ij} + r_{ij}\}} \leq E W_{ij}1_{\left\{W_{ij}' \geq \nu_{ij} + r_{ij}\right\}} +  E  \left|W_{ij} - W_{ij}'\right|1_{\{W_{ij}' \geq \nu_{ij} + r_{ij}\}} \leq C\nu_{ij}\exp\left(-c\nu_{ij}^{2\epsilon}\right).
\end{equation}
Similarly,
\begin{equation}\label{eq244}
	\begin{split}
		& E \left|W_{ij}^2 - W_{ij}'^2\right|1_{\{W_{ij}' \geq \nu_{ij} + r_{ij}\}} \leq \left( E \left(W_{ij}^2 - W_{ij}'^2\right)^2 E 1_{\{W_{ij}' \geq \nu_{ij} + r_{ij}\}}\right)^{\frac{1}{2}}\\
		\leq& \left[\left(2 E  W_{ij}^4 + 2\nu_{ij}^4\right)\cdot 3\exp\left(-c\nu_{ij}^{2\epsilon}\right)\right]^{\frac{1}{2}}
		\leq \left[\left(C\nu_{ij}^4 + 2\nu_{ij}^4\right)\cdot 3\exp\left(-c\nu_{ij}^{2\epsilon}\right)\right]^{\frac{1}{2}}
		\leq C\nu_{ij}^2\exp\left(-c\nu_{ij}^{2\epsilon}\right).
	\end{split}
\end{equation}
\eqref{eq242} and \eqref{eq244} imply that
\begin{equation}\label{eq218}
	\begin{split}
		E  W_{ij}'^21_{\{W_{ij}' \geq \nu_{ij} + r_{ij}\}} \leq  E \left|W_{ij}^2 - W_{ij}'^2\right|1_{\{W_{ij}' \geq \nu_{ij} + r_{ij}\}} +  E  W_{ij}^21_{\{W_{ij}' \geq \nu_{ij} + r_{ij}\}} \leq C\nu_{ij}^2\exp\left(-c\nu_{ij}^{2\epsilon}\right).
	\end{split}
\end{equation}
Also, by \eqref{eq215} and Cauchy-Schwarz inequality and notice that $\nu_{ij}^2 \geq \left(\frac{\nu_{i} + \alpha_i + 1}{\alpha + 1}\right)^2 = \zeta_i^2$,
\begin{equation}\label{eq219}
	\begin{split}
		\left( E  z_i^2 1_{\left\{W_{ij}' \geq \nu_{ij} + r_{ij}\right\}}\right)^2 \leq  E  z_i^4  E  1_{\left\{W_{ij}' \geq \nu_{ij} + r_{ij}\right\}}
		\leq 2\zeta_{i}^4{\mathrm{pr}}\left(W_{ij}' \geq \nu_{ij} + r_{ij}\right)
		\leq 6\nu_{ij}^4\exp\left(-c\nu_{ij}^{2\epsilon}\right),
	\end{split}
\end{equation}
\begin{equation}\label{eq220}
	\begin{split}
		\left( E  z_i 1_{\left\{W_{ij}' \geq \nu_{ij} + r_{ij}\right\}}\right)^2 \leq  E  z_i^2  E  1_{\left\{W_{ij}' \geq \nu_{ij} + r_{ij}\right\}}
		\leq 2\zeta_{i}^2{\mathrm{pr}}\left(W_{ij}' \geq \nu_{ij} + r_{ij}\right)
		\leq 6\nu_{ij}^2\exp\left(-c\nu_{ij}^{2\epsilon}\right).
	\end{split}
\end{equation}
Combine (\ref{eq216}), (\ref{eq217}), (\ref{eq218}), (\ref{eq219}) and (\ref{eq220}) together, we know that for $\nu_{ij} \geq C(\delta) = C(\epsilon)$,
\begin{equation}\label{eq203}
	\begin{split}
		&\left|E f(W_{ij}')1_{\{W \geq \nu_{ij} + r_{ij}\}}\right|\\
		\leq& C\nu_{ij}\exp\left(-c\nu_{ij}^{2\epsilon}\right) + \frac{C}{\nu_{ij}^2}\nu_{ij}^2\exp\left(-c\nu_{ij}^{2\epsilon}\right) + \frac{C}{\nu_{ij}^2}\nu_{ij}^2\exp\left(-c\nu_{ij}^{2\epsilon}\right) + C\nu_{ij}\exp\left(-c\nu_{ij}^{\epsilon}\right)\\
		\leq& \frac{\zeta_i}{\nu_{ij}}.
	\end{split}
\end{equation}
Combine (\ref{eq201}), (\ref{eq202}) and (\ref{eq203}), we know that $\exists C(\delta) > 0$ such that for all $\nu_{ij} \geq C(\delta)$,
\begin{equation}\label{eq247}
	\left| E  f(W_{ij}')\right| \leq E \left|f(W_{ij}')1_{\{\nu_{ij} - r_{ij} < W_{ij}' \leq \nu_{ij} + r_{ij}\}}\right| + \left| E  f(W_{ij}')1_{\{W_{ij}' \leq \nu_{ij} - r_{ij}\}}\right| + \left|E f(W_{ij}')1_{\{W \geq \nu_{ij} + r_{ij}\}}\right| \leq 8\frac{\zeta_i}{\nu_{ij}}.
\end{equation}
Next, we would like to bound $\left| E \left(\frac{W_{ij} + z_i - \nu_{ij}}{\nu_{ij}}- \frac{\left(W_{ij} + z_i - \nu_{ij}\right)^2}{2\nu_{ij}^2}\right)\right|$. Actually,
\begin{equation*}
	\begin{split}
		&\left| E \left(\frac{W_{ij} + z_i - \nu_{ij}}{\nu_{ij}}- \frac{\left(W_{ij} + z_i - \nu_{ij}\right)^2}{2\nu_{ij}^2}\right)\right|\\
		=& \left| E \frac{z_i}{\nu_{ij}} -  E \frac{(W_{ij} - \nu_{ij})^2}{2\nu_{ij}^2} -  E \frac{z_i^2}{2\nu_{ij}^2} -  E \frac{z_i(W_{ij} - \nu_{ij})}{2\nu_{ij}^2}\right|\\
		\leq& \left|\frac{\frac{\nu_i + \alpha_i + 1}{2(\alpha_i + 1)}}{\nu_i X_{ij}} - \frac{\nu_i X_{ij} + \nu_i^2\frac{X_{ij}(1 - X_{ij})}{\alpha_i + 1}}{2\nu_i^2X_{ij}^2}\right| +  E \frac{z_i(W_{ij} - \nu_{ij})}{2\nu_{ij}^2} +  E \frac{z_i^2}{2\nu_{ij}^2}\\
		=& \frac{1}{2(\alpha_i + 1)} + \frac{ E  z_i(W_{ij} - \nu_{ij})}{2\nu_{ij}^2} + \frac{ E  z_i^2}{2\nu_{ij}^2}.
	\end{split}
\end{equation*}
Note that $ E  \zeta_i(W_{ij} - \nu_{ij}) = \zeta_i E (W_{ij} - \nu_{ij}) = 0$ and $ E  z_i^2 \leq 2\zeta_i^2$ if $\nu_{ij} \geq C(\delta) \geq C$, we have
\begin{equation*}
	\begin{split}
		&\left| E \left(\frac{W_{ij} + z_i - \nu_{ij}}{\nu_{ij}}- \frac{\left(W_{ij} + z_i - \nu_{ij}\right)^2}{2\nu_{ij}^2}\right)\right|\\
		\leq& \frac{1}{2(\alpha_i + 1)} + \frac{ E (z_i -\zeta_i)(W_{ij} - \nu_{ij})}{2\nu_{ij}^2} + \frac{ E  z_i^2}{2\nu_{ij}^2}\\
		\leq& \frac{1}{2(\alpha_i + 1)} + \frac{1}{2\nu_{ij}^2}\left( E \left(\frac{N_i - \nu_i}{2(\alpha_i + 1)}\right)^2\right)^{1/2}\left( E (W_{ij} - \nu_{ij})^2\right)^{1/2} + \frac{\zeta_{i}^2}{\nu_{ij}^2}\\
		\leq& \frac{1}{2(\alpha_i + 1)} + \frac{1}{2\nu_{ij}^2}\left(\frac{\nu_i}{4(\alpha_i + 1)^2}\right)^{1/2}\left(\nu_i X_{ij} + \nu_i^2\frac{X_{ij}(1 - X_{ij})}{\alpha_i + 1}\right)^{1/2} + \frac{\zeta_i^2}{\nu_{ij}^2}.
	\end{split}
\end{equation*}
Since $\frac{1}{2(\alpha_i + 1)} = \frac{\zeta_i}{\nu_i + \alpha_i + 1} \leq \frac{\zeta_i}{\nu_{i}} \asymp \frac{\zeta_i}{p\nu_{ij}}$, $\frac{\nu_{i}}{4(\alpha_i + 1)^2} \leq \frac{\zeta_i}{2(\alpha_i + 1)} \leq \zeta_i$ and $\nu_i X_{ij} + \nu_i^2\frac{X_{ij}(1 - X_{ij})}{\alpha_i + 1} \leq \nu_i X_{ij} + \nu_i^2\frac{X_{ij}}{\alpha_i + 1} = 2\nu_{ij}\zeta_i$, we have
\begin{equation}\label{eq246}
	\begin{split}
		&\left| E \left(\frac{W_{ij} + z_i - \nu_{ij}}{\nu_{ij}}- \frac{\left(W_{ij} + z_i - \nu_{ij}\right)^2}{2\nu_{ij}^2}\right)\right|
		\leq  C\frac{\zeta_i}{p\nu_{ij}} + \frac{C}{\nu_{ij}^2}\left(\zeta_{i}\right)^{1/2}\left(\nu_{ij}\zeta_i\right)^{1/2} + \frac{\zeta_{i}^2}{\nu_{ij}^2}
		\leq \frac{\zeta_i}{\nu_{ij}}.
	\end{split}
\end{equation}
The last inequality holds since $\frac{\nu_{ij}}{\zeta_i} \geq \frac{\bar{\nu}}{p\zeta_{\max}} \geq C(\delta) \geq C$ and $\nu_{ij} \geq Cp\zeta_i \geq C$.\\ 
Denote $\mathcal{T}_{ij} = \left[\frac{X_{ij}}{10}, 10X_{ij}\right]$, by Lemma \ref{lm:trancated poisson tail bound} and \ref{lm:trancated Beta tail bound},
\begin{equation}\label{eq228}
	\begin{split}
		{\mathrm{pr}}\left(W_{ij} \neq W_{ij}'\right) \leq& {\mathrm{pr}}\left(q_{ij} \notin \mathcal{T}_{ij}\right) +  E \left[{\mathrm{pr}}\left(W_{ij} \neq W_{ij}'| q_{ij}\right)1_{\left\{q_{ij} \in \mathcal{T}_{ij}\right\}}\right]\\
		\leq& {\mathrm{pr}}\left(q_{ij} < \frac{1}{10}X_{ij}\right) + {\mathrm{pr}}\left(q_{ij} > 10X_{ij}\right) +  E \left[\left({\mathrm{pr}}\left(W_{ij} < \frac{1}{10}\nu_{i}q_{ij}| q_{ij}\right)\right)1_{\left\{q_{ij} \in \mathcal{T}_{ij}\right\}}\right]\\
		& +  E \left[\left({\mathrm{pr}}\left(W_{ij} > 10\nu_{i}q_{ij}| q_{ij}\right)\right)1_{\left\{q_{ij} \in \mathcal{T}_{ij}\right\}}\right]\\
		\leq& 4\exp\left(-cp\alpha_iX_{ij}^2\right) + 2 E \exp\left(-c\nu_{i}q_{ij}\right)1_{\left\{q_{ij} \in \mathcal{T}_{ij}\right\}}\\
		\leq& 4\exp\left(-c\frac{\alpha_i}{p}\right) + 2\exp\left(-c\nu_{ij}\right).
	\end{split}
\end{equation}
Since $\nu_{ij} \geq \zeta_i^{1 + \delta} \gtrsim \left(\frac{\nu_{i}}{\alpha_i}\right)^{1 + \delta} \gtrsim \left(\nu_{ij}\right)^{1 + \delta}\left(\frac{p}{\alpha_i}\right)^{1 + \delta}$, we have 
\begin{equation}\label{eq224}
	\frac{\alpha_i}{p} \gtrsim \left(\nu_{ij}\right)^{\frac{\delta}{1 + \delta}}.
\end{equation}
Therefore,
\begin{equation*}
	{\mathrm{pr}}\left(W_{ij} \neq W_{ij}'\right) \leq 6\exp\left(-c\left(\nu_{ij}\right)^{\frac{\delta}{1 + \delta}}\right),
\end{equation*}
which means
\begin{equation}\label{eq221}
	\begin{split}
		E  |W_{ij} - W_{ij}'| =&  E |W_{ij} - W_{ij}'|1_{\{W_{ij} \neq W_{ij}'\}} =  E |W_{ij} - \nu_{ij}|1_{\{W_{ij} \neq W_{ij}'\}}\\
		\leq& \left( E (W_{ij} - \nu_{ij})^2{\mathrm{pr}}\left(W_{ij} \neq W_{ij}'\right)\right)^{\frac{1}{2}}\\
		\leq& \left(2\left( E  W_{ij}^2 + \nu_{ij}^2\right)6\exp\left(-c\left(\nu_{ij}\right)^{\frac{\delta}{1 + \delta}}\right)\right)^{\frac{1}{2}}\\
		\leq& C\nu_{ij}\exp\left(-c(\nu_{ij})^{\frac{\delta}{1 + \delta}}\right).
	\end{split}
\end{equation}	
In the first and second inequality, we used Cauchy-Schwarz inequality; the third inequality comes from \eqref{eq245}.

Similarly,
\begin{equation}\label{eq222}
	\begin{split}
		E  |W_{ij}^2 - W_{ij}'^2| =&  E |W_{ij}^2 - W_{ij}'^2|1_{\{W_{ij} \neq W_{ij}'\}} =  E |W_{ij}^2 - \nu_{ij}^2|1_{\{W_{ij} \neq W_{ij}'\}}\\
		\leq& \left( E (W_{ij}^2 - \nu_{ij}^2)^2{\mathrm{pr}}\left(W_{ij} \neq W_{ij}'\right)\right)^{\frac{1}{2}}\\
		\leq& \left(2\left( E  W_{ij}^4 + \nu_{ij}^4\right)6\exp\left(-c\left(\nu_{ij}\right)^{\frac{\delta}{1 + \delta}}\right)\right)^{\frac{1}{2}}\\
		\leq& C\nu_{ij}^2\exp\left(-c(\nu_{ij})^{\frac{\delta}{1 + \delta}}\right).
	\end{split}
\end{equation}
\begin{equation}\label{eq223}
	\begin{split}
		E  (W_{ij} - W_{ij}')^2 =&  E (W_{ij} - W_{ij}')^21_{\{W_{ij} \neq W_{ij}'\}} =  E (W_{ij} - \nu_{ij})^21_{\{W_{ij} \neq W_{ij}'\}}\\
		\leq& \left( E (W_{ij} - \nu_{ij})^4{\mathrm{pr}}\left(W_{ij} \neq W_{ij}'\right)\right)^{\frac{1}{2}}\\
		\leq& \left(8\left( E  W_{ij}^4 + \nu_{ij}^4\right)6\exp\left(-c\left(\nu_{ij}\right)^{\frac{\delta}{1 + \delta}}\right)\right)^{\frac{1}{2}}\\
		\leq& C\nu_{ij}^2\exp\left(-c(\nu_{ij})^{\frac{\delta}{1 + \delta}}\right).
	\end{split}
\end{equation}
Combine \eqref{eq246}, \eqref{eq222} and \eqref{eq223} together,
\begin{equation}\label{eq248}
	\begin{split}
		&\left| E \left(\frac{W_{ij}' + z_i - \nu_{ij}}{\nu_{ij}}- \frac{\left(W_{ij}' + z_i - \nu_{ij}\right)^2}{2\nu_{ij}^2}\right)\right|\\
		\leq &\left| E \left(\frac{W_{ij} + z_i - \nu_{ij}}{\nu_{ij}}- \frac{\left(W_{ij} + z_i - \nu_{ij}\right)^2}{2\nu_{ij}^2}\right)\right| + 2\left| E \frac{W_{ij} - W_{ij}'}{\nu_{ij}}\right|\\ &+ \left| E \frac{W_{ij}^2 - W_{ij}'^2}{2\nu_{ij}^2}\right| + \left| E \frac{z_i(W_{ij}' - W_{ij})}{\nu_{ij}^2}\right|\\
		\leq& \frac{\zeta_i}{\nu_{ij}} + C\exp\left(-c(\nu_{ij})^{\frac{\delta}{1 + \delta}}\right) + \frac{\left( E  z_i^2 E  \left(W_{ij} - W_{ij}'\right)^2\right)^{\frac{1}{2}}}{\nu_{ij}^2}\\
		\leq& \frac{\zeta_i}{\nu_{ij}} + C\exp\left(-c(\nu_{ij})^{\frac{\delta}{1 + \delta}}\right) + \frac{C\zeta_i\nu_{ij}\exp\left(-c(\nu_{ij})^{\frac{\delta}{1 + \delta}}\right)}{\nu_{ij}^2} \leq 2\frac{\zeta_{i}}{\nu_{ij}}.
	\end{split}
\end{equation}
In the second inequality we also used Cauchy-Schwarz inequality; the third inequality comes from \eqref{eq223} and $ E  z_i^2 \leq 2\zeta_i^2$ if $\nu_{i} \geq C$.

\eqref{eq247} and \eqref{eq248} together imply that
\begin{equation*}
	\begin{split}
		\left| E \left(W_{ij}' + z_i\right) - \log \nu_{ij}\right| \leq&  \left| E \left(\frac{W_{ij}' + z_i - \nu_{ij}}{\nu_{ij}}- \frac{\left(W_{ij}' + z_i - \nu_{ij}\right)^2}{2\nu_{ij}^2}\right)\right| + \left| E  f(W_{ij}')\right|
		\leq 10\frac{\zeta_i}{\nu_{ij}}.
	\end{split}
\end{equation*}
\qed
\medskip

{\noindent Proof of Lemma \ref{lm:beta-binomial_2}.}
\begin{equation*}
	\log^2(\nu_{ij} + x) = \log^2\nu_{ij} + 2\frac{\log \nu_{ij}}{\nu_{ij}}x + \frac{1-\log \nu_{ij}}{\nu_{ij}^2}x^2 + \frac{2\log(\nu_{ij} +\xi)-3}{3(\nu_{ij}+\xi)^3}x^3,
\end{equation*}
where $\xi$ is  a real number between 0 and $x$.

Let $x = t - \nu_{ij} + z_i$, we have
\begin{equation*}
	\begin{split}
		\log^2\left(t + z_i\right) =& \log^2 \nu_{ij} + 2\frac{\log \nu_{ij}}{\nu_{ij}}\left(t - \nu_{ij} + z_i\right)+ \frac{1-\log \nu_{ij}}{\nu_{ij}^2}\left(t - \nu_{ij} + z_i\right)^2\\ &+ \frac{2\log(\nu_{ij} +\xi_t) - 3}{3(\nu_{ij}+\xi_t)^3}\left(t - \nu_{ij} + z_i\right)^3,
	\end{split}
\end{equation*}
where $\xi_t$ is  a real number between 0 and $t - \nu_{ij} + z_i$.

Let $g(t) = \log^2(t + z_i) - \log^2\nu_{ij} - 2\frac{\log \nu_{ij}}{\nu_{ij}}(t - \nu_{ij} + z_i) - \frac{1-\log \nu_{ij}}{\nu_{ij}^2}(t - \nu_{ij} + z_i)^2$, then for any $\nu_{ij} - r_{ij} < t < \nu_{ij} + r_{ij}$ (where $r_{ij} = \left(\nu_{ij}\zeta_{i}\right)^{\frac{1}{2} + \frac{\delta}{24 + 12\delta}}$ defined in the proof of Lemma \ref{lm:beta-binomial_1}), $\nu_{ij} \geq C(\delta)$ , we have
\begin{equation*}
	\begin{split}
		|g(t)| \leq& \frac{2\log(\nu_{ij} + \xi_t)}{3(\nu_{ij} + \xi_t)^3} \left|t - \nu_{ij} + z_i\right|^3
		\leq \frac{2\log\left(\nu_{ij} - r_{ij}\right)}{3\left(\nu_{ij} - r_{ij}\right)^3}\left(r_{ij} + z_i\right)^3
		\leq \frac{\log\nu_{ij}}{\nu_{ij}^3}\cdot 4\left(r_{ij}^3 + z_i^3\right)\\
		=& 4\frac{\log \nu_{ij}}{\nu_{ij}^3}\left(\nu_{ij}\zeta_{i}\right)^{\frac{3}{2} + \frac{\delta}{8 + 4\delta}} + 4\frac{\log \nu_{ij}}{\nu_{ij}^3}z_i^3 \leq 4\frac{\zeta_i}{\nu_{ij}}\frac{\log(\nu_{ij})\nu_{ij}^{\frac{4 + 3\delta}{(8 + 4\delta)(\delta + 1)}}}{\nu_{ij}^{\frac{4 + \delta}{8 + 4\delta}}} + 4\frac{\log \nu_{ij}}{\nu_{ij}^3}z_i^3\\
		=& 4\frac{\log\nu_{ij}}{\nu_{ij}^{\frac{\delta}{4(1 + \delta)}}}\frac{\zeta_i}{\nu_{ij}} + 4\frac{\log \nu_{ij}}{\nu_{ij}^3}z_i^3.
	\end{split}
\end{equation*}
The second inequality holds since $\xi_t \geq t - \nu_{ij} \geq -r_{ij}$ and $\frac{\log t}{t^3}$ is a decreasing function of $t \geq e^{\frac{1}{3}}$; the third inequality comes from $(u + v)^3 \leq 4(u^3 + v^3)$ for $u, v \geq 0$ and $r_{ij} \leq \nu_{ij}^{1 - \frac{5\delta}{12 + 12\delta}}$; the last inequality is due to $\nu_{ij} \geq \zeta_{i}^{1 + \delta}$.\\
Since $\nu_{ij} \gtrsim \zeta_i^{1 + \delta}$ and $ E  z_i^3 \leq 2\zeta_i^3$ for $\nu_{i} \geq C$, we know that if $\nu_{i} > C$,
\begin{equation*}
	\left|E g(W_{ij}')1_{\{\nu_{ij} - r_{ij} < W_{ij}' < \nu_{ij} + r_{ij}\}}\right| \leq 4\frac{\log\nu_{ij}}{\nu_{ij}^{\frac{\delta}{4(1 + \delta)}}}\frac{\zeta_i}{\nu_{ij}} + 8\frac{\log \nu_{ij}}{\nu_{ij}^3}\zeta_i^3 \leq 4\frac{\log\nu_{ij}}{\nu_{ij}^{\frac{\delta}{4(1 + \delta)}}}\frac{\zeta_i}{\nu_{ij}} + 8\frac{\log \nu_{ij}}{\nu_{ij}^\frac{2\delta}{1 + \delta}}\frac{\zeta_i}{\nu_{ij}}.
\end{equation*}
Therefore $\exists C_1(\delta) > 0$, for all $\nu_{ij} \geq C_1(\delta)$, 
\begin{align}\label{eq204}
	\left|E g(W_{ij}')1_{\{\nu_{ij} - r_{ij} < W_{ij}' < \nu_{ij} + r_{ij}\}}\right| \leq 2\frac{\zeta_i}{\nu_{ij}}.
\end{align}
For $t \leq \nu_{ij} - r_{ij}$, $\nu_{ij} \geq C_2(\delta)$, we have
\begin{equation*}
	\begin{split}
		\left|g(t)\right| \leq& \max\{\log^2\left(t + z_i\right), \log^2 \nu_{ij}\} + 2\frac{\log \nu_{ij}}{\nu_{ij}}\left|\frac{t - \nu_{ij} + z_i}{\nu_{ij}}\right| + \left|\frac{1 - \log \nu_{ij}}{\nu_{ij}^2}\right| \left(t - \nu_{ij} + z_i\right)^2\\
		\leq& \log^2 (\nu_{ij} + z_i) + 2\frac{\log \nu_{ij}}{\nu_{ij}}\left(\frac{\nu_{ij}}{\nu_{ij}} + \frac{z_i}{\nu_{ij}}\right) + \frac{\log \nu_{ij}}{\nu_{ij}^2}\left(\nu_{ij} + z_i\right)^2\\
		\leq& \left(\log \nu_{ij} + \frac{z_i}{\nu_{ij}}\right)^2 + 2\frac{\log \nu_{ij}}{\nu_{ij}} + 2\frac{\log \nu_{ij}}{\nu_{ij}^2}z_i + \frac{\log \nu_{ij}}{\nu_{ij}^2}\left(\nu_{ij} + z_i\right)^2\\
		\leq& \log^2 \nu_{ij} + 2\frac{\log \nu_{ij}}{\nu_{ij}} + \log \nu_{ij} + 6\frac{\log \nu_{ij}}{\nu_{ij}}z_i + 2\frac{\log \nu_{ij}}{\nu_{ij}^2}z_i^2.
	\end{split}
\end{equation*}
Denote $\epsilon = \frac{\delta}{24 + 12\delta}$, similarly to \eqref{eq202}, for $\nu_{ij} \geq C_2(\delta)$, 
\begin{equation}\label{eq249}
	\begin{split}
		&\left|E g(W_{ij}')1_{\left\{W_{ij}' \leq \nu_{ij} - r_{ij}\right\}}\right|\\ =&\left|E \left(\log^2 \nu_{ij} + 2\frac{\log \nu_{ij}}{\nu_{ij}} + \log \nu_{ij}\right)1_{\left\{W_{ij}' \leq \nu_{ij} - r_{ij}\right\}}\right| + 6\frac{\log \nu_{ij}}{\nu_{ij}}( E  z_i^2)^{\frac{1}{2}}\left( E  1_{\left\{W_{ij}' \leq \nu_{ij} - r_{ij}\right\}}\right)^{\frac{1}{2}}\\ &+ 2\frac{\log \nu_{ij}}{\nu_{ij}^2}( E  z_i^4)^{\frac{1}{2}}\left( E  1_{\left\{W_{ij}' \leq \nu_{ij} - r_{ij}\right\}}\right)^{\frac{1}{2}}\\ \leq& \left(\log^2 \nu_{ij} + 2\frac{\log \nu_{ij}}{\nu_{ij}} + \log \nu_{ij}\right){\mathrm{pr}}\left(W_{ij}' \leq \nu_{ij} - r_{ij}\right) + C\frac{\log \nu_{ij}}{\nu_{ij}}\zeta_i\left({\mathrm{pr}}\left(W_{ij}' \leq \nu_{ij} - r_{ij}\right)\right)^{\frac{1}{2}}\\ &+ C\frac{\log \nu_{ij}}{\nu_{ij}^2}\zeta_i^2\left({\mathrm{pr}}\left(W_{ij}' \leq \nu_{ij} - r_{ij}\right)\right)^{\frac{1}{2}}\\
		\leq& \left(\log^2 \nu_{ij} + 2\frac{\log \nu_{ij}}{\nu_{ij}} + \log \nu_{ij}\right)\cdot 3\exp\left(- c\nu_{ij}^{2\epsilon}\right) + C\left(\frac{\log \nu_{ij}}{\nu_{ij}}\zeta_i + \frac{\log \nu_{ij}}{\nu_{ij}^2}\zeta_i^2\right)\exp\left(- c\nu_{ij}^{2\epsilon}\right)\\
		\leq& C\frac{\zeta_i}{\nu_{ij}}.
	\end{split}
\end{equation}

$\forall t > \nu_{ij} + r_{ij}$ and $\nu_{ij} \geq C_3(\delta)$, by Cauchy-Schwarz inequality,
\begin{equation*}
	\begin{split}
		g(t) \leq& \left|\log^2 \left(t + z_i\right) - \log^2 \nu_{ij}\right| + \left|2\frac{\log \nu_{ij}}{\nu_{ij}}\left(t - \nu_{ij} + z_i\right)+ \frac{1-\log \nu_{ij}}{\nu_{ij}^2}\left(t - \nu_{ij} + z_i\right)^2\right|\\
		\leq& \log^2 \left(t + z_i\right) + \frac{t - \nu_{ij} + z_i}{\nu_{ij}}\left|2\log \nu_{ij} + \frac{1 - \log \nu_{ij}}{\nu_{ij}}\left(t - \nu_{ij} + z_i\right)\right|\\
		\leq& \log^2 \left(t + z_i\right) + \frac{t - \nu_{ij} + z_i}{\nu_{ij}}\left(2\log \nu_{ij} + \frac{\log \nu_{ij}}{\nu_{ij}}\left(t - \nu_{ij} + z_i\right)\right)\\
		\leq& \left(\log t + \frac{z_i}{t}\right)^2 + 2(\log \nu_{ij}) \frac{t + z_i}{\nu_{ij}} + \frac{\log \nu_{ij}}{\nu_{ij}^2}(t + z_i)^2\\
		\leq& 2\log^2 t + 2\frac{z_i^2}{t^2} + 2\frac{\log \nu_{ij}}{\nu_{ij}}t + 2\frac{\log \nu_{ij}}{\nu_{ij}}z_i +2 \frac{\log \nu_{ij}}{\nu_{ij}^2}t^2 + 2\frac{\log \nu_{ij}}{\nu_{ij}^2}z_i^2\\
		\leq& t + 2\frac{\log \nu_{ij}}{\nu_{ij}}t +2 \frac{\log \nu_{ij}}{\nu_{ij}^2}t^2 + 2\frac{z_i^2}{\nu_{ij}^2} + 2\frac{\log \nu_{ij}}{\nu_{ij}}z_i + 2\frac{\log \nu_{ij}}{\nu_{ij}^2}z_i^2.
	\end{split}
\end{equation*}
Use the same method in Lemma \ref{lm:beta-binomial_1}, we have
\begin{equation}\label{eq226}
	\begin{split}
		\left|E g(W_{ij}')1_{\{W_{ij}' \geq \nu_{ij} + r_{ij}\}}\right| \leq C\frac{\zeta_i}{\nu_{ij}}.
	\end{split}
\end{equation}
Combine \eqref{eq204}, \eqref{eq249} and the previous inequality together, for $\nu_{ij} \geq C(\delta) = \max\{C_1(\delta), C_2(\delta), C_3(\delta)\}$, we have
\begin{equation}\label{eq250}
	\begin{split}
		\left|E g(W_{ij}')\right| =&\left|E g(W_{ij}')1_{\{\nu_{ij} - r_{ij} < W_{ij}' < \nu_{ij} + r_{ij}\}}\right| +  \left|E g(W_{ij}')1_{\{W_{ij}' \leq \nu_{ij} - r_{ij}\}}\right| + \left|E g(W_{ij}')1_{\{W_{ij}' \geq \nu_{ij} + r_{ij}\}}\right|\\\leq& C\frac{\zeta_i}{\nu_{ij}}.
	\end{split}
\end{equation}
In addition, for $\nu_{ij} \geq C(\delta) \geq C$, we have 
\begin{equation*}
	\begin{split}
		&\left| E  \left[2\frac{\log \nu_{ij}}{\nu_{ij}}(W_{ij} - \nu_{ij} + z_i) + \frac{1 - \log \nu_{ij}}{\nu_{ij}^2}(W_{ij} - \nu_{ij} + z_i)^2\right]\right|\\ \leq& \left|2\frac{\log \nu_{ij}}{\nu_{ij}} E  z_i + \frac{1 - \log \nu_{ij}}{\nu_{ij}^2} E \left(W_{ij} - \nu_{ij}\right)^2\right| + \left|\frac{1 - \log \nu_{ij}}{\nu_{ij}^2} E  z_i^2\right| + \left|2\frac{1 - \log \nu_{ij}}{\nu_{ij}^2} E  (W_{ij} - \nu_{ij})z_i\right|\\
		\leq& \left|2\frac{\log \nu_{ij}}{\nu_{ij}}\frac{\nu_i + \alpha_i + 1}{2\left(\alpha_i + 1\right)} + \frac{1 - \log \nu_{ij}}{\nu_{ij}^2}\left(\nu_iX_{ij} + \nu_i^2\frac{X_{ij}(1 - X_{ij})}{\alpha_i + 1}\right)\right| + \log (\nu_{ij})\cdot 2\frac{\zeta_i^2}{\nu_{ij}^2}\\
		&+ 2\frac{\log \nu_{ij}}{\nu_{ij}^2}\left( E \left(W_{ij} - \nu_{ij}\right)^2\right)^{\frac{1}{2}}\left( E  z_i^2\right)^{\frac{1}{2}}\\
		=& \left|2\frac{\log\nu_{ij}}{\nu_{ij}}\zeta_{i} + \frac{1 - \log\nu_{ij}}{\nu_{ij}}\left(2\zeta_{i} - \frac{\nu_{ij}}{\alpha_i + 1}\right)\right| + 2\log(\nu_{ij})\frac{\zeta_{i}^2}{\nu_{ij}^2} + 2\frac{\log \nu_{ij}}{\nu_{ij}^2}\left(\nu_iX_{ij} + \nu_i^2\frac{X_{ij}(1 - X_{ij})}{\alpha_i + 1}\right)^{\frac{1}{2}}\left( E  z_i^2\right)^{\frac{1}{2}}\\
		\leq& 2\frac{\zeta_i}{\nu_{ij}} + \frac{\log\nu_{ij}}{\alpha_i + 1} + 2\log(\nu_{ij})\frac{\zeta_{i}^2}{\nu_{ij}^2} + 2\frac{\log \nu_{ij}}{\nu_{ij}^2}\left(2\nu_{ij}\zeta_i\right)^{\frac{1}{2}}\sqrt{2}\zeta_i\\
		\leq& 2\frac{\zeta_i}{\nu_{ij}} + \log(\nu_{ij})\frac{\nu_{i} + \alpha_i + 1}{\nu_{i}(\alpha_i + 1)} + 2\log(\nu_{ij})\frac{\zeta_{i}^2}{\nu_{ij}^2} + 4\frac{\log \nu_{ij}}{\nu_{ij}^2}\nu_{ij}\zeta_i\\
		\leq& 2\frac{\zeta_i}{\nu_{ij}} + 2\log(\nu_{ij})\frac{\zeta_i}{\nu_{i}} + 2\log(\nu_{ij})\frac{\zeta_{i}^2}{\nu_{ij}^2} + 4\frac{\log \nu_{ij}}{\nu_{ij}}\zeta_i \leq 10\log(\nu_{ij})\frac{\zeta_{i}}{\nu_{ij}}.
	\end{split}
\end{equation*}
In the second inequality, we used Cauchy-Schwarz inequality and $ E  z_i^2 \leq 2\zeta_{i}^2$ for $\nu_{i} \geq C$; the fourth inequality and the fifth inequality come from $\nu_{ij} \geq \zeta_i$.
By (\ref{eq221}), (\ref{eq222}) and (\ref{eq223}) and the previous inequality, we have
\begin{equation}\label{eq225}
	\begin{split}
		&\left| E  \left[2\frac{\log \nu_{ij}}{\nu_{ij}}(W_{ij}' - \nu_{ij} + z_i) + \frac{1 - \log \nu_{ij}}{\nu_{ij}^2}(W_{ij}' - \nu_{ij} + z_i)^2\right]\right|\\
		\leq& \left| E  \left[2\frac{\log \nu_{ij}}{\nu_{ij}}(W_{ij} - \nu_{ij} + z_i) + \frac{1 - \log \nu_{ij}}{\nu_{ij}^2}(W_{ij} - \nu_{ij}+ z_i)^2\right]\right| + \frac{4\log(\nu_{ij}) - 2}{\nu_{ij}} E \left|W_{ij} - W_{ij}'\right|\\ & + 2\frac{\log(\nu_{ij}) - 1}{\nu_{ij}^2}\left| E \left[(W_{ij} - W_{ij}')z_i\right]\right| + \frac{\log(\nu_{ij}) - 1}{\nu_{ij}^2}\left| E (W_{ij}^2 - W_{ij}'^2)\right|\\
		\leq& \left| E  \left[2\frac{\log \nu_{ij}}{\nu_{ij}}(W_{ij} - \nu_{ij} + z_i) + \frac{1 - \log \nu_{ij}}{\nu_{ij}^2}(W_{ij} - \nu_{ij}+ z_i)^2\right]\right|\\ &+ 4\frac{\log\nu_{ij}}{\nu_{ij}} E \left|W_{ij} - W_{ij}'\right|
		+ 2\frac{\log\nu_{ij}}{\nu_{ij}^2}\left[ E  (W_{ij} - W_{ij}')^2  E  z_i^2\right]^{\frac{1}{2}} + \frac{\log\nu_{ij}}{\nu_{ij}^2}\left| E \left(W_{ij}^2 - W_{ij}'^2\right)\right|\\
		\leq& 10\log(\nu_{ij})\frac{\zeta_{i}}{\nu_{ij}} + C\log(\nu_{ij})\exp\left(-c\left(\nu_{ij}\right)^{\frac{\delta}{1 + \delta}}\right) + C\log(\nu_{ij})\frac{\zeta_{i}}{\nu_{ij}}\exp\left(-c\left(\nu_{ij}\right)^{\frac{\delta}{1 + \delta}}\right)\\ &+ C\log(\nu_{ij})\exp\left(-c\left(\nu_{ij}\right)^{\frac{\delta}{1 + \delta}}\right)\\
		\leq& C\log(\nu_{ij})\frac{\zeta_{i}}{\nu_{ij}}.
	\end{split}
\end{equation}
Combine (\ref{eq250}) and (\ref{eq225}) together, we have
\begin{equation}
	\left|E \log^2(W_{ij}' + z_i) - \log^2 \nu_{ij}\right| \leq C\log(\nu_{ij})\frac{\zeta}{\nu_{ij}}.
\end{equation} \qed
\medskip

{\noindent Proof of Lemma \ref{lm:sub-Gaussian_diri-mul}.}
For convenience, denote $D_{ij} = \log\left(W'_{ij} + z_i\right) -  E \log\left(W'_{ij} + z_i\right)$, by Lemma \ref{lm:beta-binomial_1}, $\left| E \log\left(W'_{ij} + z_i\right) - \log\nu_{ij}\right| \leq 10\frac{\zeta_i}{\nu_{ij}}$. By lemma \ref{lm:tail bound}, for $t \geq 1$, note that $\frac{\zeta_i}{\nu_{ij}} < 1$,
\begin{equation}\label{eq205}
	\begin{split}
		&{\mathrm{pr}}\left(\left|\sqrt{\frac{\nu_{ij}}{\zeta_i}}D_{ij}\right| > t\right)\\  =& {\mathrm{pr}}\left(\log\left(W_{ij}' +z_i\right) > E \log\left(W_{ij}'+z_i\right) + \sqrt{\frac{\zeta_i}{\nu_{ij}}}t\right) +  {\mathrm{pr}}\left(\log\left(W_{ij}' + z_i\right) < E \log\left(W_{ij}'+z_i\right) - \sqrt{\frac{\zeta_i}{\nu_{ij}}}t\right)\\
		\leq& {\mathrm{pr}}\left(\log\left(W'_{ij}+z_i\right) > \log\nu_{ij} - 10\frac{\zeta_i}{\nu_{ij}}  + \sqrt{\frac{\zeta_i}{\nu_{ij}}}t\right) + {\mathrm{pr}}\left(\log\left(W'_{ij}+z_i\right) < \log \nu_{ij} + 10\frac{\zeta_i}{\nu_{ij}} - \sqrt{\frac{\zeta_i}{\nu_{ij}}}t\right)\\
		=& {\mathrm{pr}}\left(W'_{ij} > \nu_{ij} e^{\sqrt{\frac{\zeta_i}{\nu_{ij}}}t - 10\frac{\zeta_i}{\nu_{ij}}}-z_i\right) + {\mathrm{pr}}\left(W'_{ij} < \nu_{ij} e^{-\sqrt{\frac{\zeta_i}{\nu_{ij}}}t + 10\frac{\zeta_i}{\nu_{ij}}} - z_i\right)\\
		\leq& {\mathrm{pr}}\left(W'_{ij} > \nu_{ij} e^{\frac{1}{2}\sqrt{\frac{\zeta_i}{\nu_{ij}}}t}-z_i\right) + {\mathrm{pr}}\left(W'_{ij} < \nu_{ij} e^{-\frac{1}{2}\sqrt{\frac{\zeta_i}{\nu_{ij}}}t} - z_i\right).
	\end{split}	
\end{equation}
We first consider ${\mathrm{pr}}\left(W'_{ij} > \nu_{ij} e^{\frac{1}{2}\sqrt{\frac{\zeta_i}{\nu_{ij}}}t}-z_i\right)$. $\forall t \geq 10$,
\begin{equation*}
	\begin{split}
		{\mathrm{pr}}\left(W_{ij}' > \nu_{ij} e^{\frac{1}{2}\sqrt{\frac{\zeta_i}{\nu_{ij}}}t}-z_i\right) \leq& {\mathrm{pr}}\left(W'_{ij} > \nu_{ij} e^{\frac{1}{2}\sqrt{\frac{\zeta_i}{\nu_{ij}}}t}-\frac{\nu_{i} + \alpha_i + 1}{100(\alpha_i + 1)}t^2\right) + {\mathrm{pr}}\left(z_i \geq \frac{\nu_{i} + \alpha_i + 1}{100(\alpha_i + 1)}t^2\right).
	\end{split}
\end{equation*}
For $t \geq 10$, $\frac{\nu_{i} + \alpha_i + 1}{100(\alpha_i + 1)}t^2 \geq \frac{\nu_{i} + \alpha_i + 1}{\alpha_i + 1} = 2 E  z_i$, by Lemma \ref{lm:poisson tail bound}, 
\begin{equation*}
	\begin{split}
		{\mathrm{pr}}\left(z_i \geq \frac{\nu_{i} + \alpha_i + 1}{100(\alpha_i + 1)}t^2\right) =& {\mathrm{pr}}\left(z_i -  E  z_i \geq \frac{\nu_{i} + \alpha_i + 1}{100(\alpha_i + 1)}t^2 -  E  z_i\right)\\ \leq& {\mathrm{pr}}\left(N_i - \nu_i \geq 2\left(\alpha_i + 1\right)\frac{\nu_{i} + \alpha_i + 1}{200(\alpha_i + 1)}t^2\right)\\
		\leq& \exp\left(-\frac{\left(\frac{1}{100}\left(\nu_{i} + \alpha_i + 1\right)t^2\right)^2}{\nu_i}\psi_{Benn}\left(\frac{\frac{1}{100}\left(\nu_{i} + \alpha_i + 1\right)t^2}{\nu_i}\right)\right).
	\end{split}
\end{equation*}
Since for all $t \geq 10$, $\frac{\frac{1}{100}\left(\nu_{i} + \alpha_i + 1\right)t^2}{\nu_i} \geq 1$, which means $\psi_{Benn}\left(\frac{\frac{1}{100}\left(\nu_{i} + \alpha_i + 1\right)t^2}{\nu_i}\right) \geq \frac{c}{\frac{1}{100}\left(\nu_{i} + \alpha_i + 1\right)t^2/\nu_i}$. Therefore,
\begin{equation}\label{eq251}
	\begin{split}
		{\mathrm{pr}}\left(z_i \geq \frac{\nu_{i} + \alpha_i + 1}{100(\alpha_i + 1)}t^2\right) \leq e^{-c(\nu_i + \alpha_i + 1)t^2}.
	\end{split}
\end{equation}
Since $e^x \geq \frac{x^2}{2}$ for all $x > 0$, for $\nu_{ij}$ large eough, we have
\begin{equation*}
	\begin{split}
		\frac{1}{2}\nu_{ij}\left(e^{\frac{1}{2}\sqrt{\frac{\zeta_i}{\nu_{ij}}}t} - 1\right) - \frac{\nu_{i} + \alpha_i + 1}{100(\alpha_i + 1)}t^2 \geq& \frac{1}{2}\nu_{ij}\frac{\left(\frac{1}{2}\sqrt{\frac{\zeta_i}{\nu_{ij}}}t\right)^2}{2} - \frac{\nu_{i} + \alpha_i + 1}{100(\alpha_i + 1)}t^2
		\geq \frac{1}{16}\zeta_it^2 - \frac{1}{50}\zeta_it^2
		\geq 0.
	\end{split}
\end{equation*}
By Lemma \ref{lm:tail bound}, for $\nu_{ij}$ large enough and $t \geq 1$,
\begin{equation*}
	\begin{split}
		&{\mathrm{pr}}\left(W'_{ij} > \nu_{ij} e^{\frac{1}{2}\sqrt{\frac{\zeta_i}{\nu_{ij}}}t}-\frac{\nu_{i} + \alpha_i + 1}{100(\alpha_i + 1)}t^2\right) = {\mathrm{pr}}\left(W'_{ij} - \nu_{ij} > \nu_{ij}\left(e^{\frac{1}{2}\sqrt{\frac{\zeta_i}{\nu_{ij}}}t} - 1\right) - \frac{\nu_{i} + \alpha_i + 1}{100(\alpha_i + 1)}t^2\right)\\
		\leq& {\mathrm{pr}}\left(W'_{ij} - \nu_{ij} > \frac{1}{2}\nu_{ij}\left(e^{\frac{1}{2}\sqrt{\frac{\zeta_i}{\nu_{ij}}}t} - 1\right) \right)\\ \leq& 2\exp\left(-c\left(\frac{1}{2}\nu_{ij}\left(e^{\frac{1}{2}\sqrt{\frac{\zeta_i}{\nu_{ij}}}t} - 1\right)\right)^2\min\left\{\frac{\alpha_i}{\nu_{i}\cdot\alpha_iX_{ij}}, \frac{\alpha_i^2}{\nu_{i}^2\cdot\alpha_iX_{ij}}\right\}\right)\\
		=& 2\exp\left(-c\left(\frac{1}{2}\nu_{ij}\left(e^{\frac{1}{2}\sqrt{\frac{\zeta_i}{\nu_{ij}}}t} - 1\right)\right)^2\min\left\{\frac{1}{\nu_{ij}}, \frac{1}{\nu_{ij}}\frac{\alpha_i}{\nu_{i}}\right\}\right)\\
		\leq& 2\exp\left(-c\left(\frac{1}{2}\nu_{ij}\left(e^{\frac{1}{2}\sqrt{\frac{\zeta_i}{\nu_{ij}}}t} - 1\right)\right)^2\frac{1}{\nu_{ij}}\frac{1}{\zeta_i}\right)
		\leq 2\exp\left(-c\left(\frac{1}{2}\nu_{ij}\cdot \frac{1}{2}\sqrt{\frac{\zeta_i}{\nu_{ij}}}t\right)^2\frac{1}{\nu_{ij}}\frac{1}{\zeta_i}\right)\\
		\leq& 2\exp(-ct^2).
	\end{split}
\end{equation*}
By \eqref{eq251} and the previous inequality, for $t \geq 1$, we have
\begin{equation}\label{eq206}
	{\mathrm{pr}}\left(W'_{ij} > \nu_{ij}e^{\frac{1}{2}\sqrt{\frac{\zeta_i}{\nu_{ij}}}t} - z_i\right) \leq \exp\left(-c(\nu_i + \alpha_i + 1)t^2\right) + 2\exp\left(-ct^2\right).
\end{equation}
Moreover, when $t > 10\sqrt{\frac{\nu_{ij}}{\zeta_i}}$, we have $\nu_{ij} e^{-5} - z_i \leq \frac{1}{100}\nu_{ij}$, thus 
\begin{equation*}
	{\mathrm{pr}}\left(W'_{ij} < \nu_{ij} e^{-\frac{1}{2}\sqrt{\frac{\zeta_i}{\nu_{ij}}}t} - z_i\right) = 0.
\end{equation*}
When $t \leq 10\sqrt{\frac{\nu_{ij}}{\zeta_i}}$, note that $\frac{1 - e^{-x}}{x}$ is decreasing, by Lemma \ref{lm:tail bound}, 
\begin{equation*}
	\begin{split}
		&{\mathrm{pr}}\left(W_{ij} < \nu_{ij} e^{-\frac{1}{2}\sqrt{\frac{\zeta_i}{\nu_{ij}}}t} - z_i\right) \leq {\mathrm{pr}}\left(W_{ij} < \nu_{ij} e^{-\frac{1}{2}\sqrt{\frac{\zeta_i}{\nu_{ij}}}t}\right)
		= {\mathrm{pr}}\left(W_{ij} - \nu_{ij}< - \nu_{ij}\left(1 - e^{-\frac{1}{2}\sqrt{\frac{\zeta_i}{\nu_{ij}}}t}\right)\right)\\
		\leq& 2\exp\left(-c\left(\nu_{ij}\left(1 - e^{-\frac{1}{2}\sqrt{\frac{\zeta_i}{\nu_{ij}}}t}\right)\right)^2\min\left\{\frac{\alpha_i}{\nu_{i}\cdot\alpha_{i}X_{ij}}, \frac{\alpha_i^2}{\nu_{i}^2\cdot\alpha_iX_{ij}}\right\}\right)\\
		\leq& 2\exp\left(-c\left(\nu_{ij}\left(1 - e^{-\frac{1}{2}\sqrt{\frac{\zeta_i}{\nu_{ij}}}t}\right)\right)^2\min\left\{\frac{1}{\nu_{ij}}, \frac{1}{\nu_{ij}}\frac{\alpha_i}{\nu_{ij}}\right\}\right)\\
		\leq& 2\exp\left(-c\frac{\left(\nu_{ij}\left(1 - e^{-\frac{1}{2}\sqrt{\frac{\zeta_i}{\nu_{ij}}}t}\right)\right)^2}{\nu_{ij}\zeta_i}\right)
		\leq 2\exp\left(-c\frac{\nu_{ij}}{\zeta_i}\frac{\left(1 - e^{-5}\right)^2}{5^2}\left(\frac{1}{2}\sqrt{\frac{\zeta_i}{\nu_{ij}}}t\right)^2\right)
		\leq 2\exp\left(-ct^2\right).  	
	\end{split}
\end{equation*}
By \eqref{eq206} and the previous inequality, and use the same method in Lemma \ref{lm:sub-Gaussian}, we can get the result.
\qed
\medskip

{\noindent Proof of Lemma \ref{lm:infinity_norm_bound_Dirichlet Multinomial}.}
\begin{equation*}
	\begin{split}
		& \left\|\bar{\A}_{W'}\beta^* - \bar{\B}_{W'}^\top y\right\|_\infty = \left\|\P\A_{W'}\beta^* - \P\B_{W'}^\top ( V\beta^* + \varepsilon)\right\|_\infty \leq \left\|\P(\A_{W'} - \B_{W'}^\top  V)\beta^*\right\|_\infty + \|\bar \B_{W'}^\top \varepsilon\|_\infty. 
	\end{split}
\end{equation*}
For $\bar{\B}_{W'}^\top\varepsilon$, we have
\begin{equation}\label{eq210}
	\begin{split}
		\|\bar{\B}_{W'}^\top \varepsilon\|_{\infty} =& \|\P\B_{W'}^\top\varepsilon\|_{\infty} \leq \|\P\left(\B_{W'} -  E  \B_{W'}\right)^\top\varepsilon\|_{\infty} + \|\P E  \B_{W'}^\top\varepsilon\|_{\infty}\\
		\leq& 2\|\left(\B_{W'} -  E  \B_{W'}\right)^\top\varepsilon\|_{\infty} + \|\P E  \B_{W'}^\top\varepsilon\|_{\infty}
	\end{split}
\end{equation}
Note that for $1 \leq i \leq p$, 
\begin{equation*}
	\begin{split}
		\left[\left(\B_{W'} -  E  \B_{W'}\right)^\top\varepsilon\right]_i = \sum_{k=1}^{n}\left(\B_{W'} -  E \B_{W'}\right)_{ki}\varepsilon_k,
	\end{split}
\end{equation*}
$\forall t \geq 0$, we have
\begin{equation}\label{eq252}
	{\mathrm{pr}} \left(\left|\sum_{k=1}^{n}\left(\B_{W'} -  E  \B_{W'}\right)_{ki}\varepsilon_k\right| > t\Big|{W'}_{lj}, N_l, 1 \leq l \leq n, 1 \leq j \leq p\right) \leq 2e^{-\frac{t^2}{2\sigma^2\sum_{k=1}^n \left(B_{W'} -  E  B_{W'}\right)_{ki}^2}}.
\end{equation}
By Lemma \ref{lm:sub-Gaussian_diri-mul}, there exists a constant $K > 0$ such that $\left\|\left(B_{W'}\right)_{ki} -  E \left(B_{W'}\right)_{ki}\right\|_{\psi_2} \leq K$. Thus $\left\|\left(\left(B_{W'}\right)_{ki} -  E \left(B_{W'}\right)_{ki}\right)^2\right\|_{\psi_1} \leq 2K^2$. Consider the centering, we can get
\begin{equation*}
	\left\|\left(\left(B_{W'}\right)_{ki} -  E \left(B_{W'}\right)_{ki}\right)^2 -  E \left(\left(B_{W'}\right)_{ki} -  E \left(B_{W'}\right)_{ki}\right)^2\right\|_{\psi_1} \leq 4K^2.
\end{equation*}
By Proposition 5.16 in \cite{vershynin2010introduction}, $\forall t \geq 0$, 
\begin{equation*}
	{\mathrm{pr}}\left(\sum_{k = 1}^n \left(\B_{W'} -  E  \B_{W'}\right)_{ki}^2 -  E \left[\sum_{k = 1}^n \left(\B_{W'} -  E  \B_{W'}\right)_{ki}^2\right] > t\right) \leq \exp\left[-c\min\left(\frac{t^2}{16nK^4}, \frac{t}{4K^2}\right)\right],
\end{equation*}
where $c > 0$ is a constant.

Choose $t = n$ and notice that $K$ is a constant, we have
\begin{equation*}
	{\mathrm{pr}}\left(\sum_{k = 1}^n \left(\B_{W'} -  E  \B_{W'}\right)_{ki}^2 \geq n +  E \sum_{k = 1}^n \left(\B_{W'} -  E  \B_{W'}\right)_{ki}^2\right) \leq e^{-cn}.
\end{equation*}
Also, apply Lemma \ref{lm:beta-binomial_1} and Lemma \ref{lm:beta-binomial_2}, and also recall that $\nu_{ki} \geq \zeta_i^{1 + \delta}$, for $\nu_{\min} \geq C(\delta)$,
\begin{equation*}
	\begin{split}
		E \left(\B_{W'} -  E  \B_{W'}\right)_{ki}^2 =&  E \left(B_{W'}\right)_{ki}^2 - \left( E \left(B_{W'}\right)_{ki}\right)^2 \leq \log^2\nu_{ki} + C\log(\nu_{ki})\frac{\zeta_i}{\nu_{ki}} - \left(\log\nu_{ki} - C\frac{\zeta_i}{\nu_{ki}}\right)^2\\
		\leq& C\frac{\log\nu_{ki}}{\nu_{ki}^{\frac{\delta}{1 + \delta}}} + 2C\frac{\log\nu_{ki}}{\nu_{ki}^{\frac{\delta}{1 + \delta}}}
		\leq C.
	\end{split}
\end{equation*}
Thus
\begin{equation*}
	{\mathrm{pr}}\left(\sum_{k = 1}^n \left(\B_{W'} -  E  \B_{W'}\right)_{ki}^2 \geq Cn\right) \leq e^{-cn}.
\end{equation*}
By \eqref{eq252} and the previous inequality, $\forall t \geq 0$, $\exists$ a constant $c > 0$,
\begin{equation*}
	\begin{split}
		&{\mathrm{pr}} \left(\left|\sum_{k=1}^{n}\left(\B_{W'} -  E  \B_{W'}\right)_{ki}\varepsilon_k\right| > t\right)\\
		=& {\mathrm{pr}} \left(\left|\sum_{k=1}^{n}\left(\B_{W'} -  E  \B_{W'}\right)_{ki}\varepsilon_k\right| > t, \sum_{k = 1}^n \left(\B_{W'} -  E  \B_{W'}\right)_{ki}^2 < Cn\right)\\ &+ {\mathrm{pr}} \left(\left|\sum_{k=1}^{n}\left(\B_{W'} -  E  \B_{W'}\right)_{ki}\varepsilon_k\right| > t, \sum_{k = 1}^n \left(\B_{W'} -  E  \B_{W'}\right)_{ki}^2 \geq Cn\right)\\
		\leq& 2e^{-\frac{ct^2}{n\sigma^2}} + e^{-cn}.
	\end{split}
\end{equation*}
Therefore
\begin{equation}\label{eq208}
	\begin{split}
		{\mathrm{pr}}\left(\left\|\left(\B_{W'} -  E  \B_{W'}\right)^\top\varepsilon\right\|_{\infty} > t\right)
		\leq& {\mathrm{pr}}\left(\exists 1 \leq i \leq p, \left|\left[\left(\B_{W'} -  E  \B_{W'}\right)^\top\varepsilon\right]_i\right| > t\right)\\
		= & {\mathrm{pr}} \left(\exists 1 \leq i \leq p, \left|\sum_{k=1}^{n}\left(\B_{W'} -  E  \B_{W'}\right)_{ki}\varepsilon_k\right| > t\right)\\
		\leq& p\left(2e^{-\frac{ct^2}{n\sigma^2}} + e^{-cn}\right).
	\end{split}
\end{equation}
Notice that
\begin{equation*}
	\begin{split}
		\left(\P E  \B_{W'}^\top\varepsilon\right)_{i} = \sum_{k = 1}^n\left( E  \B_{W'}\P\right)_{ki}\varepsilon_k = \sum_{k = 1}^n\left( E \phi_1(W'_{ki}) - \frac{1}{p}\sum_{j = 1}^{p} E \phi_1(W'_{kj})\right)\varepsilon_k,
	\end{split}
\end{equation*}
$\forall t \geq 0$,
\begin{equation*}
	{\mathrm{pr}}\left(\sum_{k = 1}^n\left( E \phi_1(W'_{ki}) - \frac{1}{p}\sum_{j = 1}^{p} E \phi_1(W'_{kj})\right)\varepsilon_k > t\right) \leq 2e^{-\frac{t^2}{2\sigma^2\sum_{k = 1}^{n}\left( E \phi_1(W'_{ki}) - \frac{1}{p}\sum_{j = 1}^{p} E \phi_1(W'_{kj})\right)^2}}.
\end{equation*}
Since
\begin{equation*}
	\begin{split}
		\left| E \phi_1(W'_{ki}) - \frac{1}{p}\sum_{j = 1}^{p} E \phi_1(W'_{kj})\right| \leq \left|\log\nu_{ki} - \frac{1}{p}\sum_{j = 1}^p\log\nu_{kj}\right| + C\frac{\zeta_i}{\nu_{ki}} + \frac{1}{p}\sum_{j = 1}^pC\frac{\zeta_i}{\nu_{kj}}
		\leq C,
	\end{split}
\end{equation*}
we know that
\begin{equation*}
	{\mathrm{pr}}\left(\left(\left|\P E  \B_{W'}'^\top\varepsilon\right)_{i}\right| > t\right) \leq 2e^{-\frac{ct^2}{n\sigma^2}}.
\end{equation*}
Hence, 
\begin{equation}\label{eq209}
	\begin{split}
		{\mathrm{pr}}\left(\|\P E  \B_{W'}'^\top\varepsilon\|_{\infty} > t\right) = {\mathrm{pr}}\left(\exists 1 \leq i \leq p, \left|\left(\P E  \B_{W'}^\top\varepsilon\right)_{i}\right| > t\right) \leq 2pe^{-\frac{ct^2}{n\sigma^2}}.
	\end{split}
\end{equation}
Combine (\ref{eq210}), (\ref{eq208}) and (\ref{eq209}) together, 
\begin{equation}\label{eq211}
	\begin{split}
		{\mathrm{pr}}\left(\left\|\bar{\B}_{W'}^\top\varepsilon\right\|_{\infty} > t\right) \leq& {\mathrm{pr}}\left(\|\left(\B_{W'} -  E  \B_{W'}\right)^\top\varepsilon\|_{\infty} > \frac{t}{4}\right) +  E \left(\|\P E  \B_{W'}^\top\varepsilon\|_{\infty} > \frac{t}{2}\right)\\
		\leq& p\left(4e^{-\frac{ct^2}{n\sigma^2}} + e^{-cn}\right).	
	\end{split}
\end{equation}
For the second part,
\begin{equation*}
	\begin{split}
		\left\|\P\B_{W'}^\top\left(\B_{W'} -  V\right)\beta^*\right\|_{\infty} \leq& \left\|\P\left(\B_{W'} -  E  \B_{W'}\right)^\top\left(\B_{W'} -  V\right)\beta^*\right\|_{\infty} + \left\|\P\left( E  \B_{W'}\right)^\top\left(\B_{W'} -  V\right)\beta^*\right\|_{\infty}\\
		\leq& 2\left\|\left(\B_{W'} -  E  \B_{W'}\right)^\top\left(\B_{W'} -  V\right)\beta^*\right\|_{\infty} + \left\|\P\left( E  \B_{W'}\right)^\top\left(\B_{W'} -  V\right)\beta^*\right\|_{\infty}.
	\end{split}
\end{equation*}
Denote $D_{ki} = \phi\left(W'_{ki}\right) -  E \phi\left(W'_{ki}\right)$, $S = \text{supp}(\beta^*) = \{1 \leq j \leq p: \beta^*_j \neq 0\}$. By Lemma \ref{lm:sub-Gaussian_diri-mul}, for all $1 \leq i \leq p$,
\begin{equation}\label{eq207}
	\begin{split}
		\left[\left(\B_{W'} -  E  \B_{W'}\right)^\top\left(\B_{W'} -  V\right)\beta^*\right]_i =& \sum_{j = 1}^{p}\sum_{k = 1}^{n}\left(\phi\left(W'_{ki}\right) -  E \phi\left(W'_{ki}\right)\right)\left(\phi\left(W'_{kj}\right) - \log\nu_{kj}\right)\beta_j^*\\
		=& \sum_{j \in S}\sum_{k = 1}^{n}\left(\phi\left(W'_{ki}\right) -  E \phi\left(W'_{ki}\right)\right)\left(\phi\left(W'_{kj}\right) - \log\nu_{kj}\right)\beta_j^*\\
		=& \sum_{j \in S}\sum_{k=1}^n D_{ki}D_{kj}\beta_j^* + \sum_{j \in S}\sum_{k=1}^n D_{ki}(D_{kj} - \log\nu_{kj})\beta_j^*.
	\end{split}
\end{equation}
$\forall j \in S$, $p \geq 1$, by Cauchy-Schwarz inequality,
\begin{equation*}
	\begin{split}
		\frac{1}{p}\left( E \left|XY\right|^p\right)^{\frac{1}{p}} \leq \frac{1}{p}\left( E \left|X\right|^{2p} E \left|Y\right|^{2p}\right)^{\frac{1}{2p}} = 2\frac{1}{\sqrt{2p}}\left( E \left|X\right|^{2p}\right)^{\frac{1}{2p}}\cdot \frac{1}{\sqrt{2p}}\left( E \left|Y\right|^{2p}\right)^{\frac{1}{2p}}.
	\end{split}
\end{equation*}
Thus
\begin{equation*}
	\begin{split}
		&\|D_{ki}D_{kj} -  E  D_{ki}D_{kj}\|_{\psi_1} \leq 2\|D_{ki}D_{kj}\|_{\psi_1} \leq 4\|D_{ki}\|_{\psi_2}\|D_{kj}\|_{\psi_2} \leq C\frac{\zeta_{\max}}{\nu_{\min}}.
	\end{split} 
\end{equation*}
By Proposition 5.16 in \cite{vershynin2010introduction}, $\forall t \geq 0$, 
\begin{equation}\label{eq253}
	\begin{split}
		{\mathrm{pr}}\left(\left|\sum_{k = 1}^n D_{ki}D_{kj} - \sum_{k = 1}^n  E  D_{ki}D_{kj}\right| \geq t\right) \leq 2\exp\left[-c\min\left(\frac{\nu_{\min}^{2}t^2}{n\zeta_{\max}^2}, \frac{\nu_{\min}t}{\zeta_{\max}}\right)\right].
	\end{split}
\end{equation}
By Lemma \ref{lm:beta-binomial_1} and Lemma \ref{lm:beta-binomial_2}, if $\nu_{ki} \geq \max\{C(\delta), \zeta_{k}^{1 + \delta}\}$,
\begin{equation*}
	\begin{split}
		E  D_{ki}^2 =&  E \phi^2\left(W'_{ki}\right) - \left( E \phi\left(W'_{ki}\right)\right)^2 \leq \log^2(\nu_{ki}) + C\log(\nu_{ki})\frac{\zeta_k}{\nu_{ki}} - \left(\log(\nu_{ki}) - 10\frac{\zeta_k}{\nu_{ki}}\right)^2\\ \leq& C\log(\nu_{ki})\frac{\zeta_k}{\nu_{ki}} + 20\log(\nu_{ki})\frac{\zeta_k}{\nu_{ki}} \leq C\log(\nu_{ki})\frac{1}{\nu_{ki}^{\frac{\delta}{1 + \delta}}} \leq C.
	\end{split}
\end{equation*}
\begin{equation*}
	\begin{split}
		\left| E  D_{ki}D_{kj}\right| \leq \left( E  D_{ki}^2\right)^{\frac{1}{2}}\left( E  D_{kj}^2\right)^{\frac{1}{2}} \leq C\left(\log \nu_{ki}\frac{\zeta_k}{\nu_{ki}}\right)^{\frac{1}{2}}\left(\log \nu_{kj}\frac{\zeta_k}{\nu_{kj}}\right)^{\frac{1}{2}} \leq C\frac{\log\nu_{\min}}{\nu_{\min}}\zeta_{\max}.
	\end{split}
\end{equation*}
Therefore, set $t = n\frac{\log\nu_{\min}}{\nu_{\min}}\zeta_{\max}$ in \eqref{eq253}, we have
\begin{equation}\label{eq254}
	{\mathrm{pr}}\left(\left|\sum_{k = 1}^nD_{ki}D_{kj}\right| \geq Cn\frac{\log\nu_{\min}}{\nu_{\min}}\zeta_{\max}\right) \leq 2e^{-cn\log\nu_{\min}}.
\end{equation}
In addition, by Proposition 5.10 in \cite{vershynin2010introduction} (note that $ E  D_{ki} = 0$), $\forall t \geq 0$, 
\begin{equation*}
	{\mathrm{pr}}\left(\left|\sum_{k = 1}^{n}D_{ki}\left( E \phi\left(W'_{kj}\right) - \log\nu_{kj}\right)\right| \geq t\right) \leq e\cdot \exp\left(-c\frac{\nu_{\min}t^2}{\zeta_{\max}\sum_{k = 1}^n\left( E \phi\left(W'_{kj}\right) - \log\nu_{kj}\right)^2}\right).
\end{equation*}
By Lemma \ref{lm:beta-binomial_1}, 
\begin{equation*}
	\sum_{k = 1}^{n}\left( E \phi\left(W'_{kj}\right) - \log\nu_{kj}\right)^2 \leq C\sum_{k = 1}^{n}\left(\frac{\zeta_{k}}{\nu_{kj}}\right)^2 \leq Cn\frac{\zeta_{\max}^2}{\nu_{\min}^2}.
\end{equation*}
Therefore,
\begin{equation*}
	{\mathrm{pr}}\left(\left|\sum_{k = 1}^{n}D_{ki}\left( E \phi\left(W'_{kj}\right) - \log\nu_{kj}\right)\right| \geq t\right) \leq e\cdot \exp\left(-c\frac{\nu_{\min}^{3}t^2}{n\zeta_{\max}^3}\right).
\end{equation*}
Choose $t = n\frac{\log\nu_{\min}}{\nu_{\min}}\zeta_{\max}$, note that $\nu_{\min} \asymp \nu_{\max} \geq \zeta_{\max}^{1 + \delta}$ and $\nu_{ki} \geq C(\delta)$ for all $k, i$, we have
\begin{equation*}
	\begin{split}
		{\mathrm{pr}}\left(\left|\sum_{k = 1}^{n}D_{ki}\left( E \phi\left(W'_{kj}\right) - \log\nu_{kj}\right)\right| \geq n\frac{\log\nu_{\min}}{\nu_{\min}}\zeta_{\max}\right) \leq e\cdot \exp\left(-cn\log^2\nu_{\min}\right).
	\end{split}
\end{equation*}
By (\ref{eq207}), \eqref{eq254} and the previous inequality, we have
\begin{equation*}
	\begin{split}
		&{\mathrm{pr}}\left(\left|\left[\left(\B_{W'} -  E  \B_{W'}\right)^\top\left(\B_{W'} -  V\right)\beta^*\right]_i\right| \geq Cn\frac{\log\nu_{\min}}{\nu_{\min}}\zeta_{\max}\|\beta^*\|_1\right)\\
		\leq& {\mathrm{pr}}\left(\left|\sum_{j \in S}\sum_{k = 1}^{n}D_{ki}D_{kj}\beta_j^*\right| \geq Cn\frac{\log\nu_{\min}}{\nu_{\min}}\zeta_{\max}\|\beta^*\|_1\right)\\
		&+ {\mathrm{pr}}\left(\left|\sum_{j \in S}\sum_{k = 1}^{n}D_{ki}\left(D_{kj} - \log\nu_{kj}\right)\beta_j^*\right| \geq Cn\frac{\log\nu_{\min}}{\nu_{\min}}\zeta_{\max}\|\beta^*\|_1\right)\\
		\leq& {\mathrm{pr}}\left(\sum_{j \in S}\left|\sum_{k = 1}^{n}D_{ki}D_{kj}\right|\left|\beta_j^*\right| \geq Cn\frac{\log\nu_{\min}}{\nu_{\min}}\zeta_{\max}\sum_{j \in S}\left|\beta_j^*\right|\right)\\
		&+ {\mathrm{pr}}\left(\sum_{j \in S}\left|\sum_{k = 1}^{n}D_{ki}\left(D_{kj} - \log\nu_{kj}\right)\right|\left|\beta_j^*\right| \geq Cn\frac{\log\nu_{\min}}{\nu_{\min}}\zeta_{\max}\sum_{j \in S}\left|\beta_j^*\right|\right)\\
		\leq& \sum_{j \in S}{\mathrm{pr}}\left(\left|\sum_{k = 1}^{n}D_{ki}D_{kj}\right| \geq Cn\frac{\log\nu_{\min}}{\nu_{\min}}\zeta_{\max}\right)\\ &+ \sum_{j \in S}{\mathrm{pr}}\left(\left|\sum_{k = 1}^{n}D_{ki}\left( E \phi\left(W'_{kj}\right) - \log\nu_{kj}\right)\right| \geq n\frac{\log\nu_{\min}}{\nu_{\min}}\zeta_{\max}\right)\\
		\leq& s(2 + e)e^{-cn\log\nu_{\min}}.
	\end{split}
\end{equation*}
Therefore, $\exists$ two large constant $C, C' > 0$,
\begin{equation}\label{eq212}
	\begin{split}
		&{\mathrm{pr}}\left(\left\|\left(\B_{W'} -  E  \B_{W'}\right)^\top\left(\B_{W'} -  V\right)\beta^*\right\|_{\infty} \geq Cn\frac{\log\nu_{\min}}{\nu_{\min}}\zeta_{\max}\|\beta^*\|_1\right)\\
		\leq& {\mathrm{pr}}\left(\exists 1 \leq i \leq p, \left|\left[\left(\B_{W'} -  E  \B_{W'}\right)^\top\left(\B_{W'} -  V\right)\beta^*\right]_i\right| \geq Cn\frac{\log\nu_{\min}}{\nu_{\min}}\zeta_{\max}\|\beta^*\|_1\right)\\
		\leq& p\cdot s(2 + e)e^{-cn\log\nu_{\min}} \leq p^{-C'}.
	\end{split}
\end{equation}
For all $1 \leq i \leq p$,
\begin{equation}\label{eq256}
	\begin{split}
		\left[\P\left( E  \B_{W'}\right)^\top\left(\B_{W'} -  V\right)\beta^*\right]_i =& \sum_{j = 1}^{p}\sum_{k = 1}^{n}\left( E \phi\left(W'_{ki}\right) - \frac{1}{p}\sum_{l = 1}^{p} E \phi\left(W'_{kl}\right)\right)\left(\phi_1\left(W'_{kj}\right) - \log\nu_{kj}\right)\beta_j^*\\
		=& \sum_{j \in S}\sum_{k = 1}^{n}\left( E \phi\left(W'_{ki}\right) - \frac{1}{p}\sum_{l = 1}^{p} E \phi\left(W'_{kl}\right)\right)\left(\phi\left(W'_{kj}\right) - \log\nu_{kj}\right)\beta_j^*.
	\end{split}
\end{equation}
By Lemma \ref{lm:beta-binomial_1},
\begin{equation*}
	\begin{split}
		\left| E \phi\left(W'_{ki}\right) - \frac{1}{p}\sum_{l = 1}^{p} E \phi\left(W'_{kl}\right)\right|
		\leq \left|\log\nu_{ki} - \frac{1}{p}\sum_{l = 1}^p\log\nu_{kl}\right| + C\frac{\zeta_k}{\nu_{ki}} + \frac{1}{p}\sum_{l = 1}^pC\frac{\zeta_k}{\nu_{kl}} \leq C.
	\end{split}
\end{equation*}
Denote $E_{ki} =  E \phi_1\left(W'_{ki}\right) - \frac{1}{p}\sum_{l = 1}^{p} E \phi_1\left(W'_{kl}\right)$, by Hoeffding-type inequality, $\forall j \in S, t \geq 0$, 
\begin{equation*}
	\begin{split}
		{\mathrm{pr}}\left(\left|\sum_{k = 1}^{n}E_{ki}\left(\phi\left(W'_{kj}\right) -  E \phi\left(W'_{kj}\right)\right)\right| \geq t\right) \leq e\cdot\exp\left(-c\frac{\nu_{\min}t^2}{\zeta_{\max}\sum_{k = 1}^{n}E_{ki}^2}\right) \leq e\cdot e^{-c\frac{\nu_{\min}t^2}{n\zeta_{\max}}},
	\end{split}
\end{equation*}
which is equivalent to
\begin{equation}\label{eq255}
	{\mathrm{pr}}\left(\left|\sum_{k = 1}^{n}E_{ki}\left(\phi\left(W'_{kj}\right) - \log\nu_{kj}\right) - \sum_{k = 1}^{n}E_{ki}\left( E \phi\left(W'_{kj}\right) - \log\nu_{kj}\right)\right| \geq t\right) \leq e\cdot e^{-c\frac{\nu_{\min}t^2}{n\zeta_{\max}}}.
\end{equation}
Since $\left| E \phi_1\left(W'_{kj}\right) - \log\nu_{kj}\right| \leq 10\frac{\zeta_{\max}}{\nu_{\min}}$ for all $1 \leq k \leq n$, we know that
\begin{equation*}
	\left|\sum_{k = 1}^{n}E_{ki}\left( E \phi_1\left(W'_{kj}\right) - \log\nu_{kj}\right)\right| \leq 10n\frac{\zeta_{\max}}{\nu_{\min}}. 
\end{equation*}
Choose $t = C\sqrt{n\log p}\frac{\sqrt{\zeta_{\max}}}{\sqrt{\nu_{\min}}}$ in \eqref{eq255}, we have
\begin{equation*}
	{\mathrm{pr}}\left(\left|\sum_{k = 1}^{n}E_{ki}\left(\phi_1\left(W'_{kj}\right) - \log\nu_{kj}\right)\right| \geq C\left\{\sqrt{n\log p}\left(\frac{\zeta_{\max}}{\nu_{\min}}\right)^{\frac{1}{2}} + n\frac{\zeta_{\max}}{\nu_{\min}}\right\}\right) \leq e^{-C''\log p}.
\end{equation*}
By \eqref{eq256} and the previous inequality,
\begin{equation*}
	\begin{split}
		&{\mathrm{pr}}\left(\left|\left[\P\left( E  \B_{W'}\right)^\top\left(\B_{W'} -  V\right)\beta^*\right]_i\right| \geq C\left\{\sqrt{n\log p}\left(\frac{\zeta_{\max}}{\nu_{\min}}\right)^{\frac{1}{2}} + n\frac{\zeta_{\max}}{\nu_{\min}}\right\}\|\beta^*\|_1\right)\\
		\leq&  {\mathrm{pr}}\left(\left|\sum_{j \in S}\sum_{k = 1}^{n}E_{ki}\left(\phi_1\left(W'_{kj}\right) - \log\nu_{kj}\right)\beta_j^*\right| \geq C\left\{\sqrt{n\log p}\left(\frac{\zeta_{\max}}{\nu_{\min}}\right)^{\frac{1}{2}} + n\frac{\zeta_{\max}}{\nu_{\min}}\right\}\|\beta^*\|_1\right)\\
		\leq& {\mathrm{pr}}\left(\sum_{j \in S}\left|\sum_{k = 1}^{n}E_{ki}\left(\phi_1\left(W'_{kj}\right) - \log\nu_{kj}\right)\right|\left|\beta_j^*\right| \geq C\left\{\sqrt{n\log p}\left(\frac{\zeta_{\max}}{\nu_{\min}}\right)^{\frac{1}{2}} + n\frac{\zeta_{\max}}{\nu_{\min}}\right\}\sum_{j \in S}\left|\beta_j^*\right|\right)\\
		\leq& \sum_{j \in S}{\mathrm{pr}}\left(\left|\sum_{k = 1}^{n}E_{ki}\left(\phi_1\left(W'_{kj}\right) - \log\nu_{kj}\right)\right| \geq C\left\{\sqrt{n\log p}\left(\frac{\zeta_{\max}}{\nu_{\min}}\right)^{\frac{1}{2}} + n\frac{\zeta_{\max}}{\nu_{\min}}\right\}\right)\\
		\leq& s\dot e^{-C''\log p}.
	\end{split}
\end{equation*}
Therefore, there exist two large constants $C, C' > 0$, 
\begin{equation}\label{eq213}
	\begin{split}
		&{\mathrm{pr}}\left(\left\|\P\left( E  \B_{W'}\right)^\top\left(\B_{W'} -  V\right)\beta^*\right\|_{\infty} \geq C\left\{\sqrt{n\log p}\left(\frac{\zeta_{\max}}{\nu_{\min}}\right)^{\frac{1}{2}} + n\frac{\zeta_{\max}}{\nu_{\min}}\right\}\|\beta^*\|_1\right)\\
		=&{\mathrm{pr}}\left(\exists 1 \leq i \leq p, \left|\left[\P\left( E  \B_{W'}\right)^\top\left(\B_{W'} -  V\right)\beta^*\right]_i\right| \geq C\left\{\sqrt{n\log p}\left(\frac{\zeta_{\max}}{\nu_{\min}}\right)^{\frac{1}{2}} + n\frac{\zeta_{\max}}{\nu_{\min}}\right\}\|\beta^*\|_1\right)\\
		\leq& ps\cdot p^{-C''} \leq p^{-C'}.
	\end{split}
\end{equation}
Combine (\ref{eq211}), (\ref{eq212}), (\ref{eq213}) and (\ref{eq214}), we have
\begin{equation*}
	\begin{split}
		&{\mathrm{pr}}\left(\left\|\bar{\A}_{W'} \beta^* - \bar{\B}_{W'}^\top y\right\|_\infty \leq C\left[\sqrt{n\log p}\left\{\sigma + \left(\frac{pn}{\nu}\zeta_{\max}\right)^{\frac{1}{2}}\|\beta^*\|_1\right\} + n\frac{\log\nu_{\min}}{\nu_{\min}}\zeta_{\max}\|\beta^*\|_1\right]\right)\\
		\leq&{\mathrm{pr}}\left(\left\|\bar{\B}_{W'}^\top\varepsilon\right\|_{\infty} > C\sqrt{n\log p\cdot\sigma^2}\right)\\ &+ {\mathrm{pr}}\left(\left\|\P\left( E  \B_{W'}\right)^\top\left(\B_{W'} -  V\right)\beta^*\right\|_{\infty} \geq C\left\{\sqrt{n\log p}\left(\frac{\zeta_{\max}}{\nu_{\min}}\right)^{\frac{1}{2}} + n\frac{\zeta_{\max}}{\nu_{\min}}\right\}\|\beta^*\|_1\right)\\
		&+ {\mathrm{pr}}\left(\left\|\left(\B_{W'} -  E  \B_{W'}\right)^\top\left(\B_{W'} -  V\right)\beta^*\right\|_{\infty} \geq Cn\frac{\log\nu_{\min}}{\nu_{\min}}\zeta_{\max}\|\beta^*\|_1\right)\\
		\leq& 4p^{-C'}.
	\end{split}
\end{equation*}
Also notice that (\ref{eq228}) holds, we have
\begin{equation*}
	\begin{split}
		&{\mathrm{pr}}\left(\left\|\bar{\A}_{W} \beta^* - \bar{\B}_{W}^\top y\right\|_\infty \leq C\left[\sqrt{n\log p}\left\{\sigma + \left(\frac{pn}{\nu}\zeta_{\max}\right)^{\frac{1}{2}}\|\beta^*\|_1\right\} + n\frac{\log\nu_{\min}}{\nu_{\min}}\zeta_{\max}\|\beta^*\|_1\right]\right)\\
		\leq&{\mathrm{pr}}\left(\left\|\bar{\A}_{W'} \beta^* - \bar{\B}_{W'}^\top y\right\|_\infty \leq C\left[\sqrt{n\log p}\left\{\sigma + \left(\frac{pn}{\nu}\zeta_{\max}\right)^{\frac{1}{2}}\|\beta^*\|_1\right\} + n\frac{\log\nu_{\min}}{\nu_{\min}}\zeta_{\max}\|\beta^*\|_1\right]\right)\\& + {\mathrm{pr}}\left(\exists i, j, W_{ij} \neq W_{ij}'\right)\\
		\leq& 4p^{-C'} + \sum_{i = 1}^{n}\sum_{j = 1}^{p}\left(4\exp\left(-c\frac{\alpha_i}{p}\right) + 2\exp\left(-c\nu_{ij}\right)\right).
	\end{split}
\end{equation*}
Noting that
$\frac{\alpha_i}{p} \asymp \frac{\nu_{ij}}{\nu_{i}/\alpha_i} \geq \frac{\nu_{ij}}{\zeta_i} \geq \frac{\nu_{\min}}{\zeta_{\max}} \geq C\log(np)$, for any $1 \leq i \leq n, 1 \leq j \leq p$,we have
\begin{equation*}
	4\exp\left(-c\frac{\alpha_i}{p}\right) \leq (np)^{-(C' + 1)} \leq \frac{1}{np}p^{-C'}, \quad
	2\exp\left(-c\nu_{ij}\right) \leq (np)^{-(C' + 1)} \leq \frac{1}{np}p^{-C'}.
\end{equation*}
Therefore
\begin{equation*}
	{\mathrm{pr}}\left(\left\|\bar{\A}_{W} \beta^* - \bar{\B}_{W}^\top y\right\|_\infty \leq C\left[\sqrt{n\log p}\left\{\sigma + \left(\frac{pn}{\nu}\zeta_{\max}\right)^{\frac{1}{2}}\|\beta^*\|_1\right\} + n\frac{\log\nu_{\min}}{\nu_{\min}}\zeta_{\max}\|\beta^*\|_1\right]\right) \leq 6p^{-C'}.
\end{equation*}
\qed
\medskip

{\noindent Proof of Lemma \ref{lm: RIP fixed}.} 
Let $ V'$ be the random matrix with entries $v_{ij}' = v_{ij} - v.$
By Lemma 5 in \cite{achlioptas2001database}, for all $x \in \bbR^p, 0 < \epsilon < 1$, 
\begin{equation}\label{eq44}
	{\mathrm{pr}}\left(\left|\frac{1}{n}\| V'x\|_2^2 - \|x\|_2^2\right| \geq \epsilon\|x\|_2^2\right) \leq 2\exp\left(-n\left(\frac{\epsilon^2}{4} - \frac{\epsilon^3}{6}\right)\right).
\end{equation}
Since $ V\P =  V'\P$, we know that for any fixed $x \in \bbR^p$,
\begin{equation}\label{eq42}
	{\mathrm{pr}}\left(\left|\frac{1}{n}\| V\P x\|_2^2 - \|\P x\|_2^2\right| \geq \epsilon\|\P x\|_2^2\right) \leq 2\exp\left(-n\left(\frac{\epsilon^2}{4} - \frac{\epsilon^3}{6}\right)\right).
\end{equation}
For any $2s$-sparse $x \in \bbR^p$, denote $S_0$ be the support set of $x$. By Cauchy-Schwarz inequality,
\begin{equation*}
	\begin{split}
		\|x\|_2^2 \geq \|\P x\|_2^2 = \|x\|_2^2 - \frac{1}{p}\left(\sum_{i = 1}^{p}x_i\right)^2 = \|x\|_2^2 - \frac{1}{p}\left(\sum_{i \in S_0}x_i\right)^2\geq \|x\|_2^2 - \frac{2s}{p}\|x\|_2^2 = (1 - \frac{2s}{p})\|x\|_2^2.
	\end{split}
\end{equation*}
If $\left|\frac{1}{n}\| V\P x\|_2^2 - \|\P x\|_2^2\right| \leq \epsilon\|\P x\|_2^2$, then 
\begin{equation*}
	\frac{1}{n}\| V\P x\|_2^2 \leq \left(1 + \epsilon\right)\|\P x\|_2^2 \leq \left(1 + \epsilon\right)\|x\|_2^2 \leq \left(1 + \frac{2s}{p} + \epsilon\right)\|x\|_2^2,
\end{equation*}
and 
\begin{equation*}
	\frac{1}{n}\| V\P x\|_2^2 \geq (1 - \epsilon)\|\P x\|_2^2 \geq \left(1 - \epsilon\right)\left(1 - \frac{2s}{p}\right)\|x\|_2^2 \geq \left(1 - \frac{2s}{p} - \epsilon\right)\|x\|_2^2.
\end{equation*}
(\ref{eq42}) and the previous two inequalities together imply
\begin{equation}\label{eq83}
	{\mathrm{pr}}\left(\left|\frac{1}{n}\| V\P x\|_2^2 - \|x\|_2^2\right| \geq \left(\frac{2s}{p} + \epsilon\right)\|x\|_2^2\right) \leq 2\exp\left(-n\left(\frac{\epsilon^2}{4} - \frac{\epsilon^3}{6}\right)\right).
\end{equation}
For any $S_0 \subset \{1, 2, \dots, p\}, \left|S_0\right| = 2s$, denote $\mathcal{A}_{S_0} = \{x: x \in \bbR^p, \|x\|_2 = 1, \text{supp}(x) \subset S\}$, by \cite{vershynin2010introduction}, we can find an $\epsilon$-net of $\mathcal{A}_{S_0}$, called $\mathcal{N}_{\epsilon}$, with $\left|\mathcal{N}_{\epsilon}\right| \leq \left(1 + \frac{2}{\epsilon}\right)^{2s}$. By Lemma \ref{lm:epsilon-net} with $\tilde{\A} = \left( V\P\right)^\top\left( V\P\right)$,
\begin{equation}\label{eq43}
	\sup_{x \in \mathcal{A}_{S_0}}\frac{1}{n}\| V\P x\|_2^2 \leq (1 - 2\epsilon)^{-1} \sup_{x \in \mathcal{N}_{\epsilon}} \frac{1}{n}\| V\P x\|_2^2.
\end{equation}
By \eqref{eq83} and the union bound, 
\begin{equation*}
	\begin{split}
		{\mathrm{pr}}\left(\exists x \in \mathcal{N}_{\frac{\epsilon}{4}}, \left|\frac{1}{n}\| V\P x\|_2^2 - \|x\|_2^2\right| \geq \left(\frac{2s}{p} + \epsilon\right)\|x\|_2^2\right) \leq 2\left(1 + \frac{8}{\epsilon}\right)^{2s}\exp\left(-n\left(\frac{\epsilon^2}{4} - \frac{\epsilon^3}{6}\right)\right),
	\end{split}
\end{equation*}
i.e., 
\begin{equation}\label{eq84}
	\begin{split}
		&{\mathrm{pr}}\left(\forall x \in \mathcal{N}_{\frac{\epsilon}{4}}, \left|\frac{1}{n}\| V\P x\|_2^2 - \|x\|_2^2\right| < \left(\frac{2s}{p} + \epsilon\right)\|x\|_2^2\right)\\ \geq& 1 - 2\left(1 + \frac{8}{\epsilon}\right)^{2s}\exp\left(-n\left(\frac{\epsilon^2}{4} - \frac{\epsilon^3}{6}\right)\right).
	\end{split}
\end{equation}
For $\epsilon \leq \frac{1}{3}$, if for any $x \in \mathcal{N}_{\frac{\epsilon}{4}}$,  $\left|\frac{1}{n}\| V\P x\|_2^2 - \|x\|_2^2\right| < \left(\frac{2s}{p} + \epsilon\right)\|x\|_2^2 = \frac{2s}{p} + \epsilon$, then (\ref{eq43}) shows us for any $x \in \mathcal{A}_{S}$, 
\begin{equation}\label{eq85}
	\frac{1}{n}\| V\P x\|_2^2 \leq \frac{1}{1 - \frac{\epsilon}{2}}\left(1 + \frac{2s}{p} + \epsilon\right) \leq 1 + \frac{2s}{p} + 2\epsilon.
\end{equation}
By the definition of $\epsilon$-net, $\exists y \in \mathcal{N}_{\frac{\epsilon}{4}}$, $\|x - y\|_2 \leq \frac{\epsilon}{4}$, 
\begin{equation*}
	\frac{1}{\sqrt{n}}\| V\P x\|_2 \geq \frac{1}{\sqrt{n}}\left(\| V\P y\|_2 - \| V\P(y - x)\|_2\right) \geq \sqrt{1 - \frac{2s}{p} - \epsilon} - \frac{\epsilon}{4}\sqrt{1 + \frac{2s}{p} + 2\epsilon} \geq 1 - \frac{2s}{p} - 2\epsilon.
\end{equation*}
Thus, we have 
\begin{equation*}
	\frac{1}{n}\| V\P x\|_2^2 \geq \left(1 - \frac{2s}{p} - 2\epsilon\right)^2 \geq 1 - \frac{4s}{p} - 4\epsilon.
\end{equation*}
Therefore, if for any $x \in \mathcal{N}_{\frac{\epsilon}{4}}$,  $\left|\frac{1}{n}\| V\P x\|_2^2 - \|x\|_2^2\right| < \left(\frac{2s}{p} + \epsilon\right)\|x\|_2^2 = \frac{2s}{p} + \epsilon$, then
\begin{equation}\label{eq403}
	1 - \frac{4s}{p} - 4\epsilon \leq \frac{1}{n}\| V\P x\|_2^2 \leq 1 + \frac{4s}{p} + 4\epsilon.
\end{equation}
By \eqref{eq84} and \eqref{eq403},
\begin{equation*}
	\begin{split}
		&{\mathrm{pr}}\left(\forall x \in \bbR^p, \text{ supp}(x) \subseteq S_0, \left|\frac{1}{n}\| V\P x\|_2^2 - \|x\|_2^2\right| \leq \left(\frac{4s}{p} + 4\epsilon\right)\|x\|_2^2\right)\\
		=& {\mathrm{pr}}\left(\forall x \in \mathcal{A}_{S_0}, \left|\frac{1}{n}\| V\P x\|_2^2 - \|x\|_2^2\right| \leq \left(\frac{4s}{p} + 4\epsilon\right)\|x\|_2^2\right)\\
		\geq&{\mathrm{pr}}\left(\forall x \in \mathcal{N}_{\frac{\epsilon}{4}}, \left|\frac{1}{n}\| V\P x\|_2^2 - \|x\|_2^2\right| \leq \left(\frac{2s}{p} + \epsilon\right)\|x\|_2^2\right)\\ \geq& 1 - 2\left(1 + \frac{8}{\epsilon}\right)^{2s}\exp\left(-n\left(\frac{\epsilon^2}{4} - \frac{\epsilon^3}{6}\right)\right),
	\end{split}
\end{equation*}
which means
\begin{equation*}
	\begin{split}
		&{\mathrm{pr}}\left(\exists x \in \bbR^p, \text{ supp}(x) \subseteq S_0, \left|\frac{1}{n}\| V\P x\|_2^2 - \|x\|_2^2\right| > \left(\frac{4s}{p} + 4\epsilon\right)\|x\|_2^2\right)\\ \leq &2\left(1 + \frac{8}{\epsilon}\right)^{2s}\exp\left(-n\left(\frac{\epsilon^2}{4} - \frac{\epsilon^3}{6}\right)\right).
	\end{split}
\end{equation*}
Therefore,
\begin{equation*}
	\begin{split}
		&{\mathrm{pr}}\left(\forall x \in \bbR^p, \|x\|_0 \leq 2s, \left|\frac{1}{n}\| V\P x\|_2^2 - \|x\|_2^2\right| < \left(\frac{4s}{p} + 4\epsilon\right)\|x\|_2^2\right)\\
		=& 1 - {\mathrm{pr}}\left(\exists S_0 \subset \{1, \dots, p\}, |S_0| = 2s,  x \in \bbR^p, \text{ supp}(x) \subseteq S_0, \left|\frac{1}{n}\| V\P x\|_2^2 - \|x\|_2^2\right| > \left(\frac{4s}{p} + 4\epsilon\right)\|x\|_2^2\right)\\
		\geq&1 - \binom{p}{2s}\cdot 2\left(1 + \frac{8}{\epsilon}\right)^{2s}\exp\left(-n\left(\frac{\epsilon^2}{4} - \frac{\epsilon^3}{6}\right)\right)\\
		\geq&1 - 2\left(\frac{ep}{2s}\right)^{2s}\left(1 + \frac{8}{\epsilon}\right)^{2s}\exp\left(-n\left(\frac{\epsilon^2}{4} - \frac{\epsilon^3}{6}\right)\right)\\
		=&1 - 2\exp\left(-n\left(\frac{\epsilon^2}{4} - \frac{\epsilon^3}{6}\right) + 2s\cdot\log\left(\frac{ep}{2s}\right) + 2s\log\left(1 + \frac{8}{\epsilon}\right)\right).
	\end{split}
\end{equation*}
Set $\epsilon = \frac{1}{160}$, since $n \geq Cs\log p$, $\exists c > 0$ such that
\begin{equation*}
	\begin{split}
		{\mathrm{pr}}\left(\forall x \in \bbR^p, \|x\|_0 \leq 2s, \left|\frac{1}{n}\| V\P x\|_2^2 - \|x\|_2^2\right| < \left(\frac{4s}{p} + \frac{1}{40}\right)\|x\|_2^2\right) \geq 1 - 2e^{-cn}.
	\end{split}
\end{equation*}
Choose $\epsilon = \frac{1}{160}$, for sufficient large $n$, we have
\begin{equation*}
	{\mathrm{pr}}\left(\forall x \in \bbR^p, \|x\|_0 \leq s, \left|\frac{1}{n}\| V\P x\|_2^2 - \|x\|_2^2\right| < \frac{1}{40}\|x\|_2^2\right) > \frac{2}{3}.
\end{equation*}
\qed
\medskip

{\noindent Proof of Lemma \ref{lm:general_1}.} 
First, we bound the variance of $W_{ij}$. By the sub-exponential tail assumption \ref{ineq:sub_exponential_tail}, 
\begin{equation}\label{ineq:variance_general}
	\begin{split}
		\var(W_{ij}) =& \int_{0}^{\infty}{\mathrm{pr}}\{(W_{ij} - \nu_{ij})^2 \geq t\}dt = \int_{0}^{\infty}Ce^{-c(t/\nu_{ij}) \wedge \sqrt{t}}dt\\ \leq& \int_{0}^{\infty}Ce^{-ct/\nu_{ij}}dt + \int_{\nu_{ij}^2}^{\infty}Ce^{-c\sqrt{t}}dt \leq C\nu_{ij}/c + \int_{\nu_{ij}^2}^{\infty}Ct^{-2}dt \leq C\nu_{ij}.
	\end{split}
\end{equation}
Therefore, we know that $F \leq C$ for some constant $C$.
Let $$f(t) = \log(t + \frac{1}{2}) - \log\nu_{ij} - \frac{t - \nu_{ij} + \frac{1}{2}}{\nu_{ij}} + \frac{(t - \nu_{ij} + \frac{1}{2})^2}{2\nu_{ij}^2}.$$ \eqref{ineq:variance_general} and the similar proof of \eqref{ineq40} together show that for $\nu_{ij} \geq C_{\epsilon}$, 
\begin{equation*}
	\left|E f(W_{ij})\right| \leq C\nu_{ij}^{-3/2 + \epsilon}.
\end{equation*}
Recall that $E(W_0) = v$, we have
\begin{equation*}
	E\{f(W_{ij})\} = E\{\log(W_{ij} + 1/2)\} - \log\nu_{ij} - \frac{1}{2\nu_{ij}} + \frac{\var(W_{ij}) + \frac{1}{4}}{2\nu_{ij}^2} = E\{\log(W_{ij} + 1/2)\} - \log\nu_{ij} + \frac{R_{ij}-1}{2\nu_{ij}} + \frac{1}{8\nu_{ij}^2},
\end{equation*}
where $R_{ij} = \frac{\var(W_{ij})}{\nu_{ij}}$ and $|R_{ij} - 1| \leq F$. 
\eqref{ineq:variance_general} and the previous two inequalities together imply that
\begin{equation*}
	\left|E\{\log(W_{ij} + 1/2)\} - \log\nu_{ij}\right| \leq \frac{F}{2\nu_{ij}} + \frac{C}{\nu_{ij}^{3/2 - \epsilon}}
\end{equation*}
for all $\nu_{ij} \geq C_{\epsilon}$.

By the sub-exponential tail assumption \ref{ineq:sub_exponential_tail}, we know that for any $0 \leq t \leq 9\nu_{ij}$,
\begin{equation*}
	{\mathrm{pr}}\left(|W_{ij}'  - \nu_{ij}| \geq t\right) \leq {\mathrm{pr}}\left(|W_{ij}  - \nu_{ij}| \geq t\right) + {\mathrm{pr}}\left(W_{ij} \neq W_{ij}'\right) \leq C\exp\left(-ct^2/\nu_{ij}\right) + C\exp\left(-c\nu_{ij}\right) \leq C\exp\left(-ct^2/\nu_{ij}\right).
\end{equation*}
Moreover, for any $t > 9\nu_{ij}$, by the definition of $W_{ij}'$, we have $\bbP\left(|W_{ij}'  - \nu_{ij}| \geq t\right) = 0$. Therefore, for any $t \geq 0$, we have
\begin{equation}\label{ineq:sub_Gaussian_W_0'}
	{\mathrm{pr}}\left(|W_{ij}'  - \nu_{ij}| \geq t\right) \leq C\exp\left(-ct^2/\nu_{ij}\right).
\end{equation}
Similar arguments of \eqref{eq90} lead us to 
\begin{equation*}
	\left|E \left\{\log\left(W_{ij}' + \frac{1}{2}\right)\right\} - \log\left\{\left(W_{ij} + \frac{1}{2}\right)\right\}\right| \leq \frac{C}{\nu_{ij}^{3/2 - \epsilon}}.
\end{equation*}
Therefore, we conclude that for any $\nu_{ij} \geq C_{\epsilon}$,
\begin{equation*}
	\left|E \left\{\log\left(W_{ij}' + \frac{1}{2}\right)\right\} - \log\nu_{ij}\right| \leq \frac{F}{2\nu_{ij}} + \frac{C}{\nu_{ij}^{3/2 - \epsilon}}.
\end{equation*}
\qed
\medskip

{\noindent Proof of Lemma \ref{lm:general_2}.} 
Let $g(t) = \log^2(t + \frac{1}{2}) - \log^2\nu_{ij} - 2\frac{\log \nu_{ij}}{\nu_{ij}}(t - \nu_{ij} + \frac{1}{2}) - \frac{1-\log\nu_{ij}}{v^2}(t - \nu_{ij} + \frac{1}{2})^2$. Then 
\begin{equation*}
	\begin{split}
		E\{g(W_{ij})\} =& E\left\{\log^2\left(W_{ij} + \frac{1}{2}\right)\right\} - \log^2\nu_{ij} - \frac{\log \nu_{ij}}{\nu_{ij}} + \frac{\log\nu_{ij} - 1}{\nu_{ij}^2}\left\{\var(W_{ij}) + \frac{1}{4}\right\}\\
		=& E\left\{\log^2\left(W_{ij} + \frac{1}{2}\right)\right\} - \log^2\nu_{ij} + (R_{\nu_{ij}} - 1)\frac{\log\nu_{ij}}{\nu_{ij}} - \frac{R_{ij}}{\nu_{ij}} + \frac{\log \nu_{ij} - 1}{4\nu_{ij}^2},
	\end{split}
\end{equation*}
where $R_{ij} = \frac{\var(W_{ij})}{\nu_{ij}}$. By similar arguments of \eqref{ineq41} and \eqref{ineq42}, one has
\begin{equation*}
	\left|E \{g(W_{ij})\}\right| \leq C/\nu_{ij}
\end{equation*}
and
\begin{equation*}
	\left|E\left\{\log^2\left(W_{ij} + \frac{1}{2}\right)\right\} - E\left\{\log^2\left(W_{ij}' + \frac{1}{2}\right)\right\}\right| \leq \frac{C}{\nu_{ij}}.
\end{equation*}
for any $v \geq C$.
\eqref{ineq:variance_general} and the precious inequalities together show that 
\begin{equation*}
	\left|E\left\{\log^2\left(W_{ij}' + \frac{1}{2}\right)\right\} - \log^2\nu_{ij}\right| \leq F\frac{\log\nu_{ij}}{\nu_{ij}} + \frac{C}{\nu_{ij}}
\end{equation*}
provided that $\nu \geq C$.

\qed
\medskip

{\noindent Proof of Lemma \ref{lm:sub-Gaussian_general}.} 
By Lemma \ref{lm:general_1} and \eqref{ineq:sub_Gaussian_W_0'}, we can prove Lemma \ref{lm:sub-Gaussian_general} essentially the same as the proof of Lemma \ref{lm:sub-Gaussian}.
\qed
\medskip

{\noindent Proof of Lemma \ref{lm:infinity_norm_bound3}.} 
By the same proof of \eqref{eq510}, we have
\begin{equation}\label{ineq43}
	{\mathrm{pr}} \left\{\left|\sum_{k=1}^{n}\left(\bar{\B}_{W'}\right)_{ki}\varepsilon_k\right| > t\Big|W'_{kj}, 1 \leq k \leq n, 1 \leq j \leq p\right\} \leq 2e^{-\frac{t^2}{2\sigma^2\sum_{k=1}^n \left(\bar{\B}_{W'}\right)_{ki}^2}}.
\end{equation}
Denote $D_{ki} = \phi(W_{ki}) -  E \phi(W_{ki}), Q_{ki} = D_{ki}^2 -  E  D_{ki}^2, M_{ki} = \left(\bar{\B}_W\right)_{ki} -  E \left(\bar{\B}_W\right)_{ki}$, $R_{ki} = M_{ki}^2 -  E  M_{ki}^2$.
By the same proof of \eqref{eq9}, we have
\begin{equation}\label{ineq44}
	\begin{split}
		{\mathrm{pr}}\left(\|\bar{\B}_{W'}^\top\varepsilon\|_{\infty} > t\right) \leq Cp\left(e^{-\frac{ct^2}{n\sigma^2}} + e^{-cn}\right).
	\end{split}	
\end{equation}
By Lemma \ref{lm:general_1} and Cauchy-Schwarz inequality, 
\begin{equation*}
	\begin{split}
		&\left(\sum_{j \in S} E \left(\phi_{1}\left(W'_{kj}\right) - \log\nu_{kj}\right)\beta_j^*\right)^2 \leq \sum_{j \in S}\left( E \phi_1\left(W'_{kj}\right) - \log\nu_{kj}\right)^2\|\beta^*\|_2^2 \leq s\left(\frac{F^2}{\nu_{\min}^{2}} + \frac{C}{\nu_{\min}^{3 - 2\epsilon}}\right)\|\beta^*\|_2^2,
	\end{split}	
\end{equation*} 
which means
\begin{equation}\label{ineq45}
	\left|\sum_{j \in S} E \left(\phi_{1}\left(W'_{kj}\right) - \log\nu_{kj}\right)\beta_j^*\right| \leq \sqrt{s}\left(\frac{F}{\nu_{\min}} + \frac{C}{\nu_{\min}^{3/2 - \epsilon}}\right)\|\beta^*\|_2.
\end{equation}
Moreover, by Lemma \ref{lm:general_1}, Lemma \ref{lm:general_2} and $\nu_{\min} \asymp \nu_{kj}$,
\begin{equation*}
	\begin{split}
		\var\{\phi_1(W_{kj}')\} =& E\{\phi_1^2(W_{kj}')\} - [E\{\phi_1(W_{kj}')\}]^2\\ \leq& \log^2(\nu_{ki}) + F\frac{\log(\nu_{kj})}{\nu_{kj}} + \frac{C}{\nu_{kj}} - \left\{\log(\nu_{kj}) - \frac{F}{\nu_{kj}} - \frac{C}{\nu_{kj}^{3/2 - \epsilon}}\right\}^2\\ \leq& C\left(F\frac{\log(\nu_{\min})}{\nu_{\min}} + \frac{1}{\nu_{\min}}\right).
	\end{split}
\end{equation*}
Therefore, we have
\begin{equation*}
	\begin{split}
		& \var\left\{\sum_{k=1}^{n}\left(\bar{\B}_{W'}\right)_{ki}\right\} = \sum_{k=1}^{n}\var\left\{\left(\bar{\B}_{W'}\right)_{ki}\right\}\\ 
		=& \sum_{k=1}^{n}\left[\left(1 - \frac{1}{p}\right)^2\var\left\{\phi_1(W_{ki}')\right\}+ \frac{1}{p^2}\sum_{j \neq i}\var\left\{\phi_1(W_{kj}')\right\}\right] \leq Cn\left(F\frac{\log(\nu_{\min})}{\nu_{\min}} + \frac{1}{\nu_{\min}}\right).
	\end{split}
\end{equation*}
and
\begin{equation*}
	\var\left(\sum_{j \in S}U_{kj}\beta_j^*\right) = \sum_{j \in S}\beta_j^{*2}\var\left(U_{kj}\right) = \sum_{j \in S}\beta_j^{*2}\var\left\{\phi_1\left(W'_{kj}\right)\right\} \leq C\left(F\frac{\log(\nu_{\min})}{\nu_{\min}} + \frac{1}{\nu_{\min}}\right)\|\beta^*\|_2^2.
\end{equation*}
Similarly to \eqref{eq8}, we can show that
\begin{equation}\label{ineq46}
	\begin{split}
		\left| E \left[\P\left(\A_{W'} - \B_{W'}^\top  V\right)\beta^*\right]_i\right|
		\leq Cn\sqrt{s}\left(\frac{F}{\nu_{\min}} + \frac{1}{\nu_{\min}^{3/2 - \epsilon}}\right)\|\beta^*\|_2 + C\frac{\sqrt{n}}{\sqrt{\nu_{\min}}}\|\beta^*\|_2.
	\end{split}
\end{equation}
By the similar proof of \eqref{eq10}, we have
\begin{equation}\label{ineq47}
	{\mathrm{pr}}\left[\|\P\left(\A_{W'} - \B_{W'}^\top  V\right)\beta^*\|_{\infty} \geq C\sqrt{n\log p}\left(\frac{p}{\bar\nu}\right)^{\frac{1}{2}}\|\beta^*\|_2 + Cn\sqrt{s}\left\{\left(\frac{p}{\bar{\nu}}\right)^{\frac{3}{2} - \epsilon} + F\frac{p}{\bar{\nu}}\right\}\|\beta^*\|_2\right] \leq e^{-C'\log p}.
\end{equation}
Then similar arguments of \eqref{ineq48} and \eqref{ineq49} can show that
\begin{equation}\label{ineq50}
	\begin{split}
		&{\mathrm{pr}}\left(\|\bar{\A}_{W'}\beta^* - \bar{\B}_{W'}^\top y\|_{\infty} > C\sqrt{n\log p\left(\sigma^2 + \frac{p}{\bar\nu}\|\beta^*\|_2^2\right) + Cn^2s\left\{\left(\frac{p}{\bar{\nu}}\right)^{3 - 2\epsilon} + F^2\left(\frac{p}{\bar{\nu}}\right)^2\right\}\|\beta^*\|_2^2}\right)\\
		\leq& {\mathrm{pr}}\left[\|\P\left(\A_{W'} - \B_{W'}^\top  V\right)\beta^*\|_{\infty} \geq C\sqrt{n\log p}\left(\frac{p}{\bar\nu}\right)^{\frac{1}{2}}\|\beta^*\|_2 + Cn\sqrt{s}\left\{\left(\frac{p}{\bar{\nu}}\right)^{\frac{3}{2} - \epsilon} + F\frac{p}{\bar{\nu}}\right\}\|\beta^*\|_2\right]\\
		& + {\mathrm{pr}}\left(\|\bar{\B}_{W'}^\top\epsilon\|_{\infty} > C\sqrt{n\log p\cdot\sigma^2}\right)\\
		\leq& 3e^{-C'\log p} = 3p^{-C'}.
	\end{split}	
\end{equation}
and 
\begin{equation}\label{ineq51}
	\begin{split}
		&{\mathrm{pr}}\left(\|\bar{\A}_{W}\beta^* - \bar{\B}_{W}^\top y\|_{\infty} > C\sqrt{n\log p\left(\sigma^2 + \frac{p}{\bar\nu}\|\beta^*\|_2^2\right) + Cn^2s\left\{\left(\frac{p}{\bar{\nu}}\right)^{3 - 2\epsilon} + F^2\left(\frac{p}{\bar{\nu}}\right)^2\right\}\|\beta^*\|_2^2}\right)\\
		\leq& {\mathrm{pr}}\left(\|\bar{\A}_{W'}\beta^* - \bar{\B}_{W'}^\top y\|_{\infty} > C\sqrt{n\log p\left(\sigma^2 + \frac{p}{\bar\nu}\|\beta^*\|_2^2\right) + Cn^2s\left\{\left(\frac{p}{\bar{\nu}}\right)^{3 - 2\epsilon} + F^2\left(\frac{p}{\bar{\nu}}\right)^2\right\}\|\beta^*\|_2^2}\right)\\
		&+ {\mathrm{pr}}\left(\exists 1 \leq i \leq n, 1 \leq j \leq p, W_{ij} \neq W'_{ij}\right)\\
		\leq& 3p^{-C'} + Cnpe^{-c\nu_{\min}} \leq 3p^{-C'} + Cnp\exp\left(-C''\log(np)\right)\leq 4p^{-C'},
	\end{split}
\end{equation}
where the second last inequality holds since $\nu_{\min} \geq C\log(np)$.

Finally, \eqref{ineq:variance_general} shows that $F^2 \leq C$, which has finished the proof of Lemma \ref{lm:infinity_norm_bound3}.
\qed
\medskip

\begin{lemma}[RIP condition for general setting]\label{lm:RIP3}
	Suppose $n \geq cs\log p$ and $|\log(\nu_{ij}) - \log(\nu_{kl})| \leq a$ for some constant $a$ for all $1 \leq i, k \leq n, 1 \leq j, l \leq p$. If $\bar{V} = V\{\I_p - (1/p)1_p1_p^\top\}$ satisfies RIP condition with constant $\delta_{2s}(\bar{V}) > 0$, , $\bar\nu \geq Cp\{s\log s + \log(np)\}$, then
	$\bar{\B}_W$ satisfies RIP condition with constants $\delta_{2s}(\bar{\B}_W) = 2\delta_{2s}(\bar{V})$, i.e.,
	\begin{equation}\label{eq314}
		n\left(1 - 2\delta_{2s}(\bar{V})\right)\|x\|_2^2 \leq x^\top \bar{\A}_W x \leq n\left(1 + 2\delta_{2s}(\bar{V})\right)\|x\|_2^2
	\end{equation}
	holds for all $2s$-sparse vector $x \in \bbR^p$, with probability $1 - 2\times9^{2s} e^{2s\log p-cn}$.
\end{lemma}
{\noindent Proof of Lemma \ref{lm:RIP3}.}
By Lemmas \ref{lm:general_1}, \ref{lm:general_2} and the similar arguments of \eqref{eq52} in Lemma \ref{lm:RIP}, 
\begin{equation}\label{eq305}
	{\mathrm{pr}}\left(\exists x, \|x\|_0 \leq 2s, \|x\|_2 = 1, \left|x^{\top}(\bar{\A}_{W'} - E \bar{\A}_{W'})x\right|\geq \frac{1}{2}\delta\left(\Gamma\right)n\right) \leq e^{2s\log p} \times 9^{2s} \times 2e^{-cn}.
\end{equation}
In addition, similarly to \eqref{ineq52}, we know that
\begin{equation*}
	\left|x^\top (E  \bar{\A}_{W'} - \bar{V}^\top\bar{V}) x\right| \leq C_1sn\left(F\frac{\log\nu_{\min}}{\nu_{\min}} + \frac{1}{\nu_{\min}}\right)\|x\|_2^2
\end{equation*}
for some constant $C_1 > 0$.
Since $\nu_{\min} \gtrsim \bar{\nu}/p \geq C(s + Fs\log s)$, we know that 
\begin{equation*}
	\left|x^\top (E  \bar{\A}_{W'} - \bar{V}^\top\bar{V}) x\right| \leq \frac{1}{2}\delta_{2s}(\bar V)n\|x\|_2^2.
\end{equation*}
Combining \eqref{eq305} and the previous inequality together, we know that
\begin{equation*}
	n\left(1 - 2\delta_{2s}\left(\bar{ V}\right)\right)\|x\|_2^2 \leq x^\top \bar{\A}_W x \leq n\left(1 + 2\delta_{2s}\left(\bar{ V}\right)\right)\|x\|_2^2
\end{equation*}
with probability $1 - 2\times9^{2s} e^{2s\log p-cn}$.  \qed
\medskip

\begin{lemma}\label{lm:epsilon-net}
	Denote $\mathcal{A}_{S_0} \triangleq \{x: x \in S^{p-1}, \|x\|_2 = 1, \text{supp}(x) \subseteq S_0\}$ ,  where $S_0$ is a subset of $\{1, \dots, p\}$. Then $\forall p \times p$ matrix $\widetilde{\A}$, 
	\begin{equation}\label{eq29}
		\sup_{x \in \mathcal{A}_{S_0}} \left|x^{\top}\widetilde{\A}x\right| \leq (1 - 2\epsilon)^{-1} \sup_{x \in \mathcal{N}_{\epsilon}} \left|x^{\top}\widetilde{\A}x\right|,
	\end{equation}
	where $\mathcal{N}_{\epsilon}$ is an $\epsilon$-net of $\mathcal{A}_{S_0}$.	
\end{lemma}
{\noindent Proof of Lemma \ref{lm:epsilon-net}.}
Suppose $S_0 = \{i_1, \dots, i_{s_0}\}$, by considering the submatrix $\tilde{\A}_{S_0 \times S_0}$ with $\left(\widetilde{\A}_{S_0 \times S_0}\right)_{jk} = \left(\widetilde{\A}\right)_{i_j, i_k}$ it is equivalent to prove
\begin{equation}\label{eq30}
	\sup_{x \in S^{s_0 - 1}}\left|x^\top\tilde{\A}_{S_0 \times S_0}x\right| \leq \left(1 - 2\epsilon\right)^{-1}\sup_{x \in \mathcal{N}_{\epsilon, s_0}}\left|x^\top\tilde{\A}_{S_0 \times S_0}x\right|,
\end{equation} 
where $\mathcal{N}_{\epsilon, s_0}$ is an $\epsilon$-net of $S^{s_0 - 1}$.
By Lemma 5.4 in \cite{vershynin2010introduction}, (\ref{eq30}) holds, therefore (\ref{eq29}) holds.
\qed
\medskip

\end{document}